\documentclass[12pt,preprint,defended,final,twoside,preprintmargins,singlespace]{IJS_cit_thesis}
\usepackage{graphicx}
\usepackage{amsmath}
\usepackage{amssymb}
\usepackage{youngtab}
\usepackage{latexsym}
\usepackage{hyperref}
\usepackage[sort]{cite}
\usepackage{rotating}

\def\Bra#1{\mathinner{\langle{#1}|}}
\def\Ket#1{\mathinner{|{#1}\rangle}}
\def\braket#1{\mathinner{\langle{#1}\rangle}}
\def\bra#1{\left<#1\right|}
\def\ket#1{\left|#1\right>}
{\catcode`\|=\active 
  \gdef\Braket#1{\left<\mathcode`\|"8000\let|\bravert {#1}\right>}}
\def\bravert{\egroup\,\vrule\,\bgroup}
\newcommand{\alg}[1]{\mathfrak{#1}}

\newcommand{\su}{\alg{su}}
\newcommand{\psu}{\alg{psu}}
\newcommand{\Sl}{\alg{sl}}
\newcommand{\so}{\alg{so}}
\newcommand{\bt}{{\tilde b}}
\newcommand{\at}{{\tilde a}}
\newcommand{\atopfrac}[2]{\genfrac{}{}{0pt}{}{#1}{#2}}
\newcommand{\tr}{\mathop{\rm tr}}

\newcommand{\be}{\begin{eqnarray}}
\newcommand{\ee}{\end{eqnarray}}
\newcommand{\bc}{\begin{center}}
\newcommand{\ec}{\end{center}}
\newcommand{\bea}{\begin{eqnarray}}
\newcommand{\eea}{\end{eqnarray}}
\newcommand{\ben}{\begin{equation}}
\newcommand{\een}{\end{equation}}
\newcommand{\ap}{\alpha^\prime}

\newcommand{\Rhat}{\widehat R}
\newcommand{\del}{\partial}
\newcommand{\Qo}{\widehat Q}
\newcommand{\pderiv}[1]{\frac{\partial}{\partial{#1}}}
\newcommand{\labphi}[2]{\phi^{\,\Yboxdim5pt\tiny\young(#1,#2)}}

\newcommand{\suphi}{\phi^{\,\Yboxdim5pt\yng(1,1)}}
\newcommand{\suchi}{\chi^{\,\Yboxdim5pt\yng(1)}}
\newcommand{\suchib}{\overline\chi^{\,\Yboxdim5pt\yng(1,1,1)}}

\newcommand{\nn}{\nonumber}

\begin{document}

\degreeaward{Doctor of Philosophy }             
\university{California Institute of Technology}    
\address{Pasadena, California}                     
\unilogo{cit_logo}                                 
\copyyear{\the\year} 
\pubnum{CALT-68-2542 \\ hep-th/0505028}
\dedication{For my parents, Frank and Sheila, and my brother, Cory.}
\title{SUPERSTRING HOLOGRAPHY AND INTEGRABILITY IN \mbox{\boldmath$AdS_5\times S^5$}}
\author{Ian Swanson}                                      
\maketitle

   \begin{frontmatter}
      \makecopyright            
      \makededication           
      \begin{acknowledgements}  
%
%
I am deeply indebted to my advisor, John Schwarz, and to my collaborator, Curt Callan.
Their patience, encouragement, mentorship and wisdom have been invaluable in my studies, and
I will always be inspired by their incisive intellect and their thirst for investigation.  
This work would not have been possible without their kind efforts.
I would also like to thank Anton Kapustin, Hirosi Ooguri and Alan Weinstein, 
who, in addition to serving on my thesis committee, have helped to create an
exciting academic atmosphere in the physics department at Caltech.
In this respect I am especially thankful to Tristan McLoughlin for his friendship, 
patience and many lucid discussions of physics.

I have also benefited from interaction with
Gleb Arutyunov,
Niklas Beisert,
Andrei Belitsky,
Louise Dolan,
Sergey Frolov,
Umut G\"ursoy,
Jonathan Heckman,
Vladimir Kazakov,
Charlotte Kristjansen,
Martin Kruczenski,
Andrei Mikhailov,
Jan Plefka,
Didina Serban,
Matthias Staudacher,
Arkady Tseytlin and
Kostya Zarembo.
In addition, let me specifically thank current and former 
members of the theory group at Caltech, including
Parsa Bonderson,
Oleg Evnin,
Andrew Frey,
Hok Kong Lee,
Sanefumi Moriyama,
Takuya Okuda,
Jong-won Park,
David Politzer,
Benjamin Rahn,
Harlan Robins,
Michael Shultz,
Xin-Kai Wu,
and particularly 
Sharlene Cartier and
Carol Silberstein for all their hard work.
I would like to acknowledge entertaining and inspiring discussions 
of physics and otherwise with
Nick Halmagyi,
Lisa Li Fang Huang,
Will Linch,
Lubo\v{s} Motl,
Joe Phillips,
Peter Svr\v{c}ek
and Xi Yin.
I am also grateful to the organizers, lecturers and participants of the 2003 TASI 
summer school and the 2004 PiTP summer program.

I am honored to have the support and faithful friendship of
Megan Eckart,
Nathan Lundblad,
Chris O'Brien,
Alex Papandrew,
Mike Prosser,
Dave Richelsoph,
Demian Smith-Llera
and Reed Wangerud.
I would especially like to acknowledge
Nelly Khidekel, who has encouraged and supported me in every facet of my 
academic and personal life.  People of her caliber are rare, and 
in having met her I consider myself fortunate beyond words.
Let me also thank my family, to whom this work is dedicated.

In addition, this work was supported in part by the 
California Institute of Technology,
the James A. Cullen Memorial Fund
and US Department of Energy grant DE-FG03-92-ER40701.

      \end{acknowledgements}
      \begin{abstract}

The AdS/CFT correspondence provides a
rich testing ground for many important topics in theoretical 
physics.  The earliest and most striking example of the 
correspondence is the conjectured duality between the energy spectrum of
type IIB superstring theory
on $AdS_5\times S^5$ and the operator anomalous dimensions of 
${\cal N}=4$ supersymmetric Yang-Mills theory
in four dimensions.  While there is a substantial amount of evidence
in support of this conjecture, direct tests have been elusive.  
The difficulty of quantizing superstring 
theory in a curved Ramond-Ramond background 
is compounded by the problem of computing 
anomalous dimensions for non-BPS operators in the 
strongly coupled regime of the gauge theory.
The former problem can be circumvented
to some extent by taking a Penrose limit of $AdS_5\times S^5$,
reducing the background to that of a pp-wave (where the string theory
is soluble).  A corresponding limit of the gauge theory was discovered by
Berenstein, Maldacena and Nastase, who obtained
successful agreement between a class of operator dimensions in this limit
and corresponding string energies in the Penrose limit.
In this dissertation we present a body of work based largely 
on the introduction of worldsheet interaction corrections to 
the free pp-wave string theory by lifting the Penrose limit of $AdS_5\times S^5$.
This provides a new class of rigorous tests of AdS/CFT that probe a truly quantum 
realm of the string theory.  
By studying the correspondence in greater detail, we stand to learn
not only about how the duality is realized on a more microscopic level,
but how Yang-Mills theories behave at strong coupling.
The methods presented here will hopefully contribute to the realization of 
these important goals.
     
      \end{abstract}
	\tableofcontents
\clearpage
   \end{frontmatter}



\numberwithin{equation}{section}
%
%
%
\extrachapterintro	              				          
Since conservation laws arise from symmetries of the Lagrangian \cite{noether}, 
an efficient way to characterize physical systems is to describe
the mathematical symmetries under which they are invariant.
From a certain perspective, the symmetries themselves may be viewed 
as paramount: a complete description of fundamental physics will likely be founded 
on an account of which symmetries are allowed by nature, under what circumstances 
these symmetries are realized and how and when these symmetries are broken.  
At the energies probed by current experiments, nature is described
at the microscopic level by a quantum field theory with certain gauge 
symmetries.  This framework is remarkably successful at describing 
particle spectra and interactions, but there are many convincing indications
that this picture breaks down near the Planck scale, where gravitational effects 
become important.  

To incorporate gravity in a way that is consistent at the quantum level,
one must make a dramatic departure from the point-particle quantum field
theory upon which the Standard Model is based.  Only by replacing the fundamental
point-particle constituents of the theory with one-dimensional
extended objects (strings) is one afforded the freedom necessary to 
accommodate gravity \cite{Green:1987sp,Green:1987mn}.  
The physical theory of these objects, or string theory,
is not only able to provide a consistent theory of quantum gravity, but also 
has a rich enough structure to give rise to the types of gauge symmetries 
observed in nature (and is free of quantum anomalies) 
\cite{Green:1987sp,Green:1987mn,Schwarz:1986ju,Schwarz:1986jv,Polchinski:1998rq,Polchinski:1998rr}.  
One fascinating
aspect of string theory, however, is that quantum consistency demands that 
the theory occupies ten spacetime dimensions (M-theory is eleven dimensional).
Since we observe only four spacetime dimensions in the universe, theorists
are charged with the task of understanding the role of the six `extra'
spatial dimensions that are predicted by string theory.  At first glance,
the idea that six spatial dimensions exist in the universe but
are somehow hidden seems fanciful.
Stated concisely, a strong hope among theorists is that the extra dimensions 
in string theory will provide a mechanism through which the gauge symmetries in 
nature are realized naturally.  

In the course of trying to describe the known symmetries of the 
vacuum, the study of string theory has led to 
the discovery of a dramatically new class of fundamental 
symmetries known as dualities.  These symmetries stand apart from more 
traditional examples in that they connect
physical theories that, at least superficially, appear to be entirely 
distinct in their formulation.  This notion of duality, or the underlying 
equivalence of two seemingly disparate physical systems, has emerged 
as a powerful tool in recent decades.  The usefulness of duality derives 
in part from the fact that dual descriptions are typically complementary, 
insofar as information that is inaccessible in one physical theory 
may often be extracted from a straightforward calculation in the
theory's dual description.  This is often realized in the form of
a strong/weak duality, whereby a small parameter useful for perturbation 
theory on one side is mapped to a large parameter on the other.
Information provided by a perturbative expansion in one theory
therefore equates to knowledge about nonperturbative physics in the
dual theory (and vice versa).

In this work we will primarily be concerned with dualities that arise
holographically, meaning that information (or degrees of freedom) existing 
in one theory with a given number of spacetime dimensions can be encoded in 
some dual theory with fewer spatial dimensions.  This is of course analogous to
an actual hologram, wherein information about the shape of an object in three
spatial dimensions can be encoded on a two-dimensional film: in addition
to recording the location in two dimensions of laser light incident on its surface, 
a hologram records the polarization of this light as it is reflected off of the object.
A major theme in holographic dualities is that the importance of the spatial
dimensions in which a theory is defined is often secondary to a proper accounting
of the degrees of freedom accessible to the theory.  This leads us to how
holography was initially recognized as an important concept in theoretical 
physics:  the black-hole entropy problem.

\section{The holographic entropy bound}
\label{holo}
As described above, the degrees of freedom
in the universe appear to be described by quantum fields living in a 
four-dimensional spacetime, at least down to 
the scales accessible to current accelerator experiments. The belief among 
theorists is that this description holds all the way down to the Planck
scale, $l_{\rm Planck}$.  The implication is that, with $l_{\rm Planck}$ serving
as an ultraviolet cutoff, the degrees of freedom 
available to the vacuum can be roughly described by a three-dimensional 
lattice theory with internal lattice spacing equal to $l_{\rm Planck}$.
With one binary degree of freedom per Planck volume, the maximum entropy of a system 
enclosed in a volume $V$ should scale in direct proportion to $V$ 
\cite{Susskind:1994vu,Susskind:1998dq,Bigatti:1999dp}.

The limitations of this simple picture can be seen by considering a thermodynamic 
system in which gravitational effects are important:  namely, a black hole.
The entropy of an isolated black hole is given by the Bekenstein-Hawking formula
\cite{Hawking:1976de,Bekenstein:1973ur}:
\be
S_{\rm BH} = \frac{A}{4G}\ .
\ee
The most striking aspect of this formula is that $S_{\rm BH}$ scales linearly
with the area $A$ of the event horizon.  
A simple thought experiment, following Bekenstein 
\cite{Bekenstein:1973ur,Bekenstein:1972tm,Bekenstein:1974ax}, 
leads to an interesting problem.  
Imagine some volume $V$ of space that contains a thermodynamic system with entropy 
$S>S_{\rm BH}$. 
If the entropy of the system is bounded by its volume, then this is a reasonable
proposal. 
The mass of the system must be no greater than the mass of a black hole whose 
horizon is the boundary of $V$, otherwise the system would be larger than $V$.  Now, if 
a thin shell of mass collapses into the system and forms a black hole whose 
horizon is precisely defined by $V$, the entropy of the new system is given by the 
Bekenstein-Hawking formula:  this process violates the second law of thermodynamics.

A striking solution to this problem, proposed by 't~Hooft \cite{'tHooft:1993gx}, 
is that nature obeys a 
\emph{holographic entropy bound}, which states that the degrees of freedom available 
to a physical system occupying a volume $V$ can be mapped to some physical theory
defined to exist strictly on the boundary $\del V$ (see also 
\cite{Susskind:1993aa,Thorn:1991fv,Susskind:1998dq,Susskind:1994vu}).  
The maximum entropy of a system is thus limited by the number
of degrees of freedom that can be mapped from the interior of the system to its boundary.  
The most striking aspect of this claim is that, while both theories must give rise to
equivalent physical predictions, the `dual' theory defined on the boundary necessarily 
exists in a fewer number of spatial dimensions than the original theory living in the bulk.

\section{Holography and string theory}
\label{holo_and_string}
The holographic principal is deeply enmeshed in the intricate relationship between
string theory and point-particle gauge theory. 
As a toy example, consider the analogy 
between the classical statistical mechanics of a $D$ dimensional system and the 
quantum dynamics of a $D-1$ dimensional system.  (This analogy was alluded to extensively
by Polyakov in \cite{Polyakov:1987ez}.)  The statement for $D=1$ is that 
the quantum transition amplitude for a point particle over some time interval $T$ 
can be interpreted as the classical partition function of a string whose length is 
determined by $T$.  Although not strictly holographic, this example captures several 
themes that are ubiquitous in gauge/string-theory dualities.

We should first take note of the types of gauge theories that will be of interest to us.
The theory of the strong nuclear force, or quantum chromodynamics (QCD), is an 
$SU(3)$ gauge theory:  it is a non-Abelian Yang-Mills theory with three colors ($N_c=3$).
QCD is known to be asymptotically free, meaning that the theory is free at high energies.
At very low energies one enters a regime where perturbation theory 
is no longer useful, and with no further advancements (such as a dual string formulation)
the only hope is that lattice
computations will one day be able to probe these regions of the theory in detail.
In 1974 't~Hooft suggested that a more general $SU(N_c)$ Yang-Mills theory would 
simplify when the rank of the gauge group (or the number of colors) $N_c$ becomes
large \cite{'tHooft:1973jz}.  
Such a simplification is intriguing, because if the theory is solved in the
large $N_c$ limit, one could study a perturbative expansion with coupling
$1/N_c = 1/3$ and perhaps learn about the non-perturbative regime of QCD.
In the course of these studies 't~Hooft noticed that when 
$1/N_c$ is interpreted as a coupling strength, the resulting Feynman graph expansion is 
topologically identical to the worldsheet genus expansion of a generic interacting 
string theory.  This was one of the early indications that Yang-Mills theory could
be realized, in certain respects, as a theory of string.

In 1997 Maldacena fused 't~Hooft's holographic principle and the $1/N_c$ expansion
in a dramatic new proposal \cite{Maldacena:1997re}. 
It was known that one can construct a four-dimensional maximally 
supersymmetric $({\cal N}=4)$
$SU(N_c)$ gauge theory by stacking $N_c$ coincident D3-branes and allowing 
open strings to stretch between pairs of branes \cite{Witten:1995im}.  The 
't~Hooft limit becomes accessible in this setting by taking the number of branes
to be large.  Since the D-branes are massive, however, a large number of them 
warp the ten-dimensional background geometry and a horizon is formed.  
The geometry in the near-horizon 
limit can be computed to be the product space of a five-dimensional anti-de-Sitter manifold 
and a five-dimensional sphere, or $AdS_5\times S^5$.  Furthermore, the branes are sources for
closed string states, and the physics in the region just exterior to the branes is
described by type IIB closed superstring theory in an $AdS_5\times S^5$ background
geometry.  According to holography, the theory on the horizon should
correspond to the physics inside the horizon.  Maldacena was thereby led to conclude
that type IIB superstring theory on $AdS_5\times S^5$ is {\it equivalent} to ${\cal N}=4$ 
supersymmetric Yang-Mills theory with $SU(N_c)$ gauge group in four spacetime dimensions!  
The conjectured equivalence of these two theories is a holographic duality.
The relationship turns out to be dual in the more traditional sense, insofar as the
coupling strengths that govern perturbative expansions in each theory are inversely 
proportional:  perturbative physics in one theory corresponds to a non-perturbative
regime in the dual theory.  
The power afforded by a conjectured duality, however, is sometimes tempered by 
the inability to directly verify the proposal.  Generically, a direct verification
would require specific knowledge of non-perturbative physics on 
at least one side of the duality.

\section{The Penrose limit}
\label{penrose}
It should be noted that there is a substantial body of evidence that stands in support 
of Maldacena's conjecture.  Most notably, the string and gauge 
theories are both invariant under the same superconformal symmetry group:
$PSU(2,2|4)$.  
Apart from the satisfaction of achieving a proof of the conjecture,
an exploration of the underlying details would be useful in its own right;
a more detailed understanding of how the AdS/CFT correspondence 
is realized on the microscopic level would be extremely valuable. 
The primary obstructions to such a program 
have been the difficulty of computing
the dimensions of non-BPS operators in the strong-coupling limit of the
gauge theory, and the unsolved problem of string quantization in the
presence of a curved, Ramond-Ramond (RR) background geometry.  In February of
2002, Berenstein, Maldacena and Nastase (BMN) found a specific set of limits
where these problems can, to some extent, be circumvented \cite{Berenstein:2002jq}.  
In this section we will briefly review how this is achieved, 
paying particular attention to the string side of the duality (relevant 
details of the gauge theory will be covered in Chapter~\ref{SYM}).

In convenient global coordinates, the ${ AdS}_5 \times S^5$
metric can be written in the form
\be
ds^2 = \widehat R^2 ( - {\rm cosh}^2 \rho~ dt^2 + d \rho^2 + {\rm sinh}^2
\rho~ d \Omega_3^2 + {\rm cos}^2  \theta~ d \phi^2 +  d \theta^2 +
{\rm sin}^2 \theta~ d \widetilde\Omega_3^2)~,
\label{adsmetric0}
\ee
where $\widehat R$ denotes the radius of both the sphere and the
AdS space. (The hat is introduced because we reserve the symbol
$R$ for $R$-charge in the gauge theory.) The coordinate $\phi$ is periodic with period
$2\pi$ and, strictly speaking, the time coordinate $t$ exhibits
the same periodicity. In order to accommodate string dynamics, it
is necessary to pass to the covering space in which time is {\sl
not} taken to be periodic. This geometry is accompanied by an RR
field with $N_c$ units of flux on the $S^5$. It is a consistent,
maximally supersymmetric type IIB superstring background provided
that 
\be
\widehat R^4 =  g_s N_c (\alpha^{\prime})^2~,
\ee
where $g_s$ is the string coupling. Explicitly, 
the AdS/CFT correspondence asserts that this
string theory is equivalent to ${\cal N} =4$ super Yang--Mills
theory in four dimensions with an $SU(N_c)$ gauge group and coupling
constant $ g_{YM}^2 = g_s $. To simplify both sides of the
correspondence, we study the duality in the simultaneous limits
$g_s\to 0$ (the classical limit of the string theory) and
$N_c\to\infty$ (the planar diagram limit of the gauge theory) with
the 't Hooft coupling $g^2_{YM}N_c$ held fixed. The holographically
dual gauge theory is defined on the conformal boundary of
$AdS_5\times S^5$, which, in this case, is $R\times S^3$.  Specifically, 
duality demands that operator dimensions in the conformally invariant
gauge theory be equal to the energies of corresponding states of
the `first-quantized' string propagating in the $AdS_5\times S^5$
background \cite{Witten:1998qj}.

The quantization problem is simplified by
boosting the string to lightlike momentum along some direction or,
equivalently, by quantizing the string in the background obtained
by taking a Penrose limit of the original geometry using the
lightlike geodesic corresponding to the boosted trajectory. The
simplest choice is to boost along an equator of the $S^5$ or,
equivalently, to take a Penrose limit with respect to the
lightlike geodesic $\phi=t,~\rho=\theta=0$. To perform lightcone
quantization about this geodesic, it is helpful to make the
reparameterizations
\begin{eqnarray}
    \cosh\rho  =  \frac{1+z^2/4}{1-z^2/4}~, \qquad
    \cos\theta  =  \frac{1-y^2/4}{1+ y^2/4}\ ,
\end{eqnarray}
and work with the metric
\be
\label{metric0}
ds^2  = \widehat R^2
\biggl[ -\left({1+ \frac{1}{4}z^2\over 1-\frac{1}{4}z^2}\right)^2dt^2
        +\left({1-\frac{1}{4}y^2\over 1+\frac{1}{4}y^2}\right)^2d\phi^2
    + \frac{dz_k dz_k}{(1-\frac{1}{4}z^2)^{2}}
    + \frac{dy_{k'} dy_{k'}}{(1+\frac{1}{4}y^2)^{2}} \biggr]\ ,
\ee
where $y^2 = \sum_{k'} y^{k'} y^{k'}$ with $k'=5,\dots,8$ and $z^2 = \sum_k z^k
z^k$ with $k=1,\dots,4$ define eight `Cartesian' coordinates
transverse to the geodesic.  
This metric is invariant under the full $SO(4,2)
\times SO(6)$ symmetry, but only translation invariance in $t$ and
$\phi$ and the $SO(4)\times SO(4)$ symmetry of the transverse
coordinates remain manifest in this form. The translation
symmetries mean that string states have a conserved energy
$\omega$, conjugate to $t$, and a conserved (integer) angular
momentum $J$, conjugate to $\phi$. Boosting along the equatorial
geodesic is equivalent to studying states with large $J$ and the
lightcone Hamiltonian will give the (finite) allowed values for
$\omega-J$ in that limit. On the gauge theory side, the $S^5$
geometry is replaced by an $SO(6)$ $R$-symmetry group, and $J$
corresponds to the eigenvalue $R$ of an $SO(2)$ $R$-symmetry
generator. The AdS/CFT correspondence implies that string energies
in the large-$J$ limit should match operator dimensions in the
limit of large $R$-charge.

On dimensional grounds, taking the $J\to\infty$ limit on string states is
equivalent to taking the $\widehat R\to\infty$ limit of the geometry
(in properly chosen coordinates). The coordinate redefinitions
\begin{eqnarray}
\label{rescale}
    t \rightarrow x^+~,
\qquad
    \phi \rightarrow x^+ + \frac{x^-}{\widehat R^2}~,
\qquad
    z_k \rightarrow \frac{z_k}{\widehat R}~,
\qquad
    y_{k'} \rightarrow \frac{y_{k'}}{\widehat R}
\end{eqnarray}
make it possible to take a smooth $\widehat R\to\infty$ limit. (The
lightcone coordinates $x^\pm$ are a bit unusual, but have been chosen
for future convenience in quantizing the worldsheet Hamiltonian.)
Expressing the metric (\ref{metric0}) in these new coordinates, we obtain the
following expansion in powers of $1/\widehat R^2$:
\begin{eqnarray}
\label{expndmet0}
ds^2 & \approx &
2\,{dx^+}{dx^-} + {dz}^2 + {dy }^2  -
        \left( {z }^2 + {y }^2 \right) ({dx}^+)^2 + O(1/\Rhat^2)\ . 
\end{eqnarray}
The leading contribution (which we will call $ds_{\rm pp}^2$)
is the Penrose limit, or pp-wave
geometry: it describes the geometry seen by the infinitely boosted string.
The $x^+$ coordinate is dimensionless, $x^-$ has dimensions of
length squared, and the transverse coordinates now have dimensions of length.

In lightcone gauge quantization of the string dynamics, one
identifies worldsheet time $\tau$ with the $x^+$ coordinate, so
that the worldsheet Hamiltonian corresponds to the conjugate
space-time momentum $p_+=\omega-J$. Additionally, one sets the
worldsheet momentum density $p_{-} =1$ so that the other
conserved quantity carried by the string, $p_-=J/\widehat R^2$, is
encoded in the length of the $\sigma$ interval (though we will later
keep $p_-$ explicit for reasons covered in Chapter~\ref{twoimp}).
Once $x^\pm$ are
eliminated, the quadratic dependence of $ds^2_{\rm pp}$ on the
remaining eight transverse bosonic coordinates leads to a
quadratic (and hence soluble) bosonic lightcone Hamiltonian $p_+$.
Things are less simple when $1/\widehat R^{2}$ corrections to the
metric are taken into account: they add quartic interactions to
the lightcone Hamiltonian and lead to nontrivial shifts in the
spectrum of the string. This phenomenon, generalized to the
superstring, will be the primary subject of this dissertation.

While it is clear how the Penrose limit can bring the bosonic
dynamics of the string under perturbative control, the RR field
strength survives this limit and causes problems for quantizing
the superstring. The Green-Schwarz (GS) action is the only practical
approach to quantizing the superstring in RR backgrounds, and we
must construct this action for the IIB superstring in the $AdS_5
\times S^5$ background \cite{Metsaev:1998it}, pass to lightcone
gauge and then take the Penrose limit. The latter step reduces the
otherwise extremely complicated action to a worldsheet theory of
free, equally massive transverse bosons and fermions
\cite{Metsaev:2001bj}. As an introduction to the issues we will
be concerned with, we give a concise summary of
the construction and properties of the lightcone Hamiltonian
$H^{\rm GS}_{\rm pp}$ that describe the superstring in this limit. This
will be a helpful preliminary to our principal goal of evaluating
the corrections to the Penrose limit of the GS action.

Gauge fixing eliminates the oscillating contributions to 
both lightcone coordinates $x^\pm$,
leaving eight transverse coordinates $x^I$ as bosonic dynamical
variables. Type IIB supergravity has two ten-dimensional
supersymmetries that are described by two 16-component
Majorana--Weyl spinors of the same ten-dimensional chirality. The
GS superstring action contains just such a set of
spinors (so that the desired spacetime supersymmetry comes out
`naturally'). In the course of lightcone gauge fixing, half of
these fermi fields are set to zero, leaving behind a complex
eight-component worldsheet fermion $\psi$. This field is further
subject to the condition that it transform in an ${\bf 8}_s$
representation under $SO(8)$ rotations of the transverse
coordinates (while the bosons of course transform as an ${\bf
8}_v$). In a 16-component notation
the restriction of the worldsheet fermions to the ${\bf
8}_s$ representation is implemented by the condition $\gamma^9
\psi =+\psi$ where $\gamma^9=\gamma^1\cdots\gamma^8$ and the
$\gamma^A$ are eight real, symmetric gamma matrices satisfying a
Clifford algebra $\{\gamma^A,\gamma^B\}=2\delta^{AB}$. Another
quantity, which proves to be important in what follows, is
$\Pi\equiv\gamma^1\gamma^2\gamma^3\gamma^4.$ One could also define
$\tilde\Pi = \gamma^5\gamma^6\gamma^7\gamma^8$, but $\Pi\psi
=\tilde\Pi\psi$ for an ${\bf 8}_s$ spinor.

In the Penrose limit, the lightcone GS superstring action takes the form
\be
\label{ppwavact}
S_{\rm pp} = \frac{1}{2\pi\alpha'} \int d\tau
\int d \sigma ({\cal L}_B+{\cal L}_F)\ ,
\ee
where
\begin{eqnarray}
\label{LBandLF}
{\cal L}_B &=& \frac{1}{2} \left[ ( \dot x^A)^2
-  (x^{\prime A})^2 - (x^A)^2\right]~,
\\
{\cal L}_F &=& i \psi^{\dagger} \dot\psi + \psi^{\dagger} \Pi
\psi +\frac{i}{2} ( \psi \psi^{\prime} + \psi^{\dagger}
\psi^{\prime\dagger} )~.
\end{eqnarray}
The fermion mass term $\psi^{\dagger} \Pi \psi$ arises from the
coupling to the background RR five-form field strength, and matches
the bosonic mass term (as required by supersymmetry). It is
important that the quantization procedure preserve supersymmetry.
However, as is typical in lightcone quantization, some of the
conserved generators are linearly realized on the $x^A$ and
$\psi^\alpha$, and others have a more complicated non-linear
realization.

The equation of motion of the transverse string coordinates is
\be
\label{ppeqn} \ddot x^{A}- x^{\prime\, \prime A} +  x^A=0\ .
\ee
The requirement that $x^A$ be periodic in the worldsheet
coordinate $\sigma$ (with period $2\pi\ap p_-$) leads to the mode
expansion
\be
\label{modexpn}
x^A(\sigma, \tau) = \sum_{n= -\infty}^{\infty}x_n^A (\tau)
e^{-ik_n\sigma}~, \qquad  k_n = \frac{n}{\alpha^{\prime} p_-}
= \frac{n \widehat R^2}{\alpha^{\prime} J}~.
\ee
The canonical momentum $p^A$ also has a mode expansion, related to that
of $x^A$ by the free-field equation $p^A=\dot x^A$. The coefficient
functions are most conveniently expressed in
terms of harmonic oscillator raising and lowering operators:
\begin{eqnarray}
\label{candaops}
x_n^A(\tau) &=&\frac{i}{\sqrt{2\omega_n p_-}}
(a_n^A e^{-i\omega_n\tau}-a_{-n}^{A\dagger} e^{i\omega_n\tau})\ ,
\\
p_n^A(\tau) &=& \sqrt{ \frac{\omega_n}{2 p_-}}
(a_n^A e^{-i\omega_n\tau}+a_{-n}^{A\dagger} e^{i\omega_n\tau})~.
\end{eqnarray}
The harmonic oscillator frequencies are determined by the equation
of motion (\ref{ppeqn}) to be
\be
\label{oscens}
\omega_n=\sqrt{1+k_n^2}=\sqrt{1+({n\Rhat^2}/{\ap J})^2}=
    \sqrt{1+({g^2_{YM}N_c n^2}/{J^2})}~,
\ee
where the mode index $n$ runs from $-\infty$ to $+\infty$.
(Because of the mass term, there is no separation into
right-movers and left-movers.) The canonical commutation relations
are satisfied by imposing the usual creation and annihilation
operator algebra:
\be
\label{candaopalg}
\left[a_m^A,
a_n^{B\dagger}\right] =
    \delta_{mn}\delta^{AB}~
\Rightarrow ~ \left[ x^A(\sigma),p^B(\sigma^\prime)\right]=
     i 2\pi\ap\delta(\sigma-\sigma^\prime)\delta^{AB}~.
\ee

The fermion equation of motion is
\be
i(\dot\psi +\psi^{\prime\dagger} ) + \Pi \psi =0\ .
\ee
The expansion of $\psi$ in terms of creation and annihilation operators
is achieved by expanding the field in worldsheet momentum eigenstates
\be
\psi(\sigma, \tau) = \sum_{n= -\infty}^{\infty}\psi_{n} (\tau) 
e^{-ik_n\sigma}~,
\ee
which are further expanded in terms of convenient positive and negative
frequency solutions of the fermion equation of motion:
\be
\label{fmodexp}
\psi_n(\tau) = \frac{1}{\sqrt{4p_-\omega_n}}
    (e^{-i\omega_n\tau}(\Pi+\omega_n-k_n) b_n
     +e^{i\omega_n\tau}(1-(\omega_n-k_n)\Pi) b^\dagger_n )~.
\ee
The frequencies and momenta in this expansion are equivalent to
those of the bosonic coordinates. In order to reproduce the
anticommutation relations
\be
\{\psi(\tau,\sigma),\psi^\dagger(\tau,\sigma^\prime)\}
    = 2\pi\ap\delta(\sigma-\sigma^\prime)~,
\ee
we impose the standard oscillator algebra
\be
\{ b_m^\alpha , b_n^{\beta\dagger} \} =
\frac{1}{2} (1 + \gamma_9)^{\alpha\beta} \delta_{m,n}~.
\ee
The spinor fields $\psi$ carry 16 components, but the ${\bf 8}_s$
projection reduces this to eight anticommuting oscillators,
exactly matching the eight transverse oscillators in the bosonic
sector. The final expression for the lightcone Hamiltonian is
\be\label{ppham}
 H^{\rm GS}_{\rm pp} = \sum_{n= -\infty}^{+\infty}\omega_n
\left( \sum_A (a_n^A)^\dagger a_n^A +
\sum_\alpha (b_n^\alpha)^\dagger b_n^\alpha \right)\ .
\ee
The harmonic oscillator zero-point energies nicely cancel between
bosons and fermions for each mode $n$. The frequencies $\omega_n$
depend on the single parameter
\be
\lambda^{\prime}= g^2_{YM}N_c/J^2~, \qquad
    \omega_n=\sqrt{1+\lambda^\prime n^2}~,
\ee
so that one can take $J$
{\sl and} $g^2_{YM}N_c$ to be simultaneously large while keeping
$\lambda^{\prime}$ fixed. If $\lambda^{\prime}$ is kept fixed and
small, $\omega_n$ may be expanded in powers of $\lambda^{\prime}$,
suggesting that contact with perturbative Yang--Mills gauge theory
is possible. 

The spectrum is generated by $8+8$ transverse oscillators acting
on ground states labeled by an $SO(2)$ angular momentum taking
integer values $-\infty<J<\infty$ (note that the oscillators
themselves carry zero $SO(2)$ charge). Any combination of
oscillators may be applied to a ground state, subject to the
constraint that the sum of the oscillator mode numbers must vanish
(this is the level-matching constraint, the only constraint not
eliminated by lightcone gauge-fixing). The energies of these
states are the sum of the individual oscillator energies
(\ref{oscens}), and the spectrum is very degenerate.\footnote{Note
that the $n=0$ oscillators raise and lower the string energy by a
protected amount $\delta p_+=1$, independent of the variable
parameters. These oscillators play a special role, enlarging the
degeneracy of the string states in a crucial way, and we will call
them `zero-modes' for short.} For example, the 256 states of the
form $A^\dagger_n B^\dagger_{-n} \vert J\rangle$ for a given mode
number $n$ (where $A^\dagger$ and $B^\dagger$ each can be any of
the 8+8 bosonic and fermionic oscillators) all have the energy
\be \label{twoimpen}
p_+= \omega-J = 2\sqrt{1+({
g^2_{YM}N_c n^2}/{J^2})}
    \sim {2+({ g^2_{YM}N_c n^2}/{J^2})+\cdots}\, .
\ee
In the weak coupling limit ($\lambda^\prime\to 0$) the degeneracy is
even larger because the dependence on the oscillator mode number $n$
goes away! This actually makes sense from the dual gauge theory point
of view where $p_+\to D-R$ ($D$ is the dimension and $R$ is the $R$-charge
carried by gauge-invariant operators of large $R$): at zero coupling,
operators have integer dimensions and the number of operators with
$D-R=2$, for example, grows with $R$, providing a basis on which string
multiplicities are reproduced.  Even more remarkably, BMN were able to show
\cite{Berenstein:2002jq} that subleading terms in a $\lambda^{\prime}$
expansion of the string energies match the first perturbative corrections
to the gauge theory operator dimensions in the large $R$-charge limit. We will
further review the details of this agreement in Chapters~\ref{SYM} and \ref{twoimp}.

More generally, we expect exact string energies in the
$AdS_5\times S^5$ background to have a joint expansion in the
parameters $\lambda^{\prime}$, defined above, and $1/J$. 
We also expect the degeneracies found in the $J\to\infty$ limit (for fixed
$\lambda^\prime$) to be lifted by interaction terms that arise in
the worldsheet Hamiltonian describing string physics at large but
finite $J$. Large degeneracies must nevertheless remain in order
for the spectrum to be consistent with the $PSU(2,2\vert 4)$
global supergroup that should characterize the exact string
dynamics. The specific pattern of degeneracies should also match
that of operator dimensions in the ${\cal N}=4$ super Yang--Mills
theory. Since the dimensions must be organized by the
$PSU(2,2\vert 4)$ superconformal symmetry of the gauge theory,
consistency is at least possible, if not guaranteed.

\section{The $1/J$ expansion and post-BMN physics}
\label{1/J}
As noted above, the matching achieved by BMN
should not be confined to the Penrose (or large-radius) limit of the bulk theory, or to the 
large $R$-charge limit of the CFT.  When the Penrose limit is lifted, 
finite-radius curvature corrections to the pp-wave geometry can be 
viewed as interaction perturbations to the free 
string theory, which, in turn, correspond to first-order
corrections, in inverse powers of the $R$-charge,
to the spectrum of anomalous dimensions in the gauge theory.  
With the hope that the underlying structure of the duality can be understood
more clearly in this perturbative context, this dissertation is dedicated to exploring 
the AdS/CFT correspondence when these effects are included.  
In this section we will briefly review the work appearing in the literature 
upon which this thesis is based.  In addition, we will also point out some of the
more important developments that have appeared as part of the large body of
research that has appeared following the original BMN paper.

In references \cite{Callan:2003xr} and \cite{Callan:2004uv}, 
it was demonstrated that the first-order curvature corrections to the 
pp-wave superstring theory precisely reproduce finite $R$-charge corrections to  
the anomalous dimensions of so-called BMN operators, and exhibit the full ${\cal N}=4$ 
extended supermultiplet structure of the dual gauge theory.  
The leading-order correction to the string theory 
gives rise to a complicated interacting theory of bosons and fermions in a curved 
RR background.  While the steps taken to quantize
the resulting theory were fairly elaborate, it was demonstrated that
they comprise a practical and correct method for defining the GS
superstring action in that background.
A detailed prescription for matching string states to gauge theory operators was
given specifically in \cite{Callan:2004uv}, 
along with a description of the procedure used to quantize the fully supersymmetric 
string theory and manage the set of second-class fermionic constraints that arise 
in lightcone gauge.

While the conjectured equivalence of the two theories emerged in this
perturbative context in a remarkable manner,
these studies also took advantage of the underlying 
duality structure of the correspondence.  In particular, finite $R$-charge
corrections to operator dimensions in the gauge theory emerge at all
orders in $1/R$ (where $R$ denotes the $R$-charge), 
but are defined perturbatively in the 't~Hooft
coupling $\lambda = g_{YM}^2 N$.  Conversely, finite-radius
corrections to string state energies appear perturbatively in inverse powers
of the radius, or, equivalently, in inverse powers of the angular momentum
$J$ about the $S^5$ (which is identified with the gauge theory $R$-charge).  
According to duality, however, the string theory should provide
a strong-coupling description of the gauge theory.  
This is realized by the fact that string energy corrections can be computed to all orders in the
so-called modified 't-Hooft coupling $\lambda' = g_{YM}^2 N/J^2$.
By studying the dilatation generator of ${\cal N}=4$ SYM theory, several groups have been 
able to compute gauge theory operator dimensions to higher loop-order
in $\lambda$ 
(see, e.g.,~\cite{Beisert:2002tn,Beisert:2003ea,Beisert:2003jb,Beisert:2003tq,Beisert:2003yb,Beisert:2003ys,Beisert:2004hm,Beisert:2004ry,Beisert:2004yq}), 
and, by expanding the corresponding 
string energy formulas in small $\lambda'$,
the one- and two-loop energy corrections can be shown to precisely match the gauge theory 
results in a highly nontrivial way.  
The three-loop terms disagree, however, and this mismatch comprises a 
longstanding puzzle in these studies.  Some investigations indicate that an
order-of-limits issue may be responsible for this disagreement, whereby the small-$\lambda$ expansion in 
the gauge theory fails to capture certain mixing interactions (known as wrapping terms)
that are mediated by the dilatation generator \cite{Serban:2004jf}.

To explore the correspondence further, and perhaps to shed light on the established 
three-loop disagreement, a complete treatment of the 4,096-dimensional space of
three-excitation string states was given in reference \cite{Callan:2004ev}, including a comparison with 
corresponding SYM operators carrying three $R$-charge impurities.
(The investigations in references \cite{Callan:2003xr} and \cite{Callan:2004uv} 
were restricted to the 256-dimensional
space of two-excitation string states, also known as two-impurity states.)
Although the interacting
theory in this larger space is much more complicated, it was found that the full ${\cal N}=4$ SYM
extended supermultiplet structure is again realized by the string 
theory, and precise agreement with the anomalous dimension spectrum in the gauge theory
was obtained to two-loop order in 
$\lambda'$.  Once again, however, the three-loop formulas disagree.

Concurrent with these studies, a new formalism emerged for computing operator dimensions in
the gauge theory.  This began when Minahan and Zarembo were able to identify 
the one-loop mixing matrix of SYM operator dimensions with the Hamiltonian of an integrable $SO(6)$
spin chain with vector lattice sites \cite{Minahan:2002ve}.  
One practical consequence of this discovery is that the quantum spin chain Hamiltonian
describing the SYM dilatation generator can be completely diagonalized by a set of
algebraic relations known as the Bethe ansatz.  
Work in the $SO(6)$ sector was extended by Beisert and Staudacher, who
formulated a Bethe ansatz for the full $PSU(2,2|4)$ superconformal symmetry of the theory
(under which the complete dilatation generator is invariant) \cite{Beisert:2003yb}.

The emergence of integrable structures in the gauge 
theory has given rise to many novel tests of AdS/CFT
(see, e.g.,~\cite{Alday:2003zb,Arutyunov:2003rg,Arutyunov:2003uj,Arutyunov:2003za,Arutyunov:2004vx,Arutyunov:2004xy,Beisert:2004ag,Dolan:2003uh,Dolan:2004ps,Kazakov:2004nh,Kazakov:2004qf,Mandal:2002fs,Staudacher:2004tk,Tseytlin:2003ii,Beisert:2005bm,Alday:2005gi,Beisert:2005mq,Hernandez:2005nf,Fischbacher:2004iu}).  
It has been suggested by Bena, Polchinski and Roiban, for instance, that the classical lightcone gauge 
worldsheet action of type IIB superstring theory in $AdS_5 \times S^5$ may 
itself be integrable \cite{Bena:2003wd}.  If both theories are indeed integrable, 
they should admit infinite towers of hidden charges that, in turn, should be equated via the
AdS/CFT correspondence, analogous to identifying the SYM dilatation generator with the string 
Hamiltonian.  Numerous investigations have been successful in matching classically conserved
hidden string charges with corresponding charges derived from the integrable structure of
the gauge theory.  Arutyunov and Staudacher, for example, were able to show that an infinite set of
local charges generated via B\"acklund transformations on certain classical extended string solutions
can be matched to an infinite tower of charges generated by a corresponding 
sector of gauge theory operators \cite{Arutyunov:2003rg}.  It is important to note, however, that these 
identifications are between the structures of {\it classically} integrable string sigma 
models and integrable {\it quantum} spin chains.  
Along these lines of investigation, Arutyunov, Frolov and Staudacher developed an
interpolation between the classical string sigma model and the quantum spin chain that
yielded a Bethe ansatz purported to capture the dynamics of an $SU(2)$ sector of the string theory 
\cite{Arutyunov:2004vx}.  
This ansatz, though conjectural, allowed the authors to extract multi-impurity string energy 
predictions in the near-pp-wave limit (at $O(1/J)$ in the curvature expansion).
Corresponding predictions were extracted in reference \cite{McLoughlin:2004dh} 
directly from the quantized string theory, 
and the resulting formulas matched the Bethe ansatz predictions to all loop-orders in $\lambda'$
in a remarkable and highly intricate fashion.  

Recently the question of quantum integrability in the string theory was addressed
in reference \cite{Swanson:2004qa}.  Using a perturbed Lax representation of a 
particular solitonic solution to the 
string sigma model, one is able to argue that the string theory admits an infinite 
tower of hidden commuting 
charges that are conserved by the quantized theory to quartic order in field fluctuations.  
In addition, a prescription for matching the eigenvalue spectra of these 
charges to dual quantities in the gauge theory can also be formulated.

At this point there is a considerable amount of evidence that both the string and gauge theories
are exactly integrable (see also \cite{McLoughlin:2005gj,Beisert:2005fw} for recent developments).  
The hope is of course that we will ultimately be led to an exact
solution to large-$N_c$ Yang-Mills theory.  Before reaching this goal, it is reasonable to 
expect that type IIB string theory on $AdS_5\times S^5$ and ${\cal N}=4$ super Yang-Mills theory will be
shown to admit identical Bethe ansatz equations, thereby proving this particular duality.
This is likely the next major step in these investigations.  
There are several intermediate problems that need to solved, however, including the known
mismatch between the string and gauge theory at three-loop order in the 't~Hooft coupling.
The resolution of these outstanding problems
will inevitably lead to a deeper understanding of both the relationship between gauge and string theory,
and the capacity of string theory itself to generate realistic models of particle physics.

\section{Overview}
\label{overview}
In this dissertation we will work in the large-$N_c$ limit, where we can
ignore string splitting and joining interactions;  the ``stringy'' effects
we are concerned with arise strictly from interactions among the bosonic and 
fermionic field excitations on the worldsheet.  In Chapter~\ref{SYM}
we will provide a brief treatment of the relevant calculations that 
are needed on the gauge theory side of the correspondence, based on 
work originally presented in \cite{Callan:2003xr}.  While the
results computed there can be found elsewhere in the literature
(see, e.g.,~\cite{Beisert:2002tn}), we present our own derivation 
for pedagogical reasons and to arrange the computation in a way that 
clarifies the eventual comparison with string theory.

As noted above, the task of calculating operator dimensions in the planar limit of 
${\cal N}=4$ super Yang-Mills theory can be vastly simplified by 
mapping the dilatation generator to the Hamiltonian of an integrable spin chain.  
These techniques are powerful at leading order in 
perturbation theory but become increasingly complicated beyond one loop in the 
't~Hooft parameter $\lambda=g_{\rm YM}^2 N_c$, where spin chains typically acquire 
long-range (non-nearest-neighbor) interactions.  
In certain sectors of the theory, moreover, higher-loop Bethe ans\"atze
do not even exist.  In Chapter~\ref{virial} 
we develop a virial expansion of the spin chain Hamiltonian
as an alternative to the Bethe ansatz methodology, a method that simplifies the 
computation of dimensions of multi-impurity operators at higher loops in $\lambda$.
We use these methods to extract numerical gauge theory predictions
near the BMN limit for comparison with corresponding results on the string theory 
side of the AdS/CFT correspondence.  For completeness, we compare our virial results
with predictions that can be derived from current Bethe ansatz technology.

In Chapter~\ref{twoimp} we compute the complete set of first curvature corrections
to the lightcone gauge string theory Hamiltonian that arise in the 
expansion of $AdS_5\times S^5$ about the pp-wave limit.
We develop a systematic quantization of the interacting
worldsheet string theory and use it to obtain the interacting spectrum
of the so-called `two-impurity' states of the string. The
quantization is technically rather intricate and we provide a detailed
account of the methods we use to extract explicit results. We give a 
systematic treatment of the fermionic states and are able to show 
that the spectrum possesses the proper extended supermultiplet structure 
(a nontrivial fact since half the supersymmetry is nonlinearly realized). 
We test holography by comparing the string energy spectrum with 
the scaling dimensions of corresponding gauge theory operators. We show
that agreement is obtained in low orders of perturbation
theory, but breaks down at third order.

Notwithstanding this third-order mismatch, we proceed with this line of 
investigation in Chapter~\ref{threeimp} by subjecting the string and gauge theories 
to significantly more rigorous tests.  Specifically, we extend 
the results of Chapter~\ref{twoimp} at $O(1/J)$ in the curvature expansion
to include string states and SYM operators 
with three worldsheet or $R$-charge impurities.  In accordance with
the two-impurity problem, we find a perfect and intricate agreement 
between both sides of the correspondence to two-loop order in $\lambda$ and, 
once again, the string and gauge theory predictions fail to agree
at third order.

In Chapter~\ref{Nimp} we generalize this analysis on the 
string side by directly computing
string energy eigenvalues in certain protected sectors of the theory for an
arbitrary number of worldsheet excitations with arbitrary mode-number assignments.
While our results match all existing gauge theory predictions to two-loop
order in $\lambda'$, we again observe a mismatch at three loops between
string and gauge theory.  We find remarkable agreement to \emph{all}
loops in $\lambda'$, however, with the near pp-wave limit of a
Bethe ansatz for the quantized string Hamiltonian given in an $\su(2)$ sector.
Based on earlier two- and three-impurity results, we also infer the full multiplet
decomposition of the $N$-impurity superstring theory with distinct
mode excitations to two loops in $\lambda'$.

In Chapter~\ref{integ2}
we build on recent explorations of the AdS/CFT correspondence that 
have unveiled integrable structures 
underlying both the gauge and string theory sides of the correspondence.  
By studying a semiclassical expansion about a class of point-like solitonic solutions to the 
classical string equations of motion on $AdS_5\times S^5$,
we take a step
toward demonstrating that integrability in the string theory survives quantum 
corrections beyond tree level.  Quantum fluctuations are chosen to align with 
background curvature corrections to the pp-wave limit of $AdS_5\times S^5$, 
and we present evidence for an infinite tower of local bosonic charges that are conserved
by the quantum theory to quartic order in the expansion.  
We explicitly compute several higher charges based on a Lax representation of the 
worldsheet sigma model and provide a prescription for matching the eigenvalue spectra of these
charges with corresponding quantities descending from the integrable structure of the 
gauge theory. 
The final chapter is dedicated to a discussion of the current status of
these studies and an overview of future directions of investigation.



%
%
\chapter{{${\cal N}=4$} super Yang-Mills theory}	                  
\label{SYM}
As discussed in the introduction, the AdS/CFT correspondence 
states that the energy spectrum of string excitations in an 
anti-de-Sitter background should be equivalent (albeit related by 
a strong/weak duality) to the spectrum of operator anomalous dimensions 
of the field theory living on the conformal boundary of that background.
Any attempt to test the validity of this statement directly must therefore
involve a computation of operator dimensions in the
gauge theory, particularly for those operators that are non-BPS.
As discussed above, this is a nontrivial task for generic gauge theory
operators, but the advent of the BMN mechanism has led to dramatic 
simplifications and insights.  
Following the appearance of the original BMN paper \cite{Berenstein:2002jq},
the field witnessed remarkable progress in understanding the dilatation generator
of ${\cal N}=4$ SYM theory (see,e.g.,~\cite{Beisert:2002tn,Beisert:2003ea,Beisert:2003jb,Beisert:2003tq,Beisert:2003yb,Beisert:2003ys,Beisert:2004hm,Beisert:2004ry,Beisert:2004yq,Alday:2003zb,Arutyunov:2003rg,Arutyunov:2003uj,Arutyunov:2003za,Arutyunov:2004vx,Arutyunov:2004xy,Beisert:2004ag,Dolan:2003uh,Dolan:2004ps,Kazakov:2004nh,Kazakov:2004qf,Mandal:2002fs,Staudacher:2004tk,Tseytlin:2003ii,Beisert:2005bm,Alday:2005gi,Beisert:2005mq,Hernandez:2005nf,Fischbacher:2004iu}).   
The review presented in this chapter
will focus on some of the major contributions to this understanding. 
Since this work is dedicated primarily to understanding the 
string theory side of the AdS/CFT correspondence, 
special preference will be given to information that contributes 
directly to our ability to interpret the dual spectrum of string excitations.
For a more comprehensive and detailed review of the gauge theory aspects
of these studies, the reader is referred to \cite{Beisert:2004ry}.

To arrange the calculation in a way that is 
more useful for our subsequent comparison with string theory, 
and to emphasize a few specific points, it is useful to rederive 
several important results.  
We will focus in Section~\ref{sec_two_imp} 
on the dimensions and multiplicities 
of a specific set of near-BPS (two-impurity) operators in the planar limit.  
Most of the information to be covered 
in this section originally appeared in \cite{Beisert:2002tn}, though we will
orient our review around a rederivation of these results first presented in 
\cite{Callan:2003xr}.  Section~\ref{gen_mult} generalizes these results to
the complete set of two-impurity, single-trace operators.  
This will set the stage for a detailed analysis of the corresponding
string energy spectrum.

\section{Dimensions and multiplicities}
\label{sec_two_imp}
As explained above, the planar large-$N_c$ 
limit of the gauge theory
corresponds to the noninteracting sector ($g_s\to 0$) of the dual 
string theory.\footnote{The Yang-Mills genus-counting parameter is $g_2 = J^2/N_c$  
\cite{Kristjansen:2002bb,Constable:2002hw}.}  In this limit the gauge theory 
operators are single-trace field monomials classified by dimension $D$ and 
the scalar $U(1)_R$ component (denoted by $R$) of the $SU(4)$ $R$-symmetry group.
We will focus in this section on the simple case of operators containing only
two $R$-charge impurities.  The classical dimension will be
denoted by $K$, and the BMN limit is reached by taking $K,R \to \infty$ such that
$\Delta_0 \equiv K-R$ is a fixed, finite integer.  The anomalous dimensions
(or $D-K$) are assumed to be finite in this limit, and the quantity
$\Delta \equiv D-R$ is defined for comparison with the
string lightcone Hamiltonian $P_+ = \omega - J$ (see Section~\ref{penrose} of the
introduction).

It is useful to classify operators in the gauge theory according to their 
representation under the exact global $SU(4)$ $R$-symmetry group.  
This is possible because the dimension operator commutes with the 
$R$-symmetry.  We therefore find it convenient to label 
the component fields with Young boxes, which clarifies the
decomposition of composite operators into irreducible 
tensor representations of $SU(4)$.  More specifically, the tensor irreps
of $SU(4)$ are represented by Young diagrams composed of at most three rows
of boxes denoted by a set of three numbers $(n_1,n_2,n_3)$ 
indicating the differences in length of successive rows.
The fields available are a gauge field, a set of
gluinos transforming as ${\bf 4}$ and ${\bf\bar 4}$ under the
$R$-symmetry group, and a set of scalars transforming as a ${\bf 6}$.
In terms of Young diagrams, the gluinos transform as two-component 
Weyl spinors in the $(1,0,0)$ fundamental $({\bf 4})$ and its
adjoint $(0,0,1)$ in the antifundamental $({\bf \bar 4})$:
\be
\suchi_{~a}~~({\bf 4})~, \qquad \qquad \suchib_{~\dot a}~~({\bf \bar 4})~.
\nn
\ee
The $a$ and $\dot a$ indices denote transformation in the $({\bf 2,1})$ or
$({\bf 1,2})$ representations of $SL(2,C)$ (the covering group of the
spacetime Lorentz group), respectively.
Likewise, the scalars appear as
\be
\suphi~~({\bf 6})~.
\nn
\ee

In the planar large-$N_c$ limit the operators of interest are those
containing only a single gauge trace.  To work through an explicit example, 
we will restrict attention for the moment to operators comprising spacetime
scalars.  It is convenient to further classify these operators under the 
decomposition 
\be
SU(4) \supset SU(2)\times SU(2)\times U(1)_R\ ,
\ee
since we are eventually interested in taking the scalar $U(1)_R$ 
component to be large (which corresponds to the large angular momentum limit 
of the string theory).  The $U(1)_R$ charge of the component fields above 
can be determined by labeling the Young diagrams attached to each field
with $SU(4)$ indices, assigning  $R=\frac{1}{2}$ to
the indices $1,2$ and $R=-\frac{1}{2}$ to the indices $3,4$:
\be
R=1:    ~\phi^{\,\Yboxdim5pt\tiny\young(1,2)}~ (Z)~, \qquad
R=0:    ~\phi^{\,\Yboxdim5pt\tiny\young(1,3)}~,
    \phi^{\,\Yboxdim5pt\tiny\young(1,4)}~,
    \phi^{\,\Yboxdim5pt\tiny\young(2,3)}~,
    \phi^{\,\Yboxdim5pt\tiny\young(2,4)}~ (\phi^A)~, \qquad
R=-1:   ~\phi^{\,\Yboxdim5pt\tiny\young(3,4)}~ (\bar Z)~, \nonumber\\
R=1/2:  ~\chi^{\,\Yboxdim5pt\tiny\young(1)}~,
    \chi^{\,\Yboxdim5pt\tiny\young(2)}~,
    \bar\chi^{\,\Yboxdim5pt\tiny\young(1,2,3)}~,
    \bar\chi^{\,\Yboxdim5pt\tiny\young(1,2,4)}~, \qquad
R=-1/2: ~\chi^{\,\Yboxdim5pt\tiny\young(3)}~,
    \chi^{\,\Yboxdim5pt\tiny\young(4)}~,
    \bar\chi^{\,\Yboxdim5pt\tiny\young(1,3,4)}~,
    \bar\chi^{\,\Yboxdim5pt\tiny\young(2,3,4)}~.
\ee
To remain consistent with the literature we have labeled the scalars 
using either $Z$ or $\bar Z$ for fields with $R=1$ or $R=-1$, respectively, 
or $\phi^A$ (with $A\in 1,\ldots,4$) for fields with zero $R$-charge.
The types of operators of interest to us are those with large naive
dimension $K$ and large $R$-charge, with the quantity $\Delta_0 \equiv K-R$
held fixed.  The number $\Delta_0$ is typically referred to as the
impurity number of the operator; as explained above, $N$-impurity SYM 
operators map to string states created by $N$ oscillators acting on the 
vacuum, subject to level matching.  Operators in the gauge theory
with zero impurity number are BPS, and their dimensions are protected.
The first interesting set of non-BPS operators are those with 
$\Delta_0 = 2$.  Restricting to spacetime scalars with $\Delta_0\le 2$,
we have
\begin{eqnarray}
\label{optypes}
\tr\big((\suphi)^K\big), & \qquad  \qquad
			& (R_{\rm max} = K ) \nonumber\\
\tr\big((\suchi\sigma_2\suchi)(\suphi)^{K-3}\big),
~\tr\big((\suchi\suphi\sigma_2\suchi)(\suphi)^{K-4}\big), ~&\ldots \qquad
			&(R_{\rm max} = K-2) \nonumber\\
\tr\big((\suchib\sigma_2\suchib)(\suphi)^{K-3}\big),
~\tr\big((\suchib\suphi\sigma_2\suchib)(\suphi)^{K-4}\big),&~ \ldots\qquad
			&(R_{\rm max} = K-2) \nonumber\\
\tr\big(\nabla_\mu\suphi\nabla^\mu\suphi(\suphi)^{K-4}\big),
    & \qquad \qquad
			&(R_{\rm max} = K-2 )\ ,
\nn\\
\end{eqnarray}
where $\nabla$ is the spacetime gauge-covariant derivative.

Starting with purely bosonic operators with no derivative insertions,
we must decompose into irreps an $SU(4)$ tensor of rank $2K$.  These irreps
are encoded in Young diagrams with $2K$ total boxes, and the goal is to 
determine the multiplicity with which each diagram appears.  
(An alternative approach, taken in \cite{Beisert:2002tn},
is to use the bosonic $SO(6)$ sector of the $R$-symmetry group.)
For the purposes of this example, we restrict to irreducible tensors in the 
expansion with $\Delta_0 = 0,2$.
For $K$ odd we have
\be
\label{phi_irrepodd}
\tr\big({\suphi}^{~K}~\big) &\to&
    1\times{\underbrace{\tiny\yng(7,7)}_{ K}} ~\oplus ~~
\Yvcentermath1
\left(\frac{K-1}{2}\right)\times{\underbrace{\tiny\yng(6,4)}_{K-1}} ~\oplus ~~
\left(\frac{K-1}{2}\right)\times{\underbrace{\tiny\yng(5,5)}_{K-2}} 
\nonumber\\
&& 
\kern-25pt   \oplus \left(\frac{K-1}{2}\right)\times{\underbrace{\tiny\yng(6,6,2)}_{K-1}} 
~\oplus ~~
\left(\frac{K-3}{2}\right)\times{\underbrace{\tiny\yng(7,5,2)}_K}
~\oplus~~ \dots\ ,
\ee
while for $K$ even we have
\be
\label{phi_irrep}
\Yvcentermath1
\tr\big({\suphi}^{~K}~\big) \to~
    1\times{\underbrace{\tiny\yng(7,7)}_{ K}} ~\oplus ~~
\Yvcentermath1
\left(\frac{K-2}{2}\right)\times{\underbrace{\tiny\yng(6,4)}_{K-1}} ~\oplus
~~ \left(\frac{K}{2}\right)\times{\underbrace{\tiny\yng(5,5)}_{K-2}}
\nonumber
\\
\Yvcentermath1
\oplus \left(\frac{K-2}{2}\right)\times{\underbrace{\tiny\yng(6,6,2)}_{K-1}} ~\oplus
~~ \left(\frac{K-2}{2}\right)\times{\underbrace{\tiny\yng(7,5,2)}_K}
~\oplus~\dots~.~~
\ee
The irreps with larger minimal values of $\Delta_0=K-R$ have
multiplicities that grow as higher powers of $K$. This is very
significant for the eventual string theory interpretation of the
anomalous dimensions, but we will not expand on this point here.

The bifermion operators (that are spacetime scalars) 
with $\Delta_0 = 2$ contain products of two gluinos and $K-3$ scalars:
\begin{eqnarray}
\label{chi_irrep} \Yvcentermath1
\tr\big(\suchi~\sigma_2~\suchi~ {(\suphi)}^{ K-3}\big) \to~
    1\times{\underbrace{\tiny\yng(5,5)}_{K-2}} ~\oplus ~~
1\times{\underbrace{\tiny\yng(6,4)}_{K-1}} ~\oplus ~\ldots~,
\\
\label{chib_irrep}
\Yvcentermath1
{\tr}\big(\suchib~\sigma_2~\suchib~
{(\suphi)}^{ K-3}\big) \to~
    1\times{\underbrace{\tiny\yng(6,6,2)}_{K-1}} ~\oplus ~~
1\times{\underbrace{\tiny\yng(5,5)}_{K-2}} ~\oplus ~\dots~.
\end{eqnarray}
Note that products of $\suchi$ and $\suchib$ cannot be made to form spacetime
scalars because they transform under inequivalent irreps of $SL(2,C)$.

Different operators are obtained by different orderings of the
component fields, but such operators are not necessarily independent
under cyclic permutations or permutations of the individual fields themselves,
subject to the appropriate statistics.  Using an obvious
shorthand notation, the total multiplicities of bifermion irreps are as
follows for $K$ odd:
\be
\label{gluino_irrep_odd}
\Yvcentermath1
\tr\big(\suchi~\sigma_2~\suchi~
{(\suphi)}^{ K-3}\big) \to
\Yvcentermath1
\left(\frac{K-3}{2}\right)\times{\underbrace{\tiny\yng(5,5)}_{K-2}} \oplus
\left(\frac{K-1}{2}\right)\times{\underbrace{\tiny\yng(6,4)}_{K-1}} 
\oplus \ldots~,
\Yvcentermath1
\\
\Yvcentermath1
{\tr}\big(\suchib~\sigma_2~\suchib~
{(\suphi)}^{ K-3}\big) \to
\left(\frac{K-3}{2}\right)\times{\underbrace{\tiny\yng(5,5)}_{K-2}} \oplus 
\left(\frac{K-1}{2}\right)\times{\underbrace{\tiny\yng(6,6,2)}_{K-1}}
    \oplus \ldots~. 
\Yvcentermath1
\ee
The results for $K$ even are, once again, slightly different:
\begin{eqnarray}
\label{gluino_irrep_even}
\Yvcentermath1
\tr\big(\suchi~\sigma_2~\suchi~
{(\suphi)}^{ K-3}\big) \to
\Yvcentermath1
\left(\frac{K-2}{2}\right)\times{\underbrace{\tiny\yng(5,5)}_{K-2}} \oplus 
\left(\frac{K-2}{2}\right)\times{\underbrace{\tiny\yng(6,4)}_{K-1}} 
\oplus \ldots~, 
\Yvcentermath1
\\
\Yvcentermath1
{\tr}\big(\suchib~\sigma_2~\suchib~
{(\suphi)}^{ K-3}\big) \to
\left(\frac{K-2}{2}\right)\times{\underbrace{\tiny\yng(5,5)}_{K-2}} \oplus 
\left(\frac{K-2}{2}\right)\times{\underbrace{\tiny\yng(6,6,2)}_{K-1}} 
\oplus \ldots~.
\Yvcentermath1
\end{eqnarray}
Since the dimension operator can only have
matrix elements between operators belonging to the same $SU(4)$
irrep, this decomposition amounts to a 
block diagonalization
of the problem.  The result of this program can be summarized 
by first noting that the decomposition can be divided into a BPS and 
non-BPS sector.  The BPS states ($\Delta_0 = 0$)
appear in the $(0,K,0)$ irrep and do not mix with the remaining non-BPS 
sectors, which yield irreps whose multiplicities scale 
roughly as $K/2$ for large $K$.  Even at this stage it is clear that
certain irreps only appear in the decomposition of certain types of
operators.  The $(2,K-4,2)$ irrep, for example, will only appear 
within the sector of purely bosonic operators (the same statement 
does not hold for the $(0,K-3,2)$ irrep).  Restricting to the $(2,K-4,2)$ 
irrep, we see that the dimension matrix cannot mix operators in the 
purely bosonic sector with bifermions, for example.  We will eventually
make these sorts of observations much more precise, as they will become 
invaluable in subsequent analyses.  The general problem involves 
diagonalizing matrices that are approximately $K/2 \times K/2$ in 
size.  The operators of interest will have large $K=R+2$ and fixed
$\Delta_0 = K-R = 2$.  As noted above, we expect that the anomalous
dimension spectrum should match the energy spectrum of string states
created by two oscillators acting on a ground state with angular 
momentum $J=R$.

As an example we will start with the basis of $K-1$ purely bosonic
operators with dimension $K$ and $\Delta_0 = 2$.  The anomalous dimensions
are the eigenvalues of the mixing matrix $d_1^{ab}$, appearing in the
perturbative expansion of the generic two-point function according to
\begin{equation}
\label{opprodexp} \langle O_a(x)O_b(0)\rangle \sim
(x)^{-2d_0}(\delta_{ab}+\ln(x^2)d_1^{ab})\ ,
\end{equation}
where $d_0$ is the naive dimension. The $\delta_{ab}$ term 
implies that the operator basis is orthonormal in
the free theory (in the large-$N_c$ limit, this is enforced by
multiplying the operator basis by a common overall normalization
constant).  The operator basis can be expressed as
\begin{eqnarray}
\label{singltrbasis}
\lbrace O^{AB}_{K,1},\ldots,O^{AB}_{K,K-1}
\rbrace =
    \lbrace \tr(ABZ^{K-2}),~\tr(AZBZ^{K-3}),~\ldots~,
\nonumber\\
        \tr(AZ^{K-3}BZ),~\tr(AZ^{K-2}B) \rbrace~,
\end{eqnarray}
where $Z$ stands for $\labphi{1}{2}$ and has $R=1$, while
$A,B$ stand for any of the four $\phi^A$ ($A=1,\ldots,4$) with
$R=0$ (the so-called $R$-charge impurities). 
The overall constant needed to orthonormalize this basis
is easy to compute, but is not needed for the present purposes.
Since the $R$-charge impurities $A$ and $B$ are $SO(4)$ vectors,
the operators in this basis are rank-two $SO(4)$ tensors.
In the language of $SO(4)$ irreps, the symmetric-traceless 
tensor descends from the $SU(4)$ irrep labeled by the
$(2,K-4,2)$ Young diagram.  Likewise, the antisymmetric tensor
belongs to the pair $(0,K-3,2)+(2,K-3,0)$,
and the $SO(4)$ trace (when completed to a full $SO(6)$ trace)
belongs to the $(0,K-2,0)$ irrep.  In what follows, we refer to
these three classes of operator as $\overline T_K^{(+)}$,
$\overline T_K^{(-)}$ and $\overline T_K^{(0)}$, respectively. If
we take $A\ne B$, the trace part drops out and the $\overline
T_K^{(\pm)}$ operators are isolated by symmetrizing and
antisymmetrizing on $A,B$.

At one-loop order in the 't~Hooft coupling $g^2_{YM}N_c$ the
action of the dilatation operator on the basis in eqn.~(\ref{singltrbasis}), 
correct to all orders in $1/K$, produces a sum of 
interchanges of all nearest-neighbor fields in the trace.
All diagrams that exchange fields at greater separation (at this
loop order) are non-planar, and are suppressed by powers of 
$1/N_c$.  As an example, we may restrict to the $A\ne B$ case.
Omitting the overall factor coming from the details of the 
Feynman diagram, the leading action of the anomalous dimension 
on the $K-1$ bosonic monomials of (\ref{singltrbasis}) has the 
following structure:
\begin{eqnarray}\label{bosopermute}
(ABZ^{K-2})\rightarrow
(BAZ^{K-2})+2(AZBZ^{K-3})+(K-3)(ABZ^{K-2})~,\nonumber\\
(AZBZ^{K-3})\rightarrow
2(ABZ^{K-2})+2(AZ^2BZ^{K-4})+(K-4)(AZBZ^{K-3})~,\nonumber\\
    \ldots\ldots\qquad\qquad \nonumber\\
(AZ^{K-2}B)\rightarrow
2(AZ^{K-3}BZ)+(K-3)(BAZ^{K-2})+(ABZ^{K-2})~.
\end{eqnarray}
Arranging this into matrix form, we have
\be
\label{bigmat}
\bigl[~{\rm
Anom~~Dim}~\bigr]_{(K-1)\times(K-1)}~\sim~
{\scriptsize
\left(
\begin{array}{ccccc}
    K-3&2&0&\ldots&1\\
    2&K-4&2&\ldots&0\\
     &	 &\ddots&  & \\
    0&\ldots&2&K-4&2\\
    1&\ldots&0&2&K-3\\
\end{array}
\right)}~.
\ee

As a final step, we must observe that the anomalous dimension
matrix in eqn.\ (\ref{bigmat}) contains contributions from the
$SU(4)$ irrep $(0,K,0)$, which corresponds to the chiral 
primary $\tr(Z^K)$.  The eigenstate associated with this operator
is $\vec X_0=(1,\ldots,1)$, with eigenvalue $K$ (the naive dimension).  
Since this operator is BPS, however, its anomalous dimension must be 
zero: to normalize the (\ref{bigmat}) we therefore subtract $K$
times the identity, leaving
\begin{equation}
\label{renormbigmat}
\bigl[~{\rm
Anom~~Dim}~\bigr]_{(K-1)\times(K-1)}~\sim~
{\scriptsize
\begin{pmatrix}
        -3&+2&0&\ldots&1\cr
        +2&-4&+2&\ldots&0\cr
	  &  &\ddots & & \cr
        0&\ldots&+2&-4&+2\cr
        +1&\ldots&0&+2&-3\cr
\end{pmatrix}}~.
\end{equation}
The zero eigenvector belonging to the $(0,K,0)$ representation
should then be dropped.  The anomalous dimensions are thus the
nonzero eigenvalues of (\ref{renormbigmat}). 
This looks very much like the lattice Laplacian for a particle
hopping from site to site on a periodic lattice. The special
structure of the first and last rows assigns an extra energy to
the particle when it hops past the origin. This breaks strict
lattice translation invariance but makes sense as a picture of the
dynamics involving two-impurity states: the impurities propagate
freely when they are on different sites and have a contact
interaction when they collide.  This picture has led people to
map the problem of finding operator dimensions onto the technically
much simpler one of finding the spectrum of an equivalent
quantum-mechanical Hamiltonian \cite{Beisert:2002ff}; this important 
topic will be reserved for later chapters.

To determine the $SU(4)$ irrep assignment of each of the eigenvalues
of (\ref{renormbigmat}), note that the set of operator monomials
is invariant under $A\leftrightarrow B$.  
For some vector $\vec C=(C_1,\ldots,C_{K-1})$ representing a given 
linear combination of monomials, this transformation sends
$C_i\to C_{K-i}$.  The matrix (\ref{renormbigmat}) itself is invariant
under $A\leftrightarrow B$, so its eigenvectors will either be even
($C_i=C_{K-i}$) or odd ($C_i=-C_{K-i}$) under the same exchange.
The two classes of eigenvalues and
normalized eigenvectors are:
\begin{eqnarray}
\label{symspect}
\lambda^{(K+)}_{n}~=~8\sin^2\left(\frac{n\pi}{K-1}\right)~, \qquad
n=1,2,\ldots, n_{max} = 
{\scriptsize
\begin{cases}
(K-3)/2~~K~{\rm odd}\cr
(K-2)/2~~K~{\rm even} 
\end{cases}}~, 
\nonumber\\
   C^{(K+)}_{n,i} =
\frac{2}{\sqrt{K-1}}\cos\left[\frac{2\pi n}{K-1}(i-\frac{1}{2})\right]~,
    \qquad i=1,\ldots,K-1~, \qquad
\end{eqnarray}
\begin{eqnarray}
\label{antisymspect}
\lambda^{(K-)}_{n}~=~8\sin^2\left(\frac{n\pi}{K}\right)~, \qquad n=1,2,\ldots,
n_{max} = 
{\scriptsize
\begin{cases}
(K-1)/2~~K~{\rm odd}\cr (K-2)/2~~K~{\rm even} 
\end{cases}}~,
\nonumber\\
   C^{(K-)}_{n,i} =
\frac{2}{\sqrt{K}}\sin\left[\frac{2\pi n}{K}(i)\right]~,
    \qquad i=1,\ldots,K-1~.\qquad
\end{eqnarray}
The eigenoperators are constructed from the eigenvectors according to
\begin{equation}
\label{eigenop}
  \overline T^{(\pm)}_{K,n} (x) =
    \sum_{i=1}^{K-1} C^{(K\pm)}_{n,i}O^{AB}_{K,i}(x) ~.
\end{equation}

By appending the appropriate overall normalization factor and adding 
the zeroth order value $\Delta_0=2$, we obtain $\Delta=D-R$.
The results are divided according to operators
belonging to the $(2,K-4,2)$ irrep ($\overline T_K^{(+)}$),
the $(0,K-3,2)+(2,K-3,0)$ irreps ($\overline T_K^{(-)}$) and
$(0,K-2,0)$ ($\overline T_K^{(0)}$).  In $SO(4)$ language, these
are the symmetric-traceless, antisymmetric and trace representations,
as described above.  We therefore have the following, exact in $K$:
\begin{eqnarray} 
\label{firstopdim}
{\Delta}(\overline T_K^{(+)}) = 2 +
        \frac{g^2_{YM} N_c}{\pi^2} \sin^2\left(\frac{n\pi}{K-1}\right)~,
& ~~ &
n=1,2,\ldots,n_{max} =
{\scriptsize
\begin{cases}
(K-3)/2~~K~{\rm odd}\cr (K-2)/2~~K~{\rm even} \end{cases}}~,
\nn
\ee
\be
    { \Delta}(\overline T_K^{(-)}) = 2 +
        \frac{g^2_{YM} N_c}{\pi^2} \sin^2\left(\frac{n\pi}{K}\right)~, 
& ~~ &
    n=1,2,\ldots, n_{max}= 
{\scriptsize
\begin{cases}(K-1)/2~~K~{\rm odd}\cr
        (K-2)/2~~K~{\rm even}\end{cases}}~,
\nn
\ee
\be
    { \Delta}(\overline T_K^{(0)}) = 2 +
        \frac{g^2_{YM} N_c}{\pi^2} \sin^2\left(\frac{n\pi}{K+1}\right)~, 
& ~ &
    n=1,2,\ldots, n_{max}= 
{\scriptsize
\begin{cases}(K-1)/2~~K~{\rm odd}\cr
        (K/2)~~~~~~~K~{\rm even}\end{cases}}~.
\nonumber\\
&&
\end{eqnarray}
The multiplicities match the earlier predictions given by the
expansion in Young diagrams in eqns.~(\ref{phi_irrepodd}) and (\ref{phi_irrep}).

We will eventually be interested in exploring the overlap of such results
with that which can be predicted by the dual string theory.  As described above, 
the central assumption introduced by Berenstein, Maldacena and Nastase is 
that the $R$-charge and the rank of the gauge group $N_c$ can be
taken to infinity such that the quantity $N_c/R^2$ remains fixed.
The perturbation expansion in the gauge theory is then controlled by 
$g_{\rm YM}^2N_c$ (which is kept small in the $g_{\rm YM}^2 \to 0$ limit,
which is the classical $g_s \to 0$ limit of the string theory), 
while worldsheet interactions in the string theory are controlled by $1/\Rhat$.
If we express the dimension formulas (\ref{firstopdim}) in terms of
$R$-charge $R$, rather than naive dimension $K$ (using $K=R+2$)
and take the limit in this way, we find
\begin{eqnarray} \label{finlopdim}
    {\Delta}(\overline T_{R+2}^{(+)})
    \rightarrow 2 +\frac{g^2_{YM} N_c}{R^2}~n^2~
            \left(1-\frac{2}{R}+O(R^{-2})\right)~,
\nonumber\\
        {\Delta}(\overline T_{R+2}^{(-)})
        \rightarrow 2 +\frac{g^2_{YM} N_c}{R^2}~n^2~
                        \left(1-\frac{4}{R}+O(R^{-2})\right)~,
\nonumber\ee\be
        {\Delta}(\overline T_{R+2}^{(0)})
        \rightarrow 2 +\frac{g^2_{YM} N_c}{R^2}~n^2~
                        \left(1-\frac{6}{R}+O(R^{-2})\right)~.
\end{eqnarray}
The key fact is that the degeneracy of the full BMN limit
(at leading order in $1/R$) is lifted at subleading order 
in $1/R$.  By including these subleading terms we generate 
an interesting spectrum that will prove to be a powerful
tool for comparison with string theory and testing the claims
of the AdS/CFT correspondence.

\section{The complete supermultiplet}
\label{gen_mult}
We have thus far reviewed the anomalous dimension
computation for a specific set of operators.  For a complete comparison 
with the string theory, we need to carry out some version of
the above arguments for all the relevant operators
with $\Delta_0 = 2$.  While this is certainly possible, we can instead 
rely on supersymmetry to determine the full spectrum of anomalous dimensions
for all single-trace, two-impurity operators.  
The extended superconformal symmetry of the gauge theory
means that operator dimensions will be organized into
multiplets based on a lowest-dimension primary ${\cal O}_D$
of dimension $D$.  Other conformal primaries within the multiplet 
can be generated by acting on super-primaries with any of eight supercharges
that increment the anomalous dimension shifts by a fixed amount but leave the 
impurity number unchanged.
We need only concern ourselves here
with the case in which ${\cal O}_D$ is a spacetime scalar (of
dimension $D$ and $R$-charge $R$). There are sixteen supercharges
and we can choose eight of them to be raising operators; there are
$2^8=256$ operators we can reach by `raising' the lowest one.
Since the raising operators increase the dimension and $R$-charge
by 1/2 each time they act, the operators at level $L$, obtained by
acting with $L$ supercharges, all have the same dimension and
$R$-charge. The corresponding decomposition of the 256-dimensional
multiplet is shown in table~\ref{table1}.
\bc
\begin{table}[ht!]
\begin{eqnarray}
{\footnotesize
\begin{array}{|l|l|l|l|l|l|l|l|l|l|}\hline
{\rm Level}& 0& 1& 2& 3& 4& 5& 6& 7& 8 \\ \hline {\rm
Multiplicity}& 1& 8& 28& 56& 70& 56& 28& 8& 1 \\ \hline {\rm
Dimension}& D& D+{1}/{2} & D+{1}& D+{3}/{2} & D+{2} & D+{5}/{2} &
D+{3} & D+{7}/{2} & D+4 \\ \hline R-{\rm charge} & R& R+{1}/{2} &
R+{1}& R+{3}/{2} & R+{2} & R+{5}/{2} & R+{3} & R+{7}/{2} & R+4 \\
\hline
\end{array} \nonumber    }
\end{eqnarray}
\caption{$R$-charge content of a supermultiplet} \label{table1}
\end{table}  
\ec
The states at each level can be classified under the
Lorentz group and the $SO(4)\sim SU(2)\times SU(2)$ subgroup of
the $R$-symmetry group, which is unbroken after we have fixed the
$SO(2)$ $R$-charge. For instance, the 28 states at level 2
decompose under $SO(4)_{Lor}\times SO(4)_R$ as
$(6,1)+(1,6)+(4,4)$. For the present, the most important point is
that, given the dimension of one operator at one level, we can
infer the dimensions of all other operators in the supermultiplet.

By working in this fashion we can generate complete anomalous
dimension spectra of all two-impurity operators.  The results
obtained in this manner agree with work originally completed
by Beisert in \cite{Beisert:2002tn}.  We will summarize these results
here, adding some further useful information that emerges
from our own $SU(4)$ analysis.  
The supermultiplet of interest is
based on the set of scalars
$\sum_A \tr\left(\phi^AZ^p\phi^AZ^{R-p}\right)$, the operator
class we have denoted by $\overline T_{R+2}^{(0)}$. According to
(\ref{firstopdim}), the spectrum of $\Delta=D-R$ eigenvalues
associated with this operator basis is
\begin{eqnarray}
\label{multopdim}
\Delta(\overline T_{R+2}^{(0)}) =
    2 + \frac{g^2_{YM} N_c}{\pi^2} \sin^2\left(\frac{n\pi}{R+3}\right)
    \rightarrow 2 +\frac{g^2_{YM} N_c}{R^2}~n^2~\left(1-\frac{6}{R}
+O(R^{-2})\right)~.
\end{eqnarray}
The remaining scalar operators $\overline T_{R+2}^{(\pm)}$ are 
included in the supermultiplet and the dimension formulas are
expressed in terms of the $R$-charge of the lowest-dimension
member. It turns out that (\ref{multopdim})
governs {\it all} the operators at {\it all} levels in the
supermultiplet.  The results of this program, carried out on the
spacetime scalar operators, are summarized in table~\ref{tableone}.

\bc
\begin{table}[ht!]
\begin{eqnarray}
{\scriptsize
\begin{array}{|l|l|l|l|l|l|}\hline
L& R& SU(4) ~ {\rm Irreps} & {\rm Operator} & \Delta-2 & {\rm Multiplicity}
\\ \hline
0 & R_0 & (0,R_0,0) & \Sigma_A\tr\left(\phi^AZ^p\phi^AZ^{R_0-p}\right) &
\frac{g^2_{YM} N_c}{\pi^2} \sin^2(\frac{n\pi}{(R_0)+3}) &
    n=1,.,\frac{R_0+1}{2} \\ \hline
2 & R_0+1 & (0,R_0,2)+c.c. & \tr\left(\phi^{[i}Z^p\phi^{j]}Z^{R_0+1-p}\right) &
\frac{g^2_{YM} N_c}{\pi^2} \sin^2(\frac{n\pi}{(R_0+1)+2}) &
    n=1,.,\frac{R_0+1}{2} \\ \hline
4 & R_0+2 & (2,R_0,2) & \tr\left(\phi^{(i}Z^p\phi^{j)}Z^{R_0+2-p}\right) &
\frac{g^2_{YM} N_c}{\pi^2} \sin^2(\frac{n\pi}{(R_0+2)+1}) &
    n=1,.,\frac{R_0+1}{2} \\ \hline
4&R_0+2&(0,R_0+2,0)\times 2&
\tr\left(\chi^{[\alpha}Z^p\chi^{\beta]}Z^{R_0+1-p}\right) &
\frac{g^2_{YM} N_c}{\pi^2} \sin^2(\frac{n\pi}{(R_0+2)+1}) &
    n=1,.,\frac{R_0+1}{2} \\ \hline
6&R_0+3&(0,R_0+2,2)+c.c.&\tr\left(\chi^{(\alpha}
Z^p\chi^{\beta)}Z^{R_0+2-p}\right) &
\frac{g^2_{YM} N_c}{\pi^2} \sin^2(\frac{n\pi}{(R_0+3)+0}) &
    n=1,.,\frac{R_0+1}{2} \\ \hline
8 & R_0+4 & (0,R_0,0) & \tr\left(\nabla_\mu Z Z^p\nabla^\mu Z Z^{R_0+2-p}
\right) &
\frac{g^2_{YM} N_c}{\pi^2} \sin^2(\frac{n\pi}{(R_0+4)-1}) &
    n=1,.,\frac{R_0+1}{2} \\ \hline
\end{array}  }
\nn\\ \nonumber  
\end{eqnarray}
\caption{Dimensions and multiplicities of spacetime scalar
operators} \label{tableone}
\end{table}
\ec

The supermultiplet contains operators that are not spacetime
scalars (i.e., that transform nontrivially under the $SU(2,2)$
conformal group) and group theory determines at what levels in
the supermultiplet they must lie. A representative sampling of
data on such operators (extracted from Beisert's paper) is
collected in table~\ref{tabletwo}. We have worked out neither the
$SU(4)$ representations to which these lowest-$\Delta$ operators
belong nor their precise multiplicities. The ellipses indicate
that the operators in question contain further monomials involving
fermion fields (so that they are not uniquely specified by their
bosonic content). This information will be useful in consistency
checks to be carried out below.

\bc
\begin{table}[ht!]
\begin{eqnarray}
\begin{array}{|l|l|l|l|l|}\hline
L & R& {\rm Operator} & \Delta-2 &  \Delta-2\to \\ \hline 2 & R_0+1 &
\tr\left(\phi^iZ^p\nabla_\mu Z Z^{R_0-p}\right)+\ldots &
\frac{g^2_{YM} N_c}{\pi^2} \sin^2(\frac{n\pi}{(R_0+1)+2}) &
\frac{g^2_{YM} N_c}{R_0^2}n^2(1-\frac{4}{R_0}) \\ \hline 4 & R_0+2  &
\tr\left(\phi^iZ^p\nabla_\mu Z Z^{R_0+1-p}\right) & \frac{g^2_{YM}
N_c}{\pi^2} \sin^2(\frac{n\pi}{(R_0+2)+1}) & \frac{g^2_{YM}
N_c}{R_0^2}n^2(1-\frac{2}{R_0}) \\ \hline 4 & R_0+2  &
\tr\left(\nabla_{(\mu}ZZ^p\nabla_{\nu)}Z Z^{R_0-p}\right) &
\frac{g^2_{YM} N_c}{\pi^2} \sin^2(\frac{n\pi}{(R_0+2)+1}) &
\frac{g^2_{YM} N_c}{R_0^2}n^2(1-\frac{2}{R_0}) \\ \hline 6 & R_0+3  &
\tr\left(\phi^iZ^p\nabla_\mu Z Z^{R_0+2-p}\right)+\ldots &
\frac{g^2_{YM} N_c}{\pi^2} \sin^2(\frac{n\pi}{R_0+3}) &
\frac{g^2_{YM} N_c}{R_0^2}n^2(1-\frac{0}{R_0}) \\ \hline 6 & R_0+3  &
\tr\left(\nabla_{[\mu}ZZ^p\nabla_{\nu]}Z Z^{R_0+1-p}\right) &
\frac{g^2_{YM} N_c}{\pi^2} \sin^2(\frac{n\pi}{R_0+3}) &
\frac{g^2_{YM} N_c}{R_0^2}n^2(1-\frac{0}{R_0}) \\ \hline
\end{array}\nonumber
\end{eqnarray}
\caption{Anomalous dimensions of some operators that are not
scalars} \label{tabletwo}
\end{table}
\ec

\ \\

\bc
\begin{table}[ht!]
\begin{eqnarray}
{\small
\begin{array}{|l|l|l|l|l|l|l|l|l|l|}\hline
{\rm Level} & 0& 1& 2& 3& 4& 5& 6& 7& 8 \\ \hline
{\rm Multiplicity} & 1& 8& 28& 56& 70& 56& 28& 8& 1 \\ \hline
\delta E\times (R^2/g_{YM}^2N_c n^2) & -{6}/{R} & -{5}/{R} &
-{4}/{R} & -{3}/{R} & -{2}/{R} & -{1}/{R} &
0  & {1}/{R} & {2}/{R} \\ \hline
\end{array} \nonumber }
\end{eqnarray}
\caption{Predicted energy shifts of two-impurity string states}
\label{smultiplicity}
\end{table}
\ec

\ \\

The complete dimension spectrum of operators with
$R$-charge $R$ at level $L$ in the supermultiplet are given by the
general formula (valid for large $R$ and fixed $n$):
\begin{eqnarray}
\label{deltalevel} \Delta^{R,L}_n 
	&=& 2 + \frac{g^2_{YM} N_c}{\pi^2}
\sin^2\left(\frac{n\pi}{R+3-L/2}\right) 
\nn\\
 &=&  2+\frac{g^2_{YM}
N_c}{R^2}~n^2\left(1-\frac{6-L}{R}+O(R^{-2})\right)~.
\end{eqnarray}
It should be emphasized that, for fixed
$R$, the operators associated with different levels are actually
coming from {\it different} supermultiplets; this is why they have
different dimensions! As mentioned before, we can also precisely
identify transformation properties under the Lorentz group and
under the rest of the $R$-symmetry group of the degenerate states
at each level. 
This again leads to useful consistency checks, and
we will elaborate on this when we analyze the eigenstates of the
string worldsheet Hamiltonian.

%
%
\chapter{A virial approach to operator dimensions}	                  
\label{virial}
In the previous chapter we reviewed how the problem 
of computing operator dimensions in the planar limit of large-$N$
${\cal N}=4$ SYM theory maps to that of diagonalizing the Hamiltonian
of certain quantum mechanical systems.  Calculating operator dimensions is 
equivalent to finding the eigenvalue spectrum of spin chain Hamiltonians, 
and various established techniques associated with integrable systems (most notably the
Bethe ansatz) have proved useful in this context (for a general review of the 
Bethe ansatz method, see \cite{Faddeev:1996iy}). The utility of this approach 
was first demonstrated by 
Minahan and Zarembo in \cite{Minahan:2002ve}. For operators with two $R$-charge 
impurities, the spin chain spectra can be computed exactly via the Bethe ansatz.  
For three- or higher-impurity operators, however, the Bethe equations have
only been solved perturbatively near the limit 
of infinite chain length \cite{Minahan:2002ve,Beisert:2003yb,Lubcke:2004dg}. Furthermore, at 
higher-loop order in $\lambda$, the spin chain Hamiltonians typically acquire 
long-range or non-nearest-neighbor interactions for which a general Bethe ansatz may not be 
available.  For example, while the action of the spin chain Hamiltonian in the 
``closed $\su(2|3)$'' sector is known to three-loop order \cite{Beisert:2003ys}, 
the corresponding long-range Bethe ansatz is not known (though it may well exist).
(See \cite{Staudacher:2004tk} for a more recent approach to deriving Bethe ansatz equations.)  
A long-range Bethe ansatz does exist for the particularly simple ``closed $\su(2)$'' 
sector of the theory \cite{Serban:2004jf,Beisert:2004hm}, and
our methods will provide a useful cross-check on these approaches
to gauge theory anomalous dimensions at higher order in the 't~Hooft parameter 
$\lambda = g_{\rm YM}^2 N_c$.

In this chapter we will present a virial approach to the spin chain systems 
of ${\cal N}=4$ SYM theory.
The generic spin chain Hamiltonian acts on single-impurity pseudoparticles 
as a lattice Laplacian and higher $N$-body interactions among pseudoparticles
are suppressed  relative to the one-body pseudoparticle energy by inverse powers of 
the lattice length ${K}$.  Surprisingly, this expansion of the spin chain
Hamiltonian is truncated at $O({K}^{-3})$ in certain subsectors of the theory, 
allowing straightforward eigenvalue calculations that are exact in the chain length for
operators with more than two $R$-charge impurities.  
Furthermore, since the goal is to eventually compare anomalous
dimensions with $1/J$ energy corrections to corresponding string states
near the pp-wave limit of $AdS_5\times S^5$, and because the string
angular momentum $J$ is proportional to the lattice length ${K}$, 
this virial expansion is precisely what is needed to devise a practical 
method for testing the AdS/CFT correspondence at any order in the gauge theory loop 
expansion for an arbitrary number of $R$-charge (or worldsheet) impurities.

We will focus on three particular closed 
sectors of the theory, each labeled by the subalgebra of the full superconformal algebra 
that characterizes the spin variables of the equivalent spin chain system. 
Specifically, there are two sectors spanned by bosonic operators and
labeled by $\su(2)$ and $\Sl(2)$ subalgebras plus an $\su(2|3)$ sector which includes 
fermionic operators. Section \ref{virialSEC2} is dedicated to an analysis of the bosonic 
$\su(2)$ closed 
sector to three-loop order in $\lambda$. In Section~\ref{virialSEC3} we analyze an 
$\su(1|1)$  subsector 
of the closed $\su(2|3)$ system to three-loop order. The spin chain Hamiltonian in the bosonic 
$\Sl(2)$ sector has previously been determined to one loop, and we analyze this system in 
Section~\ref{virialSEC4}.

\section{The $\protect\su(2)$ sector}
\label{virialSEC2}
Single-trace operators in the closed $\su(2)$ sector are constructed from
two complex scalar fields of ${\cal N}=4$ SYM, typically 
denoted by $Z$ and $\phi$.  Under the 
$SO(6) \simeq U(1)_R\times SO(4)$ decomposition of the full
$SU(4)$ $R$-symmetry group, the $Z$ fields are charged under the
scalar $U(1)_R$ component and $\phi$ is a particular scalar field
carrying zero $R$-charge.  The basis of length-${K}$ operators
in the planar limit is constructed from single-trace monomials with $I$ 
impurities and total $R$-charge equal to ${K}-I$: 
\be
\label{SO6basis3}
\tr(\phi^I Z^{{K}-I})\ , \qquad  
\tr(\phi^{I-1}Z\phi Z^{{{K}-I}-1})\ , \qquad 
\tr(\phi^{I-2}Z\phi^2 Z^{{{K}-I}-1})\ , \qquad \ldots~.
\ee
The statement that this sector of operators is ``closed'' means simply
that the anomalous dimension operator can be diagonalized on this
basis, at least to leading order in large $N_c$ \cite{Beisert:2003tq,Beisert:2003jj}.

The heart of the spin chain approach is the proposition that there
exists a one-dimensional spin system whose Hamiltonian can be identified 
with the large-$N_c$ limit of the anomalous dimension operator acting
on this closed subspace of operators \cite{Minahan:2002ve}. Since the anomalous dimensions
are perturbative in the 't~Hooft coupling $\lambda$, it is natural 
to expand the $\su(2)$ spin chain Hamiltonian in powers of $\lambda$ as well:
\be
H_{\su(2)} = I + \sum_n \left( \frac{\lambda}{8 \pi^2} \right)^n H_{\su(2)}^{(2n)}\ .
\ee
Comparison with the gauge theory has shown that successive terms in the
expansion of the Hamiltonian have a remarkably simple structure:
the one-loop-order Hamiltonian $H^{(2)}_{\su(2)}$ is built out of 
permutations of pairs of nearest-neighbor fields and, at $n^{\rm th}$ order, the 
Hamiltonian permutes among themselves fields that are at most $n$ lattice 
sites apart. This is a universal structure that leads to remarkable 
simplifications in the various closed sectors of the theory \cite{Beisert:2003yb}.

Beisert, Kristjansen and Staudacher \cite{Beisert:2003tq} have introduced the
following useful notation for products of permutations acting on operators 
separated by an arbitrary number of lattice sites:
\be
\{ n_1,n_2,\dots\} = \sum_{k=1}^{K} P_{k+n_1,k+n_1+1} P_{k+n_2,k+n_2+1}\cdots\ ,
\ee
where $P_{i,j}$ simply exchanges fields on the $i^{\rm th}$ and $j^{\rm th}$ lattice sites 
on the chain. The upshot of the gauge theory analysis is that the 
equivalent spin chain Hamiltonian for the $\su(2)$ sector can be written
in a rather compact form in terms of this notation. The result, correct to 
three-loop order, is (see \cite{Beisert:2003tq} for details)
\be
H_{\su(2)}^{(2)} & = & 2\left( \{ \}-\{0\} \right)~,
\label{Hsu2_1}
\\
H_{\su(2)}^{(4)} & = & 2\bigl( -4\{\} + 6\{0\} - (\{0,1\} + \{1,0\}) \bigr)~,
\label{Hsu2_2}
\\
H_{\su(2)}^{(6)} & = & 4\bigl[ 15\{\}-26\{0\} +6\left(\{0,1\}+\{1,0\}\right)
		+\{0,2\} 
\nn\\
&&
	-\left(\{0,1,2\}+\{2,1,0\}\right) \bigr]\ .
\label{Hsu2_3}
\ee
(Note that $\{ \}$ is just the identity operator.)
The form of the three-loop term $H_{\su(2)}^{(6)}$ was first conjectured in 
\cite{Beisert:2003tq} based on integrability restrictions and BMN scaling;
this conjecture was later corroborated by direct field-theoretic methods in 
\cite{Beisert:2003ys} (see also \cite{Beisert:2003jb} for relevant discussion
on this point). Our goal is to develop practical methods for finding the
eigenvalue spectrum of the spin chain Hamiltonian for various interesting cases.

\subsection{One-loop order}
We start at one-loop order with $H_{\su(2)}^{(2)}$ in eqn.~(\ref{Hsu2_1}), which
provides a natural `position-space' prescription for constructing 
matrix elements in an $I$-impurity basis of operators.  As an explicit
example, we consider first the basis of two-impurity operators of length ${K}=8$:
\be
\tr(\phi^2 Z^6)~, \qquad 
\tr(\phi Z \phi Z^5)~,  \qquad 
\tr(\phi Z^2 \phi Z^4)~,  \qquad 
\tr(\phi Z^3 \phi Z^3)~.
\ee 
It is easy to see that the one-loop Hamiltonian mixes the four elements of 
this basis according to the matrix
\be
H_{\su(2)}^{(2)} = 
{\scriptsize
\left(
\begin{array}{cccc}
2 & -2 & 0 & 0 \\
-2 & 4 & -2 & 0 \\
0 & -2 & 4 & -2\sqrt{2} \\
0 & 0 & -2\sqrt{2} & 4 
\end{array}\right)}\ .
\label{L6PS}
\ee
This matrix generalizes to arbitrary ${K}$ and it is simple to show that
the two-impurity one-loop eigenvalues of $H_{\su(2)}^{(2)}$ are given by the formula 
\cite{Beisert:2002tn}
\be
E_{\su(2)}^{(2)} = {8}\, \sin^2 \left(\frac{\pi n}{{K}-1}\right)~, \qquad
n = 0,\ldots,n_{\rm max} = 
	\biggl\{ \genfrac{}{}{0pt}{0}{ ({K}-2)/2, 
	\quad  {K}\ {\rm even} }{ ({K}-3)/2, \quad  {K}\ {\rm odd} }\ .
\label{su2EIG}
\ee
Note that if the denominator ${K}-1$ were replaced by 
${K}$, the above expression would agree with the usual lattice Laplacian energy for a 
lattice of length ${K}$. The difference amounts to corrections to the free Laplacian of 
higher order in $1/{K}$ and we will seek to understand the physical origin of such 
corrections in what follows.

To compare gauge theory predictions with $1/J$ corrections to the 
three-impurity spectrum of the string theory on $AdS_5\times S^5$,
we need to determine the large-${K}$ behavior of the three-impurity
spin chain spectrum.  We are primarily interested in systems with few 
impurities compared to the length of the spin chain and we expect that
impurity interaction terms in the Hamiltonian will be suppressed by
powers of the impurity density (i.e.,~inverse powers of the lattice length).
This suggests that we develop a virial expansion of the spin 
chain Hamiltonian in which the leading-order term in $1/{K}$ gives the 
energy of free pseudoparticle states on the lattice (labeled by lattice
momentum mode numbers as in the two-impurity spectrum eqn.~(\ref{su2EIG})) and higher $1/{K}$
corrections come from $N$-body interactions described by vertices $V_N$. 
A reasonable guess about how the $N$-body interactions should scale with $1/{K}$
suggests that we can write the one-loop-order energy for $I$ impurities
in the form 
\be
E(\{n_i\}) = I + \frac{\lambda}{2\pi^2}\sum_{i=1}^I \sin^2\frac{n_i\pi}{{K}}
+ \sum_{N=2}^{2I} \frac{\lambda}{{K}^{2N-1}} {V_{N-\rm body}(n_1,\ldots,n_I)} + \cdots\ ,
\label{Nbody}
\ee
where the leading-order contribution $I$ measures the naive dimension
minus $R$-charge, the next term is the lattice Laplacian energy of $I$
non-interacting pseudoparticles and the $1/{K}$ corrections account for 
interactions between pseudoparticles (which may depend on the lattice 
momenta mode numbers $n_i$). In the many-body approach, one would try
to derive such energy expressions by rewriting the Hamiltonian in terms of 
creation/annihilation operators $b_{n_i}$, $b_{n_i}^\dag$ for the
pseudoparticles (commuting or anticommuting as appropriate). The $N$-body 
interaction vertex would generically be written in terms of the $b,b^\dag$ as
\be
V_N = \sum_{n_i,m_i}~\delta_{n_1+\cdots+n_N,m_1+\cdots+m_N}
	f_N(\{n_i\},\{m_i\})\prod_{i=1}^{N} b_{n_i}^\dag \prod_{i=1}^{N} b_{m_i}\ ,
\ee
where $f_N(\{n_i\},\{m_i\})$ is some function of the lattice momenta 
and the Kronecker delta enforces lattice momentum conservation. One has to
determine the functions $f_N$ by matching the many-body form of the Hamiltonian
to exact spin chain expressions such as eqn.~(\ref{Hsu2_1}). We will see that,
once the Hamiltonian is in many-body form, it is straightforward to obtain
a density expansion of the higher-impurity energy eigenvalues.

The discussion so far has been in the context of one-loop gauge theory
physics, but the logic of the virial expansion should be applicable
to the general case. To include higher-loop order physics we must do
two things: a) generalize the functions $f_N(\{n_i\},\{m_i\})$ defining 
the multi-particle interaction vertices to power series in $\lambda$ and
b) allow the free pseudoparticle kinetic energies themselves to become
power series in $\lambda$. We will be able to carry out the detailed construction 
of the higher-loop virial Hamiltonian in a few well-chosen cases.
To match this expansion at $n$-loop order in $\lambda$ to the corresponding
loop order (in the modified 't~Hooft coupling $\lambda' = g_{\rm YM}^2 N_c / J^2$) 
in the string theory, we need to determine the Hamiltonian to $O({K}^{-(2n+1)})$ 
in this virial expansion.  (The first curvature correction to the pp-wave string theory 
at one loop, for example, appears at $O(\lambda'/J)$ or, in terms of gauge theory 
parameters, at $O(\lambda/{K}^3)$.) Auspiciously, it will turn out that this virial 
expansion in the $\su(2)$ sector is truncated at small orders in $1/{K}$, allowing 
for simple eigenvalue calculations that are exact in ${K}$ (although perturbative
in $\lambda$).

The first step toward obtaining the desired virial expansion is to recast the spin 
chain Hamiltonian $H_{\su(2)}$, which is initially expressed in terms of permutation 
operators, in terms of a creation and annihilation operator algebra. 
We begin by introducing the spin operators
\be
S^\pm = \frac{1}{2}\left( \sigma_x \pm i \sigma_y \right)~, \qquad
S^z = \frac{1}{2} \sigma_z\ ,
\ee
where $\vec \sigma$ are the Pauli matrices and $S^\pm_j,\ S^z_j$ act on 
a two-dimensional spinor space at the $j^{\rm th}$ lattice site in the chain.
In this setting the $Z$ and $\phi$ fields are understood to be modeled
by up and down spins on the lattice. The nearest-neighbor permutation operator 
$P_{i,i+1}$ can be written in terms of spin operators as
\be
P_{i,i+1} = S_i^+S_{i+1}^- + S_i^-S_{i+1}^+ + 2S_i^zS_{i+1}^z + \frac{1}{2}\ , 
\ee
and the one-loop Hamiltonian in eqn.~(\ref{Hsu2_1}) can be written as
\be
H_{\su(2)}^{(2)} = - \sum_{j=1}^{K} 
	\left( S^+_j S^-_{j+1} + S^-_j S^+_{j+1}\right)
	-2 \sum_{j=1}^{K} S_j^z S_{j+1}^z + \frac{1}{2}\ .
\label{HS1}
\ee
A Jordan-Wigner transformation can now be used to express the spin generators
in terms of anti-commuting creation and annihilation operators (anticommuting
because each site can be either unoccupied ($Z$) or occupied once ($\phi$)).
A pedagogical introduction to this technique can be found in \cite{Nagaosa:1999uc}.
The explicit transformation is
\be
\label{JordWig}
S_j^+ & = & b_j^\dag K(j) = K(j) b_j^\dag~, \nn\\
S_j^- & = & K(j) b_j = b_j K(j)~,  \nn\\
S_j^z & = & b_j^\dag b_j - {1}/{2}\ ,
\ee
where the Klein factors
\be
K(j) = \exp \left( i\pi \sum_{k=1}^{j-1} b_k^\dag b_k \right)
\ee
serve to ensure that spin operators on different
sites commute, despite the anticommuting nature of the $b_j$.
The functions $K(j)$ are real, Abelian and, for $j\leq k$, 
\be
[ K(j),{\bf S}_k ] = 0\ . 
\ee
The operators $b_j^\dag$ and $b_j$ can therefore be written as
\be
b_j^\dag = S_j^+ K(j)~, \qquad b_j = S_j^- K(j)\ ,
\ee
and we easily verify that they satisfy the standard anticommutation relations
\be
\{ b_j, b_k^\dag \} = \delta_{jk}~, \qquad \{b_j^\dag,b_k^\dag\} = \{b_j,b_k\} = 0\ .
\ee
Cyclicity on the lattice requires that ${\bf S}_{{K}+1} = {\bf S}_1$, a condition
that can be enforced by the following  boundary condition on the creation and annihilation operators:
\be
b_{{K}+1} = (-1)^{I+1} b_1~, \qquad I \equiv \sum_{j=1}^{K} b_j^\dag b_j\ ,
\label{BC}
\ee
where the integer $I$ counts the number of spin chain impurities.
In this chapter we will be primarily interested in analyzing spin chains with three
impurities.  The two-impurity problem can usually be solved more directly and, although
the techniques presented here are certainly applicable, going to four impurities introduces 
unnecessary complications.
We will henceforth impose the boundary conditions in eqn.~(\ref{BC}) for 
odd impurity number only. We can use all of this to re-express eqn.~(\ref{HS1}) 
in creation and annihilation operator language, with the result
\be
H_{\su(2)}^{(2)} =   \sum_{j=1}^{{K}} \left(
	b_{j}^\dag b_{j} + b_{j+1}^\dag b_{j+1} - b_{j+1}^\dag b_{j} - b_{j}^\dag b_{j+1}
	+ 2\,b_{j}^\dag b_{j+1}^\dag b_{j} b_{j+1}  \right)\ .
\label{Hsu2PS}
\ee
Converting to momentum space via the usual Fourier transform
\be
b_j = \frac{1}{\sqrt{{K}}}\sum_{p=0}^{{K}-1} e^{ -\frac{2\pi i j}{{K}} p }~\bt_p\
\ee
yields
\be
H_{\su(2)}^{(2)} & = & 4\sum_{p=0}^{{K}-1} \sin^2\left( \frac{\pi p}{{K}} \right) \bt_p^\dag \bt_p
	+ \frac{2}{{K}} \sum_{p,q,r,s=0}^{{K}-1} e^{\frac{2\pi i (q-s)}{{K}}} 
	\bt_p^\dag \bt_q^\dag \bt_r \bt_s\, \delta_{p+q,r+s}\ .
\label{Hsu2MS}
\ee
This is a rather standard many-body Hamiltonian: it acts on a Fock space
of momentum eigenstate pseudoparticles, contains a one-body pseudoparticle 
kinetic energy term and a two-body pseudoparticle interaction (the latter
having the critical property that it conserves the number of pseudoparticles).
Note that the Hamiltonian terminates at two-body interactions, a fact that will
simplify the virial expansion of the energy spectrum. This termination is a 
consequence of the fact that the one-loop Hamiltonian contains only 
nearest-neighbor interactions and that lattice sites can only be once-occupied.

Because the pseudoparticle (or impurity) number is conserved by the interaction,
three-impurity eigenstates of the Hamiltonian must lie in the space spanned by
\be
\bt_{k_1}^\dag \bt_{k_2}^\dag \bt_{k_3}^\dag \ket{{K}}~, \qquad
	k_1 + k_2 + k_3 = 0 \mod {K}\ ,
\ee
where the ground state $\ket{{K}}$ is identified with the zero-impurity operator $\tr(Z^{K})$
and the condition of vanishing net lattice momentum arises from 
translation invariance on the spin chain (which in turn arises from the cyclicity
of the  single-trace operators in the operator basis). As a concrete example, 
the basis of three-impurity states of the ${K}=6$ $\su(2)$ spin chain is
\be
\bt_0^\dag \bt_1^\dag \bt_5^\dag \ket{{K}}~, \qquad 
\bt_0^\dag \bt_2^\dag \bt_4^\dag \ket{{K}}~, \qquad 
\bt_1^\dag \bt_2^\dag \bt_3^\dag \ket{{K}}~, \qquad 
\bt_3^\dag \bt_4^\dag \bt_5^\dag \ket{{K}}\ ,
\ee
and the matrix elements of the Hamiltonian (\ref{Hsu2MS}) 
in this basis are easily computed:
\be
H_{\su(2)}^{(2)} = 
{\scriptsize
\left(
\begin{array}{cccc}
\frac{1}{3} & -1 &  \frac{1}{3}& \frac{1}{3} \\
-1 & 3 & -1 & -1 \\
\frac{1}{3} & -1 &\frac{19}{3}  & \frac{1}{3} \\
\frac{1}{3} & -1 & \frac{1}{3} &  \frac{19}{3}
\end{array}\right)}\ .
\label{L6MS}
\ee
The first-order perturbation theory corrections to the 
three-impurity operator anomalous dimensions are the eigenvalues of
this matrix.

\begin{figure}[htb]
\begin{center}
\includegraphics[width=3.5596in,height=2.2in,angle=0]{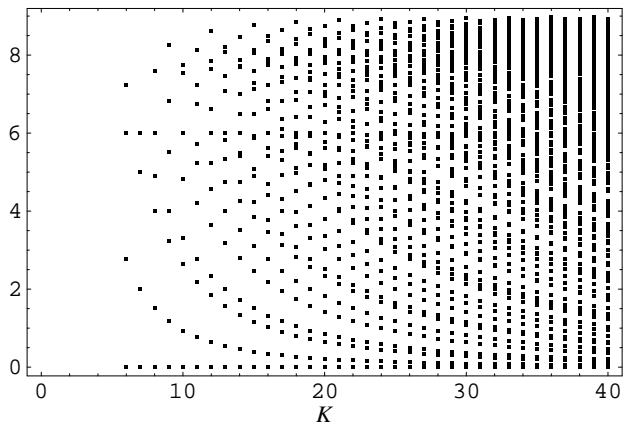}
\caption{One-loop $\su(2)$ spin chain spectrum vs.~lattice length ${K}$ ($6\leq {K} \leq 40$)}
\label{SO63aSTR}
\end{center}
\end{figure}

The construction and diagonalization of the Hamiltonian matrix on
the degenerate basis of three-impurity operators can easily be carried
out for larger ${K}$. The results of doing this\footnote{Using the position- 
or momentum-space formalism is purely a matter of 
convenience.  In practice we have found that for all sectors the momentum-space 
treatment is computationally much more efficient. The large-${K}$ extrapolations
of both methods can be checked against each other, and we of course find
that they are in agreement.}
for lattice sizes out to ${K}=40$ are displayed in figure~\ref{SO63aSTR}.  
According to eqn.~(\ref{Nbody}), we expect the eigenvalues of $H_{\su(2)}^{(2)}$ 
to scale for large ${K}$ according to
\be
\label{scaling3}
E_{{K}}(\lbrace k_i \rbrace) =  \frac{\lambda}{{K}^2}{E^{(1,2)}
	(\lbrace k_i \rbrace)} + 
\frac{\lambda}{{K}^3}{E^{(1,3)}(\lbrace k_i \rbrace)} + O(\lambda {K}^{-4})\ .
\label{expG13}
\ee
The scaling coefficients $E^{(1,2)}_{\su(2)}$ and $E^{(1,3)}_{\su(2)}$ can easily be extracted from 
the data displayed in figure~\ref{SO63aSTR} by fitting the spectral curves to 
large-order polynomials in $1/{K}$ (a similar treatment was used in \cite{Beisert:2003ea}). 
The results of this procedure are recorded for several 
low-lying levels in the spectrum (excluding zero eigenvalues) in table~\ref{NUM_SU(2)_1Loop}. 
\begin{table}[ht!]
\begin{eqnarray}
\begin{array}{|cccc|}
\hline
E_{\su(2)}^{(1,2)} & E_{\su(2)}^{(1,3)} & E_{\su(2)}^{(1,3)}/E_{\su(2)}^{(1,2)}  
                                  & {\rm Lattice\ Momenta\ }(k_1,k_2,k_3) \\
\hline
1+2.6\times 10^{-9}	&2-4.9\times 10^{-7}	&2-5.0\times 10^{-7}	& (1,0,-1)	 \\
3+4.6\times 10^{-9}	&7-8.8\times 10^{-7}    &7/3-3.0\times 10^{-7} & (1,1,-2)	 \\	
3+4.6\times 10^{-9}	&7-8.8\times 10^{-7}	&7/3-3.0\times 10^{-7}	&  (-1,-1,2)	 \\
4+6.0\times 10^{-9}	&8-1.1\times 10^{-6}    &2-2.9\times 10^{-7}   &  (2,0,-2)	 \\
7+3.2\times 10^{-8}	&14-7.1\times 10^{-6}	&2-1.0\times 10^{-6}	&  (1,2,-3) 	 \\
7+3.2\times 10^{-8}	&14-7.1\times 10^{-6}	&2-1.0\times 10^{-6}	&  (-1,-2,3)	 \\
9+2.2\times 10^{-7}	&18-5.1\times 10^{-5}	&2-5.7\times 10^{-6}	&  (3,0,-3)	 \\
12+5.7\times 10^{-5}	&28+3.8\times 10^{-3}	&7/3-1.4\times 10^{-3}	&  (2,2,-4)	 \\
12+5.7\times 10^{-5} 	&28+3.8\times 10^{-3}	&7/3-1.4\times 10^{-3}	&  (-2,-2,4)	 \\
13-5.6\times 10^{-5}	&26-3.8\times 10^{-3}	&2+1.3\times 10^{-3}	&  (1,3,-4) 	 \\
13-5.6\times 10^{-5}	&26-3.8\times 10^{-3}	&2+1.3\times 10^{-3}	&  (-1,-3,4)	 \\
\hline
\end{array} \nonumber
\end{eqnarray}
\caption{Scaling limit of three-impurity
	$\su(2)$ numerical spectrum at one loop in $\lambda$}
\label{NUM_SU(2)_1Loop}
\end{table}
Subtracting the small errors, we claim that we have the
following simple predictions for the large-${K}$ $\su(2)$ expansion coefficients 
$E_{\su(2)}^{(1,3)}$ and $E_{\su(2)}^{(1,2)}$: 
\be
E_{\su(2)}^{(1,2)} = (k_1^2+k_2^2+k_3^2)/2~, & \qquad & k_1+k_2+k_3=0~,\nonumber\nonumber\\
E_{\su(2)}^{(1,3)}/E_{\su(2)}^{(1,2)}  =  2~, & \qquad & (k_1\neq k_2\neq k_3)~,
\nonumber\\
E_{\su(2)}^{(1,3)}/E_{\su(2)}^{(1,2)} = \frac{7}{3}~, & \qquad &(k_1 = k_2,\ k_3=-2k_1)\ .
\label{su2num}
\ee
Note the slight annoyance that we must distinguish the case where all mode indices are 
unequal from the case where two indices are equal and different from the third.
The last column of table~\ref{NUM_SU(2)_1Loop} displays the choice of indices
$\{k_i\}$ that best fit each spectral series.  As the lattice momenta
increase, higher-order $1/{K}$ corrections to the spectrum become stronger
and more data will be required to maintain a given level of 
precision of the polynomial fit.  
This effect can be seen directly in the extrapolated eigenvalues in 
table~\ref{NUM_SU(2)_1Loop}.

We also note that the spectrum in table~\ref{NUM_SU(2)_1Loop} exhibits 
a degeneracy of eigenstates whose momentum labels are related by an overall
sign flip (a symmetry that is implemented on the operator basis by a parity 
operator $P$ that reverses the ordering of all fields within the trace). 
This degeneracy among ``parity pairs'' of gauge theory operators
was observed in \cite{Beisert:2003tq}, where it 
was shown that it arises as a consequence of integrability (which can, in turn, 
be used to constrain the form of the Hamiltonian at higher loop order 
\cite{Beisert:2003jb}). See \cite{Swanson:2004mk} 
for further discussion on the implications of this degeneracy.

To corroborate these results we turn to the one-loop 
Bethe ansatz for the Heisenberg spin chain. The Bethe ansatz for chains of 
spins in arbitrary representations of arbitrary simple Lie groups was developed
some time ago \cite{Ogievetsky:1986hu} (see also \cite{Saleur:1999cx} for an extension to
supersymmetric spin chains) and applied only recently to the specific
case of the dilatation operator of ${\cal N}=4$ SYM \cite{Minahan:2002ve,Beisert:2003yb}.
In the notation of \cite{Beisert:2003yb}, the Bethe equations are expressed in 
terms of the so-called Bethe roots 
(or rapidities) $u_{i}$ associated with the various impurity insertions 
in the single-trace ground state $\tr (Z^{K})$.  In a one-dimensional dynamical 
interpretation, the impurities are pseudoparticle excitations and the roots 
parameterize in some fashion the lattice momenta of the pseudoparticles. 
The index $i$ in the Bethe root $u_i$ runs over the total number $I$ of 
impurities.  A second index $q_i = 1,\ldots ,7$ is used to associate each of the
$I$ Bethe roots with a particular simple root of the $\Sl(4|4)$ symmetry
algebra associated with ${\cal N}=4$ SYM.   
The Bethe ansatz then takes the form (see \cite{Beisert:2003yb} and references therein 
for further details)
\be
\left( \frac{u_{i} + \frac{i}{2}V_{q_i}}{u_{i}-\frac{i}{2}V_{q_i} }\right)^{K}
	= \prod_{j\neq i}^{I}
	\left(
	\frac{u_{i}-u_{j}+\frac{i}{2}M_{q_i,q_j}}
	     {u_{i}-u_{j}-\frac{i}{2}M_{q_i,q_j}}
	\right)\ ,
\label{BAEFull}
\ee
where $V_{q_i}$ denotes the ${q_i}^{\rm th}$ Dynkin coefficient of the 
spin representation and $M$ is the Cartan matrix of the algebra.
To be slightly more specific, if $\alpha_{q_i}$ are the root vectors
associated with the nodes of the Dynkin diagram and $\mu$ is the
highest weight of the spin representation, then the Dynkin coefficient (for a bosonic algebra) 
is $V_{q_i} = 2\,\alpha^{(q_i)}\cdot \mu / (\alpha^{(q_i)})^2$ and the elements of
the Cartan matrix are $M_{q_i,q_j} = 2\,\alpha^{(q_i)}\cdot \alpha^{(q_j)} / (\alpha^{(q_j)})^2$
(note that diagonal elements $M_{q_i,q_i}=2$). 
(For superalgebras see, e.g.,~\cite{LSA1,LSA2}.)
Furthermore, since the spin chain systems of interest to us are cyclic and 
carry no net momentum (analogous to the level-matching condition in the string 
theory), the Bethe roots $u_{i}$ are subject to the additional constraint
\be
1 = \prod_{i}^{I}\left(
	\frac{u_{i} + \frac{i}{2}V_{q_i}}{u_{i}-\frac{i}{2}V_{q_i}}
	\right)\ .
\label{BAE2}
\ee
Having found a set of Bethe roots $u_{i}$ that solve the above
equations, the corresponding energy eigenvalue 
(up to an overall additive constant; see, e.g.,~\cite{Beisert:2003yb}) is given by
\be
E 
= \sum_{j=1}^{I} \left( \frac{V_{q_j}}{u_j^2+V_{q_j}^2/4}\right)\ .
\label{betheenergy1L}
\ee
In the current application all impurities are of the same type 
(i.e.,~carry the same Dynkin label), so the index $q_i$ can be ignored. 
It is worth noting, however, that the Dynkin coefficient $V_{q_i}$ can vanish, in which
case the associated Bethe roots do not contribute directly to the energy.

The Bethe equations are typically exactly soluble for the case of
two identical impurities (i.e.,~two Bethe roots $u_{1}$, $u_{2}$ associated
with the same simple root of the algebra). The two-impurity $\su(2)$ Bethe equations, 
for example, yield solutions
that reproduce the familiar two-impurity anomalous dimension formula
noted above in eqn.~(\ref{su2EIG}) (see \cite{Minahan:2002ve,Beisert:2003yb} 
for further examples).   For three and higher impurities, however, 
exact solutions are not known. Since we are 
ultimately interested in comparing with string theory predictions at large 
values of the $S^5$ angular momentum $J$, an alternate approach is to solve 
the Bethe equations perturbatively in small $1/{K}$. Experience
shows that, in the limit where we can neglect interactions between
excitations (or impurities), the Bethe roots are simply the inverse
of the conserved momentum carried by the impurities. 
With a little work, one can show that the Bethe ansatz conditions,
eqns.~(\ref{BAEFull}, \ref{BAE2}), can be solved order-by-order in a 
large-${K}$ expansion:
\be
u_{i} = \frac{1}{2\pi k_{i}}\left({K} + A_{i}\sqrt{{K}} + B_{i} + \cdots\right)\ ,
\label{uexp}
\ee
where $0<k_i<{K}$ is the usual integer lattice momentum. The half-integer powers 
of ${K}$ may or may not be present in eqn.~(\ref{uexp}): they are needed to deal with special kinematic 
situations (such as when a pair of impurities has the same lattice momentum) where
the integral power expansion would be singular.
The eigenvalues of the spin chain 
(or the anomalous dimensions of the corresponding gauge theory operator) 
are then obtained as a power series in $1/{K}$ by substituting the expansion
of the Bethe roots into eqn.~(\ref{betheenergy1L}).
This is the approach introduced by Minahan and Zarembo
for the $\so(6)$ spin chain in \cite{Minahan:2002ve}.
Since we wish to carry out similar calculations at higher orders in $\lambda$, 
we will review this methodology at one-loop order for the specific case of three 
identical impurities in the $\su(2)$ spin chain.  
(Since the $\su(2)$ chain is a subsector of the $\so(6)$ system studied in
\cite{Minahan:2002ve}, the three-impurity Bethe ansatz predictions derived here
are of course implied by the all-impurity $\so(6)$ anomalous dimension formula 
derived in \cite{Minahan:2002ve} at one loop.)

We now apply this to the closed $\su(2)$ sector where the Dynkin diagram has
a single node, the Cartan matrix is $M_{\su(2)} = 2$ and the Dynkin coefficient
of the fundamental representation is $V_{\su(2)} = 1$. Consequently, the Bethe equations
(\ref{BAEFull}, \ref{BAE2}) reduce to
\be
&&\kern-25pt
	\left(\frac{u_{i}+i/2}{u_{i}-i/2}\right)^{K}
	= \prod_{j\neq i}^{I}\left(
	\frac{u_{i} - u_{j} + i}{u_{i}-u_{j} - i}
	\right)\ ,
\label{BAESU2_1}
\\
&&
\kern+10pt	1 = \prod_{i}^{I}\left(
	\frac{u_{i} + {i}/{2}}{u_{i}- {i}/{2}}
	\right)\ .
\label{BAESU2_2}
\ee
With three or more pseudoparticle excitations, bound-state solutions
can arise that satisfy the second equation (\ref{BAESU2_2}).
These solutions are characterized as having
pseudoparticle states sharing the same lattice momenta 
(e.g.,~$k_i = k_j$ for the $i^{\rm th}$ and $j^{\rm th}$ roots).
The generic solutions to the Bethe equations can therefore be
loosely divided into those that do or do not contain bound 
states.  For three impurities with no bound states present
($k_1 \neq k_2 \neq k_3$), eqn.~(\ref{BAESU2_2}) states that
$k_3 = -k_1 -k_2$. The strategy of \cite{Minahan:2002ve} can then be
used to obtain a systematic expansion of $\su(2)$ Bethe roots in powers
of ${K}^{-1}$, with the result
\be
u_1 & = & \frac{{K}-4}{2\pi k_1} + \frac{3k_1}{\pi(k_1-k_2)(2k_1+k_2)} + O({K}^{-1})~,
\nn\\
u_2 & = & \frac{({K}-4)k_1^2+({K}-4)k_1k_2-2({K}-1)k_2^2}{2\pi k_2(k_1^2+k_1 k_2 -2k_2^2)} + O({K}^{-1})~,
\nn\\
u_3 & = & -\frac{({K}-1)k_1^2-(8-5{K})k_1k_2+2({K}-1)k_2^2}{2\pi (k_1+k_2)(2k_1+k_2)(k_1+2k_2)} + O({K}^{-1})  \ .
\ee
Substituting these roots into the energy formula in eqn.~(\ref{BAE2}) gives the following
expression for the anomalous dimension of the $\su(2)$ three-impurity operator
at one loop:
\be
E_{\su(2)}^{(2)}(k_1,k_2)& = &\frac{8 \pi^2}{{K}^3}\left(k_1^2 + k_1 k_2 + k_2^2\right)\left({K}+2\right)
	+ O({K}^{-4})~,
\nn\\
&&\kern+140pt 
(k_1 \neq k_2 \neq k_3)\ .
\label{BAsu2}
\ee
This is in perfect agreement with the results of eqn.~(\ref{su2num}) and the 
numerical gauge theory results in table~\ref{NUM_SU(2)_1Loop}.
When a single bound state is present the Bethe roots must be altered.  Taking,
for example, $k_1 = k_2$, the cyclic constraint in eqn.~(\ref{BAESU2_2})
sets $k_3 = -2k_1$, and the Bethe roots are
\be
u_1  & = & \frac{-7+3i\sqrt{{K}}+3{K}}{6\pi k_1}+ O({K}^{-1/2})~,
\nn\\
u_2  & = &  -\frac{7+3i\sqrt{{K}}-3{K}}{6\pi k_1}+ O({K}^{-1/2})~,
\nn\\
u_3  & = &   \frac{4-3{K}}{12\pi k_1}+ O({K}^{-1/2})\ .
\ee
In this case the anomalous dimension is predicted to be
\be
E_{\su(2)}^{(2)}(k_1) &=& \frac{8\pi^2}{{K}^3} k_1^2 (3{K}+7)+ O({K}^{-4})~, 
\qquad 
(k_1 = k_2,\ k_3=-2k_1)\ ,
\label{BAsu2NN}
\ee
which is again in agreement with the results of eqn.~(\ref{su2num})
and table~\ref{NUM_SU(2)_1Loop} (note that the fractional powers of
${K}^{-1}$ have obligingly canceled out of the final expression for
the energy).

\subsection{Two- and three-loop order}
A similar analysis can be performed on the two-loop $\su(2)$ spin chain Hamiltonian. 
As before, we use the Jordan-Wigner transformation restricted to an odd-impurity
basis of operators to rewrite the two-loop Hamiltonian (\ref{Hsu2_2}) in terms of 
position-space fermionic oscillators, obtaining a result similar to eqn.~(\ref{Hsu2PS}):
\be
H_{\su(2)}^{(4)} & = & 
	\sum_{j=1}^{K} \biggl\{
	-\frac{1}{2}\Bigl[
	b_{j+2}^\dag b_{j} 
	+b_{j}^\dag b_{j+2}
	-4 \Bigl( b_{j+1}^\dag b_{j} 
	+ b_{j}^\dag b_{j+1}\Bigr)
	\Bigr]
	-3\,  b_{j}^\dag b_{j} 	
	-4\, b_{j}^\dag b_{j+1}^\dag b_{j} b_{j+1}
\nn\\
&&	+ b_{j+1}^\dag b_{j+2}^\dag b_{j} b_{j+1}
	+ b_{j}^\dag b_{j+1}^\dag b_{j+1} b_{j+2}
	+ b_{j}^\dag b_{j+2}^\dag b_{j} b_{j+2}
	\biggr\}\ .
\ee
Passing to momentum space, we obtain the two-loop analogue of eqn.~(\ref{Hsu2MS}):
\be
H_{\su(2)}^{(4)} & = & -8 \sum_{p=0}^{{K}-1} \sin^4\left(\frac{p\pi}{{K}}\right) \bt_p^\dag \bt_p
\nn\\
&&	+\frac{1}{{K}}\sum_{p,q,r,s=0}^{{K}-1} \left(
	e^{\frac{2\pi i (q+r)}{{K}}}
	+e^{\frac{-2\pi i (p+s)}{{K}}}
	+e^{\frac{4\pi i (q-s)}{{K}}}
	-4\,e^{\frac{2\pi i (q-s)}{{K}}}
	\right)
	 \bt_p^\dag \bt_q^\dag \bt_r \bt_s\, \delta_{p+q,r+s}\ .
\nn\\
&&
\label{Hsu2MS2}
\ee
Although the two-loop Hamiltonian includes ``long-range'' interactions among non-neighboring 
lattice sites, the momentum-space Hamiltonian (\ref{Hsu2MS2}) conveniently terminates
at two-body interaction terms. An equally important point is that, for fixed momenta
$p,q,\ldots$, the one-body (two-body) operators scale as ${K}^{-4}$ (${K}^{-5}$) for large
${K}$ (the corresponding scalings for the one-loop Hamiltonian were ${K}^{-2}$ (${K}^{-3}$)).
This special relation between density scaling and power of coupling constant is critical
for matching to string theory.

We deal with the problem of finding the eigenvalues of the combined one- and two-loop 
Hamiltonian via Rayleigh-Schr\"odinger perturbation theory: at each value of the lattice 
length ${K}$ we treat the one-loop operator $H_{\su(2)}^{(2)}$ as a zeroth-order Hamiltonian 
and regard $H_{\su(2)}^{(4)}$ as a first-order perturbation. The $O(\lambda^2)$ corrections 
to the spectrum of $H_{\su(2)}^{(2)}$ are then found by taking expectation values of 
the perturbation $H_{\su(2)}^{(4)}$ in the (numerically-determined) eigenvectors of $H_{\su(2)}^{(2)}$. 
This is the recipe for non-degenerate first-order perturbation theory and we might worry
that the previously-noted parity-pair degeneracy of the eigenvalues of  $H_{\su(2)}^{(2)}$
would force us to use the rules of degenerate perturbation theory. As discussed in 
\cite{Beisert:2003tq,Callan:2004ev,Swanson:2004mk}, however, parity degeneracy can be traced to 
the existence of a higher Abelian charge that is conserved to at least three-loop 
order. This charge can be used to show that the formulas of non-degenerate perturbation 
theory can be used without modification. The basic observation is that conservation of
the Abelian charge guarantees that the matrix element of $H_{\su(2)}^{(4)}$ between two 
degenerate eigenstates of $H_{\su(2)}^{(2)}$ with different eigenvalues of the higher 
Abelian charge vanishes: this eliminates the vanishing energy-denominator singularities 
that would otherwise invalidate the non-degenerate first-order perturbation theory formulas 
(and similar arguments apply to the higher-order cases).

Using this method, we have evaluated the $O(\lambda^2)$ corrections to the spectrum
of anomalous dimensions for lattice sizes from ${K}=6$ to ${K}=40$. As before, we fit the 
spectral data to a power series in $1/{K}$ to read off the leading scaling
coefficients of the low-lying eigenvalues. As mentioned in the discussion of the
two-loop Hamiltonian (\ref{Hsu2MS2}), we expect the two-loop eigenvalues to have
the following scaling behavior in $1/{K}$:
\be
\label{scaling23}
E_{ {K}}^{(2)}(\lbrace k_i \rbrace) =  \frac{\lambda^2}{{K}^4}{E^{(2,4)}
	(\lbrace k_i \rbrace)} + 
\frac{\lambda^2}{{K}^5}{E^{(2,5)}(\lbrace k_i \rbrace)} + O(\lambda^2 {K}^{-6})\ .
\ee
The numerical data confirm that the eigenvalues scale at least as fast as ${K}^{-4}$.
The resulting numerical values for the leading scaling coefficients of low-lying eigenvalues, 
$E_{\su(2)}^{(2,4)}$ and $E_{\su(2)}^{(2,5)}$, are presented in table~\ref{NUM_SU(2)_2Loop}. 
\begin{table}[ht!]
\begin{eqnarray}
\begin{array}{|cccc|}
\hline
E_{\su(2)}^{(2,4)} & E_{\su(2)}^{(2,5)} & E_{\su(2)}^{(2,5)}/E_{\su(2)}^{(2,4)}  
                                  & (k_1,k_2,k_3) \\
\hline
-0.25-4.6\times 10^{-9}		&-2+8.0\times 10^{-7}	&8-3.4\times 10^{-6}	& (1,0,-1)	 \\
-2.25-1.4\times 10^{-6}		&-19+2.6\times 10^{-4}  &76/9+1.2\times 10^{-4} & (1,1,-2)	 \\	
-2.25-1.4\times 10^{-6}		&-19+2.6\times 10^{-4}	&76/9+1.2\times 10^{-4}	&  (-1,-1,2)	 \\
-4+8.3\times 10^{-7}		&-32-1.1\times 10^{-4}  &8+3.0\times 10^{-5}   &  (2,0,-2)	 \\
-12.25-9.9\times 10^{-6}	&-98+2.3\times 10^{-3}	&8-2.0\times 10^{-4}	&  (1,2,-3) 	 \\
-12.25-9.9\times 10^{-6}	&-98+2.3\times 10^{-3}	&8-2.0\times 10^{-4}	&  (-1,-2,3)	 \\
-20.25+3.2\times 10^{-3}	&-161.4			&7.97			&  (3,0,-3)	 \\
-36-2.8\times 10^{-3}		&-304.6			&8.46			&  (2,2,-4)	 \\
-36-2.8\times 10^{-3} 		&-304.6			&8.46			&  (-2,-2,4)	 \\
-42.25+4.9\times 10^{-3}	&-337.0			&7.97			&  (1,3,-4) 	 \\
-42.25+4.9\times 10^{-3}	&-337.0			&7.97			&  (-1,-3,4)	 \\
\hline
\end{array} \nonumber
\end{eqnarray}
\caption{Scaling limit of three-impurity
	$\su(2)$ numerical spectrum at two loops in $\lambda$}
\label{NUM_SU(2)_2Loop}
\end{table}
We thus have the following simple predictions for the two-loop large-${K}$ expansion coefficients: 
\be
E_{\su(2)}^{(2,4)} = -(k_1^2+k_2^2+k_3^2)^2/16~, & \qquad & k_1+k_2+k_3=0~,\nonumber\\
E_{\su(2)}^{(2,5)}/E_{\su(2)}^{(2,3)}  =  8~, & \qquad & (k_1\neq k_2\neq k_3)~,
\nonumber\\
E_{\su(2)}^{(2,5)}/E_{\su(2)}^{(2,3)} = \frac{76}{9}~, & \qquad &(k_1 = k_2,\ k_3=-2k_1)\ .
\label{su2num2}
\ee
Once again, the decline in precision as one goes to higher energies is expected. As a consistency
check we note that this time we have no freedom to choose the momenta $(k_1,k_2,k_3)$
associated with each state: they have been fixed in the one-loop matching exercise.


The three-loop $\su(2)$ Hamiltonian (\ref{Hsu2_3}) can be dealt with in a similar
fashion. The position space operator version of this Hamiltonian is too long 
to record here, but its momentum space version is fairly compact:
\be
H_{\su(2)}^{(6)} & = & 32 \sum_{p=0}^{{K}-1} \sin^6\left(\frac{p\pi}{{K}}\right)\bt_p^\dag \bt_p
	+\frac{1}{2{K}}\sum_{p,q,r,s=0}^{{K}-1}\biggl\{
	-10\, e^{\frac{2\pi i(q+r)}{{K}}} 
	+ e^{\frac{2\pi i(2q+r)}{{K}}}
	+ e^{\frac{2\pi i(q+2r)}{{K}}}
\nn\\
&&	+ e^{\frac{2\pi i(q-3s)}{{K}}}
	+ e^{\frac{2\pi i(2q-2r-3s)}{{K}}}
	+ e^{\frac{2\pi i(3q-2r-3s)}{{K}}}
	+ e^{\frac{2\pi i(q-r-3s)}{{K}}}
	+ e^{\frac{2\pi i(2q-r-3s)}{{K}}}
\nn\\
&&	- e^{\frac{2\pi i(q-2s)}{{K}}}
	-10\, e^{\frac{2\pi i(q-r-2s)}{{K}}}
	- e^{\frac{2\pi i(2q-r-2s)}{{K}}}
	- e^{\frac{2\pi i(3q-r-2s)}{{K}}}
	- e^{\frac{2\pi i(q+r-2s)}{{K}}}
\nn\\
&&	+29\, e^{\frac{2\pi i(q-s)}{{K}}}
	-10\, e^{\frac{4\pi i(q-s)}{{K}}}
	+ e^{\frac{6\pi i(q-s)}{{K}}}
	- e^{\frac{2\pi i(2q-s)}{{K}}}
	+ e^{\frac{2\pi i(3q-s)}{{K}}}
\nn\\
&&	- e^{\frac{2\pi i(q+r-s)}{{K}}}
	+ e^{\frac{2\pi i(2q+r-s)}{{K}}}
	+ e^{\frac{2\pi i(q+2r-s)}{{K}}}
	\biggr\}
	\bt_p^\dag \bt_q^\dag \bt_r\bt_s\,\delta_{p+q,r+s} 
\nn\\
&&	+ \frac{1}{{K}^2}\sum_{p,q,r,s,t,u=0}^{{K}-1}\biggl\{
	e^{\frac{2\pi i (q+3r-2t-3u)}{{K}}}
	+ e^{\frac{2\pi i (q+2r-s-2t-3u)}{{K}}}
\nn\\
&&\kern+30pt
	+ e^{\frac{2\pi i (2q+3r-t-3u)}{{K}}}
	+ e^{\frac{2\pi i (q+2r+s-u)}{{K}}}
	\biggr\}
	\bt_p^\dag \bt_q^\dag \bt_r^\dag
	\bt_s \bt_t \bt_u\, \delta_{p+q+r,s+t+u} \ .
\ee
It contains at most three-body operators and a careful examination of
terms shows that, for fixed momenta, the one-body operators scale as
${K}^{-6}$, the two-body operators as ${K}^{-7}$ and so on. We therefore
expect the leading scaling coefficients in the $O(\lambda^3)$ eigenvalues
to be $E_{\su(2)}^{(3,6)}$ and $E_{\su(2)}^{(3,7)}$, to use a by-now-familiar
notation. To find the eigenvalues to this order, we continue with the
Rayleigh-Schr\"odinger perturbation theory strategy: the $O(\lambda^3)$
correction to any eigenvalue is the sum of the matrix element of $H_{\su(2)}^{(6)}$
in the appropriate eigenvector of $H_{\su(2)}^{(2)}$ plus the second-order
sum-over-states contribution of $H_{\su(2)}^{(4)}$. These two pieces
can easily be computed numerically from the explicit Hamiltonian operators
at a fixed ${K}$. Parity degeneracy and conservation of the higher Abelian
charge mentioned above continue to hold, and we can again use 
non-degenerate perturbation theory formulas to compute the eigenvalue 
corrections. We have generated numerical eigenvalue data for lattices
from ${K}=6$ to ${K}=40$ and the large-${K}$ scaling coefficients of the low-lying
states extracted from those data are given in table~\ref{NUM_SU(2)_3Loop}. 
\begin{table}[ht!]
\begin{eqnarray}
\begin{array}{|cccc|}
\hline
E_{\su(2)}^{(3,4)} & E_{\su(2)}^{(3,7)} & E_{\su(2)}^{(3,7)}/E_{\su(2)}^{(3,6)}  
                                  & (k_1,k_2,k_3) \\
\hline
0.1250		&2.0003		&16.003		& (1,0,-1)	 \\
4.125		&58.03  	&14.07 		& (1,1,-2)	 \\	
4.125		&58.03		&14.07		&  (-1,-1,2)	 \\
7.999		&128.2		&16.03		&  (2,0,-2)	 \\
49.62		&713.3		&14.37		&  (1,2,-3) 	 \\
49.62		&713.3		&14.37		&  (-1,-2,3)	 \\
91.15		&1,454		&15.96		&  (3,0,-3)	 \\
263.8		&3,739		&14.17		&  (2,2,-4)	 \\
263.8 		&3,739		&14.17		&  (-2,-2,4)	 \\
\hline
\end{array} \nonumber
\end{eqnarray}
\caption{Scaling limit of three-impurity
	$\su(2)$ numerical spectrum at three loops in $\lambda$}
\label{NUM_SU(2)_3Loop}
\end{table}

A modified Bethe ansatz for the $\su(2)$ sector of the gauge theory, possibly 
incorporating all orders of higher-loop physics, has been proposed 
in \cite{Serban:2004jf,Beisert:2004hm}.\footnote{The long-range ansatz based on the 
Inozemtsev spin chain in \cite{Serban:2004jf} suffers from improper 
BMN scaling at four-loop order, a problem that is surmounted in \cite{Beisert:2004hm}. 
For further insights into the importance of BMN scaling, see \cite{Fischbacher:2004iu}. }
It is an instructive exercise and a useful consistency check on this bold proposal 
to verify that it reproduces the higher-loop scaling coefficients for three-impurity 
anomalous dimensions that we have just computed by virial methods 
(and displayed in tables ~\ref{NUM_SU(2)_1Loop},~\ref{NUM_SU(2)_2Loop} 
and \ref{NUM_SU(2)_3Loop}). For completeness, we briefly summarize the
new ansatz, referring the reader to \cite{Beisert:2004hm} for a detailed account.
In the new ansatz, the momenta $p_i$ of the excitations (closely related to the 
Bethe roots) become functions of $\lambda$ (as well as ${K}$ 
and mode numbers) and are determined by a modified version of eqns.~(\ref{BAESU2_1}, \ref{BAESU2_2}):
\be
e^{i {K} p_i} & = &\prod_{j\neq i}^I 
	\frac{\varphi(p_i)-\varphi(p_j)+i}{\varphi(p_i)-\varphi(p_j)-i}~, \qquad 
\sum_{i=1}^I p_i = 0\ .
\label{LRBA}
\ee
Dependence on $\lambda$ 
enters through the phase function $\varphi(p_i)$, which is defined in terms of the 
excitation momenta $p_i$ as follows:
\be
\varphi(p_i) & \equiv & \frac{1}{2}\cot\left({p_i}/{2}\right)
	\sqrt{1+\frac{\lambda}{\pi^2}\sin^2\left({p_i}/{2}\right)}\ .
\ee
The energy eigenvalue corresponding to a particular root of these equations is given in 
terms of the excitation momenta $p_i$ by the formula
\be
E_{\su(2)} = \sum_{i=1}^I \frac{8\pi^2}{\lambda}\left(
	\sqrt{1+\frac{\lambda}{\pi^2}\sin^2 \left(p_i/2\right)} -1\right)\ .
\ee
Finding exact solutions of these equations is even more difficult than before,
but we can follow the previous strategy of developing an expansion in powers of 
$1/{K}$ about non-interacting impurities on an infinite lattice. This is achieved 
by expanding the excitation momenta $p_i$ according to
\be
p_i = \frac{2\pi k_i}{{K}} + \sum_{n=1}\frac{p_i^{(n)}}{{K}^{\frac{n+2}{2}}}\ ,
\ee
where the integers $k_i$ (subject to the cyclicity constraint $\sum_i k_i = 0$) 
characterize the non-interacting state about which the expansion is developed.
The appearance of half-integer powers of ${K}^{-1}$ in this expansion is needed
to accommodate bound-state solutions to the Bethe equations that arise when
some of the momenta $k_i$ are equal. Solutions to the Bethe equation (\ref{LRBA}) 
will determine the expansion coefficients $p_{i}^{(n)}$ in terms of the mode 
numbers $k_i$ and ultimately lead to expansions of the energies as power 
series in ${K}^{-1}$, with coefficients that are functions of $\lambda/{K}^2$.

Explicit results for the ${K}^{-1}$ expansion of gauge theory operators of arbitrary 
impurity number, derived by the above method, were presented in 
\cite{Arutyunov:2004vx}.\footnote{It is important to note that the focus of this
chapter is a \emph{different} Bethe ansatz, designed to match the spectrum of
the string theory: the gauge theory Bethe ansatz results are derived for comparison
purposes.}
As usual, expressions are different depending on whether all momenta are unequal or some
subset of them is equal.  For all mode numbers $k_i$ unequal the $I$-impurity energy formula 
in \cite{Arutyunov:2004vx} is
\be
\label{abanearBMN}
E_{\su(2)} & = & {K}-I+\sum_{i=1}^I
	\left(\sqrt{1+\lambda'\,k_i^2}
	-\frac{\lambda'}{{K}-I} \frac{I\, k_i^2}
	{\sqrt{1+\lambda'\,k_i^2}} 
	\right)
\nn\\
&&	-\frac{\lambda'}{{K}-I}
	\sum_{\textstyle\atopfrac{i,j=1}{i\neq j}}^I \frac{2k^2_i k_j}{k^2_i-k^2_j}
	\left(
	k_j+k_i\sqrt{\frac{1+\lambda'\,k_j^2}{1+\lambda'\,k_i^2}} \right)
		+O({K}^{-2})\ ,
\label{LRBAeng}
\ee
where we have used $\lambda'=\lambda/J^2 = \lambda/({K}-I)^2$ for convenience 
($J={K}-I$ is the total $R$-charge). 
To compare with our virial results, we must further expand in
$\lambda$; expanding to first and second order yields the following scaling coefficients 
(valid for all $k_i$ unequal):
\be
E_{\su(2)}^{(1,2)} = k_1^2 + k_1k_2 + k_2^2~, & \qquad &
E_{\su(2)}^{(1,3)} = 2(k_1^2 + k_1k_2 + k_2^2)~, \nn\\
E_{\su(2)}^{(2,4)} = -\frac{1}{4}(q^2 + qr + r^2)^2~, &\qquad&
E_{\su(2)}^{(2,5)} = -2(q^2 + qr + r^2)^2\ . 
\ee
These one- and two-loop coefficients match the numerical results presented in 
tables~\ref{NUM_SU(2)_1Loop} and \ref{NUM_SU(2)_2Loop} and the analytic string
formulas in eqns.~(\ref{su2num}, \ref{su2num2}). It is harder to
write down a general formula for the many cases in which subsets of momenta are
equal but the solution for the particular case of three impurities with a two-excitation 
bound state $(k_1 = k_2 = n,\ k_3=-2n)$ was also presented in \cite{Arutyunov:2004vx}:
\be
E_{\su(2)}&=&{K}-3+2 \sqrt{1+\lambda'\,n^2}+\sqrt{1+\lambda'\,4 n^2} 
\nn\\
&&\kern-50pt	
	-\frac{\lambda'\,n^2}{{K}-3}
	\left(
	\frac{1}{1 + \lambda'\,n^2} + \frac{6}{{\sqrt{1 + \lambda'\,n^2}}} +
	\frac{12}{{\sqrt{1 + \lambda'\,4 n^2}}} -
  	\frac{8}{{\sqrt{1 + \lambda'\,n^2}}\,{\sqrt{1 + \lambda'\,4 n^2}}}
	\right)\ .
\nn\\
&&
\label{LRBAnn}
\ee
To compare with the virial results, one must again expand the energy in powers of 
$\lambda$. Doing so yields the following one- and two-loop bound-state scaling coefficients:
\be
E_{\su(2)}^{(1,2)} = 3\,n^2~, & \qquad &
E_{\su(2)}^{(1,3)} = 7\,n^2~, \nn\\
E_{\su(2)}^{(2,4)} = -\frac{9}{4}\,n^4~, &\qquad&
E_{\su(2)}^{(2,5)} = -19\,n^4\ .
\ee 
We easily verify that this agrees with numerical virial results to two-loop order.

The three-loop coefficients obtained by expanding the energy formulas
in eqns. (\ref{LRBAeng}, \ref{LRBAnn}) are given by
\be
E_{\su(2)}^{(3,6)} & = & \frac{1}{16}\Bigl(
	2\,{{k_1}}^6 + 6\,{{k_1}}^5\,{k_2} + 15\,{{k_1}}^4\,{{k_2}}^2 + 
    20\,{{k_1}}^3\,{{k_2}}^3 
\nn\\
&&	+ 15\,{{k_1}}^2\,{{k_2}}^4 + 6\,{k_1}\,{{k_2}}^5 + 
    	2\,{{k_2}}^6\Bigr)~,
\nn\\
E_{\su(2)}^{(3,7)} & = &
	\frac{1}{4}\Bigl(
	8\,{{k_1}}^6 + 24\,{{k_1}}^5\,{k_2} + 51\,{{k_1}}^4\,{{k_2}}^2 + 
    62\,{{k_1}}^3\,{{k_2}}^3 
\nn\\
&&	+ 51\,{{k_1}}^2\,{{k_2}}^4 + 24\,{k_1}\,{{k_2}}^5 + 
    	8\,{{k_2}}^6\Bigr)\ ,
\ee
for $(k_1\neq k_2\neq k_3)$, and 
\be
E_{\su(2)}^{(3,6)} = \frac{33}{8}\,n^6~, \qquad E_{\su(2)}^{(3,7)} = 58\,n^6\ ,
\ee
for the bound-state solution with $(k_1=k_2=n,\ k_3=-2n)$. The numerical values 
of these $O(\lambda^3)$ coefficients are tabulated for several low-lying states 
in the spectrum in table~\ref{BA3}. The correspondence with table~\ref{NUM_SU(2)_3Loop}, 
which displays the three-loop expansion coefficients extracted from numerical
diagonalization of the three-loop Hamiltonian, is good.  At this order in the loop 
expansion higher-order $1/{K}$ corrections to the spectrum are more important 
(compared to the one- and two-loop cases), and the numerical extrapolation is
less reliable (especially as the lattice momenta increase). The precision can always 
be improved by including data from larger lattices in the extrapolation.  
We emphasize that this discussion concerns the different methods of calculation 
of operator dimensions in the $\su(2)$ sector only. It seems to
us to give useful further evidence that the long-range Bethe ansatz for the $\su(2)$
sector of the gauge theory \cite{Beisert:2004hm} is exact. 
\begin{table}[ht!]
\begin{eqnarray}
\begin{array}{|cccc|}
\hline
E_{\su(2)}^{(3,6)} & E_{\su(2)}^{(3,7)} &E_{\su(2)}^{(3,7)}/E_{\su(2)}^{(3,6)}
					&  (k_1,k_2,k_3) \\
\hline
0.125	& 2	& 16 		& (1,0,-1)	 \\
4.125	& 58	& 14.06		& (1,1,-2)	\\
4.125	& 58	& 14.06		& (-1,-1,2)	\\
8	& 128 	& 16		& (2,0,-2)	 \\
49.625  & 713   & 14.37  	& (1,2,-3)	 \\
49.625  & 713   & 14.37  	& (-1,-2,3)	 \\
91.125  & 1,458 & 16		& (3,0,-3)	\\
264	& 3,712	& 14.06		& (2,2,-4)	\\
264	& 3,712	& 14.06		& (-2,-2,4)	\\
\hline
\end{array} \nonumber
\end{eqnarray}
\caption{Three-impurity $\su(2)$ spectrum from the long-range 
	Bethe ansatz at three loops}
\label{BA3}
\end{table}

\section{A closed $\protect\su(1|1)$ subsector of $\protect\su(2|3)$ }
\label{virialSEC3}
The three-impurity string theory analysis of \cite{Callan:2004ev} identified 
a fermionic sector of the theory that is diagonalized by string states composed 
of fermionic excitations projected onto particular four-dimensional subspaces 
(which transform in an $SU(2)^2\times SU(2)^2$ notation 
as a $({\bf 2,1;2,1})$ or $({\bf 1,2;1,2})$ of $SO(4)\times SO(4)$) 
and symmetrized in their $SO(4)\times SO(4)$ indices. 
It was also shown that this three-impurity 
subsector of the theory decouples at all orders in $\lambda$.  

On the gauge theory side this subsector corresponds to an $\su(1|1)$ 
subgroup of the closed $\su(2|3)$ sector studied by Beisert in 
\cite{Beisert:2003ys,Beisert:2003jj}.
(Supersymmetric integrable $\su(n|m)$ spin chains have previously been studied in 
certain condensed-matter applications; see, e.g.,~\cite{liuwang}.)
In the present setting the fields of $\su(2|3)$ consist of three complex scalars $\phi_a$ and 
two complex fermions $\psi_\alpha$.  In the closed $\su(1|1)$ subspace we are restricted
to a single scalar denoted by $Z$ and a single fermion labeled by $\psi$.
Just as in the $\su(2)$ sector, we use the fermionic position-space oscillators
$b_j^\dag,\ b_j$ to create or annihilate fermionic $\psi$ insertions in a ground
state composed of ${K}$ scalars:
\be
\Ket{{K}} = \tr(Z^{K})~, \qquad 
b_j^\dag \ket{{K}} = \tr( Z_1\cdots Z_{j-1} \psi Z_{j+1} \cdots Z_{K} )\ .
\label{oscsu11}
\ee

In \cite{Beisert:2003ys}, Beisert gave the action of the Hamiltonian on the $\su(2|3)$ 
spin chain to three-loop order.\footnote{Beisert's three-loop Hamiltonian was restricted
in \cite{Beisert:2003ys} to the bosonic sector, but the author has since provided us with 
the complete version.}  In the notation of \cite{Beisert:2003ys}, the action of the Hamiltonian on basis 
states can be represented in terms of special permutation operators denoted by
\be
\genfrac{\{\}}{}{0pt}{0}{A_1 \ldots A_N }{ B_1 \ldots B_N }\ ,
\nn
\ee
which replace all occurrences of the upper sequence of fields 
$A_1\ldots A_N$ in the trace by the lower sequence $B_1\ldots B_N$.  
Restricting Beisert's $\su(2|3)$ Hamiltonian to the $\su(1|1)$ subsector 
at one-loop order yields
\be
H_{\su(1|1)}^{(2)} =  
	\genfrac{\{\}}{}{0pt}{0}{Z\psi }{ Z\psi  } 
	+  \genfrac{\{\}}{}{0pt}{0}{\psi Z }{ \psi Z}
	-  \genfrac{\{\}}{}{0pt}{0}{Z\psi }{ \psi Z} 
	-  \genfrac{\{\}}{}{0pt}{0}{\psi Z }{ Z\psi}
	+ 2  \genfrac{\{\}}{}{0pt}{0}{\psi\psi }{ \psi\psi}\ .
\label{BsrtH2}
\ee
In terms of the position-space oscillators of eqn.~(\ref{oscsu11}), the $\su(1|1)$
Hamiltonian can be assembled by inspection and takes the form
\be
H_{\su(1|1)}^{(2)} = \sum_{j=1}^{K}\left(
	b_j^\dag b_j + b_{j+1}^\dag b_{j+1} - b_{j+1}^\dag b_j - b_j^\dag b_{j+1}
	\right)\ .
\ee
There are no higher-body interaction terms at
this order in $\lambda$.  This fact can be checked by computing 
\be
\braket{ {K}| b_{i+1}b_i (H_{\su(1|1)}^{(2)}) b_i^\dag b_{i+1}^\dag | {K}} = 2\ ,
\ee
which reproduces the two-body matrix element given by the last term
in eqn.~(\ref{BsrtH2}).  In momentum space we obtain
\be
H_{\su(1|1)}^{(2)} = 4\sum_{p=0}^{{K}-1}\sin^2\left(\frac{p\pi}{{K}}\right)
	\bt_p^\dag \bt_p\ .
\label{su11oneloop3}
\ee

The two-loop $\su(1|1)$ momentum-space Hamiltonian can be extracted in the same manner
(the position-space version is too long to print here):
\be
H_{\su(1|1)}^{(4)} & = & -8\sum_{p=0}^{{K}-1}\sin^4\left(\frac{p\pi}{{K}}\right)
	\bt_p^\dag \bt_p
	+\frac{1}{4{K}}\sum_{p,q,r,s=0}^{{K}-1}\biggl\{
	e^{\frac{2\pi i (q-2r)}{{K}}}
	+ e^{\frac{2\pi i (2q-r)}{{K}}}
	-4\, e^{\frac{2\pi i (q-r)}{{K}}}
\nn\\
&&\kern-20pt
	-2\, e^{\frac{2\pi i (q-2r-s)}{{K}}}
	-2\, e^{\frac{2\pi i (q+s)}{{K}}}
	+ e^{\frac{2\pi i (q-r+s)}{{K}}}
	+ e^{\frac{2\pi i (2q-2r-s)}{{K}}}
	\biggr\} \bt_p^\dag \bt_q^\dag \bt_r \bt_s\, \delta_{p+q,r+s}\ .
\ee
Finally, the complete three-loop Hamiltonian for this subsector is
\be
H_{\su(1|1)}^{(6)} & = & 32\sum_{p=0}^{{K}-1}\sin^6\left(\frac{p\pi}{{K}}\right)
	\bt_p^\dag \bt_p
	-\frac{1}{16}\sum_{p,q,r,s=0}^{{K}-1}e^{\frac{60\pi i(q-r)}{{K}}}\biggl\{
	2\, e^{-\frac{2\pi i (27q-29r)}{{K}}}
\nn\\
&&	+2\, e^{-\frac{2\pi i (28q-29r)}{{K}}}
	-4\, e^{-\frac{2\pi i (27q-28r)}{{K}}}
	+37\, e^{-\frac{2\pi i (29q-28r)}{{K}}}
	-6\, e^{-\frac{2\pi i (29q-27r)}{{K}}}
\nn\\
&&	+8\, e^{-\frac{56\pi i (q-r)}{{K}}}
	-72\, e^{-\frac{58\pi i (q-r)}{{K}}}
	-6\, e^{-\frac{2\pi i (29q-29r-2s)}{{K}}}
	-40\, e^{-\frac{2\pi i (29q-30r-s)}{{K}}}
\nn\\
&&	+37\, e^{-\frac{2\pi i (29q-29r-s)}{{K}}}
	-8\, e^{-\frac{2\pi i (29q-28r-s)}{{K}}}
	+8\, e^{-\frac{2\pi i (27q-28r+s)}{{K}}}
	+2\, e^{-\frac{2\pi i (28q-28r+s)}{{K}}}
\nn\\
&&	-40\, e^{-\frac{2\pi i (29q-28r+s)}{{K}}}
	-4\, e^{-\frac{2\pi i (27q-27r+s)}{{K}}}
	+8\, e^{-\frac{2\pi i (29q-27r+s)}{{K}}}
	+2\, e^{-\frac{2\pi i (27q-27r+2s)}{{K}}}
\nn\\
&&	+8\, e^{-\frac{2\pi i (29q-30r-2s)}{{K}}}
	\biggr\}\bt_p^\dag \bt_q^\dag \bt_r \bt_s\, \delta_{p+q,r+s}
	+\frac{1}{16}\sum_{p,q,r,s,t,u=0}^{{K}-1}\biggl\{
	2\, e^{\frac{2\pi i (q+2r-3s-2t )}{{K}}}
\nn\\
&&	- e^{\frac{2\pi i (q+3r-3s-2t )}{{K}}}
	-4\, e^{\frac{2\pi i (q+2r-3s-t )}{{K}}}
	- e^{\frac{2\pi i (2q+3r-3s-t )}{{K}}}
	+8\, e^{\frac{2\pi i (q+2r-2s-t )}{{K}}}
\nn\\
&&	+2\, e^{\frac{2\pi i (2q+3r-2s-t )}{{K}}}
	-4\, e^{\frac{2\pi i (q+2r-3s-2t-u )}{{K}}}
	+2\, e^{\frac{2\pi i (q+3r-3s-2t-u )}{{K}}}
	+2\, e^{\frac{2\pi i (q+2r-2s+u )}{{K}}}
\nn\\
&&\kern+00pt
	-4\, e^{\frac{2\pi i (q+2r-s+u )}{{K}}}
	-4\, e^{\frac{2\pi i (q+2r-2s-t+u )}{{K}}}
	\biggr\} \bt_p^\dag \bt_q^\dag \bt_r^\dag \bt_s \bt_t \bt_u\,
	\delta_{p+q+r,s+t+u}\ .
\ee
We note that $H_{\su(1|1)}^{(2)}$, $H_{\su(1|1)}^{(4)}$ and $H_{\su(1|1)}^{(6)}$
terminate at one-body, two-body and three-body interactions, respectively.
This will permit us to obtain the exact ${K}$-dependence of successive 
terms in the $\lambda$ expansion of energy eigenvalues.

As in the $\su(2)$ sector, we can use non-degenerate perturbation theory 
to extract the ${K}^{-1}$ scaling coefficients of the $\su(1|1)$ eigenvalue spectrum 
up to three loops in $\lambda$. The scaling coefficients extrapolated from
numerical diagonalization of lattices up to ${K}=40$ are recorded  
for one-loop, two-loop and three-loop orders in tables~\ref{NUM_fermi_1Loop3}, 
\ref{NUM_fermi_2Loop3} and \ref{NUM_fermi_3Loop3}, respectively. The same increase 
in leading power of ${K}^{-1}$ with corresponding order in $\lambda$ 
that was noted in the $\su(2)$ sector is found here as well 
(we use the same notation for the scaling coefficients as before in order to keep 
track of these powers). It should also be noted 
that, because the impurities in this sector are fermions symmetrized on all
group indices, the lattice momenta of all pseudoparticles must be different.
These results amount to the following
predictions for the one-loop and two-loop scaling coefficients:
\be
E_{\su(1|1)}^{(1,2)} = (k_1^2+k_1k_2+k_2^2)~, &\qquad& E_{\su(1|1)}^{(1,3)} = 0~, \nn\\
E_{\su(1|1)}^{(2,4)} -\frac{1}{4}(k_1^2+k_1k_2+k_2^2)^2~,  &\qquad&  
E_{\su(1|1)}^{(2,5)} = -(k_1^2+k_1k_2+k_2^2)^2 \ .
\label{su11ext}
\ee
We again have the usual caveat that data
on larger and larger lattices are required to maintain a fixed precision as
one goes to higher and higher energy levels.
\begin{table}[ht!]
\begin{eqnarray}
\begin{array}{|cccc|}
\hline
E_{\su(1|1)}^{(1,2)} & E_{\su(1|1)}^{(1,3)} & E_{\su(1|1)}^{(1,3)}/E_{\su(1|1)}^{(1,2)}  
		& (k_1,k_2,k_3)  \\
\hline
1+1.3 \times 10^{-10} 	&	-1.9\times 10^{-8}	& -1.9\times 10^{-8} 	&(1,0,-1) \\
4-1.0\times 10^{-7} 	&	1.8\times 10^{-5}	& 4.6\times 10^{-6}	&	(2,0,-2)  \\
7-2.5\times 10^{-7}  	&	4.4\times 10^{-5}	  & 6.3\times 10^{-6} 	&	(1,2,-3)  \\
7-2.5\times 10^{-7}  	&	4.4\times 10^{-5} 	  & 6.3\times 10^{-6}	&	(-1,-2,3)  \\
9-3.9\times 10^{-7} 	&	7.9\times 10^{-5}	   & 8.7\times 10^{-6}	&	(3,0,-3)  \\
13-4.0\times 10^{-6}  	&	8.2\times 10^{-4}	  & 6.3 \times 10^{-5}	&	(1,3,-4)    \\
13-4.0\times 10^{-6}  	&	8.2\times 10^{-4}	   & 6.3 \times 10^{-5} &	(-1,-3,4)    \\
16-2.0\times 10^{-5}  	&	4.1\times 10^{-3}	  & 2.6 \times 10^{-4}	&	(4,0,-4)      \\
19-3.5\times 10^{-5}  	&	7.3\times 10^{-3}	  & 3.8 \times 10^{-4}	&	(2,3,-5)    \\
19-3.5\times 10^{-5}  	&	7.3\times 10^{-3}	  & 3.8 \times 10^{-4}	&	(-2,-3,5)    \\
\hline
\end{array} \nonumber
\end{eqnarray}
\caption{Scaling limit of one-loop numerical spectrum of three-impurity 
	$\su(1|1)$ subsector}
\label{NUM_fermi_1Loop3}
\end{table}
\begin{table}[ht!]
\begin{eqnarray}
\begin{array}{|cccc|}
\hline
E_{\su(1|1)}^{(2,4)} & E_{\su(1|1)}^{(2,5)} & E_{\su(1|1)}^{(2,5)}/E_{\su(1|1)}^{(2,4)}  
		& (k_1,k_2,k_3)   \\
\hline
-0.25	  	  &-0.99999 &	  	3.99995  & (1,0,-1)   \\
-4.00006	  &-15.990 & 		3.998  &(2,0,-2)    \\	
-12.251		  &-48.899 &	  	3.992  &(1,2,-3)    \\
-12.251		  &-48.899 &	  	3.992  &(-1,-2,3)    \\
-20.25		  &-80.89 &	  	3.995   & (3,0,-3)   \\
-42.25		  &-168.2 &	  	3.98  &(1,3,-4)    \\
-42.25		  &-168.2 &	  	3.98  &(-1,-3,4)    \\
-64.00		  &-254.6 & 	  	3.98  &(4,0,-4)    \\
-90.26		  &-359.3 &	  	3.98   &(2,3,-5)    \\
-90.26		  &-359.8 &  		3.99  &(-2,-3,5)   \\
\hline
\end{array} \nonumber
\end{eqnarray}
\caption{Scaling limit of two-loop numerical spectrum of three-impurity 
	$\su(1|1)$ subsector }
\label{NUM_fermi_2Loop3}
\end{table}

The scaling limit of the three-loop ratio $E_{\su(1|1)}^{(3,7)}/E_{\su(1|1)}^{(3,6)}$
is recorded for the first few low-lying states in the spectrum in 
table~\ref{NUM_fermi_3Loop3}. 
\begin{table}[ht!]
\begin{eqnarray}
\begin{array}{|cc|}
\hline
 E_{\su(1|1)}^{(3,7)}/E_{\su(1|1)}^{(3,6)}  
		& (k_1,k_2,k_3)   \\
\hline
  		-86.41  & (1,0,-1) \\
		-85.71  &(2,0,-2) \\	
	  	-83.74  &(1,2,-3) \\
	  	-83.74  &(-1,-2,3) \\
	  	-101.9   & (3,0,-3) \\
	  	-96.01  &(1,3,-4) \\
	  	-96.01  &(-1,-3,4) \\
	  	-158.1 &(4,0,-4) \\
\hline
\end{array} \nonumber
\end{eqnarray}
\caption{Scaling limit of three-loop numerical spectrum of three-impurity 
	$\su(1|1)$ fermionic subsector }
\label{NUM_fermi_3Loop3}
\end{table}

The extrapolated gauge theory results in eqn.~(\ref{su11ext}) for the one-loop 
coefficients $E_{\su(1|1)}^{(1,3)}$ and $E_{\su(1|1)}^{(1,2)}$ should be checked 
against the predictions of the general one-loop Bethe ansatz \cite{Minahan:2002ve,Beisert:2003yb}
applied to the $\su(1|1)$ sector (as far as we know, no higher-loop Bethe
ansatz is available here). To apply the general Bethe ansatz equation of 
eqn.~(\ref{BAEFull}), 
we note that the $\su(1|1)$ Dynkin diagram is just a single fermionic node: 
the Cartan matrix is empty and the single Dynkin label is $V_{\su(1|1)}=1$ \cite{LSA1,LSA2}.
We therefore obtain the simple one-loop Bethe equation
\be
\left( \frac{u_{i} + \frac{i}{2} }{u_{i}-\frac{i}{2}}\right)^{K} = 1\ .
\label{BAEFull2}
\ee
Rather remarkably, eqn.~(\ref{BAEFull2}) 
can be solved exactly for arbitrary impurity number! The general $\su(1|1)$ 
Bethe roots are
\be
u_{i} = \frac{1}{2}\cot \left(\frac{k_i \pi}{{K}}\right)~,
\ee
and the energy eigenvalues computed from eqn.~(\ref{betheenergy1L}) are
\be
E_{\su(1|1)} = 4\sum_{i=1}^I\sin^2\left(\frac{\pi k_i}{{K}}\right)\ ,
\label{BAsu11}
\ee
with the usual condition $\sum k_i = 0 \mod {K}$ from eqn.~(\ref{BAE2}).  
This is just the sum of free lattice Laplacian energies and clearly matches 
the energies one would obtain from the one-loop $\su(1|1)$ Hamiltonian 
of eqn.~(\ref{su11oneloop3}) (since the latter has no interaction terms). 
No expansion in $1/{K}$ was necessary in this argument, but it is straightforward to
expand the energies in $1/{K}$ and verify the numerical results obtained 
in table~\ref{NUM_fermi_1Loop3} and eqn.~(\ref{su11ext}).

\section{The $\protect\Sl(2)$ sector}
\label{virialSEC4}
As noted in \cite{Callan:2004ev}, integrable $\Sl(2)$ spin chains have previously
been the subject of several studies involving, among other interesting problems, 
high-energy scattering amplitudes in non-supersymmetric QCD 
(see, e.g.,~\cite{Belitsky:2003ys} and references therein).
The $\Sl(2)$ closed sector of ${\cal N}=4$ SYM was studied in \cite{Beisert:2003jj}, and
the spin chain Hamiltonian in this sector is presently known to one loop in $\lambda$.
(For more recent progress, see ref.~\cite{Staudacher:2004tk}.)

The constituent fields in this sector are $SO(6)$ bosons, $Z$, carrying a single unit
of $R$-charge ($Z = \phi_5 + i\phi_6$ or, in the language of Chapter~\ref{SYM}, 
$\phi^{\,\Yboxdim5pt\tiny\young(1,2)}$),
and each lattice site on the $\Sl(2)$ spin chain is occupied 
by a single $Z$ field acted on by any number of the spacetime covariant derivatives
$\nabla \equiv \nabla_1+i\nabla_2$.  
The total $R$-charge of a particular operator is therefore equal to the 
lattice length ${K}$, and an $I$-impurity operator basis is spanned by single-trace
operators carrying all possible distributions of $I$ derivatives among the
${K}$ lattice sites:
\be
{\rm Tr}\left(\nabla^I Z~Z^{{{K}}	-1} \right)\ ,&\quad&  
{\rm Tr}\left(\nabla^{I-1}Z~\nabla Z~Z^{{{K}}	-2} \right)\ , \nn\\
	~{\rm Tr}\left(\nabla^{I-1}Z~Z\nabla Z~Z^{{{K}}	-3} \right)\ ,&\quad& \ldots~.
\ee
The integer $I$ counts the total number of derivatives in the operator and, since any 
number of impurities can occupy the same lattice site, one can think of $n$ derivative 
insertions at the $i^{\rm th}$ lattice site as $n$ bosonic oscillator excitations at 
the $i^{\rm th}$ lattice position:
\be
(a_i^\dag)^n \ket{{K}} \sim {\rm Tr} \left( Z^{i-1} \nabla^n Z  Z^{{{K}}-i} \right)\ ,
\ \ldots~.
\ee
The ground state $\ket{{K}}$ is represented by a length ${K}$ chain with 
no derivative insertions: $\ket{{K}} = {\rm Tr} \left(Z^{{K}} \right)$.

The one-loop $\Sl(2)$ spin chain Hamiltonian (corresponding to the dilatation 
operator in this sector) was constructed in \cite{Beisert:2003jj} and was defined 
by its action on basis states rather than directly expressed as an operator:
\be
H_{\Sl(2)}^{(2)}  &=& 
	\sum_{j=1}^{{K}} H^{\Sl(2)}_{j,j+1}\ , 
\nn\\
H^{\Sl(2)}_{1,2} (a_1^\dag)^j (a_2^\dag)^{n-j}\ket{{K}}  &=&  
	\sum_{j'=0}^{n}\left[
	\delta_{j=j'}\left(h(j)+h(n-j)\right)
	-\frac{\delta_{j\neq j'}}{|j-j'|}\right]
	(a_1^\dag)^{j'}(a_2^\dag)^{n-j'}\ket{{{K}}}
\nn\\
&&
\label{sl2int3}
\ee
(where $h(n)=1+\ldots+1/n$ are the harmonic numbers).
In other words, $H_{\Sl(2)}^{(2)}$ is a sum over the position-space Hamiltonian
$H_{j,j+1}^{\Sl(2)}$, which acts on the $j^{\rm th}$ and $(j+1)^{\rm th}$ (neighboring)
lattice sites; the action of $H_{j,j+1}^{\Sl(2)}$ can be summarized by the
explicit form given for $H_{1,2}^{\Sl(2)}$ above.  Since it
is only defined by its action on the state $(a_1^\dag)^j (a_2^\dag)^{n-j}\ket{{K}}$,
it is difficult to immediately translate $H_{\Sl(2)}^{(2)}$ to momentum space.
However, it is possible to expand it in powers of fields and use eqn.~(\ref{sl2int3})
to iteratively determine the expansion coefficients. The virial argument 
furthermore tells us that higher powers in the fields will determine 
higher powers of ${K}^{-1}$ in the expansion of the energy. For our current
purposes, it suffices to know the Hamiltonian expanded out to terms of
fourth order in the fields and this truncation of the 
Hamiltonian can easily be constructed by inspection:
\be
H_{\Sl(2)}^{(2)} & = & -\sum_{j=1}^{K} \biggl[
	\Bigl(a_{j+1}^\dag-2a_j^\dag+a_{j-1}^\dag\Bigr)
	\Bigl(a_j - \frac{1}{2}a_j^\dag a_j^2 \Bigr)
\nn\\
&&	+\frac{1}{4}\Bigl({a_{j+1}^{\dag\ 2}}-2{a_{j}^{\dag\ 2}}+{a_{j-1}^{\dag\ 2}}\Bigr)
	a_j^2 \biggr]\ + \cdots~.
\ee
Transformation to momentum space gives
\begin{eqnarray}
&&\kern-25pt	H_{\Sl(2)}^{(2)}  = 
	~{\sum_{p=0}^{{K}-1}}4\sin^{2}\frac{p \pi}{{K}}~\at_{p}^{\dag}\at_{p} \nn\\
&&\kern+10pt 	+ \frac{1}{{K}} {\sum_{p,q,r,s=0}^{{K}-1}}\delta_{p+q,r+s}
		\left(-\sin^{2}\frac{p\pi}{{K}}
		-\sin^{2}\frac{q\pi}{{K}}+\sin^{2}\frac{(p+q)\pi}{{K}}\right)  
		\at_{p}^{\dag}\at_{q}^{\dag }\at_{r}\at_{s}~
		+\cdots~.
\nn\\
&&
\label{hamilsltwofin3}
\end{eqnarray}
This Hamiltonian acts on an $I$-impurity Fock space spanned by the generic states
\be
\at_{k_1}^\dag \at_{k_2}^\dag \at_{k_3}^\dag \cdots \ket{{K}}\ ,
\ee
with lattice momenta labeled by 
$k_i = 0,\dots,{K}-1$, and subject to the constraint $\sum_{i} k_i = 0 \mod {K}$.  
Numerically diagonalizing this Hamiltonian on a range of lattice sizes, we obtain
data from which we extract the numerical predictions for the one-loop coefficients 
$E_{\Sl(2)}^{(1,2)}$ and $E_{\Sl(2)}^{(1,3)}$ presented in table~\ref{NUM_SL(2)_1Loop3}.
We arrive at the following predictions for the scaling coefficients 
\be
\label{SL2pred_ch3}
E_{\Sl(2)}^{(1,2)} = (k_1^2 +k_1 k_2+ k_2^2 )~, \qquad
E_{\Sl(2)}^{(1,3)}/E_{\Sl(2)}^{(1,2)} = -2~, \qquad k_1\neq k_2\neq k_3~,\nonumber\\
E_{\Sl(2)}^{(1,2)} = 3 n^2~, \qquad 
E_{\Sl(2)}^{(1,3)}/E_{\Sl(2)}^{(1,2)} = -7/3~, \qquad k_1=k_2=n, k_3=-2n\ ,
\ee
and we can easily verify that the agreement with table~\ref{NUM_SL(2)_1Loop3}
is excellent.
\begin{table}[ht!]
\begin{eqnarray}
\begin{array}{|cccc|}
\hline
E_{\Sl(2)}^{(1,2)} & E_{\Sl(2)}^{(1,3)} & E_{\Sl(2)}^{(1,3)}/E_{\Sl(2)}^{(1,2)}  
                                  & (k_1,k_2,k_3) \\
\hline
1+1.2\times 10^{-9}	&-2-3.1\times 10^{-7}	&-2-3.1\times 10^{-7}	& (1,0,-1)	 \\
3-7.6\times 10^{-9}	&-7+1.9\times 10^{-6}   &-7/3+6.3\times 10^{-7} & (1,1,-2)	 \\	
3-7.6\times 10^{-9}	&-7+1.9\times 10^{-6}	&-7/3+6.3\times 10^{-7}	&  (-1,-1,2)	 \\
4-2.8\times 10^{-7}	&-8+6.9\times 10^{-6}   &-2+1.7\times 10^{-6}   &  (2,0,-2)	 \\
7-2.9\times 10^{-7}	&-14+7.1\times 10^{-5}	&-2+1.0\times 10^{-5}	&  (1,2,-3) 	 \\
7-2.9\times 10^{-7}	&-14+7.1\times 10^{-5}	&-2+1.0\times 10^{-5}	&  (-1,-2,3)	 \\
9-4.1\times 10^{-7}	&-18+1.0\times 10^{-4}	&-2+1.0\times 10^{-5}	&  (3,0,-3)	 \\
12+8.4\times 10^{-7}	&-28-1.5\times 10^{-4}	&-7/3-1.2\times 10^{-5}	&  (2,2,-4)	 \\
12+8.4\times 10^{-7} 	&-28-1.5\times 10^{-4}	&-7/3-1.2\times 10^{-5}	&  (-2,-2,4)	 \\
13-7.0\times 10^{-6}	&-26+1.7\times 10^{-3}	&-2+1.3\times 10^{-4}	&  (1,3,-4) 	 \\
13-7.0\times 10^{-6}	&-26+1.7\times 10^{-3}	&-2+1.3\times 10^{-4}	&  (-1,-3,4)	 \\
16-1.4\times 10^{-6}	&-32+3.9\times 10^{-4}	&-2+2.4\times 10^{-5}	&  (4,0,-4)	 \\
19-7.5\times 10^{-6}	&-38+2.2\times 10^{-3}	&-2+1.1\times 10^{-4}	&  (2,3,-5)	 \\
19-7.5\times 10^{-6}	&-38+2.2\times 10^{-3}	&-2+1.1\times 10^{-4}	&  (-2,-3,5)	 \\
21-3.4\times 10^{-6}	&-42+8.8\times 10^{-4}	&-2+4.2\times 10^{-5}	&  (1,4,-5)	 \\
21-3.4\times 10^{-6}	&-42+8.8\times 10^{-4}	&-2+4.2\times 10^{-5}	&  (-1,-4,5)	 \\
\hline
\end{array} \nonumber
\end{eqnarray}
\caption{Scaling limit of numerical spectrum of three-impurity
	$\Sl(2)$ sector at one loop}
\label{NUM_SL(2)_1Loop3}
\end{table}

The extrapolated predictions can again be checked against those of 
the corresponding one-loop Bethe ansatz equations.
In the $\Sl(2)$ sector the highest weight is $-1/2$: the
Dynkin diagram therefore has coefficient $V_{\Sl(2)} = -1$ and the Cartan
matrix is $M_{\Sl(2)}=2$.  The Bethe equations (\ref{BAEFull}, \ref{BAE2}) 
thus reduce to
\be
&&\kern-25pt
	\left(\frac{u_{i}-i/2}{u_{i}+i/2}\right)^{K}
	= \prod_{j\neq i}^{n}\left(
	\frac{u_{i} - u_{j} + i}{u_{i}-u_{j} - i}
	\right)~,
\label{BAESL2_1}
\\
&&
\kern+10pt	1 = \prod_{i}^{n}\left(
	\frac{u_{i} - {i}/{2}}{u_{i}+ {i}/{2}}
	\right)\ .
\label{BAESL2_2}
\ee
Apart from a crucial minus sign, this is identical to the $\su(2)$ Bethe equation (\ref{BAESU2_2}).
In the absence of bound states, eqn.~(\ref{BAESL2_1}) is satisfied by the following Bethe
roots:
\be
u_1 & = & -\frac{ 2(1+{K})k_1^2 - (4+{K})k_1 k_2 - (4+{K})k_2^2}{2 \pi k_1(k_2^2+k_1 k_2-2k_1^2)}+O({K}^{-1})~,
\nn\\
u_2 & = & -\frac{ 2(1+{K})k_2^2 - (4+{K})k_1 k_2 - (4+{K})k_1^2}{2 \pi k_2(k_1^2+k_1 k_2-2k_2^2)}+O({K}^{-1})~,
\nn\\
u_3 & = & -\frac{2(1+{K})k_1^2 + (8+5{K})k_1 k_2 + 2(1+{K})k_2^2}{2\pi (k_1+k_2)(2k_1+k_2)(k_1+2k_2)}
	+O({K}^{-1})\ .
\ee
Using eqn.~(\ref{betheenergy1L}), we obtain
\be
E_{\Sl(2)}^{(2)}(k_1,k_2) &=& \frac{\lambda}{{K}^3}\left(k_1^2 + k_1 k_2 + k_2^2\right)\left({K}-2\right)
	+ O({K}^{-4})~,
\nn\\
&&\kern+120pt (k_1 \neq k_2 \neq k_3)\ .
\ee
For the bound state characterized by $k_1=k_2=n$ and $k_3=-2n$, 
the Bethe roots are
\be
u_1  &=&  \frac{7-3\sqrt{{K}}+3{K}}{6\pi n}+O({K}^{-1/2})~, \nn\\
u_2 &=& \frac{7+3\sqrt{{K}}+3{K}}{6\pi n}+O({K}^{-1/2})~,  \nn\\
u_3 &=& -\frac{4+3{K}}{12\pi n}+O({K}^{-1/2})\ ,
\ee
with spin chain energy
\be
E_{\Sl(2)}^{(2)}(n) = \frac{\lambda n^2}{{K}^3}(3{K}-7)
	+ O({K}^{-4})~,  \qquad (k_1 = k_2 = n,\ k_3=-2n)\ .
\ee
These results again agree with the numerical results in table~\ref{NUM_SL(2)_1Loop3}.

\section{Discussion}
In this chapter we have demonstrated that the virial expansion of the ${\cal N}=4$ SYM 
spin chain Hamiltonian for small impurity number provides a simple and reliable
method for computing exact anomalous dimensions of multi-impurity operators 
at small scalar $R$-charge (chain length) and estimating with great precision the 
near-BMN scaling behavior of these dimensions as the $R$-charge becomes large.  
The latter application, which is suited to direct comparison 
of gauge theory predictions with corresponding results on the string side of
the AdS/CFT correspondence, works well for three-impurity operators 
to three-loop order in $\lambda$  in the $\su(2)$ sector 
(the order to which the $\su(2)$ Hamiltonian is known definitively).  
Specifically, the numerical predictions from the virial approach 
for the near-BMN scaling coefficients 
($E_{\su(2)}^{(1,2)},\ E_{\su(2)}^{(1,3)},\ 
E_{\su(2)}^{(2,4)},\ E_{\su(2)}^{(2,5)},\ 
E_{\su(2)}^{(3,6)}$ and $E_{\su(2)}^{(3,7)}$) 
match corresponding results from the $\su(2)$ long-range Bethe ansatz to three-loop order,
and will eventually be shown to agree with near-plane-wave 
string theory predictions to two loops (the disagreement with string theory
at three loops is by now an expected outcome in these studies; this will also be demonstrated below).
We also find convincing agreement near the BMN limit between the virial approach
and the Bethe ansatz results at one-loop order in the closed $\Sl(2)$ and $\su(1|1)$
subsectors.  As a side result we have found 
in the $\su(1|1)$ sector an {\it exact} (in chain length) agreement between the Bethe ansatz 
and the virial expansion for one-loop operator dimensions with arbitrary 
impurity number (this was only possible because the Bethe equations can be
solved exactly in this subsector for any number of impurities).
There are currently no higher-loop Bethe ans\"atze for the $\su(1|1)$
system, however, so in this sense our numerical predictions go beyond 
the current state of Bethe ansatz technology (see \cite{Minahan:2004ds} for further 
developments of higher-loop gauge theory physics in non-$\su(2)$ sectors).  
Recent progress in developing reliable Bethe equations in the $\Sl(2)$ sector
beyond one-loop order has been made by Staudacher in \cite{Staudacher:2004tk}.
It would be very interesting to find a general long-range Bethe equation 
appropriate for ${\cal N}=4$ SYM at higher loop-order in $\lambda$, 
both for comparison with string predictions and with the virial approach 
studied here.  (For recent developments in this direction, see reference \cite{Beisert:2005fw}.)

%
%
\chapter{A curvature expansion of $AdS_5\times S^5$}      
\label{twoimp}
In order to address specifically stringy aspects of the duality,
it is typically necessary to consider simplifying limits of the
canonical $AdS_5\times S^5$ background. As described above, Metsaev \cite{Metsaev:2001bj} 
showed that, in a certain plane-wave geometry supported by a constant 
RR flux, lightcone gauge worldsheet string theory 
reduces to a free theory with the novel feature  
that the worldsheet bosons and fermions acquire a mass.
This solution was later shown to be a Penrose limit of the familiar 
${AdS}_5 \times S^5$ supergravity solution \cite{Berenstein:2002jq},
and describes the geometry near a null geodesic boosted 
around the equator of the $S^5$ subspace. The energies of Metsaev's
free string theory are thus understood to be those of a string in 
the full ${AdS}_5 \times S^5$ space, in the limit that the states 
are boosted to large angular momentum about an equatorial circle in the $S^5$. 
Corrections to the string spectrum that arise if the string is given a large, but finite,
boost can be computed. Comparison of the resulting interacting spectrum with corrections 
(in inverse powers of the $R$-charge) to the dimensions of the corresponding 
gauge theory operators largely (but not completely) confirms expectations
from AdS/CFT duality (see \cite{Callan:2003xr,Beisert:2003jb} for discussion). 
The purpose of this chapter is to
describe in fairly complete detail the methods used to obtain the results
presented in \cite{Callan:2003xr} (but only outlined in that paper). Some aspects of
the purely bosonic side of this problem were studied by Parnachev and 
Ryzhov \cite{Parnachev:2002kk}. Although we find no disagreement with them, 
our approach differs from theirs in certain respects, most notably
in taking full account of supersymmetry.

The approach is to take the GS superstring action on $AdS_5 \times S^5$,
constructed using the formalism of Cartan forms and superconnections on the 
$SU(2,2|4)/(SO(4,1)\times SO(5))$ coset superspace \cite{Metsaev:2000bj}, 
expand it in powers of the background curvature and
finally eliminate unphysical degrees of freedom
by lightcone gauge quantization. We treat the resulting interaction 
Hamiltonian in first-order degenerate perturbation theory to find the
first corrections to the highly-degenerate pp-wave spectrum. The 
complexity of the problem is such that we are forced to resort to
symbolic manipulation programs to construct and diagonalize the
perturbation matrix. In this chapter we give a proof of principle
by applying our methods to the subspace of two-impurity excitations
of the string. We show that the spectrum organizes itself into correct 
extended supersymmetry multiplets whose energies match well 
(if not perfectly) with what is known about gauge theory anomalous 
dimensions.

In Section~\ref{twoimpSEC2} we introduce the problem by considering the 
bosonic sector of the theory alone. We comment on some 
interesting aspects of the theory that arise when
restricting to the point-particle (or zero-mode) subsector.
In Section~\ref{twoimpSEC3} we review the construction of the GS superstring action 
on $AdS_5 \times S^5$ as a nonlinear sigma model on the 
$SU(2,2|4)/(SO(4,1)\times SO(5))$ coset superspace. In Sections~\ref{twoimpSEC4} and 
\ref{twoimpSEC5}
we perform a large-radius expansion on the relevant objects in the theory,  
and carry out the lightcone gauge reduction, thereby
extracting explicit curvature corrections to the pp-wave Hamiltonian.
Section~\ref{twoimpSEC6} presents results on the curvature-corrected energy spectrum,
further expanded to linear order in the modified 't Hooft coupling 
$\lambda^\prime = g_{YM}^2 N_c/J^2$; results from corresponding gauge theory
calculations (at one loop in $\lambda = g_{YM}^2 N_c$) are summarized 
and compared with the string theory. In Section~\ref{twoimpSEC7} we extend the string 
theory analysis to higher orders in $\lambda^\prime $, and compare 
results with what is known about gauge theory operator dimensions 
at higher-loop order. 

\section{Strings beyond the Penrose limit}
\label{twoimpSEC2}
To introduce the computation of finite-$J$ corrections
to the pp-wave string spectrum, we begin by discussing the construction of the
lightcone gauge worldsheet Hamiltonian for the bosonic string in the
full $AdS_5 \times S^5$ background. The problem is much more complicated
when fermions are introduced, and we will take up that aspect of the
calculation in a later section. A study of the purely bosonic problem
gives us the opportunity to explain various strategic points in a simpler
context.

In convenient global coordinates, the ${AdS}_5 \times S^5$
metric can be written in the form
\be
\label{adsmetricPre}
ds^2 = \Rhat^2 ( - {\rm cosh}^2 \rho\ dt^2 + d \rho^2 + {\rm sinh}^2
\rho\ d \Omega_3^2 + {\rm cos}^2  \theta\ d \phi^2 +  d \theta^2 +
{\rm sin}^2 \theta\ d \tilde\Omega_3^2)~,
\ee
where $\Rhat$ denotes the radius of both the sphere and the AdS space,
and $d\Omega_3^2$, $d\tilde\Omega_3^2$ denote separate three-spheres.
The coordinate $\phi$ is periodic with period $2\pi$ and, strictly speaking, 
so is the time coordinate $t$. In order to
accommodate string dynamics, it is necessary to pass to the
covering space in which time is {\sl not} taken to be periodic.
This geometry, supplemented by an RR field with $N_c$ units of flux
on the sphere, is a consistent, maximally supersymmetric type IIB
superstring background, provided that $\Rhat^4 =  g_s N_c
(\alpha^{\prime})^2$ (where $g_s$ is the string coupling). 

In its initial stages, development of the AdS/CFT correspondence
focused on the supergravity approximation to string theory in
$AdS^5\times S^5$. Recently, attention has turned to the problem of
evaluating truly stringy physics in this background and studying its
match to gauge theory physics. The obstacles to such a program, of
course, are the general difficulty of quantizing strings in curved
geometries, and the particular problem of defining the superstring in
the presence of RR background fields. As noted above, the string
quantization problem is partly solved by looking at the dynamics
of a string that has been boosted to lightlike momentum along
some direction, or, equivalently, by quantizing the string in the
background obtained by taking the Penrose limit of the original
geometry using the lightlike geodesic corresponding to the boosted
trajectory. The simplest choice is to boost along the equator of 
the $S^5$ or, equivalently, to take the Penrose limit with respect 
to the lightlike geodesic $\phi=t,~\rho=\theta=0$ and to quantize
the system in the appropriate lightcone gauge.

To quantize about the lightlike geodesic at $\rho=\theta=0$, 
it is helpful to make the reparameterizations
\begin{eqnarray}
    \cosh\rho  =  \frac{1+z^2/4}{1-z^2/4}~, \qquad
    \cos\theta  =  \frac{1-y^2/4}{1+ y^2/4}\ ,
\end{eqnarray}
and work with the metric
\be
\label{metric4}
ds^2  =  \Rhat^2
\biggl[ -\left({1+ \frac{1}{4}z^2\over 1-\frac{1}{4}z^2}\right)^2dt^2
        +\left({1-\frac{1}{4}y^2\over 1+\frac{1}{4}y^2}\right)^2d\phi^2
+ \frac{d z_k dz_k}{(1-\frac{1}{4}z^2)^{2}}
    + \frac{dy_{k'} dy_{k'}}{(1+\frac{1}{4}y^2)^{2}} \biggr]~.
\ee
The $SO(8)$ vectors spanning the eight directions
transverse to the geodesic are broken into two $SO(4)$ subgroups 
parameterized by $z^2 = z_k z^k$ with $k=1,\dots,4$, and $y^2 = y_{k'} y^{k'}$ 
with $k'=5,\dots,8$.
This form of the metric is well-suited for the present calculation: the spin
connection, which will be important for the superstring action, turns out to
have a simple functional form and the $AdS_5$ and $S^5$ subspaces appear
nearly symmetrically.  This metric has the full $SO(4,2) \times SO(6)$
symmetry associated with $AdS_5 \times S^5$, but only the translation symmetries
in $t$ and $\phi$ and the $SO(4)\times SO(4)$ symmetry of the transverse
coordinates remain manifest.
The translation symmetries mean that string states have a conserved energy
$\omega$, conjugate to $t$, and a conserved (integer) angular momentum $J$,
conjugate to $\phi$. Boosting along the equatorial geodesic is equivalent
to studying states with large $J$, and the lightcone Hamiltonian gives
eigenvalues for $\omega-J$ in that limit. On the gauge theory side, the $S^5$
geometry is replaced by an $SO(6)$ ${R}$-symmetry, and $J$ corresponds to the
eigenvalue of an $SO(2)$ ${R}$-symmetry generator. The AdS/CFT correspondence
implies that string energies in the boosted limit should match operator
dimensions in the limit of large ${R}$-charge (a limit in which perturbative 
evaluation of operator dimensions becomes legitimate).

On dimensional grounds, taking the $J\to\infty$ limit on the string states is
equivalent to taking the $\Rhat\to\infty$ limit on the metric (in the
right coordinates). The coordinate redefinitions
\begin{eqnarray}
\label{rescalePre4}
    t \rightarrow x^+ - \frac{x^-}{2 \Rhat^2}~,
\qquad
    \phi \rightarrow x^+ + \frac{x^-}{2 \Rhat^2}~,
\qquad
    z_k \rightarrow \frac{z_k}{\Rhat}~,
\qquad
    y_{k'} \rightarrow \frac{y_{k'}}{\Rhat}\ 
\end{eqnarray}
make it possible to take a smooth $\Rhat\to\infty$ limit. Expressing
the metric (\ref{metric4}) in these new coordinates, we obtain the
following expansion in powers of $1/\Rhat^2$:
\begin{eqnarray}
\label{expndmet}
ds^2 & \approx &
2\,{dx^+}{dx^-} + {dz }^2 + {dy }^2  -
        \left( {z }^2 + {y }^2 \right) ({dx}^+)^2  
\nonumber \\
& &     
	+ \left[ 2\left( z^2  - y^2  \right) dx^- dx^+
    + z^2 dz^2 - y^2 dy^2 -
    \left( {z }^4 - {y }^4 \right) (dx^+)^2 \right]
    \frac{1}{2 \Rhat^2} \nonumber \\
& &     + O(1/\Rhat^4)\ .
\end{eqnarray}
The leading $\Rhat$-independent part is the well-known pp-wave metric. 
The coordinate $x^+$ is dimensionless, $x^-$ has dimensions of
length squared, and the transverse coordinates now have dimensions of length.
Since it is quadratic in the eight transverse bosonic coordinates, the pp-wave limit
leads to a quadratic (and hence  soluble) Hamiltonian for the bosonic string.
The $1/\Rhat^{2}$ corrections to the metric are what will eventually concern
us: they will add quartic interactions to the lightcone Hamiltonian and
lead to first-order shifts in the energy spectrum of the string.

After introducing lightcone coordinates $x^\pm$ according to (\ref{rescalePre4}),
the general $AdS_5\times S^5$ metric can be cast in the form
\be 
\label{genmetric}
ds^2 = 2 G_{+-} dx^+ dx^- + G_{++} dx^+ dx^+ + G_{--} dx^- dx^- + G_{AB}
dx^A dx^B\ ,
\ee
where $x^A$ ($A = 1,\ldots,8$) labels the eight transverse directions, the metric components
are functions of the $x^A$ only, and the components $G_{+A}$ and $G_{-A}$
are not present. This simplifies even further for the pp-wave metric, 
where $G_{--} =0$ and $G_{+-} = 1$. 
We will use (\ref{genmetric}) as the starting point for constructing 
the lightcone gauge worldsheet Hamiltonian 
(as a function of the transverse $x^A$ and their
conjugate momenta $p_A$) and for discussing its expansion about the free
pp-wave Hamiltonian.

The general bosonic Lagrangian density has a simple expression in
terms of the target space metric:
\be
{\cal L} = \frac{1}{2}h^{ab} G_{\mu\nu}
\partial_{a}x^{\mu} \partial_{b}x^{\nu}\ ,
\ee
where $h$ is built out of the worldsheet metric $\gamma$ according to
$h^{ab}=\sqrt{-{\rm det}\, \gamma}\gamma^{ab}$ and the indices $a,b$ 	
label the worldsheet coordinates $\sigma,\tau$. Since
${\rm det} \, h = -1$, there are only two independent components of
$h$. The canonical momenta (and their inversion in terms of velocities) are
\begin{eqnarray}
\label{canmomenta}
p_{\mu} = h^{\tau a} G_{\mu\nu} \partial_{a} x^{\nu}\ ,
\qquad
\dot x^{\mu} = \frac{1 }{h^{\tau\tau} } G^{\mu\nu} p_{\nu} -
\frac{h^{\tau\sigma} }{h^{\tau\tau} } x^{\prime\mu}\ .
\end{eqnarray}
The Hamiltonian density ${H} = p_{\mu} \dot x^{\mu} - {\cal L}$ is
\be\label{hamilton}
{H} = \frac{1 }{2 h^{\tau\tau} } ( p_{\mu} G^{\mu\nu} p_{\nu}
+ x^{\prime \mu} G_{\mu\nu} x^{\prime \nu} ) -
\frac{h^{\tau\sigma}}{h^{\tau\tau} } (x^{\prime\mu}p_{\mu})\ .
\ee
As is usual in theories with general coordinate invariance (on the
worldsheet in this case), the Hamiltonian is a sum of constraints times
Lagrange multipliers built out of metric coefficients
(${1 }/{h^{\tau\tau} }$ and ${h^{\tau\sigma}}/{h^{\tau\tau}}$).

One can think of the dynamical system we wish to solve as
being defined by ${\cal L} = p_{\mu} \dot x^{\mu} - {H}$
(a phase space Lagrangian) regarded as a function of the coordinates
$x^{\mu}$, the momenta $p_{\mu}$ and the components $h^{ab}$ of
the worldsheet metric. To compute the quantum path integral, the
exponential of the action constructed from this Lagrangian is functionally
integrated over each of these variables. For a spacetime geometry like
(\ref{genmetric}), one finds that with a suitable gauge choice
for the worldsheet coordinates $(\tau,\sigma)$, the
functional integrations over all but the transverse (physical)
coordinates and momenta can be performed, leaving an
effective path integral for these physical variables.
This is the essence of the lightcone approach to quantization.

The first step is to eliminate integrations over $x^+$ and
$p_-$ by imposing the lightcone gauge conditions $x^+ = \tau$ and
$p_-= {\rm const}$.  (At this level of analysis, which is
essentially classical, we will not be concerned
with ghost determinants arising from this gauge choice.)
As noted above, integrations over the worldsheet metric
cause the coefficients ${1 }/{h^{\tau\tau} }$ and
${h^{\tau\sigma}}/{ h^{\tau\tau} }$ to act as Lagrange multipliers,
generating delta functions that impose two constraints:
\begin{eqnarray}
\label{hamconstrnts}
    x^{\prime -} p_- + &&\kern-18pt x^{\prime A} p_A = 0~,
\nonumber \\
    G^{++} p_+^2 + 2 G^{+-} p_+ p_- +  G^{--} p_-^2+ p_A &&\kern-20pt G^{AB} p_B +
    x^{\prime A} G_{AB} x^{\prime B} + G_{--}
    \frac{(x^{\prime A} p_A)^2 }{p_-^2} = 0\ . \nn\\
&&
\end{eqnarray}
When integrations over $x^-$ and $p_+$ are performed, the
delta functions imposing constraints serve to evaluate $x^-$ and $p_+$ in
terms of the dynamical transverse variables (and the constant
$p_-$). The first constraint is linear in $x^-$ and yields
$x^{\prime -}=-x^{\prime \, A} p_A/p_-$. Integrating this over
$\sigma$ and using the periodicity of $x^-$ yields the standard
level-matching constraint, without any modifications. The second
constraint is quadratic in $p_+$ and can be solved explicitly
for $p_+= - {H}_{\rm LC}(x^A,p_A)$. The remaining transverse
coordinates and momenta have dynamics that follow from the
phase space Lagrangian
\be
{\cal L}_{\rm ps} = p_+ + p_- \dot x^- + p_A \dot x^A \sim p_A \dot
x^A - {H}_{\rm LC}(x^A,p_A)\ ,
\ee
where we have eliminated the $p_-$ term by integrating by parts in time
and imposing that $p_-$ is constant. The essential result is that
$- p_+= {H}_{\rm LC}$ is the Hamiltonian that generates evolution of
the physical variables $x^A, ~ p_A$ in worldsheet time $\tau$. This is,
of course, dynamically consistent with the lightcone gauge identification
$x^+=\tau$ (which requires worldsheet and target space time translation
to be the same).

We can solve the quadratic constraint equation (\ref{hamconstrnts})
for $p_+= -{H}_{\rm LC}$ explicitly, obtaining the uninspiring result
\be
\label{hamiltonianLC}
{H}_{\rm LC} = -\frac{p_- G_{+-}}{ G_{--} } -
\frac{p_- \sqrt{G}}{ {G_{--}} } \sqrt{ 1 + \frac{G_{--}}{p_-^2}
( p_A G^{AB} p_B + x^{\prime A} G_{AB} x^{\prime B} ) +
\frac{G_{--}^2}{p_-^4}(x^{\prime A} p_A)^2 }\ ,
\nn\\
\ee
where
\be
G \equiv G_{+-}^2 - G_{++} G_{--}\ .
\ee
This is not very useful as it stands, but we can put it in more manageable
form by expanding it in powers of $1/\Rhat^{2}$. 
We can actually do slightly better by observing
that the constraint equation (\ref{hamconstrnts}) becomes a linear
equation for $p_+$ if $G_{--}=0$ (which is equivalent to $G^{++}=0$).
Solving the linear equation for $p_+$ gives
\be \label{Hlimit}
{H}_{\rm LC} = \frac{p_- G_{++} }{2 G_{+-} } + \frac{ G_{+-}}{ 2p_- }
( p_A G^{AB} p_B + x^{\prime A} G_{AB} x^{\prime B} )~,
\ee
a respectable non-linear sigma model Hamiltonian. 
In the general $AdS_5\times S^5$
metric (\ref{adsmetricPre}) we cannot define a convenient set of coordinates such that $G_{--}$ 
identically vanishes.
Using (\ref{rescalePre4}), however, we can find coordinates where
$G_{--}$ has an expansion which begins at $O(1/\Rhat^{4})$, while the
other metric coefficients have terms of all orders in $1/\Rhat^{2}$. Therefore,
if we expand in $1/\Rhat^{2}$ and keep terms of at most $O(1/\Rhat^{2})$, we may
set $G_{--}=0$ and use (\ref{Hlimit}) to construct the expansion of the
lightcone Hamiltonian to that order. The leading $O(\Rhat^{0})$ terms in
the metric reproduce (as they should) the bosonic pp-wave Hamiltonian
\begin{eqnarray}\label{ppHam}
{H}_{\rm LC}^{\rm pp} & = & \frac{1}{2}\left[ (\dot p^A)^2
    + ({x'}^A)^2 + (x^A)^2 \right]\ ~ ,
\end{eqnarray}
(choosing $p_-=1$ for the conserved worldsheet momentum density). 
The $O(1/\Rhat^{2})$ terms
generate a perturbing Hamiltonian density that is quartic in fields
and quadratic in worldsheet time and space derivatives:
\be\label{pertham}
{H}_{\rm LC}^{\Rhat^{-2}} =
\frac{1}{4\Rhat^2}(y^{2}  p_z^{2}- z^{2} p_y^{2})
+\frac{1}{4\Rhat^2}((2 z^2- y^2)({z}^{\prime})^2  -
            (2 y^2-z^2)({y}^{\prime})^2)~.
\ee
This is the bosonic part of the perturbing Hamiltonian we wish to derive.
If we express it in terms of the creation and annihilation operators of
the leading quadratic Hamiltonian (\ref{ppHam}) we can see that its
matrix elements will be of order $1/J$, as will be the first-order
perturbation theory shifts of the string energy eigenvalues. We
defer the detailed discussion of this perturbation theory until we
have the fermionic part of the problem in hand.
Note that this discussion implies that if we wanted to determine
the perturbed energies to higher orders in $1/\Rhat^{2}$, we would
have the very unpleasant problem of dealing with the square root form
of the Hamiltonian (\ref{hamiltonianLC}).


We have to this point been discussing a perturbative approach to
finding the effect of the true geometry of the $AdS_5\times S^5$
background on the string spectrum. Before proceeding with this program,
however, it is instructive to study a different limit in which the
kinematics are unrestricted (no large-$J$ limit is taken) but only
modes of the string that are independent of the worldsheet coordinate
(the zero-modes of the string) are kept in the Hamiltonian. This is
the problem of quantizing the superparticle of the underlying
supergravity in the $AdS_5\times S^5$ background, a problem that has 
been solved many times (for references, see \cite{Aharony:1999ti}).
A remarkable fact, which seems not to have been explicitly observed
before, is that the spectrum of the zero-mode Hamiltonian is {\em exactly}
a sum of harmonic oscillators: the curvature corrections we
propose to compute actually vanish on this special subspace. This fact
is important to an understanding of the full problem, so we
will make a brief digression to explain the solution to this
toy problem.

The quantization of the superparticle in a supergravity background
is equivalent to finding the eigensolutions of certain Laplacians,
one for each spin that occurs in the superparticle massless
multiplet. The point of interest to us can be made by analyzing
the dynamics of the scalar particle and its associated scalar
Laplacian, which only depends on the background metric. 
With apologies, we will adopt another
version of the $AdS_5\times S^5$ metric, chosen because the scalar
Laplacian is very simple in these coordinates:
\be
\label{zm_metric} 
ds^2 &=& -dt^2(\Rhat^2+  z^2)+d\phi^2(\Rhat^2- y^2)  
\nonumber\\
&&    
	+ dz^j\left(\delta_{jk}-\frac{z^j z^k}{\Rhat^2+ z^2}\right)dz^k +
    dy^{j'}\left(\delta_{j'k'}+\frac{y^{j'}y^{k'}}{\Rhat^2- y^2}\right)dy^{k'}~.
\ee 
As before, the coordinates $z^k$ and $y^{k'}$ parameterize the
two $SO(4)$ subspaces, and the indices $j,k$ and $j',k'$ run over
$j,k=1,\ldots,4,$ and $j',k' = 5,\ldots,8$.
This is a natural metric for
analyzing fluctuations of a particle (or string) around the
lightlike trajectory $\phi=t$ and $\vec z=\vec y=0$. Because the
metric components depend neither on $t$ nor on $\phi$, and because
the problem is clearly separable in $\vec z$ and $\vec y$, it
makes sense to look for solutions of the form
$\Phi=e^{-i\omega t}e^{i J\phi}F(\vec z)G(\vec y)$. The
scalar Laplacian for $\phi$ in the above metric then reduces to
\bea
\label{scalrlap} 
\Bigl[-\frac{\omega^2}{\Rhat^2+\vec z^2}+\frac{J^2}{\Rhat^2-\vec y^2}-
\pderiv{x^j}\Bigl(\delta^{jk}+\frac{z^j z^k}{\Rhat^2}\Bigr)\pderiv{z^k}
\qquad\qquad\nonumber\\
- \pderiv{y^{j'}}\Bigl(\delta^{j'k'}-\frac{y^{j'} y^{k'}}{\Rhat^2}\Bigr)
\pderiv{y^{k'}}~ \Bigr]F(z)G(y)=0~. 
\eea 
The radius $\Rhat$ disappears
from the equation upon rescaling the transverse coordinates by $z\to
z/\Rhat$ and $y\to y/\Rhat$, so we can set $\Rhat=1$ in what follows and use
dimensional analysis to restore $\Rhat$ if it is needed. The scalar
Laplacian is essentially the lightcone Hamiltonian constraint
(\ref{hamconstrnts}) for string coordinates $z^k, y^{k'}$ and string
momenta $p_z^k=-i \pderiv{z^k}$ and $p_y^{k'}=-i \pderiv{y^{k'}}$
(projected onto their zero modes). This implies that we
can use the structure of the Laplacian to correctly order
operators in the string Hamiltonian.

The periodicity $\phi\equiv\phi+2\pi$ means that the angular momentum $J$
is integrally quantized. The allowed values of $\omega$ then follow
from the solution of the eigenvalue problem posed by (\ref{scalrlap}).
As the trial function $\Phi$ indicates, (\ref{scalrlap})
breaks into separate problems for $\vec z$ and $\vec y$:
\bea\label{separated}
{H}_{AdS_5} F(\vec z) = 
	\left[ p^z_j(\delta^{jk}+z^j z^k)p^z_k +
    \omega^2\frac{z_k z^k}{1+(z_k z^k)^2} \right] F(\vec z)
        = A(\omega)  F(\vec z)~, \nonumber\\
{H}_{S^5} G(\vec y) = \left[ p^y_{j'}(\delta^{j'k'}-y^{j'} y^{k'})p^y_{k'}+
    J^2\frac{y_{k'} y^{k'} }{1-(y_{k'}y^{k'})^2} \right] G(\vec y) = B(J) G(\vec y)\ ,
\eea
where $\omega^2 - J^2 = A + B$.
The separation eigenvalues $A,B$ depend on their respective parameters
$\omega,J$, and we determine the energy eigenvalues $\omega$ by finding
the roots of the potentially complicated equation
$\omega^2 - J^2 - A - B=0$.  The scalar Laplacian (\ref{scalrlap})
is equivalent to the constraint equation (\ref{hamconstrnts})
projected onto string zero modes, and we are once again seeing
that the constraint doesn't directly give the Hamiltonian but rather an
equation (quadratic or worse) to be solved for the Hamiltonian.

The ${H}_{S^5}$ equation is just a
repackaging of the problem of finding the eigenvalues of the $SO(6)$
Casimir invariant (another name for the scalar Laplacian on $S^5$)
and ${H}_{AdS_5}$ poses the corresponding problem for $SO(4,2)$. The
$SO(6)$ eigenvalues are obviously discrete, and the $SO(4,2)$ problem
also turns out to be discrete when one imposes the condition of
finiteness at $z^2\to\infty$ on the eigenfunctions (this is a natural restriction
in the context of the AdS/CFT correspondence; for a detailed discussion
see \cite{Aharony:1999ti}). Thus we expect $\omega$ to have a purely discrete
spectrum, with eigenvalues labeled by a set of integers. The simplest way
to solve for the spectrum is to expand $F(\vec z)$ and $G(\vec y)$ in
$SO(4)$ harmonics (since this symmetry is explicit), recognize that
the radial equation is, in both cases, an example of Riemann's
differential equation and then use known properties of the hypergeometric
function to find the eigenvalues and eigenfunctions of (\ref{separated}).
Since it takes three integers to specify an $SO(4)$ harmonic and one
to specify a radial quantum number, we expect each of the two separated
equations to have a spectrum labeled by four integers. The exact
results for the separation eigenvalues turn out to be remarkably simple:
\be
\label{eignvls}
A & = & 2\omega\sum_1^4\left(n_i+\frac{1}{2}\right) -
        \left[\sum_1^4\left(n_i+\frac{1}{2}\right)\right]^2 +4~,
    \qquad n_i=0,1,2,\ldots~,
\nonumber\\
B & = & 2J\sum_1^4\left(m_i+\frac{1}{2}\right) +
        \left[\sum_1^4\left(m_i+\frac{1}{2}\right)\right]^2 +4~,
            \qquad m_i=0,1,2,\ldots~.
\ee
Different eigenfunctions correspond to different choices of the collection
of eight integers $\{n_i,m_i\}$, and the fact that the energies depend only
on $\Sigma n_i$ and $\Sigma m_i$ correctly accounts for the degeneracy of
eigenvalues.  The special form of $A$ and $B$ means that the equation
for the energy eigenvalue, $\omega^2-J^2-A-B=0$, can be factored as
\be
&&
\left[ \omega -J-\sum_1^4\left(n_i+\frac{1}{2}\right)-
    \sum_1^4\left(m_i+\frac{1}{2}\right)\right] \nn\\
&&\kern+90pt
	\times
	\left[ \omega+J-\sum_1^4\left(n_i+\frac{1}{2}\right)+
    \sum_1^4\left(m_i+\frac{1}{2}\right)\right] = 0 ~.
\ee
For obvious reasons, we retain the root that assigns only positive
values to $\omega$, the energy conjugate to the global time $t$:
\be
\omega-J~=~\sum_1^4\left(n_i+\frac{1}{2}\right)~+~ 
	\sum_1^4\left(m_i+\frac{1}{2}\right) ~.
\ee
From the string point of view, $\omega$ catalogs the eigenvalues 
of the string worldsheet Hamiltonian restricted to the zero-mode subspace. 
Quite remarkably, it is an exact sum of harmonic oscillators, 
independent of whether $J$ (and $\omega$) are large or not. This is
simply to say that the eigenvalues of the string Hamiltonian restricted
to the zero-mode sector receive no curvature corrections and could
have been calculated from the pp-wave string Hamiltonian (\ref{ppHam}).
We have only shown this for the massless bosons of the theory, but
we expect the same thing to be true for all the massless fields
of type IIB supergravity. The implication for a perturbative account 
of the string spectrum is that states created using only zero-mode 
oscillators (of any type) will receive no curvature corrections. This 
feature will turn out to be a useful consistency check on our 
quantization procedure. It is of course not true for a general
classical background and is yet another manifestation of the special
nature of the $AdS_5\times S^5$ geometry.

\section{GS superstring action on $AdS_5 \times S^5$}
\label{twoimpSEC3}

The $AdS_5\times S^5$ target space can be realized as the coset superspace 
\be
G/H = \frac{SU(2,2|4)}{ SO(4,1)\times SO(5)  }\ .
\ee
The bosonic reduction of this coset is precisely
$SO(4,2)\times SO(6)/SO(4,1)\times SO(5)\equiv AdS_5\times S^5$. 
To quantize the theory, we will expand the action about a classical trajectory 
that happens to be invariant under the stabilizer group
$H$. There is a general strategy for constructing a non-linear 
sigma model on a super-coset space in terms of the Cartan one-forms 
and superconnections of the super-coset manifold.
In such a construction, the symmetries of the stabilizer subgroup
remain manifest in the action while the remaining symmetries are 
nonlinearly realized 
(see, e.g.,~\cite{Metsaev:2000bj,Kallosh:1998zx,Metsaev:2000yf,Kallosh:1998ji, Metsaev:1999gz, Metsaev:1998it}).
Metsaev and Tseytlin \cite{Metsaev:2000yf} carried out this construction
for the $AdS_5\times S^5$ geometry, producing a $\kappa$-symmetric,
type IIB superstring action possessing the full $PSU(2,2|4)$ 
supersymmetry of $AdS_5\times S^5$. Their action is conceptually simple, 
comprising a kinetic term and a Wess--Zumino term built out of Cartan 
(super)one-forms on the super-coset manifold in the following way 
(this form was first presented in \cite{Grisaru:1985fv}):
\begin{eqnarray}
\label{genlagrangian}
{\cal S} & = & 
	-\frac{1}{2}\int_{\partial M_3} d^2\sigma\ h^{ab} L_a^\mu L_b^\mu
	+ i\int_{M_3} s^{IJ} L^\mu \wedge \bar L^I \Gamma^\mu \wedge L^J\ .
\end{eqnarray}
Repeated upper indices are summed over a Minkowskian inner product.
The indices $a,b$ are used to indicate the worldsheet coordinates
$(\tau,\sigma)$, and we use the values $a,b=0$ to indicate the
worldsheet time direction $\tau$, and $a,b=1$ to specify the $\sigma$ direction.
The matrix $s^{IJ}$ is defined by $s^{IJ} \equiv {\rm diag}(1,-1)$, where
$I,J = 1,2$.
The Wess-Zumino term appears as an integral over a three-manifold $M_3$, while
the kinetic term is integrated over the two-dimensional boundary $\partial M_3$.
The left-invariant Cartan forms are
defined in terms of the coset space representative $G$ by
\begin{eqnarray}
G^{-1}dG = 
	L^\mu P^\mu + L^\alpha  &&\kern-18pt  \bar{Q}_\alpha +  \bar{L}^\alpha Q_\alpha
	+ {1\over 2} L^{\mu\nu}J^{\mu\nu}~,
\nonumber \\
L^N = dX^M L^N_M~,
\qquad 
L_a^N =  L^N_M  &&\kern-18pt  \partial_a X^M~,
\qquad
X^M = (x^\mu,\theta^\alpha,\bar{\theta}^\alpha)\ .
\end{eqnarray}

The explicit expansion of this action in terms of independent fermionic degrees
of freedom is rather intricate. One starts with two 32-component Majorana-Weyl
spinors in ten dimensions: $\theta^I$, where $I=1,2$ labels the two spinors.
In a suitably chosen representation for the $32\times 32$ ten-dimensional
gamma matrices $\Gamma^\mu$, the Weyl projection reduces to picking
out the upper 16 components of $\theta$ and the surviving spinors can
combined into one complex 16-component spinor $\psi$:
\begin{eqnarray}
\theta^I  =  \left( { \theta^\alpha \atop 0} \right)^I~,
\qquad\qquad
\label{psi}
\psi^\alpha  =  \sqrt{2} \left[ (\theta^\alpha)^1 + i (\theta^\alpha)^2 \right]\ .
\end{eqnarray}
The following representation for $\Gamma^\mu$ (which has the desired property that 
$\Gamma_{11}=({\bf 1}_8,-{\bf 1}_8)$) allows us to express their action on 
$\psi$ in terms of real $16\times 16$ $\gamma$-matrices:
\begin{eqnarray}
\label{16gamma}
\Gamma^\mu = \left( \begin{array}{cc}
    0   &   \gamma^\mu  \\
    \bar\gamma^\mu &    0
    \end{array}  \right)~,
& \qquad &
\gamma^\mu \bar\gamma^\nu
+ \gamma^\nu \bar\gamma^\mu = 2\eta^{\mu\nu}~,
\nonumber \\
	\gamma^\mu = (1,\gamma^A, \gamma^9)~,
& \qquad &
	\bar\gamma^\mu = (-1,\gamma^A, \gamma^9)\ .
\end{eqnarray}
The indices  $\mu,\nu,\rho = 0,\dots,9$ denote $SO(9,1)$ vectors,
and  we will denote the corresponding spinor indices by 
$\alpha,\beta,\gamma,\delta = 1,\dots,16$ (we also use the convention 
that upper-case indices $A,B,C,D = 1,\dots,8$ indicate vectors of $SO(8)$,
while $i,j,k = 1,\dots,4$ ($i',j',k' = 5,\dots,8$) indicate vectors
from the $SO(3,1) \cong SO(4)$ ($SO(4)$) subspaces associated with
$AdS_5$ and $S^5$ respectively). The matrix $\gamma^9$ is formed by 
taking the product of the eight $\gamma^A$. A representation of
$\gamma^A$ matrices that will be convenient for explicit calculation
is given in Appendix A. We also note that in the course of quantization
we will impose the fermionic lightcone gauge fixing condition 
$\bar\gamma^9 \psi = \psi$. This restricts the worldsheet fermions
to lie in the $8_s$ representation of $SO(8)$ (and projects out the
$8_c$ spinor), thus reducing the number of independent components
of the worldsheet spinor from 16 to 8. The symmetric matrix
\be
\label{Pidef}
\Pi \equiv \gamma^1 \bar\gamma^2 \gamma^3 \bar\gamma^4\ 
\ee
appears in a number of places in the expansion of the action, so
we give it an explicit definition. Since $\Pi^2=1$, it has eigenvalues 
$\pm 1$, which turn out to provide a useful sub-classification of the 8
components of the $8_s$ worldsheet spinor into two groups of 4. 
The quantity $\tilde\Pi = \Pi \gamma_9$ also appears, but does not
require a separate definition because $\Pi \psi = \tilde\Pi \psi$ 
for spinors satisfying the lightcone gauge restriction to the
$8_s$ representation.

Kallosh, Rahmfeld and Rajaraman presented in \cite{Kallosh:1998zx}  
a general solution to the supergravity constraints (Maurer-Cartan equations) 
for coset spaces exhibiting a superconformal isometry 
algebra of the form
\begin{eqnarray}
\label{genalgebra}
\left[ B_\mu, B_\nu \right] & = & f_{\mu\nu}^\rho B_\rho~, \nonumber \\
\left[ F_\alpha, B_\nu \right] & = & f_{\alpha \nu}^\beta F_\beta~, \nonumber \\
\{ F_\alpha, F_\beta \} & = & f_{\alpha\beta}^\mu B_\mu\ ,
\end{eqnarray}
with $B_\mu$ and $F_\alpha$ representing bosonic and fermionic generators,
respectively.  In terms of these generators, the Cartan forms $L^\mu$ and superconnections $L^\alpha$
are determined completely by the structure constants
$f_{\alpha \mu}^J$ and $f_{\alpha\beta}^\mu$:
\begin{eqnarray}
\label{kallosh1}
L_{at}^\alpha & = &
	\left( \frac{\sinh t{\cal M}}{\cal M} \right)^\alpha_\beta 
	({\cal D}_a \theta)^\beta~,
 \\
\label{kallosh2}
L_{at}^\mu & = & e^\mu_{\phantom{\mu}\nu}\partial_a x^\nu + 2\theta^\alpha f_{\alpha\beta}^\mu
	\left( \frac{\sinh^2 (t {\cal M}/2)}{{\cal M}^2}\right)^\beta_\gamma 
	({\cal D}_a \theta)^\gamma~,
 \\
({\cal M}^2)^\alpha_\beta & = & -\theta^\gamma f^\alpha_{\gamma\mu} 
	\theta^\delta f^\mu_{\delta\beta}\ .
\end{eqnarray}
The dimensionless parameter $t$ is used here to define ``shifted'' Cartan
forms and superconnections where, for example, $L_a^\mu = {L_{at}^\mu}|_{t=1}$.
In the case of $AdS_5 \times S^5$, the Lagrangian takes the form
\begin{eqnarray}
\label{lagrangiank4}
{\cal L}_{\rm Kin} & = & -\frac{1}{2} h^{ab} L_a^\mu L_b^\mu~,
\\
\label{lagrangianwz4}
{\cal L}_{\rm WZ} & = & -2i\epsilon^{ab} \int_0^1 dt\, L_{at}^\mu s^{IJ}
    \bar\theta^I \Gamma^\mu L_{bt}^J\ .
\end{eqnarray}

In the context of eqns.~(\ref{kallosh1}, \ref{kallosh2}), 
it will be useful to choose a 
manifestation of the spacetime metric that yields a compact form of 
the spin connection.  The form appearing in eqn.~(\ref{metric4}) 
is well suited to this requirement; the $AdS_5$ and $S^5$ subspaces
are represented in (\ref{metric4}) nearly symmetrically, and the
spin connection is relatively simple:
\begin{eqnarray}
\omega^{t\,z_k}_{\phantom{t\,z_k}\,t}  =
         {z_k \over 1-{1\over 4}z^2}~, & \qquad &
\omega^{z_j\,z_k}_{\phantom{z_j\,z_k}\,z_j}  =
        {{1\over 2} z_k \over 1-{1\over 4}z^2}~,
\nonumber \\
\omega^{\phi\,y_{k'}}_{\phantom{\phi\,y_{k'}}\,\phi}  =
        -{y_{k'} \over 1+{1\over4}y^2}~, & \qquad &
\omega^{y_{j'}\,y_{k'}}_{\phantom{y_{j'}\,y_{k'}}\,y_{j'}}  =
        -{{1\over2}y_{k'}\over 1+{1\over4}y^2}\ .
\end{eqnarray}
%
Upon moving to the lightcone coordinate system in (\ref{rescalePre4}), the $x^+$
direction remains null ($G_{--} = 0$) to $O(1/\Rhat^4)$ in this
expansion.

By introducing dimensionless contraction parameters $\Lambda$ and
$\Omega$ \cite{Blau:2002mw}, one may express the $AdS_5 \times S^5$ isometry
algebra keeping lightcone directions explicit:
\begin{eqnarray}
\left[P^+,P^k\right] = {\Lambda^2 \Omega^2}J^{+k}~,
    & \quad &
\left[P^+,P^{k'}\right] = -{\Lambda^2 \Omega^2}J^{+k'}~, \nonumber \\
\left[P^+,J^{+k}\right] = -{\Lambda^2}P^{k}~,
    & \quad &
\left[P^+,J^{+k'}\right] = {\Lambda^2}P^{k'}~, \nonumber \\
\left[P^-,P^{A}\right] = { \Omega^2}J^{+A}~,
    & \quad &
\left[P^-,J^{+A}\right] = P^{A}~, \nonumber \\
\left[P^j,P^k\right] = {\Lambda^2 \Omega^2}J^{jk}~,
    & \quad &
\left[P^{j'},P^{k'}\right] = -{\Lambda^2 \Omega^2}J^{j'k'}~, \nonumber \\
\left[J^{+j},J^{+k}\right] = {\Lambda^2}J^{jk}~,
    & \quad &
\left[J^{+j'},J^{+k'}\right] = -{\Lambda^2}J^{j'k'}~, \nonumber \\
\left[P^j,J^{+k}\right] = -\delta^{jk}(P^+ - \Lambda^2\,P^-)~,
    & \quad &
\left[P^{r},J^{+s}\right] = -\delta^{rs}(P^+ + \Lambda^2\,P^-)~, \nonumber \ee\be
\left[P^{i},J^{{j}{k}}\right] = \delta^{ij}P^{k} - \delta^{ik}P^j~,
    & \quad &
\left[P^{i'},J^{{j'}{k'}}\right] = \delta^{i'j'}P^{k'}- \delta^{i'j'}P^{k'}~,  \nonumber \\
\left[J^{+{i}},J^{{j}{k}}\right] = \delta^{ij}J^{+k} - \delta^{ik}J^{+j}~,
    & \quad &
\left[J^{+{i'}},J^{j'k'}\right] = \delta^{i'j'}J^{+k'}- \delta^{i'j'}J^{+k'}~,  \nonumber \\
\left[J^{ij},J^{kl}\right]
    = \delta^{jk}J^{il} + 3\ {\rm terms}~,
    & \quad &
\left[J^{i'j'},J^{k'l'}\right]
    = \delta^{j'k'}J^{i'l'} + 3\ {\rm terms}\ .
\end{eqnarray}
The bosonic sector of the algebra relevant to (\ref{genalgebra}) takes the form
\begin{eqnarray}
\left[J^{ij},Q_{\alpha}\right] & = &
    {1\over2}Q_{\beta}(\gamma^{ij})^\beta_{\ \alpha}~,
    \nonumber \\
\left[J^{i'j'},Q_{\alpha}\right] & = &
    {1\over2}Q_{\beta}(\gamma^{i'j'})^\beta_{\ \alpha}~,
    \nonumber \\
\left[J^{+i},Q_{\alpha}\right] & = &
    {1\over2}Q_{\beta}(\gamma^{+i}-\Lambda^2\gamma^{-i})^\beta_{\ \alpha}~,
    \nonumber \\
\left[J^{+i'},Q_{\alpha}\right] & = &
    {1\over2}Q_{\beta}(\gamma^{+i'}+\Lambda^2\gamma^{-i'})^\beta_{\ \alpha}~, 
    \nonumber \\
\left[P^\mu,Q_{\alpha}\right] & = &
    {i \Omega\over 2}Q_{\beta}(\Pi \gamma^{+}\bar{\gamma}^\mu)^\beta_{\ \alpha}
    -{i\Lambda^2 \Omega\over 2}Q_{\beta}(\Pi \gamma^{-}\bar{\gamma}^\mu)^\beta_{\ \alpha}\ .
\end{eqnarray}
The fermi-fermi anticommutation relations are
\begin{eqnarray}
\{Q_\alpha,\bar Q_\beta\} & = &
    -2i\gamma^\mu_{\alpha\beta}P^\mu
    -{2 \Omega}(\bar\gamma^k \Pi)_{\alpha\beta}J^{+k}
    -{2 \Omega}(\bar\gamma^{k'}\Pi)_{\alpha\beta}J^{+k'} \nonumber \\
& &     + { \Omega}(\bar\gamma^+\gamma^{jk}\Pi)_{\alpha\beta}J^{jk}
    + {\Omega}(\bar\gamma^+\gamma^{j'k'}\Pi)_{\alpha\beta}J^{j'k'} \nonumber \\
& &     -{\Lambda^2 \Omega}(\bar\gamma^-\gamma^{jk}\Pi)_{\alpha\beta}J^{jk}
    +{\Lambda^2  \Omega}(\bar\gamma^-\gamma^{j'k'}\Pi)_{\alpha\beta}J^{j'k'}\ .
\end{eqnarray}
This form of the superalgebra has the virtue that one can
easily identify the flat space $(\Omega\rightarrow 0)$
and plane-wave $(\Lambda\rightarrow 0)$ limits. 
The Maurer-Cartan equations in this coordinate system take the form
\begin{eqnarray}
\label{diffeq1}
dL^\mu & = & -L^{\mu\nu}L^\nu - 2i\bar{L}\bar{\gamma}^\mu L~, \nonumber \\
dL^\alpha & = & -{1\over 4}L^{\mu\nu}(\gamma^{\mu\nu})^\alpha_{\ \beta}L^\beta
    + {i\Omega \over 2}L^\mu(\Pi\gamma^+\bar{\gamma}^\mu)^\alpha_{\ \beta}L^\beta
    - {i\Lambda^2\Omega \over 2}L^\mu(\Pi\gamma^-\bar{\gamma}^\mu)^\alpha_{\ \beta}L^\beta ~,
\nonumber \\
d\bar{L}^\alpha & = & -{1\over 4}L^{\mu\nu}(\gamma^{\mu\nu})^\alpha_{\ \beta}\bar{L}^\beta
    - {i\Omega\over 2}L^\mu(\Pi\gamma^+\bar\gamma^\mu)^\alpha_{\ \beta}\bar{L}^\beta
    + {i\Lambda^2\Omega \over 2}L^\mu(\Pi\gamma^-\bar\gamma^\mu)^\alpha_{\ \beta}\bar{L}^\beta\ ,
\end{eqnarray}
where wedge products (\ref{genlagrangian}) 
are understood to be replaced by the following rules:
\be
    L^\mu L^\nu = -L^\nu L^\mu~,
\qquad  L^\mu L^\alpha = -L^\alpha L^\mu~,
\qquad  L^\alpha L^\beta = L^\beta L^\alpha\ .
\ee
Upon choosing a parameterization of the
coset representative $G$
\begin{eqnarray}
G(x,\theta) = f(x)g(\theta)~, \qquad
g(\theta) = \exp(\theta^\alpha \bar{Q}_\alpha + \bar{\theta}^\alpha Q_\alpha)\ ,
\end{eqnarray}
one derives a set of coupled differential equations for the shifted 
Cartan forms and superconnections:
\begin{eqnarray}
\partial_t L_t & = & d\theta + {1\over 4}L_t^{\mu\nu}\gamma^{\mu\nu}\theta
    - {i\Omega\over 2}L_t^\mu \Pi\gamma^+ \bar\gamma^\mu\theta
    + {i\Lambda^2\Omega \over 2}L_t^\mu \Pi\gamma^-\bar\gamma^\mu \theta ~,
\nonumber \\
\partial_t L_t^\mu & = & -2i\theta \bar\gamma^\mu \bar L_t
    - 2i\bar\theta\bar\gamma^\mu L_t ~,
\nonumber \\
\partial_t L_t^{-i} & = & 
	{2\Omega}(\theta \bar\gamma^i\Pi \bar L_t)
	- {2\Omega}(\bar\theta \bar\gamma^i\Pi L_t) ~,
\nonumber \\
\partial_t L_t^{-r} & = & 
	{2\Omega}(\theta \bar\gamma^{r}\Pi \bar L_t)
	- {2\Omega}(\bar\theta \bar\gamma^{r}\Pi L_t) ~,
\nonumber \\
\partial_t L_t^{ij} & = &
	-{2\Omega}(\theta \bar\gamma^+ \gamma^{ij} \Pi \bar L_t)
    + {2\Omega} (\bar\theta \bar \gamma^+ \gamma^{ij} \Pi L_t)
\nn\\
&&    + {2\Lambda^2\Omega} (\theta \bar\gamma^- \gamma^{ij} \Pi \bar L_t)
    - {2\Lambda^2\Omega}(\bar\theta \bar\gamma^- \gamma^{ij} \Pi L_t )~,
\nonumber \\
\partial_t L_t^{i'j'} & = &
    -{2\Omega}(\theta \bar\gamma^+ \gamma^{i'j'} \Pi \bar L_t)
    + {2\Omega} (\bar\theta \bar \gamma^+ \gamma^{i'j'} \Pi L_t)
    - {2\Lambda^2\Omega} (\theta \bar\gamma^- \gamma^{i'j'} \Pi \bar L_t)~,
\nonumber \\
& &     + {2\Lambda^2\Omega}(\bar\theta \bar\gamma^- \gamma^{i'j'} \Pi L_t)\ .
\label{diffeq}
\end{eqnarray}
These coupled equations are subject to the following boundary conditions:
\begin{eqnarray}
L_\pm(t=0) = 0~, \qquad L_{t=0}^\mu = e^\mu~, \qquad L_{t=0}^\pm = e^\pm~,  \nonumber \\
L_{t=0}^{\mu\nu} = \omega^{\mu\nu}~, \qquad L_{t=0}^{-\mu} = \omega^{-\mu}\ .
\end{eqnarray}  
The generators $J^{-\mu}$ and $J^{kk'}$ are not present in the superalgebra, so
the conditions
\be
L^{+\mu} = 0~, \qquad L^{kk'} = 0
\ee
are imposed as constraints.

To employ the general solution to the Maurer-Cartan equations 
(\ref{kallosh1}, \ref{kallosh2}), the relevant sectors of the superalgebra
may be rewritten in the more convenient 32-dimensional notation
(setting $\Lambda = 1$ and $\Omega = 1$):
\begin{eqnarray}
\label{superalg}
\left[ Q_I , P^\mu \right] & = &
    \frac{i}{2} \epsilon^{I J}Q_J \Gamma_* \Gamma^\mu~,
\nonumber \\
\left[ Q_I , J^{\mu\nu} \right] & = &
    -\frac{1}{2} Q_I  \Gamma^{\mu\nu}~,
\nonumber \\
\{ (Q_I)^\mu, (Q_J)_\mu \} & = &
    -2i\delta_{IJ}\Gamma^0 \Gamma^\rho P_\rho
    +  \epsilon^{IJ}\left( -\Gamma^0 \Gamma^{jk}\Gamma_* J_{jk}
    + \Gamma^0 \Gamma^{j'k'} {\Gamma'}_* J_{j'k'} \right)\ ,
\end{eqnarray}
where
\begin{eqnarray}
\Gamma_* \equiv i\Gamma_{01234}~,
\qquad
{\Gamma'}_* \equiv i\Gamma_{56789}\ .
\end{eqnarray}
The Cartan forms and superconnections
then take the following form:
\be
L_{bt}^J  &=&  \frac{\sinh t{\cal M}}{{\cal M}} {\cal D}_b \theta^J~,
    \nn\\
L_{at}^\mu  &=&  e^\mu_{\phantom{\mu}\rho}\partial_a x^\rho
    - 4i\bar\theta^I \Gamma^\mu \left( \frac{\sinh^2 (t{\cal M}/2)}{{\cal M}^2}
    \right){\cal D}_a \theta^I\ ,
\label{sol}
\ee
where the covariant derivative is given by
\begin{eqnarray}
({\cal D}_a \theta)^I
    & = & \left( \partial_a \theta + {1\over 4}
    \left(\omega^{\mu\,\nu}_{\phantom{\mu\,\nu}\,\rho}\,\partial_a x^\rho \right)
    \Gamma^{\mu\nu} \theta\right)^I
    -{i\over 2}\epsilon^{IJ}
    e^\mu_{\phantom{\mu}\,\rho}\,\partial_a x^\rho \Gamma_* \Gamma^\mu \theta^J\ .
\end{eqnarray}
The object ${\cal M}$ is a $2\times 2$ matrix, which, for convenience, is 
defined in terms of its square:
\begin{eqnarray}
({\cal M}^2)^{IL} & = &
    -\epsilon^{IJ}(\Gamma_* \Gamma^\mu \theta^J \bar\theta^L \Gamma^\mu)
    +\frac{1}{2}\epsilon^{KL}(-\Gamma^{jk}\theta^I \bar\theta^K \Gamma^{jk}\Gamma_*
        + \Gamma^{j'k'}\theta^I \bar\theta^K \Gamma^{j'k'} {\Gamma'}_*)\ . \nn\\
\end{eqnarray}

At this point, the GS action on $AdS_5\times S^5$ (\ref{lagrangiank4}, \ref{lagrangianwz4}) 
may be expanded to arbitrary 
order in fermionic and bosonic fields.  In the present calculation, 
the parameters $\Omega$ and $\Lambda$ remain set to unity, and 
the action is expanded in inverse powers of the target-space radius $\Rhat$, 
introduced in the rescaled lightcone coordinates in eqn.~(\ref{rescalePre4}). 
The fact that supersymmetry must be protected
at each order in the expansion determines a rescaling prescription
for the fermions.  Accordingly, the eight transverse bosonic directions $x^A$
and the corresponding fermionic fields $\psi^\alpha$ receive a rescaling
coefficient proportional to $\Rhat^{-1}$.
The first curvature correction
away from the plane-wave limit therefore occurs at quartic order in both
bosonic and fermionic fluctuations.
The particular lightcone coordinate system chosen in (\ref{rescalePre4}), however,
gives rise to several complications.  The $x^\pm$ coordinates given by 
\be
\label{lccoords}
t  =  x^+ - \frac{x^-}{2\Rhat^2}~, \qquad\qquad
\phi  =  x^+ + \frac{x^-}{2\Rhat^2}
\ee
have conjugate momenta (in the language of BMN)
\be
-p_+ & = & i\partial_{x^+} = i(\partial_t + \partial_\phi) = \Delta - J~, \\
-p_- & = & i\partial_{x^-} = \frac{i}{2\Rhat^2}(\partial_\phi - \partial_t) 
	= -\frac{1}{2\Rhat^2}(\Delta + J)\ ,
\ee
with $\Delta = E = i\partial_t$ and $J = -i\partial_\phi$. 
The lightcone Hamiltonian is ${H} = -p_+$, so with $\Delta = J-p_+$ one
may schematically write
\be
p_- & = & \frac{1}{2\Rhat^2}(2J - p_+) \nn\\
& = & \frac{J}{\Rhat^2}+ \frac{{H}}{2\Rhat^2} \nn\\
& = & \frac{J}{\Rhat^2}\left( 1 + \frac{1}{2J}\sum N \omega \right)\ .
\ee
This result appears to be incorrect in the context of 
the lightcone gauge condition $\partial_\tau t = p_-$.  
To compensate for this, one must set the constant worldsheet density 
$p_-$ equal to something different from 1 (and non-constant) if the parameter 
length of the worldsheet is to be proportional to $J$. 
This operation introduces an additional $O(1/\Rhat^2)$ shift in the energy 
of the string oscillators.
This is acceptable because, in practice, we wish to consider only degenerate subsets of 
energy states for comparison between the gauge theory and string theory results.  
Because of the compensation between corrections to
$J$ and the Hamiltonian contribution from $p_-$, the eigenvalues of $J$ will remain
constant within these degenerate subsets.  Therefore, while it may seem incorrect to introduce
operator-valued corrections to $p_-$, one could proceed pragmatically 
with the intent of restricting oneself to these degenerate subsets.
When such a program is carried out, however, the resulting theory is subject to 
normal-ordering ambiguities; we instead use a coordinate
system that is free of these complications.  

A different choice of lightcone coordinates allows us
to avoid this problem completely. 
By choosing
\be
\label{newcoords}
t & = & x^+~, \nn\\
\phi & = & x^+ + \frac{x^-}{\Rhat^2}\ ,
\ee
we have
\be
\label{RJdefs}
-p_+ & = & \Delta - J~, \\
-p_- & = & i\partial_{x^-} = \frac{i}{\Rhat^2}\partial_\phi = -\frac{J}{\Rhat^2}\ ,
\ee
such that $p_-$ appears as a legitimate expansion parameter in the theory.
In this coordinate system, the curvature expansion of the metric becomes
\be
ds^2 & = & 2dx^+ dx^- - (x^A)^2 (dx^+)^2 + (dx^A)^2 
\nonumber \\
& & 	+ \frac{1}{\Rhat^2}\left[ 
	-2y^2 dx^+ dx^- + \frac{1}{2}(y^4-z^4) (dx^+)^2 + (dx^-)^2
	+\frac{1}{2}z^2 dz^2 - \frac{1}{2}y^2 dy^2 \right] 
\nonumber \\
\label{expmetric2}
& & 	+ O\left(\Rhat^{-4}\right).
\ee
The operator-valued terms in $p_-$ that appear under the first
coordinate choice (\ref{lccoords}) are no longer present.
However, it will be shown that this new coordinate system induces 
correction terms to the spacetime curvature of the worldsheet metric.  
Furthermore, the appearance of a nonvanishing $G_{--}$ component,
and the loss of many convenient symmetries between terms associated with the
$x^+$ and $x^-$ directions bring some additional complications into the analysis.
The advantage is that the results will be unambiguous in the end (and free
from normal-ordering ambiguities).

\section{Curvature corrections to the Penrose limit}
\label{twoimpSEC4}

In this section we expand the GS superstring action on $AdS_5 \times S^5$
in powers of $1/\Rhat^2$.  We begin by constructing various quantities
including combinations of Cartan one-forms relevant to the worldsheet 
Lagrangian.  Spacetime curvature corrections to the 
worldsheet metric will be calculated by 
analyzing the $x^-$ equation of motion and the covariant gauge constraints
order-by-order.  

We introduce the notation
\begin{eqnarray}
\Delta_n^\mu & \equiv & \bar\theta^I \Gamma^\mu {\cal D}_0^n \theta^I~, \\
{\Delta'}_n^\mu & \equiv & \bar\theta^I \Gamma^\mu {\cal D}_1^n \theta^I~,
\end{eqnarray}
where the covariant derivative is expanded in powers of $(1/\Rhat)$:
\begin{eqnarray}
{\cal D}_a & = & {\cal D}_a^0 + \frac{1}{\Rhat}{\cal D}_a^1
    + \frac{1}{\Rhat^2}{\cal D}_a^2 + O(\Rhat^{-3})\ .
\end{eqnarray}
Terms in the Wess-Zumino Lagrangian are encoded using a similar notation:
\be
\Box_n^\mu & \equiv & s^{IJ}\bar\theta^I\Gamma^\mu{\cal D}_0^n \theta^J~, \\
{\Box'}_n^\mu & \equiv & s^{IJ}\bar\theta^I\Gamma^\mu{\cal D}_1^n \theta^J~.
\ee
The subscript notation $(\Delta_n^\mu)_{\theta^4}$ will be used
to indicate the quartic fermionic term involving ${\cal M}^2$:
\be
(\Delta_n^\mu)_{\theta^4} \equiv \frac{1}{12}\bar\theta^I({\cal M}^2){\cal D}_0^n\theta^I~.
\ee
For the present, it will be convenient to remove an overall factor of $\Rhat^2$ from the
definition of the vielbeins $e^\mu_{\phantom{\mu}\nu}$.  
In practice, this choice makes it easier to recognize terms that contribute to the 
Hamiltonian at the order of interest, and, in the end, 
allows us to avoid imposing an additional rescaling operation on the fermions. 
We proceed by keeping terms to $O(1/\Rhat^4)$, with the understanding that 
an extra factor of $\Rhat^2$ must be removed in the final analysis.  
The covariant derivative
\be
{\cal D}_a\theta^I = \partial_a \theta^I + \frac{1}{4}\partial_a x^\mu \omega^{\nu\rho}_\mu
	\Gamma_{\nu\rho}\theta^I - \frac{i}{2}\epsilon^{IJ}\Gamma_*\Gamma_\mu
	e^\mu_{\phantom{\mu}\nu}\partial_a x^\nu \theta^J
\ee
may then be expanded to $O(1/\Rhat^2)$ (we will not need $O(1/\Rhat^3)$ terms, 
because the covariant derivative always appears left-multiplied by a spacetime spinor
$\bar\theta$):
\be
{\cal D}_0\theta^I & = &  
	\biggl[\partial_0\theta^I - p_-\epsilon^{IJ}\Pi\theta^J\biggr] 
	+\frac{1}{\Rhat}\biggl[ \frac{p_-}{4}  
	\left( z_j\Gamma^{-j} 
	- y_{j'}\Gamma^{-j'}\right)\theta^I
	+\frac{1}{4}\epsilon^{IJ}\Gamma^-\Pi(\dot x^A\Gamma^A)\theta^J\biggr] 
\nonumber \\
&&\kern-00pt	
	+\frac{1}{\Rhat^2}
	\biggl[\frac{1}{4}(\dot z_j z_k\Gamma^{jk} 
	- \dot y_{j'}y_{k'}\Gamma^{j'k'})\theta^I
	+ \frac{p_-}{4}\epsilon^{IJ}\Pi(y^2-z^2)\theta^J 
\nn\\
&&	- \frac{1}{2}\epsilon^{IJ}(\dot x^-) \Pi\theta^J \biggr] 
	+O(\Rhat^{-3})~, \\
{\cal D}_1\theta^I &= & 
	\partial_1\theta^I 
	+\frac{1}{4\Rhat}\epsilon^{IJ}\Gamma^-\Pi ({x'}^A \Gamma^A)\theta^J \nonumber \ee\be
& & 	+\frac{1}{\Rhat^2}\left[\frac{1}{4}(z'_jz_k\Gamma^{jk}-y'_{j'}y_{k'}\Gamma^{j'k'})\theta^I
	-\frac{1}{2}\epsilon^{IJ}({x'}^-)\Pi\theta^J\right] +O(\Rhat^{-3}) ~.
\ee
Note that we have not rescaled the spinor field $\theta$ in the above expansion.  
This allows us to isolate the bosonic scaling contribution from the covariant derivative 
when combining various terms in the Lagrangian. Subsequently, the fermionic rescaling is
performed based on the number of spinors appearing in each term (two spinors for each
$\Delta^\mu$ or $\Box^\mu$, and four for each $(\Delta^\mu)_{\theta^4}$). 
The worldsheet derivative notation is given by $\partial_\tau x = \partial_0 x = \dot x$ 
and $\partial_\sigma x = \partial_1 x = x'$.

The various sectors of the worldsheet Lagrangian 
are assembled keeping $x^-$ and its derivatives explicit;
these will be removed by imposing the covariant gauge 
constraints.  From the supervielbein and superconnection
\be
L_{at}^\mu & = & 
	 e^\mu_{\phantom{\mu}\nu}\partial_a x^\nu
  	  - 4i\bar\theta^I \Gamma^\mu \left( \frac{\sinh^2 (t{\cal M}/2)}{{\cal M}^2}
  	  \right){\cal D}_a \theta^I \nonumber \\
& \approx & 
	e^\mu_{\phantom{\mu}\nu}\partial_ax^\nu 
	- i\bar\theta^I \Gamma^\mu\left(t^2 + \frac{t^4{\cal M}^2}{12}\right){\cal D}_a\theta^I~, \\
L_{at}^I & = & \frac{\sinh t{\cal M}}{{\cal M}}{\cal D}_a\theta^I 
	\approx \left( t+\frac{t^3}{6}{\cal M}^2\right){\cal D}_a\theta^I~, 
\ee
we form the following objects:
\be
\label{L0L04}
L_0^\mu L_0^\mu & = & 
	\frac{1}{\Rhat^2}\left\{
	2p_-\dot x^- - p_-^2 (x^A)^2 + (\dot x^A)^2 - 2ip_-\Delta_0^-\right\} 
\nonumber \\
& & 	+ \frac{1}{\Rhat^4}\biggl\{
	(\dot x^-)^2 - 2p_-y^2\dot x^- + \frac{1}{2}(\dot z^2 z^2 - \dot y^2 y^2)
	+ \frac{p_-^2}{2}(y^4-z^4) 
\nn\\
&&	-2i\biggl[\frac{1}{2}\dot x^-\Delta_0^- + p_-\Delta_2^-
	+p_-(\Delta_0^-)_{\theta^4} 
\nn\\
&& 	- \frac{p_-}{4}(y^2-z^2)\Delta_0^- +\dot x^A \Delta_1^A
	\biggr] \biggr\} +O(\Rhat^{-6})~, 
\\
\label{L1L14}
L_1^\mu L_1^\mu & = & 
	\frac{1}{\Rhat^2}({x'}^A)^2 
	+ \frac{1}{\Rhat^4}\biggl\{
	\frac{1}{2}({z'}^2 z^2 - {y'}^2 y^2) + ({x'}^-)^2 
\nn\\
&&	- 2i{x'}^A {\Delta'}_1^A
	-i{x'}^-{\Delta'}_0^- \biggr\}+O(\Rhat^{-6})~, \\
\label{L0L14}
L_0^\mu L_1^\mu & = & 
	\frac{1}{\Rhat^2}\left\{ p_-{x'}^- + \dot x^A{x'}^A - ip_-{\Delta'}_0^-\right\}
\nonumber \\
& &  	+ \frac{1}{\Rhat^4}\biggl\{ {x'}^- \dot x^- - p_-y^2{x'}^- + \frac{1}{2}(z^2 \dot z_k z'_k
	-y^2 \dot y_{k'}{y'}_{k'}) 
\nn\\
&& 	-i p_-{\Delta'}_2^- -i p_-({\Delta'}_0^-)_{\theta^4} 
	-i \frac{p_-}{4}(z^2 - y^2){\Delta'}_0^- 
 	- \frac{i}{2}\dot x^-{\Delta'}_0^-
	-i\dot x^A {\Delta'}_1^A 
\nn\\
&&	-i {x'}^A \Delta_1^A - \frac{i}{2}{x'}^-\Delta_0^- \biggr\}
	+O(\Rhat^{-6})~.
\ee

It will be advantageous
to enforce the lightcone gauge condition $x^+ = \tau$ at all orders in 
the theory.\footnote{ This differs from the approach presented in \cite{Parnachev:2002kk}. }   
When fermions are included, this choice allows us to 
keep the $\kappa$-symmetry condition $\Gamma^+ \theta = 0$  exact.
In the pp-wave limit, keeping the worldsheet metric flat 
in this lightcone gauge is consistent with the equations of motion.  
Beyond leading order, however, we are forced to consider curvature 
corrections to the worldsheet metric that appear in both the 
conformal gauge constraints and the worldsheet Hamiltonian.
In the purely bosonic case described in Section~\ref{twoimpSEC2} above, these corrections
are kept implicit by defining gauge constraints
in terms of canonical momenta.    
In the supersymmetric theory, we must explicitly calculate these corrections.
The strategy is to expand the $x^-$ equations of motion in rescaled coordinates
(\ref{newcoords}) and solve for the components of the worldsheet 
metric order-by-order.  By varying $x^-$ in the full Lagrangian we obtain
\be
\label{xmeom}
\frac{\delta {\cal L}}{\delta \dot x^-} & = & 
\frac{1}{2}h^{00} \left\{ \frac{2p_-}{\Rhat^2} + \frac{1}{\Rhat^4}\left[2\dot x^-
	-2p_-y^2 - i\bar\theta^I\Gamma^- \partial_0 \theta^I
	+2ip_-\bar\theta^I\Gamma^- \epsilon^{IJ}\Pi\theta^J\right]\right\} 
\nonumber \\
& & 	+\frac{i}{2\Rhat^4}s^{IJ}\bar\theta^I \Gamma^-\partial_1\theta^J+ O(\Rhat^{-6}).
\ee
The worldsheet metric is taken to be flat at leading order, so there is no contribution
from $L_0^\mu L_1^\mu$ here.  
To obtain corrections to $h^{ab}$ entirely in terms of physical variables,
however, we must eliminate all instances of $x^-$ 
(or its derivatives) from the above variation.
We can solve the conformal gauge constraints at leading order 
to remove $\dot x^-$ from (\ref{xmeom}).  These constraints are obtained by varying the
Lagrangian with respect to the worldsheet metric itself:
\be
\label{Tab}
T_{ab} = L_a^\mu L_b^\mu - \frac{1}{2}h_{ab}h^{cd}L_c^\mu L_d^\mu~,
\ee
yielding a symmetric traceless tensor with two independent components.
To leading order in $1/\Rhat$, we find 
\be
\label{T00}
T_{00} & = & \frac{1}{2}(L_0^\mu L_0^\mu + L_1^\mu L_1^\mu) + \dots = 0
\nonumber \\
	& = & \frac{1}{2\Rhat^2}\left( 2p_-\dot x^- - p_-^2(x^A)^2 + (\dot x^A)^2
		- 2ip_- \Delta_0^- + ({x'}^A)^2 \right) + O(\Rhat^{-4})~, 
\nn\\
\label{T01}
T_{01} & = & L_0^\mu L_1^\mu + \dots = 0
\nonumber \\
	& = & p_-{x'}^- + \dot x^A {x'}^A - ip_-{\Delta'}_0^- + O(\Rhat^{-4})~.
\ee
Expanding $\dot x^-$ and ${x'}^-$ in the same fashion,
\be
\dot x^- = \sum_n \frac{a_n}{\Rhat^n}~, \qquad {x'}^- = \sum_n \frac{a'_n}{\Rhat^n}~,
\ee
we use (\ref{T00}) and (\ref{T01}) to obtain
\be
\label{a0}
a_0 & = & \frac{p_-}{2}(x^A)^2 - \frac{1}{2p_-}\left[(\dot x^A)^2 + ({x'}^A)^2\right]
	+ i\bar\theta^I\Gamma^- \partial_0 \theta^I
	-ip_-\epsilon^{IJ} \bar\theta^I\Gamma^- \Pi\theta^J~, 
 \\
\label{a1}
a'_0 & = & -\frac{1}{p_-}\dot x^A {x'}^A + i\bar\theta^I\Gamma^-\partial_1\theta^I~.
\ee
By substituting back into (\ref{xmeom}), and performing the analogous operation
for the ${x'}^-$ variation, these leading-order solutions provide the following 
expansions for the objects that enter into the $x^-$ equation of motion:
\be
\label{var1}
\frac{\delta {\cal L}}{\delta \dot x^-} & = & 
	\frac{1}{2}h^{00}\left\{ \frac{2p_-}{\Rhat^2} 
	+ \frac{1}{\Rhat^4}\left[p_-(z^2-y^2) - \frac{1}{p_-}\left[(\dot x^A)^2 + ({x'}^A)^2\right]
	+ i\bar\theta^I\Gamma^-\partial_0\theta^I \right] \right\} \nonumber \\ 
& & 	+ \frac{i}{2\Rhat^4}s^{IJ} \bar\theta^I\Gamma^-\partial_1\theta^J + O(\Rhat^{-6})~,
\nonumber \\
\frac{\delta {\cal L}}{\delta {x'}^-} & = & 
	\frac{h^{01}p_-}{\Rhat^2} + \frac{h^{11}}{\Rhat^4}\left( -\frac{1}{p_-}\dot x^A {x'}^A
	+ \frac{i}{2}\bar\theta^I\Gamma^- \partial_1\theta^I \right)
	-\frac{i}{2\Rhat^4}s^{IJ}\bar\theta^I\Gamma^-\partial_0\theta^J+ O(\Rhat^{-6})~.
\nn\\
&&
\ee
It is obvious from these expressions that the $x^-$ equation of motion will not
be consistent with the standard choice of flat worldsheet metric 
($h^{00}=-h^{11}=1,h^{01}=0$). We therefore expand $h^{ab}$ in powers
of $\Rhat^{-1}$, taking it to be flat at leading order and allowing the higher-order 
terms (the $\tilde h^{ab}$) to depend on the physical variables in some way:
\be
h^{00}  =  -1 + \frac{\tilde h^{00}}{\Rhat^2} + O(\Rhat^{-4})~, &\qquad&
h^{11}  =  1 + \frac{\tilde h^{11}}{\Rhat^2} + O(\Rhat^{-4})~, 
\nn\\
&&\kern-80pt
	h^{01}  =  \frac{\tilde h^{01}}{\Rhat^2} + O(\Rhat^{-4})~. 
\ee
Using (\ref{a0}) and (\ref{a1}), we find that the specific metric choice
\be
\label{h004}
\tilde h^{00} & = & \frac{1}{2}(z^2-y^2) 
	- \frac{1}{2p_-^2}\left[(\dot x^A)^2 + ({x'}^A)^2\right] 
	+ \frac{i}{2p_-}\bar\theta^I\Gamma^-\partial_0\theta^I
	-\frac{i}{2p_-}s^{IJ}\bar\theta^I\Gamma^-\partial_1\theta^J~, \nn\\
&& \\
\label{h014}
\tilde h^{01} & = & \frac{1}{p_-^2}\dot x^A {x'}^A 
	- \frac{i}{2p_-}\bar\theta^I \Gamma^- \partial_1\theta^I
	+\frac{i}{2p_-}s^{IJ}\bar\theta^I\Gamma^-\partial_0\theta^J~
\ee
simplifies the expressions of (\ref{var1}) to
\be
\frac{\delta {\cal L}}{\delta \dot x^-} = 1 + O(\Rhat^{-4})~, \qquad
	\frac{\delta {\cal L}}{\delta {x'}^-} = O(\Rhat^{-4})~.
\ee
The $x^-$ equation of motion is then consistent with the standard 
lightcone gauge choice $\dot x^+ = p_-$ to $O(1/\Rhat^2)$
(with no corrections to $p_-$, which must remain constant).
Note that $\tilde h^{00} = -\tilde h_{00}$ and $\tilde h_{00} = \tilde h_{11}$.
The fact that these curvature corrections have bi-fermionic contributions is
ultimately due to the presence of a non-vanishing $G_{--}$ term in the expanded 
metric (\ref{expmetric2}).  

Since the worldsheet metric is known to $O(1/\Rhat^2)$, $x^-$ 
can now be determined to this order from the covariant gauge constraints
(\ref{Tab}).  By invoking the leading-order solutions (\ref{T00}, \ref{T01}), 
we can simplify the equations to some extent:
\be
\label{tcorr00}
T_{00} & = & \frac{1}{2}\left( L_0^\mu L_0^\mu + L_1^\mu L_1^\mu \right)
	+ \frac{\tilde h^{00}}{\Rhat^2}L_1^\mu L_1^\mu + O(\Rhat^{-3}) = 0~, \\
T_{01} & = & L_0^\mu L_1^\mu - \frac{\tilde h_{01}}{\Rhat^2}L_1^\mu L_1^\mu +O(\Rhat^{-3}) = 0~.
\ee
Equation (\ref{tcorr00}) may be expanded to solve for $a_2$, the first subleading
correction to $\dot x^-$:
\be
\label{a2}
T_{00} & = & 2p_- a_2 + a_0^2 - 2p_- y^2 a_0 + {a'}_0^2 + \frac{1}{2}(\dot z^2 z^2 - \dot y^2 y^2)
	+ \frac{p_-^2}{2}(y^4 - z^4) 
\nn\\
&&
\kern-30pt
	+ \frac{1}{2}({z'}^2 z^2 - {y'}^2 y^2) 
 	+ (z^2 - y^2)({x'}^A)^2
	-\frac{1}{p_-^2}\left[(\dot x^A)^2 + ({x'}^A)^2\right]({x'}^A)^2
\nn\\
&&
\kern-30pt
	+\frac{i}{p_-}({x'}^A)^2\bar\theta^I\Gamma^-\partial_0\theta^I
 	-\frac{i}{p_-}({x'}^A)^2 s^{IJ}\bar\theta^I\Gamma^-\partial_1\theta^J
	-ia_0 \Delta_0^- - 2ip_-\Delta_2^- 
\nn\\
&&
\kern-30pt
	- 2ip_-(\Delta_0^-)_{\theta^4}
	+\frac{ip_-}{2}(y^2-z^2)\Delta_0^- 
	- 2i(\dot x^A\Delta_1^A + {x'}^A\Delta_1^A)
	-ia'_0{\Delta'}_0^- = 0~.
\ee

The remaining independent component $T_{01}$ is the current associated with
translation symmetry on the closed-string worldsheet.  Enforcing the
constraint $T_{01} = 0$ is equivalent to imposing the level-matching condition
on physical string states.  This condition can be used to fix 
higher-order corrections to ${x'}^-$, as is required by conformal 
invariance on the worldsheet.  However, since our goal is to examine curvature
corrections to the pp-wave limit using first-order perturbation theory,
we will only need to enforce the level-matching condition on 
string states that are eigenstates of the pp-wave theory.
We therefore need only consider the equation $T_{01} = 0$ 
to leading order in the expansion, which yields (\ref{a1}) above.
If we were interested in physical eigenstates of the
geometry corrected to $O(1/\Rhat^2)$ (i.e.,~solving the theory exactly to
this order), we would be forced to solve $T_{01}=0$ to $O(1/\Rhat^2)$.

With solutions to the $x^-$ equations of motion and an expansion
of the worldsheet metric to the order of interest, we may proceed
with expressing the Hamiltonian as the generator of lightcone
time translation: $p_+ = \delta{\cal L}/\delta\dot x^+$.
It is helpful to first vary 
$\Delta^\mu$ with respect to $\partial_0 t$ and $\partial_0\phi$:
\be
\frac{\delta \Delta^\mu}{\delta(\partial_0 t)}
	& = & \bar\theta^I\Gamma^\mu\left[
	-\frac{1}{2\Rhat^3}z_j\Gamma^{0j}\theta^I 
	- \frac{1}{2}\epsilon^{IJ}\Pi\left(\frac{1}{\Rhat^2}+\frac{z^2}{2\Rhat^4}\right)\theta^J\right]
	+ O(\Rhat^{-6})~,
\\
\frac{\delta \Delta^\mu}{\delta(\partial_0 \phi)}
	& = & \bar\theta^I\Gamma^\mu\left[
	-\frac{1}{2\Rhat^3}y_{j'}\Gamma^{9j'}\theta^I - \frac{1}{2}\epsilon^{IJ}\Pi
	\left(\frac{1}{\Rhat^2} - \frac{y^2}{2\Rhat^4}\right)\theta^J\right]
	+ O(\Rhat^{-6})~.
\ee
The kinetic term in the Lagrangian (\ref{lagrangiank4}) yields
\be
\frac{\delta {\cal L}_{\rm Kin}}{\delta \dot x^+} & = & 
	\frac{1}{\Rhat^2}\left\{
	p_-(x^A)^2 - \dot x^- + i\Delta_0^- 
	- ip_-\bar\theta^I\Gamma^-\epsilon^{IJ}\Pi\theta^J \right\} 
\nonumber \\
& & 	+\frac{1}{\Rhat^4}\biggl\{-\frac{p_-}{2}(y^4-z^4) + y^2\dot x^- + i\Delta_2^-
	+ i(\Delta_0^-)_{\theta^4} + \frac{i}{4}(z^2-y^2)\Delta_0^-
\nonumber \\
& & 	-\frac{ip_-}{2}(z^2-y^2)\bar\theta^I\Gamma^-\epsilon^{IJ}\Pi\theta^J
	-\frac{ip_-}{12}\bar\theta^I\Gamma^-({\cal M}^2)^{IJ}\epsilon^{JL}\Pi\theta^L
\nonumber \\
& & 	+\frac{i}{4}\dot x^A\bar\theta^I\Gamma^A\left(z_k\Gamma^{-k}-y_{k'}\Gamma^{-k'}\right)\theta^I
	-\frac{i}{2}(\dot x^-)\bar\theta^I\Gamma^-\epsilon^{IJ}\Pi\theta^J
	+\biggl[ -\frac{1}{2}(z^2-y^2)
\nonumber \\
& & 	+\frac{1}{2p_-^2}\left[ (\dot x^A)^2 + ({x'}^A)^2\right]
	-\frac{i}{2p_-}\bar\theta^I\Gamma^-\partial_0\theta^I
	+\frac{i}{2p_-}s^{IJ}\bar\theta^I\Gamma^-\partial_1\theta^J\biggr]
	\biggl[ p_-(x^A)^2  
\nonumber \\
& & 	- \dot x^-  + i\Delta_0^- 
	- ip_-\bar\theta^I\Gamma^-\epsilon^{IJ}\Pi\theta^J\biggr]
 	+\biggl[\frac{1}{p_-^2}\dot x^A{x'}^A
	-\frac{i}{2p_-}\bar\theta^I\Gamma^-\partial_1\theta^I
\nonumber \\
& & 	+\frac{i}{2p_-}s^{IJ}\bar\theta^I\Gamma^-\partial_0\theta^J\biggr]
	\left({x'}^- - i{\Delta'}_0^-\right) \biggr\}+ O(\Rhat^{-6})~,
\ee
while the Wess-Zumino term (\ref{lagrangianwz4}) gives
\be
\frac{\delta {\cal L}_{\rm WZ}}{\delta \dot x^+} & = & 
	\frac{i}{\Rhat^2}s^{IJ}\bar\theta^I\Gamma^-\partial_1\theta^J
	+\frac{1}{\Rhat^4}\biggl\{
	\frac{i}{4}s^{IJ}\bar\theta^I\Gamma^-(z'_jz_k\Gamma^{jk}-y'_{j'}y_{k'})\theta^J
\nn\\
&&	+\frac{i}{12}s^{IJ}\bar\theta^I\Gamma^-({\cal M}^2)^{JL}\partial_1\theta^L
	-\frac{i}{4}(y^2-z^2)s^{IJ}\bar\theta^I\Gamma^-\partial_1\theta^J
\nn\\
&&	+\frac{i}{4}{x'}^A s^{IJ}\bar\theta^I\Gamma^A(y_{j'}\Gamma^{-j'}-z_j\Gamma^{-j})\theta^J
	\biggr\}+ O(\Rhat^{-6})~. 
\ee
The variation is completed prior to any gauge fixing (with the worldsheet metric
held fixed).  After computing the variation, 
the lightcone coordinates $x^\pm$ and the worldsheet metric corrections
$\tilde h^{00}, \tilde h^{01}$ are to be replaced with dynamical 
variables according to the 
$x^-$ equations of motion and the gauge conditions $x^+ = \tau$ and $T_{ab} = 0$.
Hence, using $a_0$ and $a_2$ determined from the
covariant gauge constraints (\ref{a0}, \ref{a2}), we remove
$x^-$ ($x^+$ has already been replaced with $p_-\tau$ in the above variations)
and restore proper powers of $\Rhat$ in the vielbeins 
(so that the desired corrections enter at $O(1/\Rhat^2)$). 
As expected, the pp-wave Hamiltonian emerges at leading order:
\be
{H}_{\rm pp} = 
	\frac{p_-}{2}(x^A)^2 + \frac{1}{2p_-}\left[(\dot x^A)^2 + ({x'}^A)^2\right]
	-ip_-\bar\theta^I\Gamma^-\epsilon^{IJ}\Pi\theta^J
	+is^{IJ}\bar\theta^I\Gamma^-\partial_1\theta^J~.
\ee
The first curvature correction to the pp-wave limit is found to be
\be
{H}_{\rm int} & = & 
	\frac{1}{\Rhat^2}
	\biggl\{
	\frac{1}{4p_-}\left[y^2(\dot z^2 - {z'}^2 - 2{y'}^2)+
	z^2(-\dot y^2 + {y'}^2+2{z'}^2) \right]
\nonumber \ee\be
& & 	+\frac{1}{8p_-^3}\left[3(\dot x^A)^2-({x'}^A)^2\right]
	\left[(\dot x^A)^2 + ({x'}^A)^2\right] 
	+\frac{p_-}{8}\left[(x^A)^2\right]^2
	-\frac{1}{2p_-^3}(\dot x^A{x'}^A)^2
\nonumber \\
& & 	-\frac{i}{4p_-}\sum_{a=0}^1\bar\theta^I (\partial_a x^A \Gamma^A )\epsilon^{IJ}\Gamma^-
	\Pi (\partial_a x^B \Gamma^B)\theta^J
	-\frac{i}{2}p_-(x^A)^2\bar\theta^I\Gamma^-\epsilon^{IJ}\Pi\theta^J
\nonumber \\
& & 	-\frac{i}{2p_-^2}(\dot x^A)^2\bar\theta^I\Gamma^-\partial_0\theta^I
	-\frac{ip_-}{12}\bar\theta^I\Gamma^-({\cal M}^2)^{IJ}\epsilon^{JL}\Pi\theta^L
	-\frac{p_-}{2}(\bar\theta^I\Gamma^-\epsilon^{IJ}\Pi\theta^J)^2
\nonumber \\
& & 	-\frac{i}{2p_-^2}(\dot x^A {x'}^A)s^{IJ}\bar\theta^I\Gamma^-\partial_0\theta^J
	-\frac{i}{4}(y^2-z^2)s^{IJ}\bar\theta^I\Gamma^-\partial_1\theta^J
\nonumber \\
& & 	+\frac{i}{4}{x'}^A s^{IJ}\bar\theta^I\Gamma^A(y_{j'}\Gamma^{-j'}-z_j\Gamma^{-j})\theta^J
	+\frac{i}{4}s^{IJ}\bar\theta^I\Gamma^-(z'_j z_k \Gamma^{jk} - y'_{j'}y_{k'}\Gamma^{j'k'})\theta^J
\nonumber \\
& & 	+\frac{i}{4p_-^2}\left[(\dot x^A)^2-({x'}^A)^2\right]s^{IJ}\bar\theta^I\Gamma^-\partial_1\theta^J
	+\frac{i}{12}s^{IJ}\bar\theta^I\Gamma^-({\cal M}^2)^{JL}\partial_1\theta^L
\nonumber \\
& & 	+\frac{1}{2}(s^{IJ}\bar\theta^I\Gamma^-\partial_1\theta^J)(\bar\theta^K\Gamma^-\epsilon^{KL}
	\Pi\theta^L) + \frac{i}{4}(x^A)^2 s^{IJ}\bar\theta^I\Gamma^-\partial_1\theta^J
	\biggr\}~.
\ee

The full Lagrangian (\ref{lagrangiank4}, \ref{lagrangianwz4}) 
can also be expressed to this order.  
In terms of the quantities found in equations
(\ref{L0L04}, \ref{L1L14}, \ref{L0L14}, \ref{h004}, \ref{h014}), 
the kinetic term ${\cal L}_{\rm Kin} = -\frac{1}{2}h^{ab}L_a^\mu L_b^\mu$ 
can be written schematically as
\be
\label{Lkin}
{\cal L}_{\rm Kin} & = & 
	\frac{1}{2} \left(L_0^\mu L_0^\mu - L_1^\mu L_1^\mu \right)_2
	+\frac{1}{2\Rhat^2} \left(L_0^\mu L_0^\mu - L_1^\mu L_1^\mu \right)_4
	- \frac{1}{2\Rhat^2}\tilde h^{00}\left(L_0^\mu L_0^\mu \right)_2 
\nn\\
& & 	+ \frac{1}{2\Rhat^2}\tilde h^{00}\left(L_1^\mu L_1^\mu \right)_2
	- \frac{1}{\Rhat^2}\tilde h^{01}\left(L_0^\mu L_1^\mu \right)_2+ O(\Rhat^{-4})~,
\ee
where external subscripts indicate quadratic or quartic order in fields.  
The Wess-Zumino term is given explicitly by:
\be
\label{WZfull}
{\cal L}_{\rm WZ} & = & 
	-2i\epsilon^{ab}\int_0^1 dt L_{at}^\mu s^{IJ}\bar\theta^I \Gamma^\mu L_{bt}^J 
\nonumber \\
& \approx &  -{ip_-}\left( s^{IJ}\bar\theta^I\Gamma^- \partial_1\theta^J\right) 
	- \frac{i}{\Rhat^2}\biggl\{ 
	p_-{\Box'}_2^- + p_-({\Box'}_0^-)_{\theta^4}
	+\frac{p_-}{4}(z^2-y^2){\Box'}_0^- 
\nonumber \\
& &  	+ \frac{1}{2}\dot x^- {\Box'}_0^-
	-\frac{1}{2}{x'}^-\Box_0^- + \dot x^A {\Box'}_1^A - {x'}^A\Box_1^A \biggr\}
	+ O(\Rhat^{-4})~.
\ee
It will be useful to recast both the Hamiltonian and Lagrangian 
in 16-component notation (details may be found in Appendix A):
\be
\label{hamiltonian164}
{H} & = &
	\frac{1}{2p_-} \left( (\dot x^A)^2
    + ({x'}^A)^2 + p_-^2(x^A)^2 \right) 
    -  p_- \psi^\dagger \Pi \psi
    + \frac{i}{2}( \psi {\psi'}
    + \psi^\dagger {\psi'}^\dagger )
\nonumber \\
& &     +\frac{1}{\Rhat^2}\biggl\{
     \frac{z^2}{4p_-} \left[{y'}^2
    + 2{z'}^2- \dot y^2 \right]
    - \frac{y^2}{4p_-} \left[{z'}^2 + 2{y'}^2
    - \dot z^2 \right]
	-\frac{1}{2p_-^3}(\dot x^A{x'}^A)^2
\nonumber \ee\be
& & 	+\frac{1}{8p_-^3}\left[3(\dot x^A)^2-({x'}^A)^2\right]
	\left[(\dot x^A)^2+({x'}^A)^2\right]
	+\frac{p_-}{8}\left[(x^A)^2\right]^2
\nonumber \\
& &     + \frac{i}{8} \psi \left(
    	z_k z'_j \gamma^{jk}
    	- y_{k'}  y'_{j'} \gamma^{j'k'}
    	+  {x'}^A (z_k \bar\gamma^{A}\gamma^k - y_{k'} \bar\gamma^{A}\gamma^{k'})
    	\right) \psi
\nonumber \\
& &     + \frac{i}{8} \psi^\dagger \left(z_k z'_j \gamma^{jk}
    	- y_{k'} y'_{j'} \gamma^{j'k'}
    	+  {x'}^A (z_k \bar\gamma^{A}\gamma^k - y_{k'} \bar\gamma^{A}\gamma^{k'})
    	\right) \psi^\dagger
\nn\\
&&	+\frac{1}{2p_-}\left( \dot z^i \dot y^{j'} + {z'}^i {y'}^{j'}\right)
	\psi^\dagger \gamma^{ij'}\Pi \psi
    + \frac{i}{8}(z^2 - y^2) (\psi \psi'
        + \psi^\dagger  {\psi'}^\dagger )
\nn\\
&&\kern00pt 	
 	+ \frac{i}{8}\left[
	\frac{1}{p_-^2}\left( (\dot x^A)^2 - ({x'}^A)^2\right) 
	+(x^A)^2\right](\psi\psi'+\psi^\dagger{\psi'}^\dagger)
\nn\\
&&	-\frac{p_-}{2}(x^A)^2(\psi^\dagger\Pi\psi)
 	- \frac{i}{4p_-^2}(\dot x^A{x'}^A)(\psi\dot\psi+\psi^\dagger\dot\psi^\dagger)
	+\frac{p_-}{48}
   	(\psi^\dagger \gamma^{jk} \psi )( \psi^\dagger \gamma^{jk} \psi)
\nn\\
&&	-\frac{p_-}{48}
    	(\psi^\dagger \gamma^{j'k'} \psi )( \psi^\dagger \gamma^{j'k'} \psi)
     - \frac{i}{192}(\psi\gamma^{jk} \psi
    	+ \psi^\dagger\gamma^{jk}\psi^\dagger)
    	(\psi^\dagger \gamma^{jk} \Pi\psi' - \psi \gamma^{jk} \Pi{\psi'}^\dagger)
\nn\\
&&	+\frac{p_-}{2}(\psi^\dagger\Pi\psi)(\psi^\dagger\Pi\psi)
     + \frac{i}{192}(\psi\gamma^{j'k'} \psi
    	+ \psi^\dagger\gamma^{j'k'}\psi^\dagger)
	(\psi^\dagger \gamma^{j'k'} \Pi\psi' - \psi \gamma^{j'k'} \Pi{\psi'}^\dagger)
\nn\\
&&\kern+00pt
	- \frac{i}{4p_-^2}(\dot x^A)^2\left[\psi\dot\psi^\dagger + \psi^\dagger\dot\psi\right]
	- \frac{i}{4}(\psi\psi'+\psi^\dagger{\psi'}^\dagger)(\psi^\dagger\Pi\psi)
\nn\\
&&	- \frac{1}{4p_-} \left[
    	(\dot z^2 - \dot y^2)
    	+ ({z'}^2 - {y'}^2 ) \right]
    	\psi^\dagger \Pi \psi
	\biggr\}+ O(\Rhat^{-4})~.
\ee
One could scale the length of the worldsheet such that all $p_-$ are absorbed into
the upper limit on worldsheet integration over $d\sigma$. To organize correction terms by their 
corresponding coupling strength in the gauge theory, however, we find it convenient to
keep factors of $p_-$ explicit in the above expression.  
The Lagrangian can be computed from (\ref{Lkin}, \ref{WZfull}), giving
\be
\label{Lkinfinal}
{\cal L}_{\rm Kin} & = & 
	p_-\dot x^- 
	- \frac{1}{2}\left[ {p_-^2} (x^A)^2 - (\dot x^A)^2 + ({x'}^A)^2 \right]
	- \frac{ip_-}{2}(\psi\dot\psi^\dag + \psi^\dag\dot\psi)
	- p_-^2 \psi\Pi\psi^\dag
\nn\\
& & 	+ \frac{1}{2\Rhat^2}\biggl\{
	(\dot x^-)^2 - 2p_-y^2 \dot x^- + \frac{1}{2}(\dot z^2 z^2 - \dot y^2 y^2)
	+ \frac{p_-^2}{2}(y^4 - z^4)
\nn\\
& & 	-\frac{ip_-}{4}(\dot z_j z_k)(\psi\gamma^{jk} \psi^\dag + \psi^\dag \gamma^{jk} \psi)
	+\frac{ip_-}{4}(\dot y_{j'} y_{k'})(\psi\gamma^{j'k'} \psi^\dag + \psi^\dag \gamma^{j'k'} \psi)
\nn\\
& & 	-\frac{ip_-}{48}(\psi\gamma^{jk}\psi^\dag)(\psi\gamma^{jk}\Pi\dot\psi^\dag 
		- \psi^\dag\gamma^{jk}\Pi\dot\psi)
\nn\\
&&	+\frac{ip_-}{48}(\psi\gamma^{j'k'}\psi^\dag)(\psi\gamma^{j'k'}\Pi\dot\psi^\dag 
		- \psi^\dag\gamma^{j'k'}\Pi\dot\psi)
\nn\\
& & 	+ \frac{i}{2}\left[ \frac{p_-}{2}(y^2 - z^2) - \dot x^-\right](\psi\dot\psi^\dag + \psi^\dag\dot\psi)
 	-p_-\left[2\dot x^- - p_-(y^2 - z^2)\right]\psi\Pi\psi^\dag
\nn\\
& & 	-\frac{p_-^2}{24}(\psi^\dag\gamma^{jk}\psi)^2
	+\frac{p_-^2}{24}(\psi^\dag\gamma^{j'k'}\psi)^2
	+\frac{ip_-}{4}(\dot x^A z_j)(\psi\gamma^A\bar\gamma^j\psi^\dag + \psi^\dag\gamma^A\bar\gamma^j\psi)
\nn\\
& & 	-\frac{ip_-}{4}(\dot x^A y_{j'})
		(\psi\gamma^A\bar\gamma^{j'}\psi^\dag + \psi^\dag\gamma^A\bar\gamma^{j'}\psi)
\nn\ee\be
&&	+\frac{1}{4}(\dot x^A\dot x^B)(\psi^\dag\gamma^A\Pi\bar\gamma^B\psi 
		- \psi\gamma^A\Pi\bar\gamma^B\psi^\dag )
\nn\\
& & 	-\frac{1}{2}({z'}^2z^2 - {y'}^2y^2)-({x'}^-)^2
	+\frac{i}{2}{x'}^-(\psi{\psi'}^\dag + \psi^\dag\psi')
\nn\\
& & 	-\frac{1}{4}({x'}^A{x'}^B)(\psi^\dag\gamma^A\Pi\bar\gamma^B\psi - \psi\gamma^A\Pi\bar\gamma^B\psi^\dag)
\nn\\
&&	-\tilde h^{00}\bigl[ 2p_-\dot x^- - p_-^2 (x^A)^2 + (\dot x^A)^2 - ({x'}^A)^2
	- ip_-(\psi\dot\psi^\dag + \psi^\dag\dot\psi)
	-2 p_-^2 \psi\Pi\psi^\dag \bigr] 
\nn\\ 
&&	-2\tilde h^{01}\bigl[ p_-{x'}^- + \dot x^A {x'}^A -\frac{ip_-}{2}(\psi{\psi'}^\dag + \psi^\dag\psi')
	\bigr] \biggr\}  + O(\Rhat^{-4})~,
\ee
and
\be
\label{Lwzfinal}
{\cal L}_{\rm WZ} & = & 
	-\frac{ip_-}{2}(\psi\psi' + \psi^\dag{\psi'}^\dag )
	-\frac{i}{\Rhat^2}\biggl\{
	\frac{p_-}{8}(z'_j z_k)(\psi\gamma^{jk}\psi + \psi^\dag\gamma^{jk}\psi^\dag)
\nn\\
& & 	-\frac{p_-}{8}(y'_{j'}y_{k'})(\psi\gamma^{j'k'}\psi + \psi^\dag\gamma^{j'k'}\psi^\dag)
	+\frac{1}{4}\left[ \dot x^- + \frac{p_-}{2}(z^2 - y^2)\right](\psi\psi' + \psi^\dag{\psi'}^\dag)
\nn\\
& & 	-\frac{1}{4}({x'}^-)(\psi\dot\psi + \psi^\dag\dot\psi^\dag )
	+ \frac{i}{8}({x'}^A \dot x^B + \dot x^A {x'}^B)(\psi^\dag\gamma^A\Pi\bar\gamma^B\psi^\dag
		- \psi\gamma^A\Pi\bar\gamma^B\psi )
\nn\\
& & 	+ \frac{p_-}{8}({x'}^A z_j )(\psi^\dag\gamma^A\bar\gamma^j\psi^\dag
		+ \psi\gamma^A\bar\gamma^j\psi )
	- \frac{p_-}{8}({x'}^A y_{j'} )(\psi^\dag\gamma^A\bar\gamma^{j'}\psi^\dag
		+ \psi\gamma^A\bar\gamma^{j'}\psi )
\nn\\
& & 	+ \frac{p_-}{8}(\psi\gamma^{jk}\psi + \psi^\dag\gamma^{jk}\psi^\dag)
		(\psi\gamma^{jk}\Pi{\psi'}^\dag - \psi^\dag\gamma^{jk}\Pi\psi') 
\nn\\
& & \kern-20pt
	- \frac{p_-}{8}(\psi\gamma^{j'k'}\psi + \psi^\dag\gamma^{j'k'}\psi^\dag)
		(\psi\gamma^{j'k'}\Pi{\psi'}^\dag - \psi^\dag\gamma^{j'k'}\Pi\psi') 
	\biggr\}+ O(\Rhat^{-4})~.
\ee
For later convenience, the Lagrangian is not fully gauge fixed, though we set $\dot x^+$
to $p_-$ for simplicity and ignore any $\ddot x^+$ that arise through
partial integration (since we will ultimately choose the lightcone gauge 
$x^+ = p_-\tau$). 
As noted above, sending $h^{00} \rightarrow -1 + {\tilde h^{00}}/{\Rhat^2}$
simply rewrites the function $h^{00}$, 
and does not amount to a particular gauge choice 
for the worldsheet metric.

\section{Quantization}
\label{twoimpSEC5}
Our goal is to calculate explicit energy corrections due to the
rather complicated perturbed Hamiltonian derived in the last section.
To explain our strategy, we begin with a review of the pp-wave 
energy spectrum in the Penrose limit. This limit is obtained by keeping
only the leading term in $\Rhat^{-1}$ in the Hamiltonian expansion of
(\ref{hamiltonian164}) and leads to linear equations of motion for the 
fields. The eight bosonic transverse string coordinates obey the equation
\begin{eqnarray}
{\ddot{x}}^A - {x''}^A + p_-^2 x^A = 0\ .
\end{eqnarray}
This is solved by the usual expansion in terms of Fourier modes
\begin{eqnarray}
x^A(\sigma,\tau) & = &
    \sum_{n=-\infty}^\infty x_n^A(\tau) e^{-i k_n \sigma}~,
\nonumber \\
\label{adef}
x_n^A(\tau) & = & \frac{i}{\sqrt{2 \omega_n}} (a_n^A e^{-i \omega_n \tau}
    - {a_{-n}^{A\dagger}} e^{i \omega_n \tau} )\ ,
\end{eqnarray}
where $k_n = n$ (integer), $\omega_n = \sqrt{ p_-^2 + k_n^2}$,
and the raising and lowering operators obey the commutation relation
$ [ a_m^A, {a_n^B}^\dagger ] = \delta_{mn}\delta^{AB}$.
The bosonic piece of the pp-wave Hamiltonian takes the form
\begin{eqnarray}
{H}_{\rm pp}^B & = & \frac{1}{p_-} \sum_{n=-\infty}^\infty \omega_n
    \left( {a_n^A}^\dagger a_n^A + 4 \right)\ .
\end{eqnarray}
The fermionic equations of motion are
\begin{eqnarray}
(\dot\psi^\dagger + \psi') + i{p_-}\Pi \psi^\dagger = 0~, \\
(\dot\psi + {\psi'}^\dagger) - i{p_-}\Pi \psi  = 0\ ,
\end{eqnarray}
where $\psi$ is a 16-component complex SO(9,1) Weyl spinor. As mentioned
earlier, $\psi$ is further restricted by a lightcone gauge fixing condition
$\bar\gamma^9\psi=\psi$ which reduces the number of spinor components to eight
(details are given in Appendix A). In what follows, $\psi$ and the various 
matrices acting on it should therefore be regarded as eight-dimensional. The
fermionic equations of motion are solved by
\begin{eqnarray}
\label{Fourierfermi}
\psi & = &
    \sum_{n=-\infty}^\infty \psi_n(\tau)  e^{-i k_n \sigma}~,
\\
\label{bdef1}
\psi_n(\tau) & = & 
	\frac{1}{2\sqrt{p_-}}\left( 
	A_n b_n e^{-i\omega_n\tau} + B_n b_{-n}^\dagger e^{i\omega_n\tau} \right)e^{-ik_n\sigma}~,
\\
\label{bdef2}
\psi_n^\dagger(\tau) & = & 
	\frac{1}{2\sqrt{p_-}}\left( 
	\Pi B_n b_n e^{-i\omega_n\tau} 
	- \Pi A_n b_{-n}^\dagger e^{i\omega_n\tau} \right) e^{-ik_n\sigma}~,
\end{eqnarray}
where we have defined
\begin{eqnarray}
\label{Amatdef}
A_n & \equiv & \frac{1}{\sqrt{\omega_n}}\left(\sqrt{\omega_n - k_n} - \sqrt{\omega_n + k_n}\Pi\right)~, \\
\label{Bmatdef}
B_n & \equiv & \frac{1}{\sqrt{\omega_n}}\left(\sqrt{\omega_n + k_n} + \sqrt{\omega_n - k_n}\Pi\right)~.
\end{eqnarray}
The anticommuting mode operators $b_n,~b_n^\dagger$ carry a spinor index that takes
eight values. In the gamma matrix representation described in Appendix A, the matrix 
$\Pi$ is diagonal and assigns eigenvalues $\pm 1$ to the mode operators.
The fermionic canonical momentum is $\rho = {ip_-}\psi^\dagger$, which implies that
the fermionic creation and annihilation operators obey the anticommutation rule
$\{ b_m^\alpha, {b_n^\beta}^\dagger \} =  \delta^{\alpha\beta}\delta_{mn}$.
The fermionic piece of the pp-wave Hamiltonian can be written in terms of
these operators as
\begin{eqnarray}
{H}_{\rm pp}^F & = & \frac{1}{p_-}\sum_{n=-\infty}^\infty
   \omega_n \left( b_n^{\alpha\dagger} b_n^\alpha - 4 \right)\ .
\end{eqnarray}
Given our earlier conventions, it is necessary to invoke 
the coordinate reflection $x^\mu \rightarrow -x^\mu$ 
(Metsaev studied a similar operation on the pp-wave Hamiltonian in
\cite{Metsaev:2001bj}).
Such a transformation is, at this stage, 
equivalent to sending $x^A \rightarrow -x^A$, $p_- \rightarrow -p_-$,
and ${H} \rightarrow -{H}$. 
In essence, this operation allows us to choose the positive-energy 
solutions to the fermionic equations of motion while maintaining
our convention that $b^{\alpha^\dag}$ represent a creation operator and $b^{\alpha}$
denote an annihilation operator. The total pp-wave Hamiltonian
\be
{H}_{\rm pp} = \frac{1}{p_-}\sum_{n=-\infty}^\infty
   \omega_n \left( {a_n^A}^\dagger a_n^A + b_n^{\alpha\dagger} b_n^\alpha  \right)\ 
\ee
is just a collection of free, equal mass fermionic and bosonic oscillators.

Canonical quantization requires that we express the Hamiltonian in terms
of physical variables and conjugate momenta. At leading order in $1/\Rhat^2$, $\dot x^A$ 
is canonically conjugate to $x^A$ and can be expanded in terms of creation and
annihilation operators.  Beyond leading order, however, the conjugate variable 
$p_A = \delta {\cal L} /\delta \dot x^A $ differs from $\dot x^A$ by terms
of $O(1/\Rhat^2)$. Substituting these $O(1/\Rhat^2)$ corrected 
expressions for canonical momenta into the pp-wave Hamiltonian
\be
{H}_{\rm pp} \sim (\dot x^A)^2 + \psi^\dagger \Pi \psi + \psi^\dagger {\psi'}^\dagger
\ee 
to express it as a function of canonical variables
will yield indirect $O(1/\Rhat^2)$ corrections to the Hamiltonian 
(to which we must add the contribution of explicit $O(1/\Rhat^2)$ 
corrections to the action). For example, bosonic momenta in 
the $SO(4)$ descending from the $AdS_5$ subspace
acquire the following corrections:
\be
p_k & = & \dot z_k + \frac{1}{\Rhat^2}\biggl\{
	\frac{1}{2} y^2 p_k + \frac{1}{2p_-^2}\left[ (p_A)^2 + ({x'}^A)^2 \right]p_k
 	- \frac{1}{p_-^2}(p_A {x'}^A ){z'}_k
	- \frac{i}{2p_-}p_k \bar\theta^I \Gamma^-\partial_0\theta^I
\nn\\
& & 	+ \frac{i}{2p_-}p_k s^{IJ}\bar\theta^I\Gamma^-\partial_1\theta^J
	- \frac{ip_-}{4}\bar\theta^I\Gamma^- z_j \Gamma_k^{\phantom{k}j}\theta^I
	- \frac{ip_-}{4}\bar\theta^I \Gamma^k \left(
		z_j\Gamma^{-j} - y_{j'}\Gamma^{-j'}\right)\theta^I
\nn\\
& & 	+ \frac{i}{4}p_A \epsilon^{IJ}\bar\theta^I\Gamma^- \left(
		\Gamma_k \Pi \Gamma^A  + \Gamma^A \Pi \Gamma_k \right ) \theta^J
  	+ \frac{i}{2p_-}{z'}_k\bar\theta^I\Gamma^-\partial_1 \theta^I
	- \frac{i}{2p_-}{z'}_k s^{IJ}\bar\theta^I\Gamma^-\partial_0\theta^J
\nn\\
& & 	+ \frac{i}{4}{x'}^A s^{IJ}\epsilon^{JK}\bar\theta^I\Gamma^-
		\left( \Gamma_k\Pi\Gamma^A - \Gamma^A \Pi \Gamma_k \right) \theta^K
	\biggr\}+O(\Rhat^{-4})~.
\ee
The leading-order relationship $p_k = \dot z_k$ has
been substituted into the correction term at $O(1/\Rhat^2)$, and the lightcone gauge choice 
$x^+ = p_-\tau$ has been fixed after the variation.

To compute fermionic momenta $\rho = {\delta {\cal L}}/{\delta \dot \psi}$,
it is convenient to work with complex 16-component spinors. 
Terms in ${\cal L}$ relevant to the fermionic momenta $\rho$ are as follows:
\be
\label{Lferm}
{\cal L} & \sim & -ip_- \left( \psi^\dagger \dot\psi \right)
	- \frac{i}{\Rhat^2}\biggl\{
	\frac{1}{4}\left[\dot x^- + \frac{p_-}{2}(z^2 - y^2)\right]
		\left(\psi\dot \psi^\dagger + \psi^\dagger\dot\psi\right)
\nn\\
&&	-\frac{p_-\tilde h^{00}}{2}\left(\psi\dot\psi^\dagger + \psi^\dagger\dot\psi\right)
	+ \frac{p_-}{96}\left(\psi\gamma^{jk}\psi^\dagger\right)
		\left(\psi\gamma^{jk}\Pi\dot\psi^\dagger - \psi^\dagger\gamma^{jk}\Pi\dot\psi\right)
\nn\\
&& 	-\frac{{x'}^-}{4}\left(\psi\dot\psi + \psi^\dagger\dot\psi^\dagger\right) 
	- (j,k \rightleftharpoons j',k')
	\biggr\}+O(\Rhat^{-4})~. 
\ee
This structure can be manipulated to simplify the
subsequent calculation.  Using partial integration,
we can make the following replacement at leading order:
\be
\label{IBP}
\frac{ip_-}{2}\left( \psi^\dagger \dot\psi + \psi\dot\psi^\dagger\right) = 
	ip_- \left(\psi^\dagger \dot \psi\right) + {\rm surface\ terms}.
\ee
Operations of this sort have no effect on the $x^-$ equation of motion or 
the preceding calculation of $\delta {\cal L} / \delta \dot x^+$, for example.
Similarly, terms in ${\cal L}$ containing the matrix $({\cal M})^2$ 
may be transformed according to
\be
-\frac{ip_-}{96} \left(\psi\gamma^{jk}\psi^\dagger\right)\left(\psi\gamma^{jk}\Pi\dot\psi^\dagger
	- \psi^\dagger\gamma^{jk}\Pi\dot\psi\right) 
 =  	\frac{ip_-}{48}
	\left(\psi\gamma^{jk}\psi^\dagger\right)
		\left(\psi^\dagger\gamma^{jk}\Pi\dot\psi\right)~.
\ee
Terms of the form
\be
\label{fierz}
\frac{1}{4}\left(\dot x^- \right)
		\left(\psi\dot \psi^\dagger + \psi^\dagger\dot\psi\right)~,
\ee
however,
cannot be treated in the same manner.  
The presence of (\ref{fierz}) ultimately 
imposes a set of second-class constraints on the theory,
and we will eventually be led to treat $\psi^\dagger$ 
as a constrained, dynamical degree of freedom in the Lagrangian.
The fermionic momenta therefore take the form
\be
\rho_\alpha & = & ip_-\psi^\dagger_\alpha + \frac{1}{\Rhat^2}\biggl\{
	\frac{i}{4}\left(\dot x^- + \frac{p_-}{2}(z^2 - y^2)\right)\psi^\dagger_\alpha
	- \frac{ip_-}{2}\tilde h^{00} \psi^\dagger_\alpha 
	- \frac{i{x'}^-}{4}\psi_\alpha 
\nn\\
& & 	- \frac{ip_-}{48}\left[ \left(\psi\gamma^{jk}\psi^\dagger\right)
		\left(\psi^\dagger\gamma^{jk}\Pi \right)_\alpha 
	- (j,k \rightleftharpoons j',k')\right]
	\biggr\}+O(\Rhat^{-4})~, 
\\
\rho^\dagger_\alpha & = & 
	\frac{1}{\Rhat^2}\biggl\{
	\frac{i}{4}\left(\dot x^- + \frac{p_-}{2}(z^2 - y^2)\right)\psi_\alpha
	- \frac{ip_-}{2}\tilde h^{00} \psi_\alpha 
 	- \frac{i{x'}^-}{4}\psi^\dagger_\alpha 
	\biggr\}+O(\Rhat^{-4})~.
\ee
Using (\ref{a0}) and (\ref{a1}) to replace $\dot x^-$ and ${x'}^-$ at leading order
(in 16-component spinor notation), and using (\ref{h004}) to implement the appropriate
curvature corrections to the $h^{00}$ component of the worldsheet metric,
we find
\be
\label{rhoeqn}
\rho	& = & 	ip_- \psi^\dagger + \frac{1}{\Rhat^2}\biggl\{
	\frac{1}{4}y^2\rho + \frac{1}{8p_-^2}\left[ (p_A^2) + ({x'}^A)^2\right] \rho
	+ \frac{i}{4p_-}(p_A {x'}^A )\psi
	+ \frac{i}{4p_-}\left( \rho\Pi\psi \right) \rho 
\nn\\
& & 	- \frac{i}{8p_-}\left( \psi\rho' + \rho\psi' \right)\psi
	+ \frac{i}{8p_-}\left(\psi\psi' - \frac{1}{p_-^2}\rho\rho'\right)\rho
\nn\\
& &  	+ \frac{i}{48p_-}\left[ \left(\psi\gamma^{jk}\rho\right)\left(\rho\gamma^{jk}\Pi\right)
	- (j,k, \rightleftharpoons j',k') \right] 
	\biggr\}+O(\Rhat^{-4})~,\ 
\\
\rho^\dagger & = & \frac{1}{\Rhat^2}\biggl\{
	\frac{i}{4}p_- y^2\psi + \frac{i}{8p_-}\left[ (p_A^2) + ({x'}^A)^2 \right]\psi
	+ \frac{1}{4p_-^2}\left( p_A {x'}^A \right)\rho
	-\frac{1}{4}\left( \rho\Pi\psi \right)\psi
\nn\\
& & 	-\frac{1}{8p_-^2}\left(\psi\rho' + \rho\psi' \right)\rho
	- \frac{1}{8}\left(\psi\psi' - \frac{1}{p_-^2}\rho\rho'\right)\psi
	\biggr\}+O(\Rhat^{-4})~.
\ee
Denoting the $O(1/\Rhat^2)$ corrections to $\rho$ in (\ref{rhoeqn}) by $\Phi$, 
the pp-wave Hamiltonian can be expressed in terms of canonical variables as
\be
{H}_{\rm pp} & =  & - p_- \psi^\dagger \Pi \psi + \frac{i}{2}\psi\psi' 
	+ \frac{i}{2}\psi^\dagger {\psi'}^\dagger
\nn\\
&&\kern-35pt
	 =  i\rho\Pi\psi + \frac{i}{2}\psi\psi' - \frac{i}{2p_-^2}\rho\rho'
	+ \frac{1}{\Rhat^2}\left\{ 
	\frac{i}{2p_-^2}\rho \Phi' + \frac{i}{2p_-^2}\Phi \rho' 
	- i \Phi \Pi \psi  \right\}~.
\ee
The $O(1/\Rhat^2)$ correction to the Hamiltonian can also be expressed in terms
of canonical variables. The overall canonical Hamiltonian can conveniently be broken
into its BMN limit $({H}_{\rm pp})$, pure bosonic $({H}_{\rm BB})$,
pure fermionic $({H}_{\rm FF})$ and boson-fermion $({H}_{\rm BF})$
interacting subsectors:
\be
\label{Hppwave}
{H}_{\rm pp} & = & 
	\frac{p_-}{2}(x^A)^2 + \frac{1}{2p_-}\left[(p_A)^2 + ({x'}^A)^2\right]
	+ {i}\rho\Pi\psi + \frac{i}{2}\psi\psi' - \frac{i}{2p_-^2} \rho \rho'~,
\nn\\
&&
\\
\label{Hpurbos}
{H}_{\rm BB} & = & \frac{1}{\Rhat^2}\biggl\{
	\frac{1}{4p_-}\left[ -y^2\left( p_z^2 + {z'}^2 + 2{y'}^2\right)
	+ z^2\left( p_{y}^2 + {y'}^2 + 2{z'}^2 \right)\right]
	+ \frac{p_-}{8}\left[ (x^A)^2 \right]^2
\nn\\
& & \kern-00pt
	- \frac{1}{8p_-^3}\left\{  \left[ (p_A)^2\right]^2 + 2(p_A)^2({x'}^A)^2 
	+ \left[ ({x'}^A)^2\right]^2 \right\}
	 + \frac{1}{2p_-^3}\left({x'}^A p_A\right)^2
	\biggr\}~,
\\
\label{hbfeqn}
{H}_{\rm FF} & = & -\frac{1}{4\Rhat^2}\biggl\{
	\frac{1}{p_-}\left(\rho\Pi\psi\right)^2
	+ \frac{1}{p_-^3}\left(\rho\Pi\psi\right)\rho\rho'
 	+ \frac{1}{2p_-^3}\left(\psi\psi' - \frac{1}{p_-^2}\rho\rho'\right)\rho\rho'
\nn\\
& & 	+\frac{1}{2p_-}\left(\psi\psi' - \frac{1}{p_-^2}\rho\rho'\right)(\rho\Pi\psi)
	+ \frac{1}{2p_-^3}\left(\psi\rho' + \rho\psi'\right)\rho'\psi
\nn\\
& & 	+ \frac{1}{12p_-^3}\left(\psi\gamma^{jk}\rho\right)
		\left(\rho\gamma^{jk}\Pi\rho'\right)
\nn\ee\be
& & 	- \frac{1}{48p_-}\left(\psi\gamma^{jk}\psi - \frac{1}{p_-^2}\rho\gamma^{jk}\rho\right)
		\left(\rho'\gamma^{jk}\Pi\psi - \rho\gamma^{jk}\Pi\psi'\right)
	- (j,k \rightleftharpoons j',k')
	\biggr\}~,	\nn\\
&&
\\
\label{hffeqn}
{H}_{\rm BF} & = & 
	\frac{1}{\Rhat^2}\biggl\{
	\frac{i}{4}z^2 \psi\psi' -\frac{i}{8p_-^2}\left[(p_A)^2 + ({x'}^A)^2\right]\psi\psi'
\nn\\
&&	+\frac{i}{4p_-^4}\left[(p_A)^2 + ({x'}^A)^2 + p_-^2(y^2 - z^2)\right]\rho\rho'
\nn\\
& & 	-\frac{i}{2p_-^2}\left( p_k^2 + {y'}^2 - p_-^2 z^2
		-\frac{1}{4}(p_A)^2 - \frac{1}{4}({x'}^A)^2
		-\frac{p_-^2}{2}y^2 \right)\rho\Pi\psi
\nn\\
& & 	+\frac{i}{4}(z'_j z_k)\left(\psi\gamma^{jk}\psi - \frac{1}{p_-^2}\rho\gamma^{jk}\rho\right)
	-\frac{i}{4}(y'_{j'} y_{k'})\left(\psi\gamma^{j'k'}\psi - \frac{1}{p_-^2}\rho\gamma^{j'k'}\rho\right)
\nn\\
& & 	-\frac{i}{8}(z'_k y_{k'} + z_k y'_{k'})
		\left(\psi\gamma^{kk'}\psi - \frac{1}{p_-^2}\rho\gamma^{kk'}\rho\right)
	+\frac{1}{4p_-}(p_k y_{k'} + z_k p_{k'} )\psi\gamma^{kk'}\rho
\nn\\
& & 	+\frac{1}{4p_-}(p_j z'_k)\left(\psi\gamma^{jk}\Pi\psi 
		+ \frac{1}{p_-^2}\rho\gamma^{jk}\Pi\rho\right)
\nn\\
&&	-\frac{1}{4p_-}(p_{j'} y'_{k'})\left(\psi\gamma^{j'k'}\Pi\psi 
		+ \frac{1}{p_-^2}\rho\gamma^{j'k'}\Pi\rho\right)
\nn\\
& & 	-\frac{1}{4p_-}(p_k y'_{k'} + z'_k p_{k'})
		\left(\psi\gamma^{kk'}\Pi\psi + \frac{1}{p_-^2}\rho\gamma^{kk'}\Pi\rho\right)
\nn\\
& & 	-\frac{1}{4p_-^3}(p_A{x'}^A)(\rho\psi' + 2 \psi\rho' )
	-\frac{i}{2p_-^2}(p_kp_{k'} - z'_k y'_{k'})\psi\gamma^{kk'}\Pi\rho
	\biggr\}~.
\ee

This Hamiltonian has one problem that we must resolve before attempting 
to extract its detailed consequences. At the end of Section~\ref{twoimpSEC2}, we argued that
when the theory is restricted to the subspace of string zero-modes (i.e.,~excitations 
of the string that are independent of the worldsheet coordinate 
$\sigma$), curvature corrections to the leading pp-wave Hamiltonian should 
vanish. The only terms in the Hamiltonian that survive in this limit are
those with no worldsheet spatial derivatives. Although ${H}_{\rm BB}$ has 
no such terms, the fermionic pieces of the Hamiltonian do. For example,
${H}_{\rm FF}$ contains a term $\Rhat^{-2}(\rho\Pi\psi)^2$ that would appear
to modify the zero-mode spectrum at $O(1/\Rhat^2)$, contrary to 
expectation. In the end, we found that this problem can be traced to the 
presence of second-class constraints involving $\dot\psi^\dag$. As it turns
out, the constrained quantization procedure needed to handle second-class
constraints has the effect, among many others, of resolving the zero-mode
paradox just outlined. To see this, we must work out the appropriate 
constrained quantization procedure. 

The set of constraints that define canonical momenta are
known as primary constraints, and take the generic form $\chi = 0$. 
Primary constraints can be categorized as either first or
second class.  Second-class constraints arise when canonical momenta
do not have vanishing Poisson brackets with the primary
constraints themselves: $\left\{ \rho_\psi,\chi_\psi \right\} \neq
0$, $\left\{ \rho_{\psi^\dagger},\chi_{\psi^\dagger} \right\} \neq
0$. (First-class constraints are characterized by the more typical
condition $\left\{ \rho_{\psi^\dagger},\chi_{\psi^\dagger}\right\}
= \left\{ \rho_\psi,\chi_\psi \right\} = 0$.)
To the order of interest, the primary constraint equations are
\be
\label{con1}
\chi_\alpha^1 & = 0 = & \rho_\alpha - ip_-\psi^\dagger_\alpha
\nn\\
& & 	- \frac{ip_-}{8\Rhat^2}\biggl[
		2y^2 + \frac{1}{p_-^2}\left[ (p_A)^2 + ({x'}^A)^2\right]
		- 2(\psi^\dagger\Pi\psi) 
		+ \frac{i}{p_-}(\psi\psi' + \psi^\dagger{\psi'}^\dagger)
	\biggr] \psi_\alpha^\dagger
\nn\\
& & 	-\frac{i}{4p_-\Rhat^2}\biggl[
		(p_A {x'}^A) - \frac{ip_-}{2}(\psi{\psi'}^\dagger + \psi^\dagger\psi')
	\biggr]\psi_\alpha
	+ \frac{ip_-}{48\Rhat^2}(\psi\gamma^{jk}\psi^\dagger)(\psi^\dagger\gamma^{jk}\Pi)_\alpha~, \nn\\
&&
\\
\label{con2}
\chi_\alpha^2 & = 0 = & \rho^\dagger_\alpha
	-\frac{i}{4p_-\Rhat^2}\biggl[
		(p_A{x'}^A)-\frac{ip_-}{2}(\psi{\psi'}^\dagger + \psi^\dagger\psi')
	\biggr]\psi^\dagger_\alpha
\nn\\
& & 	- \frac{ip_-}{4\Rhat^2}\biggl[
		y^2 + \frac{1}{2p_-^2}\left[(p_A)^2 + ({x'}^A)^2 \right] - (\psi^\dagger\Pi\psi)
		+ \frac{i}{2p_-}(\psi\psi' + \psi^\dagger{\psi'}^\dagger)
	\biggr]\psi_\alpha\ .
\nn\\
&&
\ee
It is clear that these constraints are second-class. In the presence of
second-class constraints, consistent quantization requires that
the quantum anticommutator of two fermionic fields be identified with their
Dirac bracket (which depends on the Poisson bracket algebra of the
constraints) rather than with their classical Poisson bracket.
The Dirac bracket is given in terms of Poisson brackets by 
(see, for example, \cite{Weinberg:1995mt}) 
\be
\{A,B\}_{\rm D} = \{A,B\}_{\rm P} - \{A,\chi_N\}_{\rm P} \left( C^{-1} \right)^{NM}
	\{\chi_M, B\}_{\rm P}~,
\ee
where
\be
C_{NM} \equiv \{\chi_N,\chi_M\}_{\rm P}\ .
\ee
The indices $N$ and $M$ denote both the spinor index $\alpha$ 
and the constraint label $a = 1,2$.
For Grassmanian fields $A$ and $B$, the Poisson bracket is defined by
\be
\{A,B\}_{\rm P} = -\left( \frac{\partial A}{\partial\psi^\alpha}\frac{\partial B}{\partial \rho_\alpha}
	+ \frac{\partial B}{\partial\psi^\alpha}\frac{\partial A}{\partial \rho_\alpha}\right) 
	-\left( \frac{\partial A}{\partial\psi^{\dagger\alpha} }\frac{\partial B}{\partial \rho^\dagger_\alpha}
	+ \frac{\partial B}{\partial\psi^{\dagger\alpha}}\frac{\partial A}{\partial \rho^\dagger_\alpha}\right)~.
\ee

As an example, the Dirac bracket $\{\rho_\alpha,\rho_\beta\}_{\rm D}$ 
is readily computed (to the order of interest) by noting
that the partial integration in (\ref{IBP}) 
introduces an asymmetry between $\psi$ and $\psi^\dag$
into the system.  Since  $\{\rho_\alpha,\rho_\beta\}_{\rm D}$ contains
\be
\{\rho_\alpha, \chi_{a\gamma} \} = O(\Rhat^{-2})~, \qquad 
	\{\chi_{b\eta},\rho_\beta \} = O(\Rhat^{-2})~,
\ee
an immediate consequence of this asymmetry is that
$\{\rho_\alpha,\rho_\beta\}_{\rm D}$ vanishes to $O(1/\Rhat^4)$.   
To compute $\{ \rho_\alpha, \psi_\beta \}_{\rm D}$, we note that 
\be
\{ \rho_\alpha, \chi_{(2\gamma)} \}_{\rm P} & = & -\delta_{\alpha\rho}
			\frac{\partial\chi_{(2\gamma)}}{\partial\psi_\rho}
	 =  O(\Rhat^{-2})~,
\ee
and, to leading order,
\be
(C^{-1})^{(2\gamma)(1\eta)} = -\frac{i}{p_-}\delta_{\gamma\eta} + O(\Rhat^{-2})~,
\ee
such that
\be
\{ \rho_\alpha, \psi_\beta \}_{\rm D} & = & 
	-\delta_{\alpha\beta} -\frac{i}{p_-}\{ \rho_\alpha, \chi_{(2\beta)}\}_{\rm P}~.
\ee
Similar manipulations are required for $\{ \psi_\alpha, \psi_\beta \}_{\rm D}$,
which does exhibit $O(1/\Rhat^2)$ corrections.
The second-class constraints on the fermionic sector of the system are removed
by enforcing
\be
\label{Dirac1}
\{ \rho_\alpha(\sigma),\psi_\beta(\sigma') \}_{\rm D} & = & 
	-\delta_{\alpha\beta}\delta(\sigma - \sigma')
	+ \frac{1}{4\Rhat^2}\delta(\sigma - \sigma')\biggl\{
	\frac{-i}{p_-}(\rho \Pi)_\alpha\psi_\beta 
	+ \frac{i}{p_-}(\rho\Pi\psi)\delta_{\alpha\beta}
\nn\\
& & 	+ \frac{i}{2p_-}\left[
		\left(\psi\psi'\delta_{\alpha\beta} 
		- \frac{1}{p_-}^2\rho\rho'\delta_{\alpha\beta}\right)
		+ \psi'_\alpha\psi_\beta
		+ \frac{1}{p_-^2}\rho'_\alpha\rho_\beta \right]
\nn\\
& & 	+ \frac{1}{2p_-^2}\left[ (p_A)^2 + ({x'}^A)^2\right]\delta_{\alpha\beta}
	+ y^2\delta_{\alpha\beta} \biggr\}
\nn\\
& & \kern-30pt
	-\frac{i}{8p_-\Rhat^2}\left( \psi_\alpha\psi_\beta 
	+ \frac{1}{p_-^2}\rho_\alpha\rho_\beta\right)
	\frac{\partial}{\partial\sigma'}\delta(\sigma-\sigma')+O(\Rhat^{-4})~,
\\
\label{Dirac2}
\{ \psi_\alpha(\sigma),\psi_\beta(\sigma') \}_{\rm D} & = & 
	\frac{i}{4p_-\Rhat^2}\delta(\sigma-\sigma')
	\biggl\{
	(\psi\Pi)_{(\alpha}\psi_{\beta)} 
	-\frac{1}{p_-^2}(p_A{x'}^A)\delta_{(\alpha\beta)}
\nn\\
& & 	+ \frac{1}{2p_-^2}\left[
		\psi'_{(\alpha}\rho_{\beta)} - \rho'_{(\alpha}\psi_{\beta)}
		+ (\psi\rho' + \rho\psi')\delta_{(\alpha\beta)}\right]
	\biggr\}
\nn\\
& & 	+ \frac{i}{8p_-^3\Rhat^2}\left( \rho_{(\alpha}\psi_{\beta)} 
	- \psi_{(\alpha}\rho_{\beta)}\right)
	\frac{\partial}{\partial\sigma'}\delta(\sigma-\sigma')+O(\Rhat^{-4})~,
\\
\label{Dirac3}
\{\rho_\alpha(\sigma),\rho_\beta(\sigma')\}_{\rm D} & = & O(\Rhat^{-4})~.
\ee
Identifying these Dirac brackets with the quantum anticommutators of the
fermionic fields in the theory naturally leads to additional 
$O(1/\Rhat^2)$ corrections to the energy spectrum.  One way to implement
these corrections is to retain the Fourier expansion of $\psi$ and $\psi^\dag$ 
given in (\ref{bdef1}, \ref{bdef2}) while transforming the fermionic 
creation and annihilation operators
\be
b_n^\alpha \to c_n^\alpha~, \qquad\qquad
b_n^{\dag\alpha} \to c_n^{\dag\alpha}\ ,
\ee
such that 
$\{ \rho(c, c^\dag), \psi(c, c^\dag) \}_{\rm P}$,
for example, satisfies (\ref{Dirac1}).  This approach amounts to 
finding $O(1/\Rhat^2)$ corrections to
$\{ c_n^\alpha , c_m^{\dag\beta} \}$ that allow the
usual anticommutators to be identified with the 
above Dirac brackets (\ref{Dirac1}-\ref{Dirac3}).  
In practice, extracting these solutions from
(\ref{Dirac1}-\ref{Dirac3}) can be circumvented by 
invoking a non-linear field redefinition 
$\psi \rightarrow \tilde\psi,\ \rho \rightarrow \tilde\rho$,
such that 
\be
\{ \rho(c,c^\dag),\psi(c,c^\dag) \}_{\rm P}
= \{ \tilde\rho(b,b^\dag),\tilde\psi(b,b^\dag) \}_{\rm P}~.
\ee
Both representations satisfy (\ref{Dirac1}), and the operators
$b_n^\alpha,b_m^{\dag\beta}$ are understood to obey the usual relations:
\be
\{ b_n^\alpha, b_m^{\dag\beta} \} = \delta^{\alpha\beta}\delta_{nm}~.
\ee
In general, the non-linear field redefinition 
$\tilde \psi(b,b^\dag) = \psi(b,b^\dag) + \dots $ 
contains corrections that are cubic in the 
fields $\rho(b,b^\dag)$, $\psi(b,b^\dag)$, $x^A(a,a^\dag)$ and $p_A(a,a^\dag)$.
Such correction terms can be written down by inspection, with matrix-valued 
coefficients to be solved for by comparing 
$\{ \tilde\rho(b,b^\dag),\tilde\psi(b,b^\dag) \}_{\rm P}$
and $\{ \tilde\psi(b,b^\dag),\tilde\psi(b,b^\dag) \}_{\rm P}$
with (\ref{Dirac1}, \ref{Dirac2}).  A straightforward computation yields
\be
\label{redef1}
\rho_\alpha \rightarrow \tilde \rho_\alpha & = & \rho_\alpha~, \\
\label{redef2}
\psi_\beta \rightarrow \tilde \psi_\beta & = & \psi_\beta 
	+\frac{i}{8p_-\Rhat^2}\biggl\{
	(\psi'\psi)\psi_\beta 
	- 2(\rho\Pi\psi)\psi_\beta
	-\frac{1}{p_-^2}(\rho'\rho)\psi_\beta
	+ \frac{2}{p_-^2}(p_A {x'}^A)\rho_\beta
\nn\\
& & 	+\frac{1}{p_-^2}\left[ (\rho'\psi)\rho_\beta - (\rho\psi')\rho_\beta\right]
	+2ip_-\left[ y^2\psi_\beta 
	+ \frac{1}{2p_-^2}\left( (p_A)^2 + ({x'}^A)^2 \right)\psi_\beta\right]
	\biggr\}~.
\nn\\
& & 
\ee
This approach to enforcing the modified Dirac bracket structure amounts 
to adding $O(1/\Rhat^2)$ correction terms to the Hamiltonian while keeping 
the standard commutation relations. It is much more convenient for calculating
matrix elements than the alternative approach of adding $O(1/\Rhat^2)$ 
operator corrections to the fermi field anticommutators $\{ b, b^\dag \}$.

By invoking the redefinitions in (\ref{redef1}, \ref{redef2}), the pieces
of the interaction Hamiltonian that involve fermions take the final forms
\be
\label{Hpurferm}
{H}_{\rm FF} & = & 
	-\frac{1}{4p_-^3 \Rhat^2}\biggl\{
	p_-^2\left[ (\psi'\psi) + \frac{1}{p_-^2}(\rho\rho')\right](\rho\Pi\psi)
	-\frac{p_-^2}{2}(\psi'\psi)^2  
	+ (\psi'\psi)(\rho'\rho)
\nn\\
& & 	- \frac{1}{2p_-^2}(\rho'\rho)^2
	+ (\rho\psi')(\rho'\psi)
	-\frac{1}{2}\left[ (\psi\rho')(\psi\rho') + (\psi'\rho)^2\right]
	+ \frac{1}{12}(\psi\gamma^{jk}\rho)(\rho\gamma^{jk}\Pi\rho')
\nn\\
& & 	-\frac{p_-^2}{48}
		\left(\psi\gamma^{jk}\psi - \frac{1}{p_-^2}\rho\gamma^{jk}\rho\right)
	\left(\rho'\gamma^{jk}\Pi\psi - \rho\gamma^{jk}\Pi\psi'\right)
	- (j,k \rightleftharpoons j',k')
	\biggr\}~,
\\
\label{Hmix}
{H}_{\rm BF} & = & 
	\frac{1}{\Rhat^2}\biggl\{
	-\frac{i}{4p_-^2}\left[
	(p_A)^2+({x'}^A)^2 + p_-^2(y^2 - z^2)\right]\left(\psi\psi'-\frac{1}{p_-^2}\rho\rho'\right)
\nn\\
& & 	-\frac{1}{2p_-^3}(p_A{x'}^A)(\rho\psi' + \psi\rho' )
	-\frac{i}{2p_-^2}\left( p_k^2 + {y'}^2 - p_-^2 z^2 \right)\rho\Pi\psi
\nn\\
& & 	+\frac{i}{4}(z'_j z_k)\left(\psi\gamma^{jk}\psi - \frac{1}{p_-^2}\rho\gamma^{jk}\rho\right)
	-\frac{i}{4}(y'_{j'} y_{k'})\left(\psi\gamma^{j'k'}\psi - \frac{1}{p_-^2}\rho\gamma^{j'k'}\rho\right)
\nn\\
& & 	-\frac{i}{8}(z'_k y_{k'} + z_k y'_{k'})
		\left(\psi\gamma^{kk'}\psi - \frac{1}{p_-^2}\rho\gamma^{kk'}\rho\right)
	+\frac{1}{4p_-}(p_k y_{k'} +  z_k p_{k'} )\psi\gamma^{kk'}\rho
\nn\\
& & 	+\frac{1}{4p_-}(p_j z'_k)\left(\psi\gamma^{jk}\Pi\psi 
		+ \frac{1}{p_-^2}\rho\gamma^{jk}\Pi\rho\right)
\nn\ee\be
&&	-\frac{1}{4p_-}(p_{j'} y'_{k'})\left(\psi\gamma^{j'k'}\Pi\psi 
		+ \frac{1}{p_-^2}\rho\gamma^{j'k'}\Pi\rho\right)
\nn\\
& & 	-\frac{1}{4p_-}(p_k y'_{k'} + z'_k p_{k'})
		\left(\psi\gamma^{kk'}\Pi\psi + \frac{1}{p_-^2}\rho\gamma^{kk'}\Pi\rho\right)
\nn\\
&&	-\frac{i}{2p_-^2}(p_kp_{k'} - z'_k y'_{k'})\psi\gamma^{kk'}\Pi\rho
	\biggr\}~.
\ee
The full Hamiltonian is the sum of these two terms plus the bosonic interaction term 
${H}_{\rm BB}$ (\ref{Hpurbos}) and the free Hamiltonian ${H}_{\rm pp}$ (\ref{Hppwave}). 
This system is quantized by imposing the standard (anti)commutator algebra for $x^A,\psi$ 
and their conjugate variables $p^A,\rho$. This will be done by expanding the field variables 
in creation and annihilation operators in a standard way.

Returning to the phenomenon that led us to explore second-class constraints in the first 
place, note that (\ref{Hpurferm}) manifestly vanishes on the subspace of string
zero-modes because all terms have at least one worldsheet spatial derivative.  
The bose-fermi mixing Hamiltonian (\ref{Hmix}) still has terms that can lead to 
curvature corrections to the string zero-mode energies, but their net effect 
vanishes by virtue of nontrivial cancellations between terms that split $SO(4)\times SO(4)$ 
indices and terms that span the entire $SO(8)$. How this comes about will be seen when we 
actually compute matrix elements of this Hamiltonian.

\section{Energy spectrum}
\label{twoimpSEC6}
To compute the energy spectrum correct to first order in $O(\Rhat^{-1})$, we
will do degenerate first-order perturbation theory on the Fock space of 
eigenstates of the free Hamiltonian ${H}_{\rm pp}$. The
degenerate subspaces of the BMN theory are spanned by fixed numbers of
creation operators with specified mode indices (subject to the 
constraint that the mode indices sum up to zero) acting on the ground 
state $\ket{J}$, where $J=p_-\Rhat^2$ is the angular momentum (assumed large)
of the string center of mass in its motion around the 
equator of the $S^5$. In this chapter we
restrict attention to ``two-impurity states'' generated by pairs of
creation operators of equal and opposite mode number. For each positive
mode number $n$, the 16 bosonic and fermionic creation operators can be combined 
in pairs to form the following 256 degenerate two-impurity states:
\be
\label{2impbasis}
a_n^{A\dagger} a_{-n}^{B\dagger} \ket{J}~, \qquad b_{n}^{\alpha\dagger}
	b_{-n}^{\beta\dagger}\ket{J}~,
\qquad a_n^{A\dagger} b_{-n}^{\alpha\dagger}\ket{J}~, \qquad
	a_{-n}^{A\dagger} b_{n}^{\alpha\dagger}\ket{J}~.
\ee
The creation operators are classified under the residual $SO(4)\times SO(4)$ 
symmetry to which the isometry group of the $AdS_5\times S^5$ target
space is broken by the lightcone gauge quantization procedure. 
The bosonic creation operators $a_n^{A\dag}$ decompose as 
$({\mathbf 4},{\mathbf 1})+ ({\mathbf 1},{\mathbf 4})$,
or, in the $SU(2)^2\times SU(2)^2$ notation introduced in
\cite{Callan:2003xr}, as $({\mathbf 2},{\mathbf 2};{\mathbf 1},{\mathbf 1}) + ({\mathbf
1},{\mathbf 1};{\mathbf 2},{\mathbf 2})$. Analogously, the fermionic 
operators $b_n^{\alpha\dag}$ decompose as $({\mathbf 2},{\mathbf
1};{\mathbf 2},{\mathbf 1}) + ({\mathbf 1},{\mathbf 2};{\mathbf1},{\mathbf 2})$ 
under the covering group. It is useful to note that the two fermion irreps are
eigenvectors, with opposite eigenvalue, of the $\Pi$ operator introduced in
(\ref{Pidef}). To find the perturbed energy spectrum, we must compute explicit 
matrix elements of ${H}_{\rm int}$ in this basis and then diagonalize 
the resulting \mbox{$256\times 256$} matrix. We will compare the perturbed 
energy eigenvalues with general expectations from $PSU(2,2|4)$ as well as with
the large ${R}$-charge limit of the anomalous dimensions of gauge theory 
operators with two ${R}$-charge defects.
Higher-impurity string states can be treated in the same way, but we
defer such questions to a later chapter. 
Our purpose here is primarily to check that our methods (choice of action, 
lightcone gauge reduction, quantization rules, etc.)~are consistent and 
correct. Due to the algebraic complexity met with at each step, this 
check is far from trivial. Once reassured on these fundamental points, 
we can go on to examine a wider range of physically interesting issues.

The first step in carrying out this program is to expand ${H}_{\rm int}$ 
in creation and annihilation operators using (\ref{adef}, \ref{bdef1}) for 
$x^A,\psi$ and the related expansions for $p^A,\rho$. As an example, we quote 
the result for $H_{\rm BB}$ (keeping only terms with two creation and two 
annihilation operators):
\begin{eqnarray}
\label{Hcorrected}
{H}_{\rm BB} & = &
    -\frac{1}{32 p_- \Rhat^2}\sum \frac{\delta(n+m+l+p)}{\xi}
    \times 
\nn\\
& & \biggl\{
    2 \biggl[ \xi^2 
	- (p_-^4 - k_l k_p k_n k_m )
     +  \omega_n \omega_m k_l k_p
      +  \omega_l \omega_p k_n k_m
    + 2 \omega_n \omega_l k_m k_p
\nn\\
& & 
    + 2 \omega_m \omega_p k_n k_l
    \biggr]
    a_{-n}^{\dagger A}a_{-m}^{\dagger A}a_l^B a_p^B
   +4 \biggl[ \xi^2 
	- (p_-^4 - k_l k_p k_n k_m )
     - 2 \omega_n \omega_m k_l k_p
\nn\\
&&
     +  \omega_l \omega_m k_n k_p
   -  \omega_n \omega_l k_m k_p
    -  \omega_m \omega_p k_n k_l
    + \omega_n \omega_p k_m k_l \biggr]
    a_{-n}^{\dagger A}a_{-l}^{\dagger B}a_m^A a_p^B
\nn\\
&&     + 2  \biggl[8 k_l k_p
    a_{-n}^{\dagger i}a_{-l}^{\dagger j}a_m^i a_p^j
     + 2 (k_l k_p +k_n k_m)  
	a_{-n}^{\dagger i}a_{-m}^{\dagger i}a_l^j a_p^j
\nn\\
&&    +(\omega_l \omega_p+ k_l k_p -\omega_n 
	\omega_m- k_n k_m)a_{-n}^{\dagger i}a_{-m}^{\dagger i}a_l^{j'} a_p^{j'}
     -4 ( \omega_l \omega_p- k_l k_p)
	a_{-n}^{\dagger i}a_{-l}^{\dagger j'}a_m^i a_p^{j'} 
\nn\\
&&
	-(i,j \rightleftharpoons i',j')
    \biggr]\biggr\}~,
\end{eqnarray}
with $\xi \equiv \sqrt{\omega_n \omega_m \omega_l \omega_p}$.
The expansion of the interaction terms involving fermi fields are too 
complicated to be worth writing down explicitly at this stage. 
Schematically, we organize the two-impurity matrix elements of the 
perturbing Hamiltonian as shown in Table \ref{blockform}.
\begin{table}[ht!]
\begin{eqnarray}
\begin{array}{|c|cccc|}
\hline
 {H}_{int} & a^{A\dagger}_n a^{B\dagger}_{-n} \ket{J} &
        b^{\alpha\dagger}_n b^{\beta\dagger}_{-n}\ket{J} &
        a^{A\dagger}_n b^{\alpha\dagger}_{-n} \ket{J} &
        a^{A\dagger}_{-n} b^{\alpha\dagger}_{n} \ket{J} \\
        \hline
\bra{J} a^{A}_n a^{B}_{-n} & {H}_{\rm BB} & {H}_{\rm BF} &0&0 \\
\bra{J} b^{\alpha}_n b^{\beta}_{-n} & {H}_{\rm BF} & {H}_{\rm
FF}&0&0\\ \bra{J} a^{A}_n b^{\alpha}_{-n} &0&0& {H}_{\rm BF} & {\cal
H}_{\rm BF} \\ \bra{J} a^{A}_{-n} b^{\alpha}_n & 0 & 0 & {H}_{\rm
BF} & {H}_{\rm BF}\\
\hline
\end{array} \nonumber
\end{eqnarray}
\caption{Structure of the matrix of first-order energy
perturbations in the space of two-impurity string states}
\label{blockform}
\end{table}

To organize the perturbation theory, it is helpful to express everything in
terms of two parameters: $J$ and $\lambda'$. 
In the duality between 
Type IIB superstring theory on $AdS_5\times S^5$ and ${\cal N}=4$
$SU(N_c)$ super Yang-Mills theory in four dimensions, we identify
\be
{\cal N}=4\ {\rm SYM} & \qquad & AdS_5\times S^5~, \nn\\
SU(N_c) & \rightleftharpoons & \int_{S^5}F_5 = N_c ~,\nn\\
g_{\rm YM}^2 N_c  & \rightleftharpoons &  {\Rhat^4}~, \nn\\
g_{\rm YM}^2 & \rightleftharpoons & g_{s} .
\ee
In the pp-wave limit, however, the AdS/CFT dictionary reads
\be
{R} & \rightleftharpoons &  p_- \Rhat^2 = J~, \nn\\
 \frac{{R}^2}{N_c} & \rightleftharpoons & g_{s} p_-^2 = g_2~, \nn\\
{R} \rightarrow \infty & \rightleftharpoons & p_- \Rhat^2,\ N_c \rightarrow \infty~.
\ee
The modified 't Hooft coupling
\be
\lambda' = \frac{g_{\rm YM}^2 N_c}{{R}^2} \rightleftharpoons 
	\frac{1}{p_-^2}
\ee
is kept fixed in the ${R},N_c \rightarrow \infty$ limit.  
(We have kept $\alpha' = \mu = 1$.)
Since the 
gauge theory is perturbative in $\lambda = g_{YM}^2 N_c$, 
and $p_-^2$ on the string side is mapped to ${R}^2/(g_{YM}^2 N_c)$, 
we will expand string energies $\omega_q$ 
in powers of $1/p_-$, keeping terms up to some low order to correspond with
the loop expansion in the gauge theory.
This type of dictionary would be incorrect in the original coordinate system
characterized by the lightcone coordinates $t = x^+ - (x^-/2\Rhat^2)$ and 
$\phi = x^+ + (x^-/2\Rhat^2)$ given in (\ref{rescalePre4}).  
In this case, one would calculate corrections
to ${R} \rightleftharpoons p_- \Rhat^2$ appearing in the perturbing Hamiltonian 
(which amount to operator-valued corrections to $p_-$). 

\subsection{Evaluating Fock space matrix elements of ${H}_{\rm BB}$}
We now proceed to the construction of the perturbing Hamiltonian matrix
on the space of degenerate two-impurity states. To convey a sense of
what is involved, we display the matrix elements of ${H}_{\rm BB}$ 
(\ref{Hpurbos}) between the bosonic two-impurity Fock space states:
\begin{eqnarray}
\label{bosonmatrix}
 \Braket{ J | a_n^A a_{-n}^B \left( {H}_{\rm BB} \right)
        a_{-n}^{C \dagger} a_n^{D \dagger} | J } &  = &
        \left( N_{\rm BB}(n^2\lambda') - 2 n^2\lambda'\right)
	\frac{\delta^{ AD}\delta^{ BC}}{J} 
\qquad\qquad\qquad\nn\\
 +  &&\kern-25pt
	  \frac{n^2\lambda'}{J(1+n^2\lambda')}      
	\left[ \delta^{ab}\delta^{cd}
        + \delta^{ad}\delta^{bc} - \delta^{ac}\delta^{bd} \right]
\nn\\
 - &&\kern-25pt
	 \frac{n^2\lambda'}{J(1+n^2\lambda')}  	 
	\left[ \delta^{a'b'}\delta^{c'd'}
        + \delta^{a'd'}\delta^{b'c'} - \delta^{a'c'}\delta^{b'd'} \right]
\nn\ee\be
 \approx  \left(n_{\rm BB}-2 \right)&&\kern-25pt   
	\frac{n^2\lambda'}{J} \delta^{ AD}\delta^{ BC}
	    + \frac{n^2\lambda'}{J}\left[ \delta^{ab}\delta^{cd}
        + \delta^{ad}\delta^{bc} - \delta^{ac}\delta^{bd} \right]
\nn\\     - \frac{n^2\lambda'}{J}
	&&\kern-25pt
	\left[ \delta^{a'b'}\delta^{c'd'}
        + \delta^{a'd'}\delta^{b'c'} - \delta^{a'c'}\delta^{b'd'} \right]
	+ O({\lambda'}^{2})\ ,
\end{eqnarray}
where lower-case $SO(4)$ indices $a,b,c,d\in 1,\dots ,4$
indicate that $A,B,C,D$ are chosen from the first $SO(4)$, 
and $a',b',c',d'\in 5,\dots ,8$ indicate the second $SO(4)$ 
$(A,B,C,D \in 5,\dots ,8)$. We have also displayed the further expansion 
of these $O(1/J)$ matrix elements in powers of $\lambda'$ (using
the basic BMN-limit energy eigenvalue condition
$\omega_n/p_- = \sqrt{1+\lambda' n^2}$).
This is to facilitate eventual contact with perturbative gauge theory 
via AdS/CFT duality. Note that ${H}_{\rm BB}$ does not mix 
states built out of oscillators from different $SO(4)$ subgroups. 
There is a parallel no-mixing phenomenon in the gauge theory: two-impurity 
bosonic operators carrying spacetime vector indices do not mix with 
spacetime scalar bosonic operators carrying ${R}$-charge vector indices. 

Due to operator ordering ambiguities, two-impurity matrix 
elements of ${H}_{\rm BB}$ can differ by contributions
proportional to $\delta^{AD}\delta^{BC}$, depending on the particular
prescription chosen \cite{Callan:2003xr}. $N_{\rm BB}(n^2\lambda')$ is an 
arbitrary function of $n^2\lambda'$, which is included to account 
for such ambiguities (we will shortly succeed in fixing it).  To
match the dual gauge theory physics, it is best to expand $N_{\rm BB}$ as a 
power series in $\lambda'$. The zeroth-order term must vanish if 
the energy correction is to be perturbative in the gauge coupling. The 
next term in the expansion contributes one arbitrary constant 
(the $n_{\rm BB}$ term) and each higher term in the $\lambda'$ expansion
in principle contributes one additional arbitrary constant to this 
sector of the Hamiltonian. Simple general considerations will
fix them all.

\subsection{Evaluating Fock space matrix elements of ${H}_{\rm FF}$}
The calculation of the two-impurity matrix elements of the parts of
${H}_{\rm int}$ that involve fermionic fields is rather involved and 
we found it necessary to employ symbolic manipulation programs to 
keep track of the many different terms. The end results are fairly 
concise, however. For ${H}_{FF}$ we find
\begin{eqnarray}
\label{fermimatrix}
\Braket{J| b_n^\alpha b_{-n}^\beta\left({H}_{\rm FF}\right)
          b_{-n}^{\gamma\dagger} b_n^{\delta\dagger}|J} & = &
    \left(N_{\rm FF}(n^2\lambda')-2 {n^2\lambda'}\right) 
	\frac{\delta^{\alpha\delta}\delta^{\beta\gamma}}{J} \qquad\qquad\qquad\nn\\
     + \frac{n^2\lambda'}{24 J(1+n^2\lambda')} &&\kern-25pt
\left[ (\gamma^{ij})^{\alpha\delta}(\gamma^{ij})^{\beta\gamma}
        + (\gamma^{ij})^{\alpha\beta}(\gamma^{ij})^{\gamma\delta}
        - (\gamma^{ij})^{\alpha\gamma}(\gamma^{ij})^{\beta\delta} \right]\nn\\
    - \frac{n^2\lambda'}{24 J(1+n^2\lambda')} &&\kern-25pt
    \left[(\gamma^{i'j'})^{\alpha\delta}(\gamma^{i'j'})^{\beta\gamma}
        + (\gamma^{i'j'})^{\alpha\beta}(\gamma^{i'j'})^{\gamma\delta}
-
(\gamma^{i'j'})^{\alpha\gamma}(\gamma^{i'j'})^{\beta\delta}\right]
\nn
\ee
\be
	&\approx&
	\left( n_{\rm FF}-2 \right) 
	\frac{n^2\lambda'}{J}\delta^{\alpha\delta}\delta^{\beta\gamma} 
     	+ \frac{n^2\lambda'}{24 J} 
\left[ (\gamma^{ij})^{\alpha\delta}(\gamma^{ij})^{\beta\gamma}
        + (\gamma^{ij})^{\alpha\beta}(\gamma^{ij})^{\gamma\delta}
        - (\gamma^{ij})^{\alpha\gamma}(\gamma^{ij})^{\beta\delta} \right]
\nn\\
&&    - \frac{n^2\lambda'}{24 J}
    \left[(\gamma^{i'j'})^{\alpha\delta}(\gamma^{i'j'})^{\beta\gamma}
        + (\gamma^{i'j'})^{\alpha\beta}(\gamma^{i'j'})^{\gamma\delta}
-
(\gamma^{i'j'})^{\alpha\gamma}(\gamma^{i'j'})^{\beta\delta}\right]
	+ O({\lambda'}^2)~.
\end{eqnarray}
This sector has its own normal-ordering function $N_{\rm FF}$, with properties
similar those of $N_{\rm BB}$ described above. The index structure of the 
fermionic matrix elements is similar to that of its bosonic counterpart
(\ref{bosonmatrix}). 

We will now introduce some useful projection operators that
will help us understand the selection rules implicit in the index structure
of (\ref{fermimatrix}).
The original 16-component spinors $\psi$ were reduced to eight components 
by the Weyl condition $\bar\gamma^9 \psi = \psi$. 
The remaining eight components are further divided into spinors 
$\tilde\psi$ and $\hat\psi$, which are even or odd under the 
action of $\Pi$:
\be
\Pi \tilde\psi = -\tilde \psi~, & \qquad & 
	\Pi \tilde b^{\dag \alpha} = -\tilde b^{\dag\alpha}~, \nn\\
\Pi \hat\psi = \hat\psi~, & \qquad & \Pi\hat b^{\dag\alpha} = \hat b^{\dag\alpha}~.
\ee
The spinors $\hat \psi$ transform in the $({\bf 1,2; 1,2})$ of
$SO(4)\times SO(4)$, while $\tilde \psi$ transform in the $({\bf 2,1;2,1})$.
This correlation between $\Pi$-parity and $SO(4)\times SO(4)$ representation
will be very helpful for analyzing complicated fermionic matrix elements. 

We denote the $SU(2)$ generators of the active factors of the $({\bf 2,1;2,1})$
irrep as $\Sigma^+$ and $\Omega^+$, where the $\Sigma $ act on the SO(4) 
descended from the $AdS_5$, and the $\Omega$ act on the $SO(4)$ coming from 
the $S^5$. The $({\bf 1,2; 1,2})$ generators are similarly labeled by 
$\Sigma^-$ and $\Omega^-$.  Each set of spinors is annihilated by its 
counterpart set of $SU(2)$  generators:
\be
\label{su2def}
\Sigma^+ \hat b^{\dag\alpha} = \Omega^+ \hat b^{\dag\alpha} & = & 0~, \nn\\
\Sigma^- \tilde b^{\dag\alpha} = \Omega^- \tilde b^{\dag\alpha} & = & 0~.
\ee
In terms of the projection operators 
\be
\Pi_+ = \frac{1}{2}(1+\Pi)~, \qquad \Pi_- = \frac{1}{2}(1-\Pi)\ ,
\ee
which select the disjoint $({\bf 1,2; 1,2}) $ and
$({\bf 2,1; 2,1}) $ irreps, respectively, we have
\be
\Pi_+ \psi = \hat \psi~, & \qquad & \Pi_+ \hat b^\alpha = \hat b^\alpha~, \nn\\
\label{PP1}
\Pi_- \psi = \tilde \psi~, & \qquad & \Pi_- \tilde b^\alpha = \tilde b^\alpha~.
\ee
The $\Pi_\pm$ projections commute with the $SO(4)$ generator matrices 
$\gamma^{ij}, \gamma^{i' j'}$, a fact that implies certain useful 
selection rules for the one-loop limit of (\ref{fermimatrix}). The rules are 
most succinctly stated using an obvious $\pm$ shorthand to indicate the 
representation content of states created by multiple fermionic creation 
operators. In brief, one finds that $++$ states connect only with $++$
and $--$ states connect only with $--$. The only subtle point
is the statement that all $++\to --$ matrix elements of (\ref{fermimatrix})
must vanish: this is the consequence of a simple cancellation between
two terms. This observation will simplify the matrix diagonalization 
we will eventually carry out.

\subsection{Evaluating Fock space matrix elements of ${H}_{\rm BF}$}
The ${H}_{\rm BF}$ sector in the Hamiltonian mediates mixing between
spacetime bosons of the two types (pure boson and bi-fermion) as well as between spacetime 
fermions (which of course contain both bosonic and fermionic oscillator excitations).  
The 64-dimensional boson mixing matrix
\be
\Braket{ J| b_{n}^\alpha b_{-n}^\beta \left( {H}_{\rm BF} \right)
	a_{-n}^{A\dagger} a_{n}^{B\dagger} | J }\ ,
\nn
\ee
is an off-diagonal block in the bosonic sector of the perturbation matrix
in Table \ref{blockform}. The same methods used earlier in this section to
reduce Fock space matrix elements involving fermi fields can be used here
to obtain the simple explicit result (we omit the details) 
\begin{eqnarray}
\label{31}
\Braket{J| b_{n}^\alpha b_{-n}^\beta \left( {H}_{\rm BF} \right)
        a_{-n}^{A\dagger} a_{n}^{B\dagger} |J} & = &
	\frac{n^2 {\lambda'}}{2J(1+n^2\lambda')}
	\biggl\{
	\sqrt{1+n^2\lambda'}\Bigl[
                \left( \gamma^{ab'} \right)^{\alpha\beta}
                - \left( \gamma^{a'b} \right)^{\alpha\beta} \Bigr]
\nn\\
& & 	+~ n\sqrt{ \lambda' }\left[
	\left( \gamma^{a'b'} \right)^{\alpha\beta}
	- \left( \gamma^{ab} \right)^{\alpha\beta}
	+ \left(\delta^{ab} - \delta^{a'b'}\right)
	\delta^{\alpha\beta} \right]
	\biggr\}
\nn\\
& \approx &
	\frac{ n^2 \lambda'}{2 J}\left[
                \left( \gamma^{ab'} \right)^{\alpha\beta}
                - \left( \gamma^{a'b} \right)^{\alpha\beta} \right]
	+O({\lambda'}^{3/2})~. 
\end{eqnarray}
The complex conjugate of this matrix element gives the additional off-diagonal 
component of the upper $128\times 128$ block of spacetime bosons.  
We note that terms in the ${H}_{\rm BF}$ sector split the $SO(8)$ group
(manifest in the pp-wave limit) into its
$SO(4)$ constituents such that states of the form 
$a_{-n}^{a'\dagger} a_{n}^{b'\dagger} \ket{J}$, for example, 
which descend strictly from the $S^5$ subspace, vanish in this subsector. 
This behavior is reproduced in the gauge theory, wherein two-boson states that are
either spacetime scalars or scalars of the ${R}$-charge group do not mix with 
bi-fermionic scalars in either irrep.  

The 128-dimensional subsector of spacetime fermions is mixed by 
matrix elements of the same Hamiltonian taken between 
fermionic string states of the general form 
$b_n^{\alpha\dagger} a_{-n}^{A\dagger} \ket{J}$. Our standard methods 
yield the following simple results for the two independent types of 
spacetime fermion mixing matrix elements:
\begin{eqnarray}
\label{22}
&&
\Braket{ J | b_{n}^\alpha a_{-n}^A \left( {H}_{\rm BF} \right)
    b_{n}^{\beta\dagger} a_{-n}^{B\dagger}|J }  = 
	N_{\rm BF}(n^2\lambda')\frac{\delta^{AB}\delta^{\alpha\beta}}{J}
\nn\\
&& 	+ \frac{n^2\lambda'}{2J(1+n^2\lambda')}\biggl\{
	\left( \gamma^{ab}\right)^{\alpha\beta}
        - \left( \gamma^{a'b'}\right)^{\alpha\beta} 
	- (3+4n^2\lambda') \delta^{ab} \delta^{\alpha\beta}
	- (5+4n^2\lambda')\delta^{a'b'}\delta^{\alpha\beta}
	\biggr\}
\nn\\
&&   \approx 
	 \frac{n^2 \lambda'}{2J}
	\biggl\{\left( \gamma^{ab}\right)^{\alpha\beta} 
        - \Bigl(  \gamma^{a'b'}  \Bigr)^{\alpha\beta}
    + \left[ (2 n_{\rm BF}-3) \delta^{ab}+ (2 n_{\rm BF}-5) \delta^{a'b'}
	 \right]\delta^{\alpha\beta} \biggr\} 
	+ O({\lambda'}^{2})~,
\nn\\
& & 
\end{eqnarray}
\begin{eqnarray}
\label{22a}
\Braket{J | b_{n}^\alpha a_{-n}^A \left({H}_{\rm BF}\right)
        b_{-n}^{\beta\dagger} a_{n}^{B\dagger}|J } & = &
\frac{n^2\lambda'}{2J\sqrt{1+n^2\lambda'}}\biggl\{
	\left( \gamma^{ab}\right)^{\alpha\beta}
        - \left( \gamma^{a'b'}\right)^{\alpha\beta} 
\nn\\
 	-\frac{n {\lambda'}^{1/2}}{\sqrt{1+n^2\lambda'}}
	&&\kern-25pt\Bigl[
	\left( \gamma^{ab'}\right)^{\alpha\beta}
        - \left( \gamma^{a'b}\right)^{\alpha\beta} \Bigr]
	-\delta^{\alpha\beta}
	\left( \delta^{ab} - \delta^{a'b'} \right)
	\biggr\}
\nn
\\
  \approx  \frac{n^2 \lambda'}{2 J}
	\biggl\{
         \left( \gamma^{ab}\right)^{\alpha\beta}&&\kern-25pt
        -  \left( \gamma^{a'b'}\right)^{\alpha\beta}
    - \left( \delta^{ab}- \delta^{a'b'} \right)\delta^{\alpha\beta} \biggr\}
	+ O({\lambda^\prime}^{3/2})\ .
\end{eqnarray}

Equation (\ref{22}) involves yet another normal-ordering function. Since these
functions have a nontrivial effect on the spectrum, we must give them specific 
values before we can calculate actual numerical eigenvalues. 
The key point is that the structure of the perturbing Hamiltonian implies 
certain relations between all the normal-ordering functions.  
Because the interaction Hamiltonian
is quartic in oscillators, normal-ordering ambiguities give rise to terms
quadratic in oscillators, appearing as constant contributions to the
diagonal matrix elements.  There are
normal-ordering contributions from each sector of the theory: 
${H}_{\rm BB}$ contributes a single term quadratic in
bosonic oscillators; ${H}_{\rm FF}$ yields a term 
quadratic in fermionic oscillators;  ${H}_{\rm BF}$
contributes one term quadratic in bosons and one quadratic in fermions.
The bosonic contributions multiply terms of the form $a^\dag a$,
which are collected into the function $N_{\rm BB}(n^2\lambda')$
with one contribution from ${H}_{\rm BB}$ and one
contribution from ${H}_{\rm BF}$.  Similarly, $N_{\rm FF}(n^2\lambda')$
collects terms multiplying $b^\dag b$, receiving one contribution 
from ${H}_{\rm FF}$ and one contribution from ${H}_{\rm BF}$.
Normal-ordering contributions from both $a^\dag a$ and $b^\dag b$ terms
are non-vanishing in the spacetime fermion subsector; all possible
normal-ordering ambiguities appear in this subspace.  The normal-ordering
function $N_{\rm BF}(n^2\lambda')$ therefore must satisfy
\be
N_{\rm BF}(n^2\lambda') = N_{\rm BB}(n^2\lambda') + N_{\rm FF}(n^2\lambda')\ .
\ee
The normal ordering functions are basically finite renormalizations
that must be adjusted so that the spectrum reflects the $PSU(2,2|4)$ 
global supersymmetry of the classical worldsheet action (a symmetry 
we want to preserve at the quantum level). 

As has been explained 
elsewhere \cite{Beisert:2002tn,Callan:2003xr} (and as we shall shortly
review), energy levels should be organized into multiplets obtained 
by acting on a ``highest-weight'' level with all possible combinations 
of the eight ${R}$-charge raising supercharges. All the states 
obtained by acting with a total of $L$ supercharges have the same 
energy and we will refer to them as states at level $L$ in the 
supermultiplet. The levels of a multiplet run from $L=0$ to $L=8$.
A careful inspection of the way the normal ordering functions contribute
to the energies of states in the two-impurity sector shows that states
at levels $L=0,8$ are shifted by $N_{\rm BB}$ only. Similarly, levels $L=2,4,6$
are shifted by $N_{\rm FF}$ or $N_{\rm BB}$ and one must have 
$N_{\rm BB} = N_{\rm FF}$ if those levels are to remain internally 
degenerate. Finally, levels $L=1,3,5,7$ are shifted by $N_{\rm BF}$ only. 
By supersymmetry, the level 
spacing must be uniform throughout the supermultiplet and this is only
possible if we also set $N_{\rm BB} = N_{\rm BF}$. But then the constraint 
$N_{\rm BF} = N_{\rm BB} + N_{\rm FF}$ can only be met by setting
$N_{\rm BB} = N_{\rm FF} = N_{\rm BF} = 0$, which then eliminates
any normal-ordering ambiguity from the string theory. This is basically
an exercise in using global symmetry conditions to fix otherwise
undetermined finite renormalizations.

\subsection{Diagonalizing the one-loop perturbation matrix}
We are now ready to diagonalize the perturbing Hamiltonian and
examine whether the resulting energy shifts have the right multiplet structure
and whether the actual eigenvalues match gauge theory expectations.
To simplify the problem, we will begin by diagonalizing the perturbation 
matrix expanded to first nontrivial order 
in both $1/J$ and $\lambda'$. Our results should, by duality, match one-loop
gauge theory calculations and we will eventually return to the problem of
finding the string spectrum correct to higher orders in $\lambda'$.
From the structure of the results just obtained for the perturbation
matrices, we can see that the general structure of the
energy eigenvalues of two-impurity states must be
\be
\label{eigenformula}
E_{\rm int}(n) =  2 + n^2 \lambda'
	\left( 1 + \frac{\Lambda}{J}+ O(J^{-2})\right)  + O(\lambda'^{2})~,
\ee
where $\Lambda$ is dimensionless and the dependence on $1/J$, $\lambda'$ 
and mode number $n$ is given by (\ref{bosonmatrix}, \ref{fermimatrix}).
The eigenvalues $\Lambda$ must meet certain conditions if the requirements of 
$PSU(2,2|4)$ symmetry are to be met, and we will state those conditions before 
solving the eigenvalue problem. 

The eigenvalues $\Lambda$ must meet certain conditions if the requirements of 
$PSU(2,2|4)$ symmetry are to be met. The eigenvalues in question are lightcone
energies and thus dual to the gauge theory quantity $\Delta=D-J$, the difference
between scaling dimension and ${R}$-charge. Since conformal invariance is part
of the full symmetry group, states are organized into conformal multiplets built 
on conformal primaries. A supermultiplet will contain several conformal primaries 
having the same value of $\Delta$ and transforming into each other under the supercharges. 
All 16 supercharges increment the dimension of an operator by $1/2$, but only eight 
of them (call them ${\cal Q}_\alpha$) also increment the ${R}$-charge by $1/2$, so as to
leave $\Delta$ unchanged. These eight supercharges act as ``raising operators'' on the 
conformal primaries of a supermultiplet: starting from a super-primary of lowest
${R}$-charge, the other conformal primaries are created by acting on it in all 
possible ways with the eight ${\cal Q}_\alpha$. Primaries obtained by acting with $L$ factors 
of ${\cal Q}_\alpha$ on the super-primary are said to be at level $L$ in the supermultiplet 
(since the  ${\cal Q}_\alpha$ anticommute, the range is $L=0$ to $L=8$). The multiplicities 
of states at the various levels are also determined: for each $L=0$ primary operator, there 
will be $C^8_L$ such operators at level $L$ (where $C^n_m$ is the binomial coefficient). 
If the $L=0$ primary has multiplicity $s$, summing over all $L$ gives $2^8 s=256 s$ 
conformal primaries in all. 

These facts severely restrict the quantity $\Lambda$ in the general expression 
(\ref{eigenformula}) above. Although the states in the degenerate multiplet all
have the same $J$, they actually belong to different levels $L$ in more than one 
supermultiplet.  A state of given $L$ is a member of a supermultiplet 
built on a ``highest-weight'' or super-primary state with ${ R}=J-L/2$. 
Since all the primaries in a supermultiplet have the same $\Delta$, the joint 
dependence of eigenvalues on $\lambda,J,L$ must be of the form 
$\Delta(\lambda,J-L/2)$. The only way the expansion of (\ref{eigenformula}) 
can be consistent with this is if $\Lambda=L+c$, where $c$ is a pure numerical 
constant (recall that $\lambda'=\lambda/J^2$). Successive members of
a supermultiplet must therefore have eigenvalues separated by exactly one
and the difference between ``top'' ($L=8$) and ``bottom'' ($L=0$) eigenvalues for
$\Lambda$ must be exactly eight. 

\subsection{Details of the one-loop diagonalization procedure.}
We now confront the problem of explicitly diagonalizing the 
first-order perturbation matrix $\Lambda$ (obtained by expanding the
relevant matrix elements to first order in $\lambda'$). 
The matrix block diagonalizes
on the spacetime boson and spacetime fermion subspaces, as indicated in 
Table~\ref{blockform}. Within these sub-blocks, there are further
block diagonalizations arising from special properties of the one-loop
form of the matrix elements of the perturbing Hamiltonian. For example, 
Fock space states built out of two bosonic creation operators that 
transform only under the internal $SO(4)$ mix only with themselves,
thus providing a $16\times 16$ dimensional diagonal sub-block. Within
such sub-blocks, symmetry considerations are often sufficient to
completely diagonalize the matrix or at least to reduce it to a
low-dimensional diagonalization problem. In short, the problem
reduces almost entirely to that of projecting the matrix elements 
of ${H}_{\rm int}$ on subspaces of the two-impurity Fock space 
defined by various symmetry properties. Determining the $SO(4)\times SO(4)$
symmetry labels of each eigenstate in the diagonalization will furthermore
enable us to precisely match string states with gauge theory 
operators. In this subsection, we record for future reference the 
detailed arguments for the various special
cases that must be dealt with in order to fully diagonalize the
one-loop perturbation and characterize the irrep decomposition. 
Although the projections onto the various
invariant subspaces are matters of simple algebra, that algebra
is too complicated to be done by hand and we have resorted to
symbolic manipulation programs. The end result of the diagonalization is
quite simple and the reader willing to accept our results on faith
can skip ahead to the end of this subsection.
 
We begin with a discussion of the action of the purely bosonic
perturbation ${H}_{\rm BB}$ on the $64$-dimensional Fock space 
created by pairs of bosonic creation operators. Part of this subspace 
connects via ${H}_{\rm BF}$ to the Fock space of spacetime bosons
created by pairs of fermionic creation operators, and we will
deal with it later. There is, however, a subspace that only connects
to itself, through the purely bosonic perturbation ${H}_{\rm BB}$. 
We will first deal with this purely bosonic block diagonalization,
leading to eigenvalues we will denote by $\Lambda_{\rm BB}$.
The eight bosonic modes lie in the $SO(4) \times SO(4)$ representations 
$({\bf 2,2;1, 1})$ and $({\bf 1,1;2,2})$ (i.e.,~they are vectors in
the $SO(4)$ subgroups descended from $AdS_5$ and $S^5$, respectively). The key fact 
about ${H}_{\rm BB}$ is that the 16-dimensional spaces spanned
by two $({\bf 2,2;1, 1})$ oscillators or by two $({\bf 1,1;2,2})$
oscillators are closed under its action (it is also true that
${H}_{\rm BF}$ annihilates both of these subspaces).
The $SO(4)$ representation content of the states created by such 
oscillator pairs is given by the formula $({\bf 2,2})\times ({\bf 2,2}) = 
({\bf 3,3})+({\bf 3,1})+({\bf 1, 3})+({\bf 1,1})$ (we use $SU(2)\times SU(2)$
notation, rather than $SO(4)$, since it is unavoidable when we discuss fermions). 
By projecting the $O(\lambda')$ part of (\ref{bosonmatrix}) 
onto these subspaces, one can directly 
read off the eigenvalues $\Lambda_{\rm BB}$, with the 
results shown in Table~\ref{bosonspectrum}. The identification
of the representations associated with particular eigenvalues is
easy to do on the basis of multiplicity. In any event, projection onto
invariant subspaces is a simple matter of symmetrization or
antisymmetrization of oscillator indices and can be done directly.
The most important point to note is that the eigenvalues are
successive even integers, a simple result and one that is consistent
with our expectations from extended supersymmetry. It will
be straightforward to match these states to gauge theory operators
and compare eigenvalues with anomalous dimensions.
\begin{table}[ht!]
\be
\begin{array}{|c|c|}\hline
 SO(4)_{AdS}\times  SO(4)_{S^5} & \Lambda_{\rm BB} \\
\hline 
({\bf 1,1;1, 1}) &  {-6  } \\ ({\bf 1,1; 3,3})
&{-2 } \\ ({\bf 1,1;3,1}) +({\bf 1,1;1,3})& {-4  } \\ \hline
\end{array}\qquad
\begin{array}{|c|c|}\hline
 SO(4)_{AdS}\times  SO(4)_{S^5} & \Lambda_{\rm BB} \\
\hline 
  ({\bf 1,1;1, 1}) & {\phantom +}2  \\
 ({\bf 3,3;1, 1}) & {-2 } \\
  ({\bf 3,1;1, 1}) + ({\bf 1,3 ;1, 1})& {\phantom +} 0  \\
\hline
\end{array} \nonumber
\ee
\caption{Energy shifts at $O(1/J)$ for unmixed bosonic modes}
\label{bosonspectrum}
\end{table}

The Fock space of spacetime bosons created by pairs of fermionic creation
operators contains a similar pair of $16\times 16$ diagonal sub-blocks.
The construction and application of the relevant projection operators
and the subsequent match-up with gauge theory operators is more complicated 
than on the bosonic side and we must develop some technical tools before 
we can obtain concrete results. 

Just as ${H}_{\rm BB}$ is closed in the two 16-dimensional spaces of bosonic 
$({\bf 1,1;2,2})$ or $({\bf 2,2;1,1})$ states, ${H}_{\rm FF}$ is 
closed on subspaces of bi-fermions spanned by a pair
of $({\bf 1,2;1,2})$ or a pair of $({\bf 2,1;2,1})$ fermionic oscillators
(i.e., $--$ or $++$ states, to use an obvious shorthand).
The complete spectrum of eigenvalues from these subsectors of the Hamiltonian 
can be computed by projecting out the $({\bf 2,1;2,1})$ and $({\bf 1,2;1,2})$
spinors in ${H}_{\rm FF}$ (\ref{fermimatrix}). To do this, it will
be helpful to express the eight-component spinors of the string theory in a 
basis which allows us to define fermionic oscillators labeled by their 
$({\bf 2,1;2,1})$ and $({\bf 1,2;1,2})$ representation content. 

The original 32-component Majorana-Weyl spinors $\theta^I$ were reduced
by the Weyl projection and a lightcone gauge condition to an eight-component 
spinor $\psi^\alpha$ (transforming in the $8_s$ of $SO(8)$). The generators 
of the four $SU(2)$ factors (\ref{su2def}) of the manifest $SO(4)\times SO(4)$ 
symmetry can be expressed as $8\times 8$ $SO(8)$ matrices as follows:
\be
\label{xplctgen}
\Sigma_1^\pm = -\frac{1}{4i}(\gamma^2\gamma^3\pm\gamma^1\gamma^4)~, & \qquad & 
	\Omega_1^\pm = \frac{1}{4i}(-\gamma^6\gamma^7\pm\gamma^5)~, \nn\\
\Sigma_2^\pm = -\frac{1}{4i}(\gamma^3\gamma^1\pm\gamma^2\gamma^4)~, & \qquad & 
	\Omega_2^\pm = \frac{1}{4i}(-\gamma^7\gamma^5\pm\gamma^6)~, \nn\\
\Sigma_3^\pm = -\frac{1}{4i}(\gamma^1\gamma^2\pm\gamma^3\gamma^4)~, & \qquad & 
	\Omega_3^\pm = \frac{1}{4i}(-\gamma^5\gamma^6\pm\gamma^7)~.
\ee
We will use the representation for the $\gamma^A$ given in Appendix A 
(\ref{cliffmat}) when we need to make these generators explicit. 
The $8_s$ spinor may be further divided 
into its $({\bf 1,2; 1,2})$ and $({\bf 2,1; 2,1})$ components $\hat \psi$ 
and $\tilde \psi$, respectively, and this suggests a useful basis change
for the string creation operators: for the $({\bf 1,2; 1,2})$ spinor, we 
define four new objects $w,x,y,z$ by 
\be
\hat b^{\dag} = w 
{\scriptsize
	\left( \begin{array}{c}
	1 \\ 0 \\ 0 \\ -1 \\ 0 \\0 \\0 \\0 
	\end{array} \right) 
+ x \left( \begin{array}{c}
	0 \\ 1\\ 1 \\ 0 \\ 0 \\0 \\0 \\0 
	\end{array} \right)
+ y \left( \begin{array}{c}
	0 \\ 0\\ 0 \\ 0 \\ 1 \\0 \\0 \\1 
	\end{array} \right)
+ z \left( \begin{array}{c}
	0 \\ 1\\1 \\ 0 \\ 0 \\1 \\-1 \\0 
	\end{array} \right)}~,
\ee
which we then organize in two different ways into two-component complex spinors:
\be
\label{20spinor}
\zeta = \left( \begin{array}{c}
	w+iy \\ z+ix \end{array} \right)~, \qquad 
\varphi = \left( \begin{array}{c}
	-z+ix \\ w-iy \end{array} \right)~~
\Leftarrow ~~ \Sigma_i^-~,
\nn\\
\bar \zeta = \left( \begin{array}{c}
	w+iy \\ -z+ix \end{array} \right)~, \qquad 
\bar \varphi = \left( \begin{array}{c}
	z+ix \\ w-iy \end{array} \right)~~
\Leftarrow ~~ \Omega_i^-~.
\ee
This organization into two-spinors is meant to show how components of 
$\hat\psi$ transform under the two $SU(2)$ factors that act nontrivially
on them. As may be verified from the explicit forms of the $SU(2)$ generators 
obtained by substituting (\ref{cliffmat}) into (\ref{xplctgen}), the two-component 
spinors $\zeta$ and $\varphi$ transform as $({\bf 1,2})$ under the first $SO(4)$ 
and the spinors $\bar\zeta$ and $\bar\varphi$ transform as $({\bf 1,2})$ under the
second $SO(4)$ of $SO(4)\times SO(4)$. The explicit realization of the two 
$SU(2)$ factors involved here is found in this way to be
\be
\Sigma_1^- = \left( \begin{array}{cc} 0 & 1/2 \\  1/2 & 0 \end{array} \right)~, & &  
\Omega_1^- = \left( \begin{array}{cc} 0 & 1/2 \\ 1/2 & 0 \end{array} \right)~, 
\nn\\ 
\Sigma_2^- = \left( \begin{array}{cc} 0 & i/2 \\ -i/2 & 0 \end{array} \right)~, & &
\Omega_2^- = \left( \begin{array}{cc} 0 & -i/2 \\ i/2 & 0 \end{array} \right)~, 
\nn\\
\Sigma_3^- = \left( \begin{array}{cc} 1/2 & 0 \\ 0 & -1/2 \end{array} \right)~, & &
\Omega_3^- = \left( \begin{array}{cc} 1/2 & 0 \\ 0 & -1/2 \end{array} \right)\ .
\ee

One may similarly decompose $({\bf 2,1; 2,1})$ spinors 
and express the corresponding generators $\Sigma^+$ and $\Omega^+$. We decompose
$\tilde\psi$ into components $\bar w,\bar x,\bar y,\bar z$ according to
\be
\tilde b^\dag = \bar w 
{\scriptsize
	\left( \begin{array}{c}
	1 \\ 0 \\ 0 \\ 1 \\ 0 \\0 \\0 \\0 
	\end{array} \right)
+ \bar x \left( \begin{array}{c}
	0 \\ 1\\ -1 \\ 0 \\ 0 \\0 \\0 \\0 
	\end{array} \right)
+ \bar y \left( \begin{array}{c}
	0 \\ 0\\ 0 \\ 0 \\ 1 \\0 \\0 \\-1 
	\end{array} \right)
+ \bar z \left( \begin{array}{c}
	0 \\ 0\\0 \\ 0 \\ 0 \\1 \\ 1 \\0
	\end{array}\right)}~,
\ee
and rearrange them into two-component complex spinors:
\be
\xi = \left( \begin{array}{c}
	\bar z + i\bar x\\\bar w + i\bar y \end{array} \right)~, \qquad 
\eta = \left( \begin{array}{c}
	\bar w - i\bar y \\ -\bar z + i\bar x \end{array} \right)~~
\Leftarrow ~~\Sigma_i^+~,
\nn\\
\bar\xi = \left( \begin{array}{c}
	- \bar z + i\bar x \\ \bar w + i\bar y \end{array} \right)~, \qquad 
\bar\eta = \left( \begin{array}{c}
	\bar w - i\bar y \\ \bar z + i\bar x \end{array} \right)~~
\Leftarrow~~ \Omega_i^+.
\ee
The corresponding explicit $({\bf 2,1;2,1})$ generators are
given by
\be
\Sigma_1^+ = \left( \begin{array}{cc} 0 & -1/2 \\ -1/2 & 0 \end{array} \right)~, && 
\Omega_1^+ = \left( \begin{array}{cc} 0 & 1/2 \\ 1/2 & 0 \end{array} \right)~,
\nn\\
\Sigma_2^+ = \left( \begin{array}{cc} 0 & i/2 \\ -i/2 & 0 \end{array} \right)~, &&  
\Omega_2^+ = \left( \begin{array}{cc} 0 & -i/2 \\ i/2 & 0 \end{array} \right)~, 
\nn\\
\Sigma_3^+ = \left( \begin{array}{cc} 1/2 & 0 \\ 0 & -1/2 \end{array} \right)~, && 
\Omega_3^+ = \left( \begin{array}{cc} 1/2 & 0 \\ 0 & -1/2 \end{array} \right)~.
\ee
These observations will make it possible to construct linear combinations of
products of components of $\psi^\alpha$ transforming in chosen irreps
of $SO(4)\times SO(4)$.

Let us now use this machinery to analyze the perturbation matrix on
spacetime bosons created by two fermionic creation operators (bi-fermions).
As explained in the discussion of (\ref{fermimatrix}), ${H}_{\rm FF}$  
is block-diagonal on the 16-dimensional $++$ or $--$ bi-fermionic subspaces. 
To project out the $({\bf 2,1;2,1})$ or $++$ block of ${H}_{\rm FF}$, 
we simply act on all indices of (\ref{fermimatrix}) with the $\Pi_+$
projection operator:
\be
\label{HFF--me}
\Braket{ J | \tilde b_n^\alpha  \tilde b_{-n}^\beta \left( {H}_{\rm FF} \right) 
	  \tilde b_{-n}^{\gamma\dagger}  \tilde b_n^{\delta\dagger}| J }   & = &   
	-2 \frac{n^2\lambda'}{J} \Pi_+^{\alpha\delta}\Pi_+^{\beta\gamma}
 	+ \frac{n^2\lambda'}{24 J}\biggl\{
	\biggl[
	(\Pi_+\gamma^{ij}\Pi_+)^{\alpha\delta}(\Pi_+\gamma^{ij}\Pi_+)^{\beta\gamma}
\nn\\
& & \kern-30pt	+ (\Pi_+\gamma^{ij}\Pi_+)^{\alpha\beta}(\Pi_+\gamma^{ij}\Pi_+)^{\gamma\delta}
 	- (\Pi_+\gamma^{ij}\Pi_+)^{\alpha\gamma}(\Pi_+\gamma^{ij}\Pi_+)^{\beta\delta}
	\biggr]
\nn\\
& & \kern-50pt	- \biggl[
	(\Pi_+\gamma^{i'j'}\Pi_+)^{\alpha\delta}(\Pi_+\gamma^{i'j'}\Pi_+)^{\beta\gamma}
	+ (\Pi_+\gamma^{i'j'}\Pi_+)^{\alpha\beta}(\Pi_+\gamma^{i'j'}\Pi_+)^{\gamma\delta}
\nn\\
& & 	- (\Pi_+\gamma^{i'j'}\Pi_+)^{\alpha\gamma}(\Pi_+\gamma^{i'j'}\Pi_+)^{\beta\delta}
	\biggr]\biggr\}~.
\ee
The $SO(4)\times SO(4)$ representation content of this subspace is specified by 
$({\bf 2,1;2,1})\times ({\bf 2,1;2,1}) = ({\bf 1,1;1,1})\oplus ({\bf 1,1;3,1})\oplus
({\bf 3,1;1,1})\oplus ({\bf 3,1;3,1})$ and we must further project onto 
individual irreducible representations in order to identify the eigenvalues.

With the tools we have built up in the last few paragraphs, we are in a position
to directly project out some of the desired irreducible representations. 
Bi-fermions of $++$ type transforming as scalars under the first $SO(4)$ (i.e.,
under $\Sigma_i^+$) are constructed by making $SU(2)$ invariants out of the 
two-component spinors $\xi$ and $\eta$. There are four such objects:
\be
\label{su2scalar}
\xi_{-n} \tau_2 \xi_n~, &\qquad & \xi_{-n} \tau_2 \eta_n~, \nn\\
\eta_{-n} \tau_2 \xi_n~, &\qquad & \eta_{-n} \tau_2 \eta_n~,
\ee
where $\tau_2$ is the second Pauli matrix. At the same time, they must
also comprise a {\bf 3} and a {\bf 1} under the second $SO(4)$ (i.e.,~under 
$\Omega_i^+$). To identify the irreducible linear combinations, 
one has to re-express the objects in (\ref{su2scalar}) in terms of the spinors 
$\bar\xi$ and $\bar\eta$ that transform simply under $\Omega_i^+$. Representative 
results for properly normalized creation operators of $++$ bi-fermion states 
in particular $SO(4)\times SO(4)$ irreps are
\be
\label{adsscalarminus}
\begin{array}{cc}
-\frac{1}{2}\left( \xi_{-n} \tau_2 \eta_n - \eta_{-n} \tau_2 \xi_n \right)
	&  \phantom{\Biggr\}}\ ({\bf 1,1;1,1})~, \quad \Lambda_{\rm FF} = -2~,
\end{array}
\nn\\
\begin{array}{cc}
	\frac{1}{2}\left( \xi_{-n} \tau_2 \eta_n + \eta_{-n} \tau_2 \xi_n \right)
	& \\
	\frac{i}{2}\left( \xi_{-n} \tau_2 \xi_n + \eta_{-n} \tau_2 \eta_n \right)
	& \\
	-\frac{1}{2}\left( \xi_{-n} \tau_2 \xi_n - \eta_{-n} \tau_2 \eta_n \right)
	&  
\end{array}
  \Biggr\}\ ({\bf 1,1;3,1})~,\quad  \Lambda_{\rm FF} = 0~.
\ee
We simply have to re-express the $\xi,\eta$ bilinears in terms of the original
spinor creation operators $\tilde b$ in order to obtain an explicit projection
of the matrix elements ({\ref{HFF--me}) onto irreducible subspaces and to obtain
the eigenvalues $\Lambda_{\rm FF}$ associated with each irrep. A parallel
analysis of states constructed by forming normalized $SU(2)$ invariants from 
$\bar\xi$ and $\bar\eta$ gives another irrep and eigenvalue:
\be
\label{sscalarminus}
\begin{array}{cc}
	\frac{1}{2}\left( \bar\xi_{-n} \tau_2 \bar\eta_n + \bar\eta_{-n} \tau_2 \bar\xi_n\right)
	&   \\
	\frac{i}{2}\left( \bar\xi_{-n} \tau_2 \bar\xi_n + \bar\eta_{-n} \tau_2 \bar\eta_n \right)
	&   \\
	-\frac{1}{2}\left( \bar\xi_{-n} \tau_2 \bar\xi_n - \bar\eta_{-n} \tau_2 \bar\eta_n \right)
	&
\end{array}
  \Biggr\}\ ({\bf 3,1;1,1})~, \qquad  \Lambda_{\rm FF} = -4  ~.
\ee
By similar arguments, whose details we will omit, one can construct the creation
operator for the normalized ${\bf (3,1;3,1)}$ or $++$ bi-fermion and find the
eigenvalue $\Lambda_{\rm FF}= -2$.

An exactly parallel analysis of 
$\Braket{ J | \hat b \hat b ({H}_{\rm FF}) \hat b^\dag \hat b^\dag | J }$
on the 16-dimensional subspace spanned by $({\bf 1,2;1,2})$ bi-fermions
yields the same eigenvalue spectrum.  
The creation operators of irreducible states (built this time out of
$\zeta$ and $\phi$) and their eigenvalues are
\be
\label{adsscalarplus}
\begin{array}{c}
 -\frac{1}{2}\left( \zeta_{-n} \tau_2 \varphi_n - \varphi_{-n} \tau_2 \zeta_n \right)
	
\end{array}
\phantom{ \Biggr\}}\ ({\bf 1,1;1,1})~,  &\qquad \Lambda_{\rm FF} = -2~,
\nn\\
\begin{array}{cc}
	\frac{1}{2}\left( \zeta_{-n} \tau_2 \varphi_n + \varphi_{-n} \tau_2 \zeta_n \right)
	&  \\
	\frac{i}{2}\left( \zeta_{-n} \tau_2 \zeta_n + \varphi_{-n} \tau_2 \varphi_n \right)
	& \\
	-\frac{1}{2}\left( \zeta_{-n} \tau_2 \zeta_n - \varphi_{-n} \tau_2 \varphi_n \right)
	&  
\end{array}
\Biggr\}\ ({\bf 1,1;1,3})~, &\qquad  \Lambda_{\rm FF} = 0~,\ee\be
\label{sscalarplus}
\begin{array}{cc}
\frac{1}{2}\left( \bar\zeta_{-n} \tau_2 \bar\varphi_n + \bar\varphi_{-n} \tau_2 \bar\zeta_n \right)
	&  \\
\frac{i}{2}\left( \bar\zeta_{-n} \tau_2 \bar\zeta_n + \bar\varphi_{-n} \tau_2 \bar\varphi_n \right)
	& \\
-\frac{1}{2}\left( \bar\zeta_{-n} \tau_2 \bar\zeta_n - \bar\varphi_{-n} \tau_2 \bar\varphi_n \right)
	&   
\end{array}
\Biggr\}\ ({\bf 1,3;1,1})~, &\qquad  \Lambda_{\rm \rm FF} = -4~.
\ee
The overall results for this sector are displayed in Table \ref{FFmult1}.
\begin{table}[ht!]
\be
\begin{array}{|c|c|}\hline
 SO(4)_{AdS}\times SO(4)_{S^5} & \Lambda_{\rm FF} \\
\hline ({\bf 1},{\bf 1};{\bf 1},{\bf 1})& {-2 }
\\ ({\bf 1},{\bf 1};{\bf 3},{\bf 1})& {\phantom +}0 \\
({\bf 3},{\bf 1};{\bf 1},{\bf 1})& {-4 } \\ ({\bf
3},{\bf 1};{\bf 3},{\bf 1})& {-2 } \\ \hline
\end{array} \qquad
\begin{array}{|c|c|}\hline
 SO(4)_{AdS}\times SO(4)_{S^5} & \Lambda_{\rm FF} \\
\hline 
({\bf 1},{\bf 1};{\bf 1},{\bf 1}) & {-2 }
\\ 
({\bf 1},{\bf 1};{\bf 1},{\bf 3}) & {\phantom +}0 \\
({\bf 1},{\bf 3};{\bf 1},{\bf 1}) & {-4 } \\ 
({\bf 1},{\bf 3};{\bf 1},{\bf 3}) & {-2 } \\ \hline
\end{array} \nonumber
\ee
\caption{Energy shifts of states created by two fermions in 
({\bf 2},{\bf 1};{\bf 2},{\bf 1}) or
({\bf 1},{\bf 2};{\bf 1},{\bf 2}) } 
\label{FFmult1}
\end{table}

To this point, we have been able to study specific projections of the ${H}_{\rm BB}$
and ${H}_{\rm FF}$ subsectors by choosing states that are not mixed by ${H}_{\rm BF}$.
We now must deal with the subspace of spacetime boson two-impurity states
that is not annihilated by ${H}_{\rm BF}$. This 64-dimensional
space is spanned by pairs of bosonic creation operators taken from 
different $SO(4)$ subgroups and pairs of fermionic creation operators of opposite
$\Pi$-parity. The representation content of these creation-operator pairs
is such that the states in this sector all belong to $({\bf 2,2;2,2})$ irreps.
This space is of course also acted on by ${H}_{\rm BB}$
and ${H}_{\rm FF}$, so we will need the matrix elements of all three 
pieces of the Hamiltonian as they act on this subspace. By applying
the appropriate projections to the general one-loop matrix elements, 
we obtain the expressions
\be
\label{bosonmatrixP}
\Braket{ J | a_n^A a_{-n}^B \left( {H}_{\rm BB} \right) a_{-n}^{C \dagger} a_n^{D \dagger} | J }   \to 
	-2 \frac{n^2 \lambda'}{J}
	\left(
	\delta^{ a d' }\delta^{ b' c } 
	+\delta^{ a' d }\delta^{ b c' } 
	+\delta^{ a d }\delta^{ b' c' } 
	+\delta^{ a' d' }\delta^{ b c } 
	\right)~,
\nn\\
\ee\be
\Braket{ J| b_{n}^\alpha b_{-n}^\beta \left( {H}_{\rm BF} \right)
	a_{-n}^{A\dagger} a_{n}^{B\dagger} | J }  &\to& 
	\frac{n^2 \lambda'}{2 J}\Bigl[ 
		\left( \Pi_+\gamma^{ab'} \Pi_-\right)^{\alpha\beta}
		- \left( \Pi_+\gamma^{a'b} \Pi_- \right)^{\alpha\beta}
\nn\\
&&\kern+10pt	+\left( \Pi_-\gamma^{ab'} \Pi_+\right)^{\alpha\beta}
		- \left( \Pi_-\gamma^{a'b} \Pi_+ \right)^{\alpha\beta}
	\Bigr]~,
\ee\be
\label{fermimatrixP}
&&\kern-25pt\Braket{ J |  b_n^\alpha  b_{-n}^\beta \left( {H}_{\rm FF} \right) 
	  b_{-n}^{\gamma\dagger}  b_n^{\delta\dagger}| J }  \to 
	-2  \frac{n^2\lambda'}{J} \left(
	\Pi_+^{\alpha\delta}\Pi_-^{\beta\gamma}
	+ \Pi_-^{\alpha\delta}\Pi_+^{\beta\gamma}
	\right)	
\nn\\
&&\kern+20pt	+ \frac{n^2\lambda'}{24 J}\biggl\{
	\Bigl[
	(\Pi_+\gamma^{ij}\Pi_+)^{\alpha\delta}(\Pi_-\gamma^{ij}\Pi_-)^{\beta\gamma}
 	+ (\Pi_+\gamma^{ij}\Pi_-)^{\alpha\beta}(\Pi_-\gamma^{ij}\Pi_+)^{\gamma\delta}
\nn\\
&&\kern+20pt	- (\Pi_+\gamma^{ij}\Pi_-)^{\alpha\gamma}(\Pi_-\gamma^{ij}\Pi_+)^{\beta\delta}
	\Bigr]
	- \Bigl[
	(\Pi_+\gamma^{i'j'}\Pi_+)^{\alpha\delta}(\Pi_-\gamma^{i'j'}\Pi_-)^{\beta\gamma}
\nn\ee\be
&&\kern+20pt	+ (\Pi_+\gamma^{i'j'}\Pi_-)^{\alpha\beta}(\Pi_-\gamma^{i'j'}\Pi_+)^{\gamma\delta}
	- (\Pi_+\gamma^{i'j'}\Pi_-)^{\alpha\gamma}(\Pi_-\gamma^{i'j'}\Pi_+)^{\beta\delta}
	\Bigr]
\nn\\
&&\kern+20pt	+ \Bigl[
	(\Pi_-\gamma^{ij}\Pi_-)^{\alpha\delta}(\Pi_+\gamma^{ij}\Pi_+)^{\beta\gamma}
	+ (\Pi_-\gamma^{ij}\Pi_+)^{\alpha\beta}(\Pi_+\gamma^{ij}\Pi_-)^{\gamma\delta}
\nn\\
&&\kern+20pt	- (\Pi_-\gamma^{ij}\Pi_+)^{\alpha\gamma}(\Pi_+\gamma^{ij}\Pi_-)^{\beta\delta}
	\Bigr]
	- \Bigl[
	(\Pi_-\gamma^{i'j'}\Pi_-)^{\alpha\delta}(\Pi_+\gamma^{i'j'}\Pi_+)^{\beta\gamma}
\nn\\
&&\kern+20pt	+ (\Pi_-\gamma^{i'j'}\Pi_+)^{\alpha\beta}(\Pi_+\gamma^{i'j'}\Pi_-)^{\gamma\delta}
	- (\Pi_-\gamma^{i'j'}\Pi_+)^{\alpha\gamma}(\Pi_+\gamma^{i'j'}\Pi_-)^{\beta\delta}
	\Bigr] \biggr\}~.
\ee
Since the 64-dimensional space must contain four copies of the $({\bf 2,2;2,2})$ irrep,
the diagonalization problem is really only $4\times 4$ and quite easy to solve.
The results for the eigenvalues appear in Table \ref{mixspectrum}. 
Collecting the above results, we present the
complete $SO(4)_{AdS}\times SO(4)_{S^5}$ decomposition of
spacetime boson two-impurity states in Table \ref{specfinal}. 
\begin{table}[ht!]
\be
\begin{array}{|c|c|}\hline
 SO(4)_{AdS}\times SO(4)_{S^5} & \Lambda_{\rm BF} \\
\hline ({\bf 2},{\bf 2};{\bf 2},{\bf 2})& -4 \\ ({\bf 2},{\bf
2};{\bf 2},{\bf 2})\times 2 & -2 \\ ({\bf 2},{\bf 2};{\bf
2},{\bf 2})& 0  \\ \hline
\end{array} \nonumber
\ee
\caption{String eigenstates in the subspace for which ${H}_{\rm BF}$ has
non-zero matrix elements} \label{mixspectrum}
\end{table}

\begin{table}[ht!]
\be
\begin{array}{|c|c|c|}\hline
	& SO(4)_{AdS}\times SO(4)_{S^5}   & 	 \Lambda   \\
\hline
{H}_{\rm BB}  & ({\bf 1},{\bf 1};{\bf 1},{\bf 1})  	 & -6 \\
		&  ({\bf 1},{\bf 1};{\bf 1},{\bf 1})  		& 2 \\
		& ({\bf 1},{\bf 1};{\bf 3},{\bf 1}) +({\bf 1},{\bf 1};{\bf 1},{\bf 3}) 	& -4 \\
		& ({\bf 3},{\bf 1};{\bf 1},{\bf 1}) +({\bf 1},{\bf 3};{\bf 1},{\bf 1})  &  0 \\
		& ({\bf 1},{\bf 1};{\bf 3},{\bf 3}) 		& -2 \\
		& ({\bf 3},{\bf 3};{\bf 1},{\bf 1}) 		& -2 \\
\hline
\end{array} \qquad
\begin{array}{|c|c|c|}\hline
	& SO(4)_{AdS}\times SO(4)_{S^5}   & 	 \Lambda   \\
\hline
{H}_{\rm FF} 	& ({\bf 1},{\bf 1};{\bf 1},{\bf 1}) 	& -2 \\
			& ({\bf 1},{\bf 1};{\bf 1},{\bf 1}) 	& -2 \\
			&  ({\bf 1},{\bf 1};{\bf 3},{\bf 1}) +({\bf 1},{\bf 1};{\bf 1},{\bf 3}) & 0 \\
			&  ({\bf 3},{\bf 1};{\bf 1},{\bf 1}) +({\bf 1},{\bf 3};{\bf 1},{\bf 1}) & -4 \\
			&  ({\bf 3},{\bf 1};{\bf 3},{\bf 1}) +({\bf 1},{\bf 3};{\bf 1},{\bf 3})	& -2 \\
\hline
 {H}_{\rm BF} 	&	({\bf 2},{\bf 2};{\bf 2},{\bf 2}) 	& 0 \\
		&  	({\bf 2},{\bf 2};{\bf 2},{\bf 2}) \times 2  		& -2 \\
			&	({\bf 2},{\bf 2};{\bf 2},{\bf 2})  		& -4 \\ \hline
\end{array} \nn
\ee
\caption{Group decomposition of the 128 two-impurity spacetime bosons }
\label{specfinal}
\end{table}

By projecting out closed subspaces of the one-loop Hamiltonian
we have successfully classified each of the energy levels in the bosonic
Fock space with an $SO(4)\times SO(4)$ symmetry label.  Similar arguments
can be applied to the fermionic Fock space, where two-impurity string states
mix individual bosonic and fermionic oscillators (we omit the details).
A summary of these results for all states, including spacetime fermions, is given in 
Table~\ref{allshifts}. The important fact to note is that the $\Lambda$ eigenvalues
and their multiplicities are exactly as required for consistency with the
full $PSU(2,2|4)$ symmetry of the theory. This is a nontrivial result since the
quantization procedure does not make the full symmetry manifest. It is also a very
satisfying check of the overall correctness of the extremely complicated set
of procedures we were forced to use. We can now proceed to a comparison with
gauge theory anomalous dimensions.

\begin{table}[ht!]
\be
\begin{array}{|c|ccccc|}\hline
{\rm Level} & 0 & 2 & 4 & 6 & 8 \\
\hline
{\rm Mult.} & 1 & 28 & 70 & 28 & 1 \\
\hline
\Lambda_{\rm Bose}   & -6 & -4 & -2 & 0 & 2 \\ \hline
\end{array}
\qquad\qquad
\begin{array}{|c|cccc|} \hline
{\rm Level} & 1 & 3 & 5 & 7 \\
\hline
{\rm Mult.} & 8 & 56 & 56 & 8  \\
\hline
\Lambda_{\rm Fermi}   & -5 & -3 & -1 & 1 \\ \hline
\end{array} \nn
\ee
\caption{First-order energy shift summary: complete two-impurity string multiplet} \label{allshifts}
\end{table}

\subsection{Gauge theory comparisons}
The most comprehensive analysis of one-loop anomalous dimensions of
BMN operators and their organization into supersymmetry multiplets
was given in \cite{Beisert:2002tn}. As stated in \cite{Callan:2003xr}, 
the above string theory calculations are in 
perfect agreement with the one-loop gauge theory predictions. For
completeness, we present a summary of the spectrum of dimensions 
of gauge theory operators along with a sampling of information about
their group transformation properties.

The one-loop formula for operator dimensions takes the generic form
\be
\Delta_n^{R} = 2 + \frac{g_{YM}^2 N_c}{{R}^2} n^2
	\left( 1 + \frac{\bar\Lambda}{{R}} + O({R}^{-2}) \right)\ .
\ee
The $O({R}^{-1})$ correction $\bar\Lambda$ 
for the set of two-impurity operators is predicted to match 
the corresponding $O(J^{-1})$ energy correction to 
two-impurity string states, labeled above by $\Lambda$.
Part of the motivation for performing the special projections on two-impurity
string states detailed above was to emerge with specific symmetry labels for
each of the string eigenstates.  String states of a certain representation content
of the residual $SO(4)\times SO(4)$ symmetry of $AdS_5\times S^5$ are
expected, by duality, to map to gauge theory operators with the same
representation labels in the $SL(2,{\bf C})$ Lorentz and $SU(4)$ 
${R}$-charge sectors of the gauge theory.  Knowing the symmetry content
of the string eigenstates therefore allows us to test this mapping
in detail.

The bosonic sector of the gauge theory, characterized by single-trace operators
with two bosonic insertions in the trace, appears in Table \ref{BBmatch}.
\begin{table}[ht!]
\begin{eqnarray}
\begin{array}{|c|c|c|}\hline
{\rm Operator} &  SO(4)_{AdS}\times SO(4)_{S^5} &\bar\Lambda\\
\hline
\Sigma_A\tr\left(\phi^AZ^p\phi^AZ^{R-p}\right)
 &({\bf 1,1;1, 1}) & -6 \\
\tr\left(\phi^{(i}Z^p\phi^{j)}Z^{R-p}\right)
 &({\bf 1,1; 3,3}) & -2\\
\tr\left(\phi^{[i}Z^p\phi^{j]}Z^{R-p}\right)
 &({\bf 1,1;3,1}) +({\bf 1,1;1,3})&-4\\
\hline
\tr\left(\nabla_\mu Z Z^p\nabla^\mu Z Z^{R-2-p}\right)
 &({\bf 1,1;1, 1})& 2\\
\tr\left(\nabla_{(\mu}ZZ^p\nabla_{\nu)}Z Z^{R-2-p}\right) &
 ({\bf 3,3;1, 1}) & -2\\
\tr\left(\nabla_{[\mu}ZZ^p\nabla_{\nu]}Z Z^{R-2-p}\right) & ({\bf
3,1;1, 1}) + ({\bf 1,3 ;1, 1}) & 0 \\ \hline
\end{array} \nonumber
\end{eqnarray}
\caption{Bosonic gauge theory operators: either spacetime or ${R}$-charge singlet.}\label{BBmatch}
\end{table}
The set of operators comprising Lorentz scalars clearly agrees with the
corresponding pure-boson string states in Table~\ref{specfinal}, 
which are scalars in $AdS_5$.  Operators containing pairs of spacetime derivatives
correspond to string theory states that are scalars of the $S^5$ subspace.
The bi-fermion sector of the string theory corresponds to the set of two-gluino operators
in the gauge theory.  A few of these operators are listed in Table~\ref{FFmatch}.
These states, which form either spacetime or ${R}$-charge scalars,
clearly agree with their string theory counterparts, which were constructed
explicitly above. The string states appearing in
the $({\bf 2,2;2,2})$ representation
(listed in Table~\ref{mixspectrum}) correspond to the operators listed in 
Table~\ref{BFmatch}.
Finally, the complete supermultiplet spectrum of two-impurity gauge
theory operators
appears in Table~\ref{smultiplicity2}.
The extended supermultiplet spectrum is in perfect agreement with the complete
one-loop string theory spectrum in Table~\ref{allshifts} above.
\begin{table}[ht!]
\begin{eqnarray}
\begin{array}{|c|c|c|}\hline
{\rm Operator} & SO(4)_{AdS}\times SO(4)_{S^5}&\bar\Lambda\\ \hline
\tr\left(\chi^{[\alpha}Z^p\chi^{\beta]}Z^{R-1-p}\right)
    &({\bf 1},{\bf 1};{\bf 1},{\bf 1})& -2 \\
\tr\left(\chi^{(\alpha}Z^p\chi^{\beta)}Z^{R-1-p}\right)
    &({\bf 1},{\bf 1};{\bf 3},{\bf 1})& 0 \\
\tr\left(\chi[\sigma_{\mu},\tilde\sigma_\nu] Z^p\chi Z^{R-1-p}\right)
    &({\bf 3},{\bf 1};{\bf 1},{\bf 1})& -4 \\
\hline
\end{array} \nonumber
\end{eqnarray}
\caption{Bosonic gauge theory operators with two gluino impurities.} \label{FFmatch}
\end{table}
\begin{table}[ht!]
\begin{eqnarray}
\begin{array}{|c|c|c|}\hline
{\rm Operator} & SO(4)_{AdS}\times SO(4)_{S^5}&\bar\Lambda\\ \hline
\tr\left(\phi^i Z^p \nabla_\mu Z Z^{R-1-p}\right) + \dots
        &({\bf 2},{\bf 2};{\bf 2},{\bf 2})&  -4 \\ \hline
\tr\left(\phi^i Z^p \nabla_\mu Z Z^{R-1-p}\right)
        &({\bf 2},{\bf 2};{\bf 2},{\bf 2})& -2 \\ \hline
\tr\left(\phi^i Z^p \nabla_\mu Z Z^{R-1-p}\right) +\dots
        &({\bf 2},{\bf 2};{\bf 2},{\bf 2})& 0 \\ \hline
\end{array} \nonumber
\end{eqnarray}
\caption{Bosonic gauge theory operators: spacetime and ${R}$-charge non-singlets } \label{BFmatch}
\end{table}
\begin{table}[ht!]
\begin{eqnarray}
\begin{array}{|l|l|l|l|l|l|l|l|l|l|}\hline
{\rm Level} & 0& 1& 2& 3& 4& 5& 6& 7& 8 \\ \hline
{\rm Multiplicity} & 1& 8& 28& 56& 70& 56& 28& 8& 1 \\ \hline
\delta E\times ({R}^2/g_{YM}^2N_c n^2) & -{6}/{R} & -{5}/{R} &
-{4}/{R} & -{3}/{R} & -{2}/{R} & -{1}/{R} &
0  & {1}/{R} & {2}/{R} \\ \hline
\end{array} \nonumber
\end{eqnarray}
\caption{Anomalous dimensions of two-impurity operators}
\label{smultiplicity2}
\end{table}

\ \\
\ \\
\ \\
\ \\
\ \\
\ \\
\ \\
\ \\

\section{Energy spectrum at all loops in $\lambda'$}
\label{twoimpSEC7}		
To make comparisons with gauge theory dimensions at one loop
in $\lambda = g_{YM}^2 N_c$, we have expanded all string
energies in powers of the modified 't Hooft coupling 
$\lambda' = g_{YM}^2 N_c/{R}^2$. The string theory analysis is exact to all
orders in $\lambda'$, however, and it is possible to extract
a formula for the $O(1/J)$ string energy corrections that is 
exact in $\lambda'$ and suitable for comparison with 
higher-order corrections to operator dimensions in the gauge theory.  
In practice, it is slightly more difficult to diagonalize the string
Hamiltonian when the matrix elements are not expanded in 
small $\lambda'$.  This is mainly because, beyond leading order,
${H}_{\rm BF}$ acquires additional terms that mix bosonic
indices in the same $SO(4)$ and also mix bi-fermionic indices in the 
same $({\bf 1},{\bf 2};{\bf 1},{\bf 2})$ or $({\bf 2},{\bf 1};{\bf 2},{\bf 1})$
representation.  Instead of a direct diagonalization of the entire 
128-dimensional subspace of spacetime bosons, for example, we find
it more convenient to exploit the `dimension reduction' that can
be achieved by projecting the full Hamiltonian onto individual
irreps. 

For example, the $({\bf 1,1;1,1})$ irrep appears four times in 
Table~\ref{specfinal} and is present at levels $L=0,4,8$ in the
supermultiplet. To get the exact eigenvalues for this irrep, we 
will have to diagonalize a $4\times 4$ matrix. The basis vectors
of this bosonic sector comprise singlets of the two $SO(4)$ subgroups 
($a^{\dag a} a^{\dag a}\ket{J}$ and $a^{\dag a'}a^{\dag a'}\ket{J}$)
plus two bi-fermion singlets constructed from the $({\bf 2,1;2,1})$ 
and $({\bf 1,2;1,2})$ creation operators 
($\hat b^{\dag\alpha} \hat b^{\dag\alpha}\ket{J}$ and 
$\tilde b^{\dag\alpha}\tilde b^{\dag\alpha}\ket{J}$).
The different Hamiltonian matrix elements that enter the $4\times 4$ 
matrix are symbolically indicated in Table~\ref{singlet}. It is a 
simple matter to project the general expressions for matrix elements 
of ${H}_{\rm BB}$, etc., onto singlet states and so obtain the matrix
as an explicit function of $\lambda', n$. The matrix can be exactly
diagonalized and yields the following energies:
\be
E_0(n,J) & = & 2\sqrt{1 + \lambda'n^2} - \frac{n^2\lambda'}{J}\left[ 2 + \frac{4}{\sqrt{1+n^2\lambda'}}\right]
	+ O(1/J^2)~,
\nn\\
E_4(n,J) & = & 2\sqrt{1 + \lambda'n^2} - \frac{2 n^2\lambda'}{J} + O(1/J^2)~,
\nn\\
E_8(n,J) & = & 2\sqrt{1 + \lambda'n^2} - \frac{n^2\lambda'}{J}\left[ 2 - \frac{4}{\sqrt{1+n^2\lambda'}}\right]
	+ O(1/J^2)~.
\ee
The subscript $L=0,4,8$ indicates the supermultiplet level to which the eigenvalue
connects in the weak coupling limit. The middle eigenvalue ($L=4$) is doubly
degenerate, as it was in the one-loop limit.

\begin{table}[ht!]
\begin{eqnarray}
\begin{array}{|c|cccc|}\hline
H_{\rm int}	& a^{\dag a}a^{\dag a}\ket{J} & a^{\dag a'}a^{\dag a'}\ket{J} & 
	\hat b^{\dag\alpha} \hat b^{\dag\alpha}\ket{J} & \tilde b^{\dag\alpha}\tilde b^{\dag\alpha}\ket{J}
\\ \hline
\bra{J}a^{a}a^{a} & {H}_{\rm BB} & {H}_{\rm BB} & {H}_{\rm BF}  & {H}_{\rm BF}
\\ 
\bra{J}a^{a'}a^{a'} & {H}_{\rm BB} & {H}_{\rm BB} & {H}_{\rm BF}  & {H}_{\rm BF}
\\ 
\bra{J}\hat b^{\alpha} \hat b^{\alpha} & {H}_{\rm BF}  & {H}_{\rm BF} & 
		{H}_{\rm FF}  & {H}_{\rm FF}
\\ 
\bra{J}\tilde b^{\alpha} \tilde b^{\alpha} & {H}_{\rm BF}  & {H}_{\rm BF} & 
		{H}_{\rm FF}  & {H}_{\rm FF}
\\ \hline
\end{array} \nn
\end{eqnarray}
\caption{Singlet projection at finite $\lambda'$ }
\label{singlet}
\end{table}

There are two independent $2\times 2$ matrices that mix states at levels 
$L=2,6$. According to Table~\ref{specfinal}, one can project out the 
antisymmetric bosonic and antisymmetric bi-fermionic states in the irrep
$({\bf 1,1;3,1})+({\bf 1,1;1,3})$ or in the irrep 
$({\bf 3,1;1,1})+({\bf 1,3;1,1})$. The results of 
eqns.~(\ref{adsscalarminus}, \ref{sscalarminus}, \ref{adsscalarplus}, \ref{sscalarplus}) 
can be used to carry out the needed 
projections and obtain explicit forms for the matrix elements of the perturbing 
Hamiltonian. The actual $2\times 2$ diagonalization is trivial to do and both
problems give the same result. The final result for the energy levels (using the
same notation as before) is
\be
E_2(n,J) & = & 2\sqrt{1 + \lambda'n^2} - \frac{n^2\lambda'}{J}\left[ 2 + \frac{2}{\sqrt{1+n^2\lambda'}}\right]
	+ O(1/J^2)~,
\nn\\
E_6(n,J) & = & 2\sqrt{1 + \lambda'n^2} - \frac{n^2\lambda'}{J}\left[ 2 - \frac{2}{\sqrt{1+n^2\lambda'}}\right]
	+ O(1/J^2)~.
\ee
We can carry out similar diagonalizations for the remaining irreps of Table~\ref{specfinal},
but no new eigenvalues are encountered: the energies already listed are the exact energies 
of the $L=0,2,4,6,8$ levels. It is also easy to see that the degeneracy structure 
of the exact levels is the same as the one-loop degeneracy.

The odd levels of the supermultiplet are populated by the 128-dimensional 
spacetime fermions, and this sector of the theory can be diagonalized directly. 
Proceeding in a similar fashion as in the bosonic sector, we find exact energy
eigenvalues for the $L=1,3,5,7$ levels (with unchanged multiplicities). We refrain
from stating the individual results because the entire supermultiplet spectrum,
bosonic and fermionic, can be written in terms of a single concise formula:
to leading order in $1/J$ and all orders in $\lambda'$, the energies of the 
two-impurity multiplet are given by
\begin{eqnarray}
\label{stringfinal}
E_L(n,J) & = & 2\sqrt{1+\lambda' n^2} 
	-\frac{n^2\lambda'}{J}\left[
	2+\frac{(4-L)}{\sqrt{1+n^2\lambda'}}\right]+O(1/J^2)~,
\end{eqnarray}
where $L=0,1,\ldots,8$ indicates the level within the supermultiplet.
The degeneracies and irrep content are identical to what we found at
one loop in $\lambda'$. This expression can be rewritten, correct
to order $J^{-2}$, as follows:
\begin{eqnarray}
\label{stringxpnd}
E_L(n,J)  \approx 2\sqrt{1+\frac{\lambda n^2}{(J-{L}/{2})^2}}
	-\frac{n^2\lambda}{(J-L/2)^3}\left[
	2+\frac{4}{\sqrt{1+{\lambda n^2}/{(J-L/2)^2}}}\right] ~.
\end{eqnarray}
This shows that, within this expansion, the joint dependence on $J$ 
and $L$ is exactly what is required for extended supersymmetry
multiplets. This is a rather nontrivial functional requirement,
and a stringent check on the correctness of our quantization 
procedure (independent of any comparison with gauge theory).

In order to make contact with gauge theory we expand (\ref{stringfinal}) in 
$\lambda'$, obtaining
\begin{eqnarray}
\label{stringfinalexp}
E_L(n,J)  & \approx& \left[ 2 + \lambda' n^2 -
	 \frac{1}{4}(\lambda' n^2)^2 + \frac{1}{8}(\lambda' n^2)^3 
	+\dots \right]  \nonumber\\ 
	&~ &\kern-70pt + \frac{1}{J}\left[ n^2\lambda'(L-6)+
	(n^2\lambda')^2\left(\frac{4-L}{2}\right)+
	(n^2\lambda')^3\left(\frac{3L-12}{8}\right) +\ldots\right]~.
\end{eqnarray}

We can now address the comparison with higher-loop results 
on gauge theory operator dimensions. Beisert, Kristjansen and 
Staudacher \cite{Beisert:2003tq} computed the two-loop correction to 
the anomalous dimensions of a convenient class of operators 
lying at level four in the supermultiplet. The operators in question
lie in a symmetric-traceless irrep of an $SO(4)$ subgroup of the
${R}$-charge and are guaranteed by group theory not to mix
with any other fields \cite{Beisert:2003tq}. The following 
expression for the two-loop anomalous dimension was found: 
\begin{eqnarray}
\label{twoloopgauge}
\delta\Delta_n^{R} & = & -\frac{g_{YM}^4 N_c^2}{\pi^4}
	\sin^4\frac{n\pi}{{R}+1}\left(
	\frac{1}{4}+\frac{\cos^2\frac{n\pi}{{R}+1}}{{R}+1}\right)~.
\end{eqnarray} 
As explained above, ${\cal N}=4$ supersymmetry ensures that the dimensions 
of operators at other levels of the supermultiplet will be obtained
by making the substitution ${R} \to {R}+2-{L}/{2}$ in the
expression for the dimension of the $L=4$ operator. Making that substitution
and taking the large-${R}$ limit we obtain a general formula for the
two-loop, large-${R}$ correction to the anomalous dimension of the general
two-impurity operator:
\begin{eqnarray}
\label{twoloopgaugeL}
\delta\Delta_n^{{R},L} & = & -\frac{g_{YM}^4 N_c^2}{\pi^4}
	\sin^4\frac{n\pi}{{R}+3-L/2}~\left(
	\frac{1}{4}+\frac{\cos^2\frac{n\pi}{{R}+3-L/2}}
		{{R}+3-L/2}\right)~\nonumber\\
	&\approx &  -\frac{1}{4}(\lambda' n^2)^2 + 
	\frac{1}{2}(\lambda' n^2)^2~ \frac{4-L}{{R}} + O(1/{R}^2) ~, 
\end{eqnarray}
Using the identification ${R}\rightleftharpoons J$ specified by 
duality, we see that this expression matches the corresponding string result
in (\ref{stringfinalexp}) to $O(1/J)$, confirming the AdS/CFT
correspondence to two loops in the gauge coupling.  

The three-loop correction to the dimension of this same class of $L=4$ gauge theory 
operators has recently been definitively determined \cite{Beisert:2003ys}. The
calculation involves a remarkable interplay between gauge theory and integrable 
spin chain models
\cite{Beisert:2003jb,Beisert:2003tq,Beisert:2003jj,Klose:2003qc}. The final
result is
\be
\delta\Delta_n^{R}  =  
	\left(\frac{\lambda}{\pi^2}\right)^3
	\sin^6 \frac{n\pi}{{R}+1}
	\left[
	\frac{1}{8} + \frac{
			\cos^2 \frac{n\pi}{{R}+1}}{4({R}+1)^2}
	\left( 3{R}+2({R}+6)\cos^2 \frac{n\pi}{{R}+1}
	\right)
	\right]\ .
\label{3LP}
\ee
If we apply to this expression the same logic applied to 
the two-loop gauge theory result (\ref{twoloopgauge}), we obtain the 
following three-loop correction to the anomalous dimension of the 
general level of the two-impurity operator supermultiplet:
\begin{eqnarray}
\label{threeloopgaugeL}
\delta\Delta_n^{{R},L} \approx  
 \frac{1}{8}(\lambda' n^2)^3 - 
	\frac{1}{8}(\lambda' n^2)^2~ \frac{8-3L}{{R}} + O(1/{R}^2) ~.
\end{eqnarray}
We see that this expression differs from the third-order contribution 
to the string result (\ref{stringfinalexp}) for the corresponding quantity.
The difference is a constant shift and one might hope to absorb it in a 
normal-ordering constant. However, our discussion of the normal-ordering 
issue earlier in the chapter seems to exclude any such freedom.

\section{Discussion}
In this chapter we have given
a detailed account of the quantization of the first curvature correction 
to type IIB superstring theory in the plane-wave limit of $AdS_5\times S^5$. 
We have presented the detailed diagonalization of the resulting perturbing 
Hamiltonian on the degenerate subspace of two-impurity states, obtaining 
string energy corrections that can be compared with higher-loop anomalous
dimensions of gauge theory operators. Beyond the Penrose limit, the 
holographic mapping between each side of the correspondence is 
intricate and nontrivial, and works perfectly to two loops in the gauge 
coupling.  The agreement, however, appears to break down at three loops.  
(Similar three-loop disagreements have appeared in semiclassical 
string analyses; see, for example, ref.~\cite{Serban:2004jf}.) This troubling issue was 
first observed in \cite{Callan:2003xr}, at which time the third-order gauge theory 
anomalous dimension was somewhat conjectural. In the intervening time, the 
third-order result (\ref{3LP}) acquired a solid basis, thus confirming 
the mismatch. Several questions arise about this mismatch: is it due to
a failure of the AdS/CFT correspondence itself?  Does it signal the need
to modify the worldsheet string action?  Is it simply that the
perturbative approach to the gauge theory anomalous dimensions is not
adequate in the relevant limits?   

Regarding this final point, 
a specific explanation has been proposed that may account for the 
disagreement with gauge theory at three loops.  The essential idea is that
certain types of gauge theory mixing terms that connect fields within single-trace 
operators are dropped in the particular limit that is taken 
in this setup.  This amounts to a plausible order-of-limits problem:  
we will leave a more detailed discussion of this proposal for Chapter \ref{con}.
Despite vigorous investigation from 
several directions, all of these questions remain open. 

%
%
\chapter{The curvature expansion: Three impurities} 		          
\label{threeimp}
Thus far, we have seen that attempts to push the original results of BMN further 
have gone in two independent directions.  In the gauge theory, the calculation of
anomalous dimensions of BMN operators has been greatly simplified by Minahan and Zarembo's
discovery that the problem can be mapped to that of computing the energies of
certain integrable spin chains \cite{Minahan:2002ve}.  Based on this development,
calculations in certain sectors of the theory
have been carried out to three loops in the 't~Hooft coupling $\lambda$
\cite{Beisert:2003tq,Beisert:2003ys}.\footnote{We note that the conjectural
three-loop computation of \cite{Beisert:2003tq} was solidified by field
theoretic methods in \cite{Beisert:2003ys}.} 
Furthermore, we have shown in the previous chapter that the quantization of 
the GS string in the $AdS_5\times S^5$ background has developed 
far enough to enable perturbative computations of the effect of worldsheet 
interactions on the spectrum of the string when it is boosted to large, but finite, 
angular momentum $J$ \cite{Callan:2003xr,Callan:2004uv,Parnachev:2002kk}. 
As demonstrated in Chapter \ref{twoimp}, these two approaches lead to different 
expansions of operator anomalous dimensions (or string eigenenergies): on the 
gauge theory side, one naturally has an expansion in the coupling constant 
$\lambda$ that is typically exact in $R$-charge; on the 
string theory side one has an expansion in inverse powers of angular momentum $J$
(the dual of gauge theory $R$-charge) that is exact in $\lambda$. 

The expansion on the string side is difficult and has so far been carried out
to ${O}(1/J)$ for two-impurity states (i.e.,~states with two 
string oscillators excited). The resulting functions of the loop expansion parameter 
$\lambda$ can be compared with the large $R$-charge expansion of two-impurity BMN operators 
in the gauge theory to provide new and stringent tests of the AdS/CFT 
correspondence.  As mentioned above, gauge theory technology
has made it possible to compute anomalous dimensions of certain two-impurity
BMN operators out to at least three-loop order. The agreement between dual quantities 
is perfect out to two-loop order but, surprisingly, seems to break down at three loops 
\cite{Callan:2003xr,Callan:2004uv}.  
Exactly what this means for the AdS/CFT correspondence is not yet clear but, 
given the circumstances, it seems appropriate to at least look for further 
data on the disagreement in the hope of finding some instructive systematics.  
The subject of this chapter is to pursue one possible line of attack in which 
we extend the calculations described above to higher-impurity string states 
and gauge theory operators.  The extension of our two-impurity results to 
higher impurities is not a straightforward matter on either side of the 
correspondence and gets more complex as the number of impurities increases. 
We focus here on the three-impurity case, where we obtain results that validate 
our methods for quantizing the GS superstring; the agreement with gauge theory
at one and two loops is impressive, though we will also confirm the previously observed
breakdown of agreement at three-loop order.

In Sections \ref{threeimp2} and \ref{threeimp3} we present the details of the 
diagonalization of the perturbing string worldsheet Hamiltonian on degenerate subspaces
of three-impurity states.  We give a
compressed discussion of general strategy, concentrating on the
aspects of the problem which are new to the three-impurity case.
An interesting new element is that the non-interacting degenerate 
subspace breaks up into several different supersymmetry multiplets 
so that the detailed accounting of multiplicities and irrep
decomposition amounts to a stringent test that the quantization
has maintained the correct nonlinearly realized superconformal
symmetries.  Section \ref{threeimp4} is devoted to the comparison of the string
theory spectrum with gauge theory anomalous dimensions. 
We employ our own gauge theory data derived in Chapter \ref{virial}
for the various higher-loop spin chains onto which
the gauge theory anomalous dimension problem has been mapped.
We find perfect agreement through two-loop order and,
once again, a breakdown at three loops. 

Overall, the three-impurity regime of the string theory
offers a much more stringent test of the duality away from
the full plane-wave limit. While we are unable to offer (via this analysis) 
a solution to the disagreement with gauge theory at three loops, we can 
confirm that the complicated interacting worldsheet theory
at $O(1/\Rhat^2)$ in the curvature expansion is properly quantized 
and correct to two loops in $\lambda$.

\section{Three-impurity spectrum: one loop in $\lambda'$}
\label{threeimp2}
The three-impurity Fock space block-diagonalizes into separate spacetime 
fermion and spacetime boson sectors. The bosonic sector contains states 
that are purely bosonic (composed of three bosonic string oscillators) and 
states with bi-fermionic components:
\be
a_q^{A\dag} a_r^{B\dag} a_s^{C\dag}\ket{J}~,
	\qquad a_q^{A\dag} b_r^{\alpha\dag} b_s^{\beta\dag}\ket{J}\ .
\label{state1}
\ee
Pure boson states are mixed by the bosonic sector of the Hamiltonian
$H_{\rm BB}$, while states with bi-fermionic excitations are mixed both by the purely
fermionic Hamiltonian $H_{\rm FF}$ and the bose-fermi sector $H_{\rm BF}$.
The sector of spacetime fermion states is composed of purely fermionic excitations 
and mixed states containing two bosonic oscillators:
\be
b_q^{\alpha\dag} b_r^{\beta\dag} b_s^{\gamma\dag}\ket{J}~,
	\qquad a_q^{A\dag} a_r^{B\dag} b_s^{\alpha\dag}\ket{J}\ .
\label{state2}
\ee
Pure fermion states are acted on  by $H_{\rm FF}$, and mixed states with 
bosonic excitations are acted on by $H_{\rm BB}$ and $H_{\rm BF}$. This block 
diagonalization of the perturbing Hamiltonian is displayed schematically 
in table~\ref{blockform3}.
\begin{table}[ht!]
\begin{eqnarray}
\begin{array}{|c|cccc|}
\hline
 H_{\rm int} & a^{A\dagger} a^{B\dagger} a^{C\dagger} \ket{J} &
        a^{A\dag} b^{\alpha\dagger} b^{\beta\dagger} \ket{J} &
        b^{\alpha\dagger} b^{\beta\dagger} b^{\gamma\dag} \ket{J} &
        a^{A\dag} a^{B\dag} b^{\alpha\dag} \ket{J} \\
        \hline
\bra{J} a^{A} a^{B} a^C & H_{\rm BB} & H_{\rm BF} &0&0 \\
\bra{J} a^A b^{\alpha} b^{\beta} & H_{\rm BF} & H_{\rm FF}+ H_{\rm BF}&0&0\\ 
\bra{J} b^{\alpha}  b^{\beta}  b^\gamma  &0&0& H_{\rm FF} & H_{\rm BF} \\ 
\bra{J} a^{A}  a^{B}  b^{\alpha}  & 0 & 0 & H_{\rm BF} & H_{\rm BB} + H_{\rm BF} \\
\hline
\end{array} \nonumber
\end{eqnarray}
\caption{Three-impurity string states}
\label{blockform3}
\end{table}

The three-impurity string states are subject to the usual level-matching 
condition on the mode indices: $q+r+s=0$. There are two generically 
different solutions of this constraint: all mode indices different ($q\neq r\neq s$)
and two indices equal (e.g.,~$q=r=n,\ s=-2n$).
In the inequivalent index case, there are $16^3=4,096$ degenerate states
arising from different choices of spacetime labels on the mode creation operators.
In the case of two equivalent indices, the dimension of the degenerate subspace
is half as large (there are fewer permutations on mode indices that generate 
linearly independent states). The two types of basis break up into irreducible 
representations of $PSU(2,2|4)$ in different ways and must be studied separately.

As in the two-impurity case, the problem of diagonalizing the perturbation
simplifies enormously when the matrix elements are expanded to leading order
in $\lambda'$.  We will take this approach here
to obtain an overview of how degeneracies are lifted by the interaction.
The generalization of the results to all loop orders in $\lambda'$ (but still 
to first non-leading order in $1/J$) will be presented in the next section.
It is once again the case that in the one-loop approximation, 
projection onto invariant subspaces under the manifest global $SO(4)\times SO(4)$ 
symmetry often diagonalizes the Hamiltonian directly (and at worst reduces 
it to a low-dimensional matrix).  Symbolic manipulation programs
were used to organize the complicated algebra and to perform explicit
projections onto invariant subspaces. 

\subsection{Inequivalent mode indices $(q\neq r\neq s)$}
In the sector of spacetime bosons, the subspace
of purely bosonic states $a_q^{A\dag} a_r^{B\dag} a_s^{C\dag}\ket{J}$ 
is 512-dimensional.  When each of the three mode indices $(q,r,s)$ are 
different, states with bi-fermionic excitations 
$a_q^{A\dag} b_r^{\alpha\dag} b_s^{\beta\dag}\ket{J}$
are inequivalent under permutation of the mode indices, and form a 
1,536-dimensional subsector. The entire bosonic sector of the three-impurity 
state space therefore contains 2,048 linearly independent states.  
The fermionic sector decomposes in a similar manner:
the subsector of purely fermionic states 
$b_q^{\alpha\dag} b_r^{\beta\dag} b_s^{\gamma\dag}\ket{J}$  
is 512-dimensional; fermionic states containing two bosonic excitations 
$a_q^{A\dag} a_r^{B\dag} b_s^{\alpha\dag}\ket{J}$ 
are inequivalent under permutation of the mode indices, and comprise
an additional 1,536-dimensional subsector. Adding this 2,048-dimensional 
fermion sector brings the dimensionality of the entire state space to 4,096.  

Our first task is to evaluate the interaction Hamiltonian matrix. 
The matrix elements needed to fill out the spacetime boson sector 
are listed in table~\ref{boseblock3}. To evaluate the entries, 
we express the Hamiltonian $H_{\rm int}$ computed in Chapter \ref{twoimp}
in terms of mode creation and 
annihilation operators, expand the result in powers of $\lambda'$ and then 
compute the indicated matrix elements between three-impurity Fock space states.
The pure-boson $(H_{\rm BB})$, pure-fermion $(H_{\rm FF})$ and bose-fermi $(H_{\rm BF})$ 
mixing sectors of $H_{\rm int}$ appear
above in eqns.~(\ref{Hpurbos}), (\ref{Hpurferm}) and (\ref{Hmix}) respectively.
We collect below all the relevant results of this exercise.
\begin{table}[ht!]
\begin{eqnarray}
\begin{array}{|c|cccc|}
\hline
 H_{\rm int} & 
	a_s^{D\dag}  a^{E\dagger}_r  a^{F\dagger}_{q}\ket{J} &
	a_s^{D\dag}  b^{\gamma\dagger}_r  b^{\delta\dagger}_{q}\ket{J} &
        a_r^{D\dag}  b^{\gamma\dagger}_q  b^{\delta\dagger}_{s}\ket{J} &
	a_r^{D\dag}  b^{\gamma\dagger}_s  b^{\delta\dagger}_{q}\ket{J}
\\   \hline
\bra{J} a^{A}_q  a^{B}_r  a^{C}_{s}      & H_{\rm BB}    & H_{\rm BF}
						 & H_{\rm BF} & H_{\rm BF} \\
\bra{J} a^{A}_q  b^{\alpha}_r  b^{\beta}_{s}& H_{\rm BF} & H_{\rm FF}+H_{\rm BF} 
						& H_{\rm BF} & H_{\rm BF} \\
\bra{J} a^{A}_s  b^{\alpha}_q  b^{\beta}_{r}& H_{\rm BF} & H_{\rm BF} 
					& H_{\rm FF}+H_{\rm BF} & H_{\rm BF} \\ 
\bra{J} a^{A}_r  b^{\alpha}_s  b^{\beta}_{q}& H_{\rm BF} & H_{\rm BF} 
					& H_{\rm BF} & H_{\rm FF}+ H_{\rm BF}\\ 
\hline
\end{array} \nonumber
\end{eqnarray}
\caption{$H_{\rm int}$ on spacetime-boson
three-impurity string states $(q\neq r\neq s)$}
\label{boseblock3}
\end{table}

We will use an obvious $(m,n)$ matrix notation to distinguish the different entries in
table~\ref{boseblock3}. The purely bosonic, 512-dimensional $(1,1)$ block has
the explicit form
\be
\Braket{J| a_q^{A} a_r^{B} a_s^{C} (H_{\rm BB})a_s^{D\dag} a_r^{E\dag} a_q^{F\dag}|J}
	& = & 
	\frac{\lambda'}{J}
	\delta^{AF}\delta^{BE}\delta^{CD}
	\left( rs+q(r+s)-q^2-r^2-s^2\right)
\nn\\	+ \frac{\lambda'}{2J}
	\biggl\{
	\delta^{AF}\Bigl[
	(r^2+s^2)
	\bigl(
	\delta^{cd}\delta^{be}&&\kern-25pt  
	-  \delta^{c'd'}\delta^{b'e'}
	\bigr)
	+ (s^2-r^2)
	\bigl(
	\delta^{be}\delta^{c'd'}  
	- \delta^{cd}\delta^{b'e'}
	\bigr)
\nn\\	+2rs\bigl(
	\delta^{bd}\delta^{ce} 
	- \delta^{bc}\delta^{de} &&\kern-25pt 
	- \delta^{b'd'}\delta^{c'e'} 
	+ \delta^{b'c'}\delta^{d'e'}
	\bigr)
	\Bigr]
	+  
	\bigl( 
	r\rightleftharpoons q,\ 
	F\rightleftharpoons E,\ 
	A\rightleftharpoons B
	 \bigr)
\nn\\
&&\kern+35pt	+  \bigl( 
	s\rightleftharpoons q,\ 
	F\rightleftharpoons D,\ 
	A\rightleftharpoons C
	 \bigr)\biggr\}\ .
\label{bosons}
\ee
The off-diagonal entries that mix purely bosonic states
$a_q^{A\dag}a_r^{B\dag}a_s^{C\dag}\Ket{J}$ with states containing bi-fermions
$a_q^{A\dag}b_r^{\alpha\dag}b_s^{\beta\dag}\Ket{J}$ are given by a separate set of 
512-dimensional matrices. The $(1,2)$ block in table~\ref{boseblock3}, for 
example, yields
\be
\Braket{J| a_q^{A} a_r^{B} a_s^{C} (H_{\rm BF}) 
	a_s^{D\dag} b_r^{\alpha\dag} b_q^{\beta\dag}|J}
	& = & 
	\frac{\lambda'}{2J}\delta^{CD} qr
	\biggl\{
	\left(\gamma^{ab'}\right)^{\alpha\beta} - 
	\left(\gamma^{a'b}\right)^{\alpha\beta}
	\biggr\}\ , 
\label{12block}
\ee
where the index $a\ (a')$ symbolizes the value of the vector index $A$, 
provided it is in the first (second) $SO(4)$.
There are six blocks in this subsector, each given by 
a simple permutation of the mode indices $(q,r,s)$ in eqn.~(\ref{12block}).  
In table~\ref{boseblock3}, these matrices occupy the $(1,2)$, 
$(1,3)$ and $(1,4)$ blocks, along with their transposes in the
$(2,1)$, $(3,1)$ and $(4,1)$ entries.  

The pure-fermion sector of the Hamiltonian, $H_{\rm FF}$, has non-vanishing
matrix elements between states containing bi-fermionic excitations.
The $H_{\rm FF}$ contribution to the $(2,2)$ block, for example, is given by
\be
\Braket{J| b_q^{\alpha} b_r^{\beta} a_s^{A} (H_{\rm FF}) 
	a_s^{B\dag} b_r^{\gamma\dag} b_q^{\delta\dag}|J}
	& = & 
	-\frac{\lambda'}{2J}(q-r)^2\delta^{AB}
	\delta^{\alpha\delta}\delta^{\gamma\beta}
\nn\\
	+\frac{\lambda'}{24 J}\delta^{AB} qr\biggl\{
	\bigl(\gamma^{ij}\bigr)^{\alpha\gamma} &&\kern-25pt
	\bigl(\gamma^{ij}\bigr)^{\beta\delta}
	- \bigl(\gamma^{ij}\bigr)^{\alpha\beta}
	\bigl(\gamma^{ij}\bigr)^{\gamma\delta}
	- \bigl(\gamma^{ij}\bigr)^{\alpha\delta}
	\bigl(\gamma^{ij}\bigr)^{\beta\gamma}
\nn\\
	-\bigl(\gamma^{i'j'}\bigr)^{\alpha\gamma} 
	\bigl(\gamma^{i'j'}\bigr)^{\beta\delta} &&\kern-25pt
	+ \bigl(\gamma^{i'j'}\bigr)^{\alpha\beta}
	\bigl(\gamma^{i'j'}\bigr)^{\gamma\delta}
	+ \bigl(\gamma^{i'j'}\bigr)^{\alpha\delta}
	\bigl(\gamma^{i'j'}\bigr)^{\beta\gamma}
	\biggr\}\ .
\label{HFF22}
\ee
A similar contribution, related to this one by simple permutations of
the mode indices $(q,r,s)$, appears in the diagonal blocks
$(3,3)$ and $(4,4)$ as well.

The bose-fermi mixing Hamiltonian $H_{\rm BF}$ makes the following contribution 
to the lower diagonal blocks $(2,2)$, $(3,3)$ and $(4,4)$ in 
table~\ref{boseblock3}:  
\be
&&
\Braket{J| b_q^{\alpha} b_r^{\beta} a_s^{A} (H_{\rm BF}) 
	a_s^{B\dag} b_r^{\gamma\dag} b_q^{\delta\dag}|J}
	 =  
	\frac{\lambda'}{2J}\biggl\{
	2s(q+r-s)\delta^{ab}\delta^{\alpha\delta}\delta^{\beta\gamma}
\nn\\	
&&\kern+30pt
	- rs 
	\Bigl[
	\bigl(\gamma^{ab}\bigr)^{\beta\gamma}
	-\bigl(\gamma^{a'b'}\bigr)^{\beta\gamma}
	\Bigr]
	-sq\Bigl[
	\bigl(\gamma^{ab}\bigr)^{\alpha\delta}
	-\bigl(    \gamma^{a'b'} \bigr)^{\alpha\delta}
	\Bigr]
\nn\\
&&\kern+30pt
	-2\Bigl[q^2+r^2+s^2- 
	s(q+r)\Bigr] 
	\delta^{a'b'}\delta^{\alpha\delta}\delta^{\beta\gamma}
	 \biggr\}	\ .
\ee
The $H_{\rm BF}$ sector also makes the following contribution to the 
off-diagonal $(2,3)$ block:
\be
\Braket{J| b_q^{\alpha} b_r^{\beta} a_s^{A} (H_{\rm BF}) 
	a_r^{B\dag} b_q^{\gamma\dag} b_s^{\delta\dag}|J}
	=  &&\kern-15pt
	-\frac{\lambda'}{2J}\delta^{\alpha\gamma}rs
	\biggl\{
	\bigl(\delta^{ab}-\delta^{a'b'}
	\bigr)\delta^{\beta\delta}
	-\bigl(\gamma^{ab}\bigr)^{\beta\delta}
	+\bigl(\gamma^{a'b'}\bigr)^{\beta\delta}
	\biggr\}\ .
\nn\\
&&
\ee
The contributions of $H_{\rm BF}$ to the remaining off-diagonal blocks $(2,3)$, 
$(2,4)$, etc.~are obtained by appropriate index permutations.


The sector of spacetime fermions decomposes in a similar fashion.
The fermion analogue of table~\ref{boseblock3} for the bosonic sector
appears in table~\ref{fermiblock3}.
\begin{table}[ht!]
\begin{eqnarray}
\begin{array}{|c|cccc|}
\hline
 H_{\rm int} & 
	b_s^{\zeta\dag}  b^{\epsilon\dagger}_r  b^{\delta\dagger}_{q}\ket{J} &
	a_s^{C\dag}  a^{D\dagger}_r  b^{\delta\dagger}_{q}\ket{J} &
        a_r^{C\dag}  a^{D\dagger}_q  b^{\delta\dagger}_{s}\ket{J} &
	a_r^{C\dag}  a^{D\dagger}_s  b^{\delta\dagger}_{q}\ket{J}
\\   \hline
\bra{J} b^{\alpha}_q  b^{\beta}_r  b^{\gamma}_{s}      
				& H_{\rm FF}    & H_{\rm BF} & H_{\rm BF} & H_{\rm BF} \\
\bra{J} b^{\alpha}_q  a^{A}_r  a^{B}_{s}& 
		H_{\rm BF} & H_{\rm BB}+H_{\rm BF} & H_{\rm BF} & H_{\rm BF} \\
\bra{J} b^{\alpha}_s  a^{A}_q  a^{B}_{r}& 
		H_{\rm BF} & H_{\rm BF} & H_{\rm BB}+H_{\rm BF} & H_{\rm BF} \\ 
\bra{J} b^{\alpha}_r  a^{A}_s  a^{B}_{q}& 
		H_{\rm BF} & H_{\rm BF} & H_{\rm BF} & H_{\rm BB}+ H_{\rm BF}\\ 
\hline
\end{array} \nonumber
\end{eqnarray}
\caption{Interaction Hamiltonian on spacetime fermion 
three-impurity states $(q\neq r\neq s)$}
\label{fermiblock3}
\end{table}
The $(1,1)$ fermion block is occupied by the pure-fermion sector of the Hamiltonian
taken between the purely fermionic three-impurity states 
$b^{\alpha\dag}_q  b^{\beta\dag}_r  b^{\gamma\dag}_{s}\Ket{J}$:
\be
\Braket{J| b_q^{\alpha} b_r^{\beta} b_s^{\gamma} (H_{\rm FF}) 
	b_s^{\zeta\dag} b_r^{\epsilon\dag} b_q^{\delta\dag}|J}
	& = & 
	-\frac{\lambda'}{J}\left[
	q^2+r^2+s^2-rs-q(r+s)\right]\delta^{\alpha\delta}
	\delta^{\beta\epsilon}\delta^{\gamma\zeta}
\nn\\	+ \frac{\lambda'}{24J}\delta^{\alpha\delta}rs
	\biggl\{
	&&\kern-25pt
	\bigl(\gamma^{ij}\bigr)^{\beta\gamma} 
	\bigl(\gamma^{ij}\bigr)^{\epsilon\zeta} 
	- \bigl(\gamma^{ij}\bigr)^{\beta\epsilon}
	\bigl(\gamma^{ij}\bigr)^{\gamma\zeta}
	+ \bigl(\gamma^{ij}\bigr)^{\beta\zeta}
	\bigl(\gamma^{ij}\bigr)^{\gamma\epsilon}
\nn\\	- &&\kern-25pt  \bigl(\gamma^{i'j'}\bigr)^{\beta\gamma}
	\bigl(\gamma^{i'j'}\bigr)^{\epsilon\zeta}
	+ \bigl(\gamma^{i'j'}\bigr)^{\beta\epsilon}
	\bigl(\gamma^{i'j'}\bigr)^{\gamma\zeta}
	- \bigl(\gamma^{i'j'}\bigr)^{\beta\zeta}
	\bigl(\gamma^{i'j'}\bigr)^{\gamma\epsilon}
\nn\\
	+ \bigl( &&\kern-28pt
	r\rightleftharpoons q,\ 
	\alpha\rightleftharpoons \beta,\ 
	\delta\rightleftharpoons \epsilon
	 \bigr)
	+  
	\bigl( 
	s\rightleftharpoons q,\ 
	\alpha\rightleftharpoons \gamma,\ 
	\delta\rightleftharpoons \zeta
	 \bigr)
	\biggr\}\ .
\nn\\&&
\ee
The off-diagonal $(1,2)$, $(1,3)$ and $(1,4)$ blocks (and their transposes)
mix purely fermionic states with $a_s^{A\dag} a_r^{B\dag} b_q^{\alpha\dag}\Ket{J}$
states:
\be
\Braket{J| b_q^{\alpha} b_r^{\beta} b_s^{\gamma} (H_{\rm BF}) 
	a_s^{A\dag} a_r^{B\dag} b_q^{\delta\dag}|J}
	& = & 
	-\frac{\lambda'}{2J}\delta^{\alpha\delta}rs
	\biggl\{
	\left(\gamma^{ab'}\right)^{\beta\gamma}
	- \left(\gamma^{a'b}\right)^{\beta\gamma}
	\biggr\}\ .
\label{12fermiblock}
\ee
The lower-diagonal $(2,2)$, $(3,3)$ and $(4,4)$ blocks receive contributions from the
pure boson sector of the Hamiltonian:
\be
&&\Braket{J| b_q^{\alpha} a_r^{A} a_s^{B} (H_{\rm BB}) 
	a_s^{C\dag} a_r^{D\dag} b_q^{\beta\dag}|J}
	=  
	-\frac{\lambda'}{2J}\delta^{\alpha\beta}
	\biggl\{
	(r-s)^2\delta^{BC}\delta^{AD}
\nn\\
&&\kern+80pt
	-(r^2+s^2)\Bigl(
	\delta^{ad}\delta^{bc}
	-\delta^{a'd'}\delta^{b'c'}
	\Bigr)
\nn\\
&&\kern+80pt
	-2rs\Bigl(
	\delta^{ac}\delta^{bd} 
	-\delta^{ab}\delta^{cd} 
	-\delta^{a'c'}\delta^{b'd'}
	+\delta^{a'b'}\delta^{c'd'}
	\Bigr)
\nn\\
&&\kern+80pt
	+(r^2-s^2)\Bigl(
	\delta^{ad}\delta^{b'c'}
	- \delta^{a'd'}\delta^{bc}
	\Bigr)
	\biggr\}\ .
\ee
In the same diagonal blocks of table~\ref{fermiblock3}, the $H_{\rm BF}$
sector contributes
\be
&&\kern-20pt
	\Braket{J| b_q^{\alpha} a_r^{A} a_s^{B} (H_{\rm BF}) 
	a_s^{C\dag} a_r^{D\dag} b_q^{\beta\dag}|J}
	 =  
	\frac{\lambda'}{8J}
	\biggl\{
	\delta^{\alpha\beta}\Bigl[
	+\bigl(
	8q(r+s)-5(r^2+s^2)-6q^2
	\bigr)\delta^{AD}\delta^{BC}
\nn\\
&&\kern+30pt
	+ (3q^2+s^2)\delta^{AD}\delta^{bc} 
	+ (3q^2+r^2)\delta^{BC}\delta^{ad} 
	+ (r^2-5q^2)\delta^{BC}\delta^{a'd'}
\nn\\
&&\kern+30pt
	+ (s^2-5q^2)\delta^{AD}\delta^{b'c'}
	\Bigr]
	-4\delta^{BC}qr\Bigl[	
	\bigl(\gamma^{ad}\bigr)^{\alpha\beta}
	- \bigl(\gamma^{a'd'}\bigr)^{\alpha\beta}
	\Bigr]
\nn\\
&&\kern+30pt
	-4\delta^{AD}qs\Bigl[
	\bigl(\gamma^{bc}\bigr)^{\alpha\beta}
	- \bigl(\gamma^{b'c'}\bigr)^{\alpha\beta}
	\Bigr]
	\biggr\}\ .
\ee
Finally, the off-diagonal blocks $(2,3)$, $(2,4)$ and $(3,4)$ (plus their
transpose entries) are given by the $H_{\rm BF}$ matrix element
\be
\Braket{J| b_q^{\alpha} a_r^{A} a_s^{B} (H_{\rm BF}) 
	a_r^{C\dag} a_q^{D\dag} b_s^{\beta\dag}|J}
	& = & 
	-\frac{\lambda'}{32 J}\delta^{AC}\biggl\{
	\delta^{\alpha\beta}
	\Bigl[
	(q-s)^2\delta^{BD}
	-(q^2+14qs+ s^2)\delta^{bd}
\nn\\	
	-( &&\kern-28pt    q^2-  18qs+  s^2)\delta^{b'd'}
	\Bigr]  
	+16 qs \Bigl[
	\bigl(\gamma^{bd}\bigr)^{\alpha\beta}
	-\bigl(\gamma^{b'd'}\bigr)^{\alpha\beta}
	\Bigr]
	\biggr\}\ .
\nn\\
&&
\ee

A significant departure from the two-impurity case is that all these matrix
elements have, along with their spacetime index structures, nontrivial
dependence on the mode indices. The eigenvalues could potentially have very
complicated mode-index dependence but, as we shall see, they do not.
This amounts to a rigid consistency check on the whole procedure that 
was not present in the two-impurity case.

\subsection{Matrix diagonalization: inequivalent modes $(q\neq r\neq s)$}
We now turn to the task of diagonalizing the one-loop approximation
to the perturbing Hamiltonian. To simplify the task, we exploit certain 
block diagonalizations that hold to leading order in $\lambda'$ 
(but not to higher orders). While we eventually want to study the spectrum 
to all orders in $\lambda'$, diagonalizing the Hamiltonian at one loop
will reveal the underlying supermultiplet structure. As an example of the
simplifications we have in mind, we infer from (\ref{12block}) that the 
matrix elements of $H_{\rm BF}$ between pure boson states 
$a_q^{A\dag} a_r^{B\dag} a_s^{C\dag}\ket{J}$ and bifermionic spacetime bosons
vanish to leading order in $\lambda'$ if all three $SO(8)$ bosonic vector 
indices lie within the same $SO(4)$, descended either from 
$AdS_5$ or $S^5$. Restricting to such states brings 
the bosonic sector of the Hamiltonian into the block-diagonal form in 
table~\ref{boseblock1}.
This leaves two 64-dimensional subspaces of purely bosonic states on which
the perturbation is block diagonal, as recorded in table~\ref{boseblock2}. 
\begin{table}[ht!]
\begin{eqnarray}
\begin{array}{|c|cc|}
\hline
 H_{\rm int} & a^{a\dagger} a^{b\dagger} a^{c\dagger} \ket{J}
		+ a^{a'\dagger} a^{b'\dagger} a^{c'\dagger} \ket{J}  &
	a^{A\dagger}  b^{\alpha\dagger}  b^{\beta\dagger}  \ket{J} 
\\   \hline
\bra{J} a^{a} a^{b} a^{c} +
\bra{J} a^{a'} a^{b'} a^{c'} & H_{\rm BB} & 0  \\
\bra{J} a^{A}  b^{\alpha}  b^{\beta} & 0  & H_{\rm FF}+H_{\rm BF}  \\
\hline
\end{array} \nonumber
\end{eqnarray}
\caption{Block-diagonal $SO(4)$ projection on bosonic three-impurity string states}
\label{boseblock1}
\end{table} 
\begin{table}[ht!]
\begin{eqnarray}
\begin{array}{|c|cc|}
\hline
 H_{\rm int} & a^{a\dagger} a^{b\dagger} a^{c\dagger} \ket{J} &
	a^{a'\dagger} a^{b'\dagger} a^{c'\dagger}\ket{J} 
\\   \hline
\bra{J} a^{a} a^{b} a^{c} & (H_{\rm BB})_{64\times 64} & 0  \\
\bra{J} a^{a'} a^{b'} a^{c'} & 0  & (H_{\rm BB})_{64\times 64}  \\
\hline
\end{array} \nonumber
\end{eqnarray}
\caption{SO(4) projection on purely bosonic states}
\label{boseblock2}
\end{table}

Since the interaction Hamiltonian has manifest $SO(4)\times SO(4)$ 
symmetry, it is useful to project matrix elements onto irreps of that group
before diagonalizing. In some cases the irrep is unique, and
projection directly identifies the corresponding eigenvalue. In the cases
where an irrep has multiple occurrences, there emerges an unavoidable matrix 
diagonalization that is typically of low dimension. In what follows,
we will collect the results of carrying out this program on the one-loop
interaction Hamiltonian.  A very important feature of the results that
appear is that all the eigenvalues turn out to have a common simple 
dependence on mode indices. More precisely, 
the expansion of the eigenvalues for inequivalent mode indices $(q,r,s)$ 
out to first non-leading order in $\lambda'$ and $1/J$ can be written as
\be
\label{lambdaexp}
E_J(q,r,s) = 3+\frac{\lambda'(q^2+r^2+s^2)}{2}
	\left(1 + \frac{\Lambda}{J} + O(J^{-2})\right)\ ,
\ee
where, as in Chapter \ref{twoimp}, $\Lambda$ is a pure number that characterizes the lifting of the 
degeneracy in the various sectors. (The notation $\Lambda_{\rm BB},\ \Lambda_{\rm BF}$ and 
$\Lambda_{\rm FF}$ will again be used to denote energy corrections arising entirely from 
the indicated sectors of the perturbing Hamiltonian.) This simple quadratic 
dependence of the eigenvalues on the mode indices does not automatically 
follow from the structure of the matrix elements themselves, but is
important for the successful match to gauge theory eigenvalues. 
In what follows, we will catalog some of the different $\Lambda$ values that 
occur, along with their $SO(4)\times SO(4)$ irreps (and multiplicities). 
When we have the complete list, we will discuss 
how they are organized into supermultiplets.

In the $SO(4)$ projection in table~\ref{boseblock2}, we will find a set 
of 64 eigenvalues for both the $SO(4)_{AdS}$ and $SO(4)_{S^5}$ subsectors.
We record this eigenvalue spectrum in table~\ref{SO4BOSE}, using a now-familiar
$SU(2)^2\times SU(2)^2$ notation. For comparison, it is displayed alongside
the projection of the two-impurity spectrum onto the same subspace (as 
found in Chapter \ref{twoimp}).
\begin{table}[ht!]
\begin{equation}
\begin{array}{|c|c|}
\hline
SO(4)_{AdS}\times SO(4)_{S^5} & \Lambda_{\rm BB} \\
\hline 
{\bf (1,1;2,2)}	&	-8	\\
{\bf [1,1;(2+4),2]}+{\bf [1,1;2,(2+4)]}	&	-6	\\
{\bf [1,1;(2+4),(2+4)]}	&	-4 \\
\hline
{\bf [(2+4),(2+4);1,1]}	&	-2	\\
{\bf [(2+4),2;1,1]}+{\bf [2,(2+4);1,1]}	&	0	\\
{\bf (2,2;1,1)}	&	2 \\
\hline
\end{array} \nonumber \qquad\qquad
\begin{array}{|c|c|}
\hline
SO(4)_{AdS}\times SO(4)_{S^5} & \Lambda_{\rm BB} \\
\hline 
{\bf (1,1;1,1)}	&	-6	\\
{\bf (1,1;3,1)}+{\bf (1,1;1,3)}	&	-4	\\
{\bf (1,1;3,3)}	&	-2 \\
\hline
{\bf (3,3;1,1)}	&	-2	\\
{\bf (3,1;1,1)}+{\bf (1,3;1,1)}	&	0	\\
{\bf (1,1;1,1)}	&	2 \\
\hline  
\end{array} \nonumber
\end{equation} 
\caption{Three-impurity energy spectrum in the pure-boson $SO(4)$ projection
(left panel) and two-impurity energy spectrum in the same projection (right panel)}
\label{SO4BOSE}
\end{table}
In the three-impurity case, the ${\bf (1,1;2,2)}$ level in the
$SO(4)_{S^5}$ subsector clearly descends from the two-impurity
singlet ${\bf (1,1;1,1)}$ in the same $SO(4)$ subgroup.  
In the same manner, the three-impurity ${\bf [1,1;(2+4),2]}+{\bf [1,1;2,(2+4)]}$
level descends from the $SO(4)_{S^5}$ antisymmetric two-impurity state
${\bf (1,1;3,1)}+{\bf (1,1;1,3)}$, and the three-impurity ${\bf [1,1;(2+4),(2+4)]}$
level is tied to the two-impurity symmetric-traceless ${\bf (1,1;3,3)}$ irrep.
In the $SO(4)_{S^5}$ subsector, each of these levels receives a shift to the 
energy of $-2$.  The total multiplicity of each of these levels is also increased 
by a factor of four when the additional ${\bf (2,2)}$ is tensored into the 
two-impurity state space.  The $SO(4)_{AdS}$ subsector follows a similar pattern:
the ${\bf (2,2;1,1)}$, ${\bf [(2+4),2;1,1]}+{\bf [2,(2+4);1,1]}$ and 
${\bf [(2+4),(2+4);1,1]}$
levels appear as three-impurity descendants of the two-impurity irrep spectrum 
${\bf (1,1;1,1)}+{\bf (3,1;1,1)}+{\bf (1,3;1,1)}+{\bf (3,3;1,1)}$.
In this subsector, however, the three-impurity energies are identical 
to those in the two-impurity theory.  

The bosonic $SO(4)$ projection has a precise fermionic analogue.  
Similar to the bosons, the $SO(9,1)$ spinors $b_q^\dag$ decompose 
as ${\bf (2,1;2,1)}+{\bf (1,2;1,2)}$ under the action of $\Pi$ parity:
\be
\label{fermipi}
\Pi \hat b_q^\dag = \hat b_q^\dag~, \qquad 
\Pi \tilde b_q^\dag = - \tilde b_q^\dag\ .
\ee
(As described above, the notation $\hat b_q^\dag$ labels ${\bf (1,2;1,2)}$ spinors with
positive eigenvalue under $\Pi$, and $\tilde b_q^\dag$ 
indicates ${\bf (2,1;2,1)}$ spinors, which are negative under $\Pi$.)
Analogous to the $SO(4)$ projection on the $SO(8)$ bosonic operators 
$a_q^{A\dag } \to a_q^{a\dag} + a_q^{a'\dag}$, 
projecting out the positive or negative eigenvalues of $\Pi$ on the 
eight-component spinor $b_q^{\alpha\dag}$ 
leaves a subspace of four-component spinors spanned by 
$\hat b_q^\dag$ and $\tilde b_q^\dag$.

We can perform a projection on the subsector in table~\ref{boseblock3}
similar to that appearing in table~\ref{boseblock2}.
In this case, instead of three bosonic impurities mixing with
a single bosonic (plus a bi-fermionic) excitation,
we are now interested in projecting out particular interactions
between a purely fermionic state and a state with one fermionic 
and two bosonic excitations. 
Using $\pm$ to denote the particular representation of the 
fermionic excitations, the off-diagonal elements given 
by (\ref{12fermiblock}) vanish for $+++ \to \pm$ and $---\to \pm$ interactions.
In other words, the pure fermion states in the $(1,1)$ block 
of table~\ref{fermiblock3} will not 
mix with states containing two bosonic excitations if 
all three fermionic oscillators lie in the same $\Pi$ projection.  
This projection appears schematically in table~\ref{fermiblock1}.
\begin{table}[ht!]
\begin{eqnarray}
\begin{array}{|c|cc|}
\hline
 H_{\rm int} & \hat b^{\alpha\dagger}\hat b^{\beta\dagger}\hat b^{\gamma\dagger} \ket{J} 
	+ \tilde b^{\alpha\dagger}\tilde b^{\beta\dagger}\tilde b^{\gamma\dagger} 
	\ket{J}& 
	a^{A\dagger} a^{B\dagger} b^{\alpha\dagger} \ket{J}  
		\\   \hline
\bra{J}\hat b^{\alpha}\hat b^{\beta}\hat b^{\gamma}+
\bra{J}\tilde b^{\alpha}\tilde b^{\beta}\tilde b^{\gamma}	  
					& H_{\rm FF} & 0  \\
\bra{J} a^{A} a^{B} b^{\alpha}	 & 0  & H_{\rm BB}+H_{\rm BF}  \\
\hline
\end{array} \nonumber
\end{eqnarray}
\caption{Block-diagonal projection on fermionic three-impurity string states}
\label{fermiblock1}
\end{table}

The $(1,1)$ pure fermion block in table~\ref{fermiblock1} 
breaks into two 64-dimensional subsectors under this
projection. By tensoring an additional ${\bf (1,2;1,2)}$ or
${\bf (2,1;2,1)}$ impurity into the two-impurity
state space, we expect to see a multiplicity structure
in this projection given by
\be
{\bf (1,2)}\times {\bf(1,2;1,2)} & = & 
	{\bf (1,2;1,2)} + {\bf [1,2;1,(2+4)]} 
\nn\\
&&	+ {\bf [1,(2+4);1,2]}
	+ {\bf [1,(2+4);1,(2+4)]}\ ,
\nn\\
{\bf (2,1)}\times {\bf(2,1;2,1)} & = & 
	{\bf (2,1;2,1)} + {\bf [2,1;(2+4),1]} 
\nn\\
&&	+ {\bf [(2+4),1;2,1]}
	+ {\bf [(2+4),1;(2+4),1]}\ , 
\ee
for a total of 128 states. The projections onto the two 64-dimensional 
$\Pi_+$ and $\Pi_-$ subspaces yield identical eigenvalues and multiplicities. 
The results for both subspaces are presented in table~\ref{Pifermi}:
\begin{table}[ht!]
{\footnotesize
\begin{equation}
\begin{array}{|c|c|}
\hline
SO(4)_{AdS}\times SO(4)_{S^5} & \Lambda_{\rm FF} \\
\hline 
{\bf (2,1;2,1)}+{\bf (1,2;1,2)}	&	-3	\\
{\bf [2,1;(2+4),1]}+{\bf [1,2;1,(2+4)]}	&	-1	\\
{\bf [(2+4),1;2,1]}+{\bf [1,(2+4);1,2]}	&	-5 \\
{\bf [(2+4),1;(2+4),1]}+{\bf [1,(2+4);1,(2+4)]}	&	-3 \\
\hline
\end{array} \nonumber
\qquad\qquad
\begin{array}{|c|c|}
\hline
SO(4)_{AdS}\times SO(4)_{S^5} & \Lambda_{\rm FF} \\
\hline 
{\bf (1,1;1,1)}+{\bf (1,1;1,1)}	&	-2	\\
{\bf (1,1;3,1)}+{\bf (1,1;1,3)}	&	0	\\
{\bf (3,1;1,1)}+{\bf (1,3;1,1)}	&	-4 \\
{\bf (3,1;3,1)}+{\bf (1,3;1,3)}	&	-2 \\
\hline
\end{array} \nonumber
\end{equation} }
\caption{Spectrum of three-impurity states (left panel) and two-impurity 
states (right panel) created by $\Pi_\pm$-projected fermionic creation operators}
\label{Pifermi} 
\end{table}
The two-impurity bi-fermion states in table~\ref{Pifermi} 
are spacetime bosons while the tri-fermion states
are spacetime fermions. For comparison purposes, we have displayed
both spectra. Note that the $O(1/J)$ energy corrections of the 
two types of state are simply displaced by $-1$ relative to each other.

This exhausts the subspaces that can be diagonalized by simple 
irrep projections. The remaining eigenvalues must be obtained by
explicit diagonalization of finite dimensional submatrices obtained
by projection onto representations with multiple occurrence. The upshot
of these more complicated eigenvalue calculations is that the first-order 
$\lambda'$ eigenvalues take on all integer values from $\Lambda=-8$ to
$\Lambda= +2$, alternating between spacetime bosons and fermions as 
$\Lambda$ is successively incremented by one unit. 

\subsection{Assembling eigenvalues into supermultiplets}

Finally, we need to understand how the perturbed three-impurity
spectrum breaks up into extended supersymmetry multiplets. This is
relatively easy to infer from the multiplicities of the perturbed
eigenvalues (and the multiplicities are a side result of the calculation 
of the eigenvalues themselves). In the last subsection, we described a 
procedure for diagonalizing the one-loop
perturbing Hamiltonian on the $4,096$-dimensional space of 
three-impurity string states with mode indices $p\ne q\ne r$. 
The complete results for the eigenvalues $\Lambda$ and their 
multiplicities are stated in table~\ref{multlist} (we use the notation of 
(\ref{lambdaexp}), while the $B$ and $F$ subscripts are used to indicate 
bosonic and fermionic levels in the supermultiplet).
\begin{table}[ht!]
\begin{eqnarray}
\begin{array}{|c|ccccccccccc|}
\hline
\Lambda	& -8	& -7	& -6    & -5	& -4	& -3	& -2 & -1 & 0 
		& 1 & 2 \\	
\hline
{\rm Multiplicity}	 & 4_B	& 40_F	& 180_B	& 480_F	& 840_B	& 1008_F
		& 840_B	& 480_F	& 180_B	& 40_F	& 4_B	 \\
\hline
\end{array} \nonumber
\end{eqnarray}
\caption{Complete three-impurity energy spectrum (with multiplicities)}
\label{multlist}
\end{table}

The $\Lambda$ eigenvalues in table~\ref{multlist} are integer-spaced, which
is consistent with supersymmetry requirements (for details, the reader is referred to
the discussion following eqn.~(\ref{eigenformula}) in Chapter \ref{twoimp} above). 
However, because the range 
between top and bottom eigenvalues is ten, rather than eight, the $4,096$-dimensional 
space must be built on more than one type of extended supermultiplet, with more
than one choice of $c$ in the general formula $\Lambda=L+c$ (where $L$ is the supermultiplet level
and $c$ is some numerical constant). 
This is to be contrasted with the two-impurity case, where the degenerate space 
was exactly $256$-dimensional and was spanned by a single superconformal
primary whose lowest member was a singlet under both Lorentz transformations 
and the residual $SO(4)$ ${R}$-symmetry. We can readily infer what 
superconformal primaries are needed to span the degenerate three-impurity
state space by applying a little numerology to table~\ref{multlist}.
The lowest eigenvalue is $\Lambda=-8$: it has multiplicity $4$ and,
according to table~\ref{SO4BOSE}, its $SO(4)\times SO(4)$ decomposition
is ${\bf (1,1;2,2)}$ (spacetime scalar, ${R}$-charge $SO(4)$ four-vector).
According to the general arguments about how the full extended supermultiplet
is built by acting on a ``bottom'' state with the eight raising operators,
it is the base of a supermultiplet of $4\times 256$ states extending
up to $\Lambda=0$. By the same token, there is a highest eigenvalue
$\Lambda=+2$: it has multiplicity $4$ and, according to table~\ref{SO4BOSE}, 
its $SO(4)\times SO(4)$ decomposition is ${\bf (2,2;1,1)}$ (spacetime vector, 
${R}$-charge singlet). Using lowering operators instead of
raising operators, we see that one derives from it a supermultiplet of
$4\times 256$ operators with eigenvalues extending from $\Lambda=-6$
to $\Lambda=+2$. The multiplicities of the $\Lambda$ eigenvalues occurring
in these two supermultiplets are of course given by binomial coefficients,
as described above.  By comparing with the \emph{total} multiplicities of 
each allowed $\Lambda$ (as listed in table~\ref{multlist}) we readily see
that what remains are $8\times 256$ states with eigenvalues running from 
$\Lambda=-7$ to $\Lambda=+1$ with the correct binomial coefficient pattern 
of multiplicities. The top and bottom states here are spacetime fermions 
and must lie in a spinor representation of the Lorentz group. It is not 
hard to see that they lie in the eight-dimensional $SO(4)\times SO(4)$ irrep 
${\bf (2,1;1,2)}+{\bf (1,2;2,1)}$. This exhausts all the states and we conclude that 
the three-impurity state space is spanned by three distinct extended 
superconformal multiplets. The detailed spectrum is given in table~\ref{3mult15} 
(where the last line records the total multiplicity at each level as given 
in table~\ref{multlist} and the first line records the two-impurity spectrum
for reference). Note the peculiar feature that certain energies are shared
by all three multiplets: this is an accidental degeneracy that does not
survive at higher loop order.
\begin{table}[ht!]
{\scriptsize
\begin{eqnarray}
\begin{array}{|c|ccccccccccc|c|}
\hline
\Lambda	& -8	& -7	& -6    & -5	& -4	& -3	& -2	
		& -1	& 0	& 1	& 2	& 	\\ \hline\hline
\Delta_0=2	& 	& 	& 1_B	& 8_F	& 28_B	& 56_F	& 70_B	
		& 56_F	& 28_B	& 8_F	& 1_B	& {\rm scalar}	\\ \hline\hline 
\Delta_0=3	 & 4 & 32	& 112	& 224	& 280	& 224	& 112	
		& 32	& 4	& 	& 	& SO(4)_{S^5} ~{\rm vector} \\ \hline
	& 	& 	& 4 	& 32	& 112	& 224	& 280	
		& 224	& 112	& 32	& 4	& SO(4)_{AdS_5} ~{\rm vector} \\ \hline
	& 	& 8	& 64	& 224	& 448	& 560	& 448 	
		& 224	& 64	& 8	& 	& {\rm spinor}	\\ \hline
\hline
{\rm Total}	 & 4_B	& 40_F	& 180_B	& 480_F	& 840_B	& 1008_F
		& 840_B	& 480_F	& 180_B	& 40_F	& 4_B	& 4,096 \\
\hline 
\end{array} \nonumber
\end{eqnarray}  }
\caption{Submultiplet breakup of the three-impurity spectrum}
\label{3mult15}
\end{table}

A complete analysis of the agreement with gauge theory anomalous 
dimensions will have to be deferred until a later 
section: the dimensions of three-impurity gauge theory operators 
are much harder to calculate than those of the two-impurity operators
and there are few results in the literature, even at one loop. However, 
it is worth making a few preliminary remarks at this point. Since there 
are three superconformal multiplets, we have only three independent 
anomalous dimensions to compute. Minahan and Zarembo \cite{Minahan:2002ve} 
found that the problem simplifies dramatically if we study the one-loop 
anomalous dimension of the special subset of single-trace operators of the form
$\tr{(\phi^I Z^J)}$ (and all possible permutations of the fields inside the
trace), where the $R$-charge is carried by an $SO(4)\times SO(4)$ 
singlet scalar field $Z$ and the impurities are insertions of a scalar field 
$\phi$ lying in the ${\bf (1,1;2,2)}$ (vector) irrep of the residual 
$SO(4)\times SO(4)$ symmetry.  More formally, these operators
are in the $SO(4)\times SO(4)$ irrep obtained by completely symmetrizing $I$
vectors in the ${\bf (1,1;2,2)}$ irrep. The crucial point is that such operators 
form a closed sector, mixing only among themselves under the anomalous 
dimension operator.  More importantly, the action of the one-loop anomalous 
dimension operator on this closed sector can be recast as the action of an integrable spin 
chain Hamiltonian of a type solvable by Bethe ansatz techniques. Although the Bethe ansatz 
is generally not analytically soluble, Minahan and Zarembo used it to 
obtain a virial expansion for the anomalous dimension in which the number $I$ of 
impurities is held fixed, while the $R$-charge $J$ is taken to be large
(see eqn.~(5.29) in \cite{Minahan:2002ve}).  In terms of the number of
spin chain lattice sites ${K}$, their result appears as
\be
\gamma_{\so(6)} = \frac{\lambda}{2{K}^3}\sum_n M_n k_n^2 
	\left( {K} + M_n + 1 \right) + O({K}^{-4})\ .
\ee
The integer $k_n$ represents pseudoparticle momenta on the spin chain, 
and is dual to the string theory worldsheet mode indices;
the quantity $M_n$ labels the number of trace impurities with identical $k_n$.
With $I$ impurities, the spin chain length is given in terms of the $R$-charge
by ${K} = J+I$, which leads to
\be
\gamma_{\so(6)} = \frac{\lambda}{2J^3} \sum_n M_n k_n^2
	\left( J - 2I +M_n + 1 \right) + O(J^{-4})\ .
\label{MZBA}
\ee
This virial expansion
is similar in character to (\ref{lambdaexp}) and, for $I=3$ (the three-impurity 
case), it matches that equation precisely with $\Lambda=-4$.

On the string theory side, three completely symmetrized 
${\bf (1,1;2,2)}$ vectors form a tensor in the ${\bf (1,1;4,4)}$ irrep; such 
an irrep can be constructed from three $SO(4)_{S^5}$ vector (bosonic) creation operators.
Table~\ref{SO4BOSE} shows that the corresponding string perturbation theory 
eigenvalue is (at one-loop order) $\Lambda=-4$ as well. We infer from 
table~\ref{3mult15} that this eigenvalue lies at level $L=4$ of the $SO(4)_{S^5}$ 
vector superconformal multiplet (and this argument takes care of the gauge 
theory/string theory comparison for all other operators in that multiplet). 

The sector described above is often 
called an $\so(6)$\footnote{This notation
is used to distinguish the protected gauge theory symmetry groups from those in the string theory.}   
sector on the gauge theory side, with reference to the subalgebra of the 
full superconformal algebra under which it is invariant.
In an $\su(2)$ subspace of the $\so(6)$, this sector becomes closed to all loop order.  
For future reference, we note that Beisert \cite{Beisert:2003jj} has identified two other 
closed sectors of operators in the gauge theory.  In addition to the bosonic $\su(2)$ sector,
a bosonic $\Sl(2)$ sector and an $\su(2|3)$ sector (of which the closed $\su(2)$ sector is a subsector) 
are also exactly closed.  
It should be noted that integrable $\Sl(2)$ spin chains were discovered some time ago in 
phenomenologically motivated studies of the scaling behavior of high-energy scattering 
amplitudes in physical, non-supersymmetric QCD \cite{Belitsky:1999bf} 
(see also \cite{Kotikov:2000pm,Kotikov:2001sc,Kotikov:2002ab,Kotikov:2003fb}).
The $\su(2|3)$ spin chain was studied more recently in \cite{Beisert:2003ys}:  
this closed sector breaks into the $\su(2)$ bosonic sector and a special fermionic subsector, 
which we denote as $\su(1|1)$ (a subalgebra of $\su(2|3)$).

In the string theory, 
the subsectors analogous to the gauge theory $\Sl(2)$ and $\su(1|1)$ 
are constructed out of
completely symmetrized $SO(4)_{AdS}$ bosons and completely symmetrized fermions 
of the same $\Pi$ eigenvalue, respectively (see Chapter \ref{twoimp} or ref.~\cite{Callan:2004uv}).  
They correspond to the central $L=4$ levels
of the remaining two supermultiplets in table~\ref{3mult15}, and a calculation
of their eigenvalues would complete the analysis of the match between 
three-impurity operators and string states at one-loop order. 
Since we eventually want to go beyond one loop, where Bethe 
ansatz technology is less well-developed, we have found it useful to employ the 
numerical methods presented in Chapter~\ref{virial} 
for evaluating spin chain eigenvalues (we refer the reader to 
Chapter~\ref{virial}, or to ref.~\cite{Callan:2004dt}, 
for a check of our results against Bethe-ansatz techniques,
including the higher-loop corrections of \cite{Beisert:2004hm}).  
This subject will be developed in a later section.

\subsection{Two equivalent mode indices $(q=r=n,~s=-2n)$}
When two mode indices are allowed to be equal, the analysis becomes
slightly more complicated. Since we are diagonalizing a
Hamiltonian that is quartic in oscillators in a basis of 
three-impurity string states, one oscillator in the ``in'' state must 
always be directly contracted with one oscillator in the ``out'' state
and, with two equal mode indices, there are many more nonvanishing 
contributions to each matrix element. While the matrix 
elements are more complicated, the state space is only half as large when
two mode indices are allowed to be equal (only half as many mode-index 
permutations on the basis states generate linearly independent states).  
As a result, the fermionic and bosonic sectors of the Hamiltonian 
are each 1,024-dimensional. By the same token, the multiplet structure
of the energy eigenstates will be significantly different from the
unequal mode index case studied in the previous subsection.  

To study this case, we make the mode index choice
\be
q=r=n~, \qquad s=-2n\ .
\ee 
The structure of matrix elements of the string Hamiltonian between spacetime bosons 
is given in table~\ref{boseblock3nn}. 
\begin{table}[ht!]
\begin{eqnarray}
\begin{array}{|c|ccc|}
\hline
 H_{\rm int} & 
	a_{-2n}^{D\dag}  a^{E\dagger}_n  a^{F\dagger}_{n}\ket{J} &
	a_{-2n}^{D\dag}  b^{\gamma\dagger}_n  b^{\delta\dagger}_{n}\ket{J} &
        a_n^{D\dag}  b^{\gamma\dagger}_n  b^{\delta\dagger}_{-2n}\ket{J} 
\\   \hline
\bra{J} a^{A}_n  a^{B}_n  a^{C}_{-2n}      & H_{\rm BB}    
		& H_{\rm BF} & H_{\rm BF}  \\
\bra{J} a^{A}_n  b^{\alpha}_n  b^{\beta}_{-2n}& H_{\rm BF} 
		& H_{\rm FF}+H_{\rm BF} & H_{\rm BF}  \\
\bra{J} a^{A}_{-2n}  b^{\alpha}_n  b^{\beta}_{n}& H_{\rm BF} 
		& H_{\rm BF} & H_{\rm FF}+H_{\rm BF}  \\ 
\hline
\end{array} \nonumber
\end{eqnarray}
\caption{Bosonic three-impurity string perturbation matrix with $(q= r=n,\  s=-2n)$}
\label{boseblock3nn}
\end{table}
This table seems to describe a $3\times 3$ block matrix with $512\times 512$ blocks 
in each subsector, giving a 1,536-dimensional state space.  However, the vector and spinor 
indices are required to run over values that generate linearly independent basis 
states. This eliminates one third of the possible index assignments, implying that
the matrix is in fact $1,024\times 1,024$. 

To evaluate the entries in table~\ref{boseblock3nn}, we express the Hamiltonians 
(\ref{Hpurbos}, \ref{Hpurferm}, \ref{Hmix}) in terms of mode creation and annihilation operators, 
expand the result in powers of $\lambda'$ and compute the indicated matrix 
elements between three-impurity Fock space states. We collect below all the relevant 
results of this exercise for this equal-mode-index case.

The purely bosonic subsector in the $(1,1)$ block is given by
\be
\Bra{J}&&\kern-20pt
	a_n^{A} a_n^{B} a_{-2n}^{C} (H_{\rm BB})a_{-2n}^{D\dag} a_n^{E\dag} 
	a_n^{F\dag}\Ket{J}
	 = 
\frac{n^2\,\lambda}{2J} \biggl\{
	5\,{\delta }^{BF}\,\delta^{cd}\delta^{ae} + 
      5\,{\delta }^{AF}\,\delta^{cd}\delta^{be} 
	- 4\,{\delta }^{BF}\,\delta^{ad}\delta^{ce} 
\nn\\
&&\kern-20pt	+     4\,{\delta }^{BF}\,\delta^{ac}\delta^{de} 
	+ 4\,{\delta }^{AF}\,\delta^{bc}\delta^{de} +
      5\,{\delta }^{BE}\,\delta^{cd}\delta^{af} 
	- 4\,{\delta }^{BE}\,\delta^{ad}\delta^{cf} + 
      4\,{\delta }^{BE}\,\delta^{ac}\delta^{df} 
\nn\\
&&\kern-20pt	
	+ 4\,{\delta }^{AE}\,\delta^{bc}\delta^{df}
	-       4\,{\delta }^{bd}
	\Bigl( {\delta}^{AF}\,{\delta }^{ce} 
	+ {\delta }^{AE}\,{\delta }^{cf}
         \Bigr)  		
	+ 3\,{\delta }^{BF}\,{\delta }^{ae}\,{\delta }^{c'd'} 
	+ 3\,{\delta }^{AF}\,{{{\delta }}}^{be}\,{{{\delta }}}^{c'd'} 
\nn\\
&&\kern-20pt
	+3\,{\delta }^{BE}\,{{{\delta }}}^{af}\,{{{\delta }}}^{c'd'} 
	-      3\,{\delta }^{BF}\,{{{\delta }}}^{cd}\,{{{\delta }}}^{a'e'}  
    	-3\,{\delta }^{AF}\,{{{\delta }}}^{cd}\,{{{\delta }}}^{b'e'}   
	- 5\,{\delta }^{BF}\,{{{\delta }}}^{c'd'}\delta^{a'e'} - 
      5\,{\delta }^{AF}\,{{{\delta }}}^{c'd'}\delta^{b'e'} 
\nn\\
&&\kern-20pt
	+ 4\,{\delta }^{BF}\,{{{\delta }}}^{a'd'}\delta^{c'e'} 
	+       4\,{\delta }^{AF}\,{{{\delta }}}^{b'd'}\delta^{c'e'} 
	- 4\,{\delta }^{BF}\,{{{\delta }}}^{a'c'}     \delta^{d'e'} - 
      4\,{\delta }^{AF}\,{{{\delta }}}^{b'c'}\delta^{d'e'} 
	- 3\,{\delta }^{BE}\,{{{\delta }}}^{cd}\,{{{\delta }}}^{a'f'} 
\nn\\
&&\kern-20pt - 
     3\,{\delta }^{AE}\,{{{\delta }}}^{cd}\,{{{\delta }}}^{b'f'} 
	- 5\,{\delta }^{BE}\,{{{\delta }}}^{c'd'}\delta^{a'f'} - 
     5\,{\delta }^{AE}\,{{{\delta }}}^{c'd'}          \delta^{b'f'}     
	+ 4\,{\delta }^{BE}\,{{{\delta }}}^{a'd'}\delta^{c'f'} + 
      4\,{\delta }^{AE}\,{{{\delta }}}^{b'd'}\delta^{c'f'} 
\nn\\
&&\kern-20pt	- 4\,{\delta }^{BE}\,{{{\delta }}}^{a'c'}\delta^{d'f'} 
	-       4\,{\delta }^{AE}\,{{{\delta }}}^{b'c'}\delta^{d'f'}  
    	+  {\delta }^{AE}\,{{{\delta }}}^{bf}\,  
	\Bigl( 5\,{{{\delta }}}^{cd} 
	+ 3\,{{{\delta }}}^{c'd'} \Bigr)  
\nn\\
&&\kern-20pt - 
      2\,{\delta }^{CD}\,\Bigl[ 9\,
	\Bigl( {\delta }^{BE}\delta^{AF} 
	+ {\delta }^{AE}\delta^{BF} \Bigr)  
	-          {{{\delta }}}^{be}\delta^{af} 
\nn\\
&&\kern-20pt	- {{{\delta }}}^{ae}\delta^{bf} 
	+ {{{\delta }}}^{ab}\delta^{ef} +       
         {{{\delta }}}^{b'e'}\delta^{a'f'} + {{{\delta }}}^{a'e'}\delta^{b'f'} 
	- {{{\delta }}}^{a'b'}\delta^{e'f'} 
	\Bigr] 
	\biggr\}\ .
\label{bosonsnn}
\ee
This matrix element exhibits the same antisymmetry between the $SO(4)_{AdS}$ 
and $SO(4)_{S^5}$ indices that is exhibited in eqn.~(\ref{bosons}).  
The off-diagonal $H_{\rm BF}$ mixing sector is essentially equivalent to its
counterpart in eqn.~(\ref{12block}):
\be
\Braket{J| a_n^{A} a_n^{B} a_{-2n}^{C} (H_{\rm BF}) 
	a_{-2n}^{D\dag} b_n^{\alpha\dag} b_n^{\beta\dag}|J}
	= 
	\frac{n^2\lambda'}{2J}\delta^{CD}
	\biggl\{
	\left(\gamma^{ab'}\right)^{\alpha\beta} - 
	\left(\gamma^{a'b}\right)^{\alpha\beta}
	\biggr\}\ .
\label{12blocknn}
\ee
The diagonal contributions from the pure fermion sector $H_{\rm FF}$
in the $(2,2)$ and $(3,3)$ blocks of table~\ref{boseblock3nn} appear as
\be
\Braket{J| b_n^{\alpha} b_n^{\beta} a_{-2n}^{A} (H_{\rm FF}) 
	a_{-2n}^{B\dag} b_n^{\gamma\dag} b_n^{\delta\dag}|J}
	& = & 
	\frac{n^2\lambda'}{24 J}\delta^{AB} \biggl\{
	\bigl(\gamma^{ij}\bigr)^{\alpha\gamma} 
	\bigl(\gamma^{ij}\bigr)^{\beta\delta}
	- \bigl(\gamma^{ij}\bigr)^{\alpha\beta}
	\bigl(\gamma^{ij}\bigr)^{\gamma\delta}
\nn\\
	- \bigl(\gamma^{ij}\bigr)^{\alpha\delta}
	\bigl(\gamma^{ij}\bigr)^{\beta\gamma}  
	-\bigl(\gamma^{i'j'}\bigr)^{\alpha\gamma} 
	\bigl(   &&\kern-30pt  \gamma^{i'j'} \bigr)^{\beta\delta} 
	+ \bigl(\gamma^{i'j'}\bigr)^{\alpha\beta}
	\bigl(\gamma^{i'j'}\bigr)^{\gamma\delta}
	+ \bigl(\gamma^{i'j'}\bigr)^{\alpha\delta}
	\bigl(\gamma^{i'j'}\bigr)^{\beta\gamma}
	\biggr\}\ .
\nn\\
&&
\label{HFF22nn}
\ee
The $H_{\rm BF}$ sector exhibits the following contribution
to the lower diagonal blocks $(2,2)$ and $(3,3)$:
\be
\Braket{J| b_n^{\alpha} b_n^{\beta} a_{-2n}^{A} (H_{\rm BF}) 
	a_{-2n}^{B\dag} b_n^{\gamma\dag} b_n^{\delta\dag}|J}
	& = & 
	\frac{n^2\lambda'}{J}\biggl\{
	-10\,\delta^{a'b'} \left(
	\delta^{\alpha\delta}\delta^{\beta\gamma}
	-\delta^{\alpha\gamma}\delta^{\beta\delta}
	\right)
\nn\\
	-8\,\delta^{ab} \left(
	\delta^{\alpha\delta}\delta^{\beta\gamma}  
	-\delta^{\alpha\gamma}\delta^{\beta\delta}
	\right)				&&\kern-25pt
	-\delta^{\alpha\gamma}
	\Bigl[
	(\gamma^{ab})^{\beta\delta} - (\gamma^{a'b'})^{\beta\delta}
	\Bigr]
	+\delta^{\alpha\delta}
	\Bigl[
	(\gamma^{ab})^{\beta\gamma} - (\gamma^{a'b'})^{\beta\gamma}
	\Bigr]
\nn\\
	+\delta^{\beta\gamma}
	\Bigl[					&&\kern-25pt
	(\gamma^{ab})^{\alpha\delta} - (\gamma^{a'b'})^{\alpha\delta}
	\Bigr]
	-\delta^{\beta\delta}
	\Bigl[
	(\gamma^{ab})^{\alpha\gamma} - (\gamma^{a'b'})^{\alpha\gamma}
	\Bigr]
	\biggr\}\ .
\nn\\
&&
\label{HBF2211}
\ee
Finally, the off-diagonal version of (\ref{HBF2211}) appears in the $(2,3)$ block
(along with its transpose in the $(3,2)$ block): 
\be
&&\Braket{J| b_n^{\alpha} b_n^{\beta} a_{-2n}^{A} (H_{\rm BF}) 
	a_n^{B\dag} b_n^{\gamma\dag} b_{-2n}^{\delta\dag}|J}
	= 
	-\frac{n^2\lambda'}{J}\biggl\{
	\delta^{a'b'} \left(
	\delta^{\alpha\delta}\delta^{\beta\gamma}
	-\delta^{\alpha\gamma}\delta^{\beta\delta}
	\right)
\nn\\
&&\kern+20pt
	-\delta^{ab} \left(
	\delta^{\alpha\delta}\delta^{\beta\gamma}  
	-\delta^{\alpha\gamma}\delta^{\beta\delta}
	\right)				
	+\delta^{\alpha\gamma}
	\Bigl[  
	(\gamma^{ab})^{\beta\delta} - (\gamma^{a'b'})^{\beta\delta}
	\Bigr]		
	-\delta^{\beta\gamma}
	\Bigl[				
	(\gamma^{ab})^{\alpha\delta} - (\gamma^{a'b'})^{\alpha\delta}
	\Bigr]
	\biggr\}\ .
\nn\\
&&
\ee

The fermionic sector perturbation matrix is displayed schematically 
in table~\ref{fermiblock3nn}. Like table~\ref{boseblock3nn}, it is 
$1,024\times 1,024$ once redundant index assignments are eliminated.
\begin{table}[ht!]
\begin{eqnarray}
\begin{array}{|c|ccc|}
\hline
 H_{\rm int} & 
	b_{-2n}^{\zeta\dag}  b^{\epsilon\dagger}_n  b^{\delta\dagger}_{n}\ket{J} &
	a_{-2n}^{C\dag}  a^{D\dagger}_n  b^{\delta\dagger}_{n}\ket{J} &
        a_n^{C\dag}  a^{D\dagger}_n  b^{\delta\dagger}_{-2n}\ket{J} 
\\   \hline
\bra{J} b^{\alpha}_n  b^{\beta}_n  b^{\gamma}_{-2n}      
						& H_{\rm FF}    
		& H_{\rm BF} & H_{\rm BF}  \\
\bra{J} b^{\alpha}_n  a^{A}_n  a^{B}_{-2n}& H_{\rm BF} 
		& H_{\rm BB}+H_{\rm BF} & H_{\rm BF}  \\
\bra{J} b^{\alpha}_{-2n}  a^{A}_n  a^{B}_{n}& H_{\rm BF} & H_{\rm BF} 
		& H_{\rm BB}+H_{\rm BF}  \\ 
\hline
\end{array} \nonumber
\end{eqnarray}
\caption{Fermionic string perturbation matrix $(q=r=n,\ s=-2n)$}
\label{fermiblock3nn}
\end{table}

The purely fermionic subsector in the $(1,1)$ block of table~\ref{fermiblock3nn}
takes the form
\be
&&\Braket{J| b_n^{\alpha} b_n^{\beta} b_{-2n}^{\gamma} (H_{\rm FF}) 
	b_{-2n}^{\zeta\dag} b_n^{\epsilon\dag} b_n^{\delta\dag}|J}
	 = 
	\frac{9n^2\lambda'}{J}\delta^{\gamma\zeta}\left(
	\delta^{\alpha\epsilon}\delta^{\beta\delta} 
	- \delta^{\alpha\delta}\delta^{\beta\epsilon}
	\right)
\nn\\&&	+\frac{n^2\lambda'}{24J}\biggl\{
	\delta^{\gamma\zeta}\Bigl[    
	(\gamma^{ij})^{\alpha\beta}(\gamma^{ij})^{\delta\epsilon} 
	- (\gamma^{ij})^{\alpha\delta}(\gamma^{ij})^{\beta\epsilon}
	+ (\gamma^{ij})^{\alpha\epsilon}(\gamma^{ij})^{\beta\delta}
	-(\gamma^{i'j'})^{\alpha\beta}(\gamma^{i'j'})^{\delta\epsilon}  
\nn\\&&	+ (\gamma^{i'j'})^{\alpha\delta}(\gamma^{i'j'})^{\beta\epsilon} 
	- (\gamma^{i'j'})^{\alpha\epsilon}(\gamma^{i'j'})^{\beta\delta}  
	\Bigr]
	-2\delta^{\alpha\delta}\Bigl[
	(\gamma^{ij})^{\beta\gamma}(\gamma^{ij})^{\epsilon\zeta}
	- (\gamma^{ij})^{\beta\epsilon}(\gamma^{ij})^{\gamma\zeta}
\nn\\&&	+ (\gamma^{ij})^{\beta\zeta}(\gamma^{ij})^{\gamma\epsilon}  
	-(\gamma^{i'j'})^{\beta\gamma}(\gamma^{i'j'})^{\epsilon\zeta}
	+ (\gamma^{i'j'})^{\beta\epsilon}(\gamma^{i'j'})^{\gamma\zeta}
	- (\gamma^{i'j'})^{\beta\zeta}(\gamma^{i'j'})^{\gamma\epsilon}
	\Bigr]
\nn\\&&	+2\delta^{\alpha\epsilon}\Bigl[
	(\gamma^{ij})^{\beta\gamma}(\gamma^{ij})^{\delta\zeta}  
	- (\gamma^{ij})^{\beta\delta}(\gamma^{ij})^{\gamma\zeta}
	+ (\gamma^{ij})^{\beta\zeta}(\gamma^{ij})^{\gamma\delta}
	-(\gamma^{i'j'})^{\beta\gamma}(\gamma^{i'j'})^{\delta\zeta}
\nn\\&&	+ (\gamma^{i'j'})^{\beta\delta}(\gamma^{i'j'})^{\gamma\zeta}  
	- (\gamma^{i'j'})^{\beta\zeta}(\gamma^{i'j'})^{\gamma\delta}
	\Bigr]
	+2\delta^{\beta\delta}\Bigl[
	(\gamma^{ij})^{\alpha\gamma}(\gamma^{ij})^{\epsilon\zeta}
	- (\gamma^{ij})^{\alpha\epsilon}(\gamma^{ij})^{\gamma\zeta}
\nn\ee\be
&&	+ (\gamma^{ij})^{\alpha\zeta}(\gamma^{ij})^{\gamma\epsilon}  
	-(\gamma^{i'j'})^{\alpha\gamma}(\gamma^{i'j'})^{\epsilon\zeta}
	+ (\gamma^{i'j'})^{\alpha\epsilon}(\gamma^{i'j'})^{\gamma\zeta}
	- (\gamma^{i'j'})^{\alpha\zeta}(\gamma^{i'j'})^{\gamma\epsilon}
	\Bigr]
\nn\\
&&	-2\delta^{\beta\epsilon}\Bigl[
	(\gamma^{ij})^{\alpha\gamma}(\gamma^{ij})^{\delta\zeta}  
	- (\gamma^{ij})^{\alpha\delta}(\gamma^{ij})^{\gamma\zeta}
	+ (\gamma^{ij})^{\alpha\zeta}(\gamma^{ij})^{\gamma\delta}
	-(\gamma^{i'j'})^{\alpha\gamma}(\gamma^{i'j'})^{\delta\zeta} 
\nn\\&&	+ (\gamma^{i'j'})^{\alpha\delta}(\gamma^{i'j'})^{\gamma\zeta}  
	- (\gamma^{i'j'})^{\alpha\zeta}(\gamma^{i'j'})^{\gamma\delta}
	\Bigr]
	\biggr\}\ .
\label{pureferminn}
\ee
The off-diagonal blocks $(1,2)$ and $(1,3)$ receive contributions
from the $H_{\rm BF}$ sector:
\be
\Braket{J| b_n^{\alpha} b_n^{\beta} b_{-2n}^{\gamma} (H_{\rm BF}) 
	a_{-2n}^{A\dag} a_n^{B\dag} b_n^{\delta\dag}|J}
	& = & 
	\frac{n^2\lambda'}{J}\biggl\{
	\delta^{\alpha\delta}\Bigl[
	\left(\gamma^{ab'}\right)^{\beta\gamma}
	- \left(\gamma^{a'b}\right)^{\beta\gamma}
	\Bigr]
\nn\\
&&	-\delta^{\delta\beta}\Bigl[
	\left(\gamma^{ab'}\right)^{\alpha\gamma}
	- \left(\gamma^{a'b}\right)^{\alpha\gamma}
	\Bigr] \biggr\}\ .
\label{12fermiblocknn}
\ee
The bosonic sector $H_{\rm BB}$ contributes to the $(2,2)$ and $(3,3)$ blocks:
\be
&&\Braket{J| b_q^{\alpha} a_r^{A} a_s^{B} (H_{\rm BB}) 
	a_s^{C\dag} a_r^{D\dag} b_q^{\beta\dag}|J}
	 = 
	-\frac{n^2\lambda'}{2J}\delta^{\alpha\beta}\biggl\{
	9\,\delta^{AD}\delta^{BC}
	+4\,\delta^{ac}\delta^{bd}
	-4\,\delta^{ab}\delta^{cd}
\nn\\
&&	-\delta^{ad}\left(
		5\,\delta^{bc}+3\,\delta^{b'c'}\right)  
	-4\,\delta^{a'c'}\delta^{b'd'}
	+4\,\delta^{a'b'}\delta^{c'd'}
	+\delta^{a'd'}\left(
		5\,\delta^{b'c'}+3\,\delta^{bc}\right)
	\biggr\}\ .
\ee
In the same lower-diagonal blocks, $H_{\rm BF}$ exhibits the contribution
\be
&&\Braket{J| b_n^{\alpha} a_n^{A} a_{-2n}^{B} (H_{\rm BF}) 
	a_{-2n}^{C\dag} a_n^{D\dag} b_n^{\beta\dag}|J}
	 =  
	-\frac{n^2\lambda'}{8J}\biggl\{
	39\,\delta^{\alpha\beta}\delta^{AD}\delta^{BC}
\nn\\
&&
	+\delta^{\alpha\beta}\delta^{AD}
	\left(\delta^{b'c'}-7\,\delta^{bc}\right)
	-4\,\delta^{\alpha\beta}\delta^{BC}
	\left(\delta^{ad}-\delta^{a'd'}\right)
	+4\,\delta^{BC}\Bigl[
	(\gamma^{ad})^{\alpha\beta}
	-(\gamma^{a'd'})^{\alpha\beta}
	\Bigr]
\nn\\
&&
	-8\,\delta^{AD}\Bigl[  
	(\gamma^{bc})^{\alpha\beta}
	-(\gamma^{b'c'})^{\alpha\beta}
	\Bigr]
	\biggr\}\ .
\ee
Finally, $H_{\rm BF}$ yields matrix elements in the off-diagonal block
$(2,3)$:
\be
\Braket{J| b_n^{\alpha} a_n^{A} a_{-2n}^{B} (H_{\rm BF}) 
	a_n^{C\dag} a_n^{D\dag} b_{-2n}^{\beta\dag}|J}
	& = & 
	-\frac{n^2\lambda'}{32J}\biggl\{
	9\,\delta^{\alpha\beta}\delta^{AC}\delta^{BD}
	+9\,\delta^{\alpha\beta}\delta^{AD}\delta^{BC}
\nn\\
&&\kern-150pt
	+\delta^{\alpha\beta}       \delta^{AC}   \left(  
	23\delta^{bd}-41\delta^{b'd'}\right) 
	+\delta^{\alpha\beta}\delta^{AD}\left(
	23\delta^{bc}-41\delta^{b'c'}\right)
\nn\\
&&\kern-150pt
	-32\delta^{AD}   \Bigl[     
	(\gamma^{bc}   )^{\alpha\beta}   
	- (\gamma^{b'c'})^{\alpha\beta}
	\Bigr]
	-32\delta^{AC}\Bigl[
	(\gamma^{bd})^{\alpha\beta} 
	- (\gamma^{b'd'})^{\alpha\beta}
	\Bigr]
	\biggr\}\ .
\ee

We can perform a full symbolic diagonalization of the $1,024\times 1,024$ bosonic
and fermionic perturbation matrices to obtain the one-loop in $\lambda'$, $O(1/J)$ 
energy corrections. They can all be expressed in terms of dimensionless
eigenvalues $\Lambda$ according to the standard formula (\ref{lambdaexp}) modified
by setting $q=r=n,\ s=-2n$:
\be
E_J(n) = 3+3n^2\lambda'\left(1+\frac{\Lambda}{J}+O(J^{-2})\right)\ .
\label{Lexpnn}
\ee
The resulting spectrum is displayed in table~\ref{nnspec}.
\begin{table}[ht!]
{\small
\begin{eqnarray}
\begin{array}{|c|ccccccccc|}
\hline
\Lambda_1~({S^5~{\rm vector}})
	& -23/3& -20/3	& -17/3	& -14/3	& -11/3	& -8/3 & -5/3	& -2/3	& 1/3	 \\
\hline
{\rm Multiplicity}	& 4_B	& 32_F	& 112_B    & 224_F & 280_B & 224_F	
								& 112_B	& 32_F	& 4_B 	\\
\hline\hline
\Lambda_2~({AdS_5~{\rm vector}})	
	& -19/3	& -16/3	& -13/3	& -10/3	& -7/3	& -4/3 & -1/3	& 2/3	& 5/3	 \\
\hline
{\rm Multiplicity}	& 4_B	& 32_F	& 112_B    & 224_F & 280_B & 224_F	
								& 112_B	& 32_F	& 4_B 	\\
\hline
\end{array} \nonumber
\end{eqnarray} }
\caption{Spectrum of three-impurity string Hamiltonian with $(q=r=n,\ s=-2n)$}
\label{nnspec}
\end{table}
The levels clearly organize themselves into two superconformal multiplets built on 
vector primary states. Note that the spinor multiplet is absent and that the degeneracy 
between multiplets that was seen in the inequivalent mode index case has been lifted. 
The spinor multiplet is absent for the following reason: it contains a representation 
at level $L=4$ arising from fermion creation operators completely symmetrized on 
$SO(4)\times SO(4)$ spinor indices; such a construct must vanish unless all the creation 
operator mode indices are different. 

If we keep track of the $SO(4)\times SO(4)$
irrep structure, we find that the symmetric-traceless bosonic $SO(4)_{S^5}$ states 
arising from the closed $\su(2)$ subsector fall into the $-11/3~[280_B]$ level. This 
is the counterpart of the $-4~[280_B]$ level in table~\ref{3mult15}. To compare
with Minahan and Zarembo's Bethe ansatz calculation of the corresponding gauge theory 
operator dimension, we must evaluate eqn.~(\ref{MZBA}) with the
appropriate choice of parameters.  In particular, $M_n=2$ 
when two mode indices are allowed to coincide and, comparing with eqn.~(\ref{Lexpnn}),
we find perfect agreement with the string theory prediction $\Lambda=-11/3$. 
States at level $L=4$ in the second multiplet in table~\ref{nnspec} correspond
to operators in the $\Sl(2)$ closed sector of the gauge theory and the eigenvalue
$\Lambda=-7/3~[280_B]$ amounts to a prediction for the one-loop anomalous
dimension of that class of gauge theory operators. As mentioned at the end of the
previous subsection, we will need to develop a numerical treatment of the
$\Sl(2)$ spin chain Hamiltonian in order to assess the agreement between the
string theory and gauge theory in this sector. 

\section{Three-impurity spectrum: all orders in $\lambda'$}
\label{threeimp3}
In the previous section, we have studied the eigenvalue spectrum of the 
string theory perturbation Hamiltonian expanded to leading order in $1/J$ 
and to one-loop order in $\lambda'$. The expansion in $\lambda'$ was for
convenience only since our expressions for matrix elements are exact in
this parameter.  We should, in principle, be able to obtain results that are 
exact in $\lambda'$ (but still of leading order in $1/J$). This is a
worthwhile enterprise since recent progress on the gauge theory side
has made it possible to evaluate selected operator anomalous dimensions
to two- and three-loop order. The simple one-loop 
calculations of the previous sections have given us an overview
of how the perturbed string theory eigenvalues are organized into 
superconformal multiplets. This provides a very useful orientation for 
the more complex all-orders calculation, to which we now turn.

\subsection{Inequivalent mode indices: $(q\neq r\neq s)$}
Our first step is to collect the exact matrix elements of the perturbing
Hamiltonian between three-impurity states of unequal mode indices. The 
block structure of the perturbation matrix in the spacetime boson sector
is given in table~\ref{boseblock3} and the exact form of the $(1,1)$ block is
\be
\Bra{J}&&\kern-20pt a_q^{A} a_r^{B} a_{s}^{C} (H_{\rm BB})a_{s}^{D\dag} a_r^{E\dag} 
	a_q^{F\dag}\Ket{J}
	 = 
	-\frac{1}{2\omega_q\omega_r\omega_s}\biggl\{
	\delta^{BE}\omega_r
	\Bigl[
	\delta^{CD}\delta^{AF}(s^2+q^2(1+2s^2\lambda'))
\nn\\
&&	- (q^2+s^2)\delta^{cd}\delta^{af}	
	-2qs (\delta^{ad}\delta^{cf}  		
	-\delta^{ac}\delta^{df})
	+(q^2-s^2)\delta^{af}\delta^{c'd'}
	-(q^2-s^2)\delta^{a'f'}\delta^{cd}
\nn\\
&&	+ (q^2+s^2)\delta^{c'd'}\delta^{a'f'}	
	+2qs (\delta^{a'd'}\delta^{c'f'}  	
	-\delta^{a'c'}\delta^{d'f'})
	\Bigr]
	+ \Bigl(
	C\rightleftharpoons B,\ 	
	D\rightleftharpoons E,\		
	s\rightleftharpoons r
	\Bigr)
\nn\\
&&	+ \Bigl(
	A\rightleftharpoons B,\ 
	F\rightleftharpoons E,\
	q\rightleftharpoons r
	\Bigr)
	\biggr\}\ ,
\label{fullbose11}
\ee
where we define $\omega_q \equiv \sqrt{q^2+1/\lambda'}$ to simplify this and 
other similar expressions.

The off-diagonal $H_{\rm BF}$ contributions to the $(1,2)$, $(1,3)$ and $(1,4)$ blocks
are yet more complicated.  To simplify the expressions, we define
\be
F_1 \equiv \sqrt{(\omega_q+q)(\omega_r-r)}~, & \qquad & 
		F_2 \equiv \sqrt{(\omega_q-q)(\omega_r+r)}~, \nn\\
F_3 \equiv \sqrt{(\omega_q-q)(\omega_r-r)}~, & \qquad & 
		F_4 \equiv \sqrt{(\omega_q+q)(\omega_r+r)}\ .
\ee
Using these functions, the matrix elements in these off-diagonal
subsectors are given by:
\be
\Bra{J}&&\kern-20pt a_q^{A} a_r^{B} a_{s}^{C} (H_{\rm BF}) 
	a_{s}^{D\dag} b_r^{\alpha\dag} b_q^{\beta\dag}\Ket{J}
	 = 
	\frac{\delta^{CD}}{32 \omega_q\omega_r J}\biggl\{
	\frac{8}{\sqrt{\lambda'}}
	(F_1-F_2)\delta^{AB}\delta^{\gamma\delta}  
\nn\\
&&	-2(q-r)(F_3+F_4)\delta^{AB}\delta^{\gamma\delta} 
	+4(q-r)(F_3+F_4)(\gamma^{ab})^{\gamma\delta}		
\nn\\
&&	-2(q+r)(F_3-F_4)(\gamma^{ab'})^{\gamma\delta}
	+(2qF_3 -2qF_4 +2rF_3 - 2rF_4 )(\gamma^{a'b})^{\gamma\delta}
\nn\\
&&	-(4qF_3+4qF_4-4rF_3-4rF_4) (\gamma^{a'b'})^{\gamma\delta} 
	+\frac{8}{\sqrt{\lambda'}}(F_2-F_1)\delta^{a'b'}  \delta^{\gamma\delta}  
\nn\\
&&	+4(q- r)   
	(F_3+F_4)\delta^{\gamma\delta}\delta^{a'b'}
	-2(q-r)(F_3+F_4)\delta^{\gamma\delta}(\delta^{ab}-\delta^{a'b'})
\nn\\
&&	-4\lambda'\omega_q\omega_r(q-r)(F_3+   F_4)\delta^{AB}\delta^{\gamma\delta}
\nn\\
&&	+4\sqrt{\lambda'}(qr-\omega_q\omega_r)
	\Bigl[
	(F_1+F_2)\Bigl(
	(\gamma^{ab'})^{\gamma\delta} - (\gamma^{a'b})^{\gamma\delta}
	\Bigr)
	-(F_1-F_2)\delta^{\gamma\delta}(\delta^{ab}-  \delta^{a'b'})
	\Bigr]   
\nn\\
&&	+2(\omega_q+\omega_r)(F_3+F_4)
	\Bigl[
	(\gamma^{ab'})^{\gamma\delta} - (\gamma^{a'b})^{\gamma\delta}
	\Bigr]
\nn\ee\be
&&	+4\sqrt{\lambda'}(r\omega_q-q\omega_r)(F_1+F_2)
	\Bigl[  
	(\gamma^{ab})^{\gamma\delta}- (\gamma^{a'b'})^{\gamma\delta}
	\Bigr]
\nn\\
&&	-4\lambda'(q-r)(F_3-F_4)(r\omega_q+q\omega_r)\delta^{AB}\delta^{\gamma\delta}
	-\lambda'\delta^{AB}
\nn\\
&&	+2\lambda'(\omega_q\omega_r-qr)(q-r)(F_3+F_4)\delta^{AB}\delta^{\gamma\delta}
	+4\sqrt{\lambda'}(\omega_q\omega_r+qr)(F_1-F_2)\delta^{AB}\delta^{\gamma\delta}
\nn\\
&&	-2\lambda'  (q-r)(\omega_q\omega_r+qr)  (F_3+F_4)\delta^{AB}\delta^{\gamma\delta}
	\biggr\}\ .
\ee
The $H_{\rm FF}$ contribution to the lower-diagonal blocks $(2,2)$, $(3,3)$ and
$(4,4)$ is
\be
\Bra{J}&&\kern-20pt b_q^{\alpha} b_r^{\beta} a_{s}^{A} (\rm H_{\rm FF}) 
	a_{s}^{B\dag} b_r^{\gamma\dag} b_q^{\delta\dag}\Ket{J}
	 =
\nn\\
&&	\frac{\delta^{AB}}{48\omega_r\omega_s J}\sqrt{\lambda'}
	\biggl\{
	2 rs \sqrt{1/\lambda'}
	\biggl[
	\Bigl(
	(\gamma^{ij})^{\alpha\gamma}(\gamma^{ij})^{\beta\delta}
	-(\gamma^{i'j'})^{\alpha\gamma}(\gamma^{i'j'})^{\beta\delta}
	\Bigr)
\nn\\
&&	-\Bigl(
	(\gamma^{ij})^{\alpha\delta}(\gamma^{ij})^{\beta\gamma}
	-(\gamma^{i'j'})^{\alpha\delta}(\gamma^{i'j'})^{\beta\gamma}
	\Bigr)
	-\Bigl(
	(\gamma^{ij})^{\alpha\beta}(\gamma^{ij})^{\gamma\delta}
	-(\gamma^{i'j'})^{\alpha\beta}(\gamma^{i'j'})^{\gamma\delta}
	\Bigr)
	\biggr]
\nn\\
&&	-12
	\Bigl[
	2\delta^{\alpha\delta}\delta^{\beta\gamma}	
	\Bigl(
	s^2\sqrt{1/\lambda'}-2rs\sqrt{\lambda'}\omega_r\omega_s
	+r^2(2s^2\sqrt{\lambda'}+\sqrt{1/\lambda'})
	\Bigr) \Bigr]
	\biggr\}\ .
\ee
The bose-fermi Hamiltonian $H_{\rm BF}$ contributes the following matrix
elements to the same lower-diagonal blocks:
\be
\Bra{J}&&\kern-20pt b_q^{\alpha} b_r^{\beta} a_{s}^{A} (H_{\rm BF}) 
	a_{s}^{B\dag} b_r^{\gamma\dag} b_q^{\delta\dag}\Ket{J}
	 = 
	-\frac{1}{2\omega_q\omega_r\omega_s}
	\biggl\{
	s\sqrt{\lambda'}\delta^{ab}\delta^{\alpha\delta}\delta^{\beta\gamma}
	\Bigl[
	s\omega_r
	( 2q^2\sqrt{\lambda'}+\sqrt{1/\lambda'} )
\nn\\
&&	+ s\omega_q
	( 2r^2\sqrt{\lambda'}+\sqrt{1/\lambda'} )
	-2\omega_q\omega_r\omega_s(q+r)\sqrt{\lambda'}
	\Bigr]
	+\delta^{a'b'}\delta^{\alpha\delta}\delta^{\beta\gamma}
	\Bigl[
	2\omega_r q^2 (1+s^2\lambda') 
\nn\\
&&	+ s^2\omega_r
	+2\omega_q r^2 (1+s^2\lambda') 
	+ s^2\omega_q
	-2s(q+r)\lambda'\omega_q\omega_r\omega_s
	\Bigr]
\nn\\
&&	+sr\omega_q\delta^{\alpha\delta}\Bigl[
	(\gamma^{ab})^{\beta\gamma} - (\gamma^{a'b'})^{\beta\gamma}
	\Bigr]
	+sq\omega_r\delta^{\beta\gamma}\Bigl[
	(\gamma^{ab})^{\alpha\delta} - (\gamma^{a'b'})^{\alpha\delta}
	\Bigr]
	\biggr\}\ .
\ee

To simplify off-diagonal elements in the $(2,3)$, $(2,4)$ and $(3,4)$
blocks, we define
\be
G_1 \equiv \sqrt{(\omega_r+r)(\omega_s-s)}~, & \qquad & 
		G_2 \equiv \sqrt{(\omega_r-r)(\omega_s+s)}~, \nn\\
G_3 \equiv \sqrt{(\omega_r-r)(\omega_s-s)}~, & \qquad & 
		G_4 \equiv \sqrt{(\omega_r+r)(\omega_s+s)}\ .
\ee
The matrix elements in these subsectors are then given by
\be
\Bra{J}&&\kern-20pt b_q^{\alpha} b_r^{\beta} a_{s}^{A} (H_{\rm BF}) 
	a_r^{B\dag} b_q^{\gamma\dag} b_{s}^{\delta\dag}\Ket{J}
	 = 
\nn\\
&&	-\frac{1}{16 (\lambda'\omega_r\omega_s)^{3/2}}\biggl\{
	\sqrt{\omega_r\omega_s}\lambda'
	\delta^{\alpha\gamma}
	\biggl[
	2\delta^{ab}\delta^{\beta\delta}\Bigl[
	(G_1+G_2)(2-2\lambda'\omega_r\omega_s)
\nn\\
&&	+(r+s)\sqrt{\lambda'}
	\Bigl(
	G_4-\lambda'G_4(r-\omega_r)(s-\omega_s)
\nn\\
&&	+G_3(-1+rs\lambda'+r\omega_s\lambda'+\omega_r(s+\omega_s)\lambda')
	\Bigr)
	\Bigr]
\nn\ee\be
&&	+2\sqrt{\lambda'}
	\Bigl[
	(r+s)(G_3-G_4)+\sqrt{\lambda'}(G_1-G_2)(r\omega_s-s\omega_r)
	\Bigr]
	\Bigl[
	(\gamma^{ab})^{\beta\delta}
	-(\gamma^{a'b'})^{\beta\delta}
	\Bigr]
\nn\\
&&	+\sqrt{\lambda'}\Bigl[
	2rs\sqrt{\lambda'}G_1-2rs\sqrt{\lambda'}G_2+(r-s)(G_3+G_4)
	+(\omega_s-\omega_r)(G_3-G_4)
\nn\\
&&	+2\omega_r\omega_s\sqrt{\lambda'}(G_2-G_1)
	\Bigr]
	\Bigl[
	(\gamma^{ab'})^{\beta\delta}
	-(\gamma^{a'b})^{\beta\delta}
	\Bigr]
	+2\delta^{a'b'}\delta^{\beta\delta}
	\Bigl[
	-2rs\sqrt{\lambda'}(G_1-G_2)
\nn\\
&&	+(r+s)\sqrt{\lambda'}
	\Bigl(
	-G_4-\lambda'G_4(r-\omega_r)(s-\omega_s)
\nn\\
&&	+G_3(1+rs\lambda'+r\omega_s\lambda'+\omega_r(s+\omega_s)\lambda')
	\Bigr)
	\Bigr] \biggr]
	\biggr\}\ .
\ee

The entries in the spacetime fermion block matrix of table~\ref{fermiblock1} 
are far too complicated to write out explicitly: they
are best generated, viewed and manipulated with computer algebra
techniques. The explicit formulas, along with a collection of the
Mathematica programs written to generate and work with them, are 
available 
on the web.\footnote{\href{http://theory.caltech.edu/~swanson/MMA1/mma1.html}{http://theory.caltech.edu/{$\sim$}swanson/MMA1/mma1.html}}

We were not able to symbolically diagonalize the complete perturbation
matrix built from the exact (in $\lambda'$) matrix elements listed
above: with the computing resources available to us, the routines for
diagonalizing the full 2,048-dimensional matrices would not terminate in 
any reasonable time. As noted in the previous section, however, gauge 
theory arguments suggest that there are three protected $SO(4)\times SO(4)$ 
irreps that do not mix with any other irreps. It is a straightforward
matter to project the perturbation matrix onto these unique protected 
irreps to obtain analytic expressions for the corresponding exact 
eigenvalues. In fact, the superconformal multiplet structure of the three-impurity 
problem is such that the energies/dimensions of all other irreps can be inferred from those 
of the three protected irreps. Hence, this method will give us exact
expressions for all the energy levels of the three-impurity problem.

Consider first the $\Sl(2)$ closed sector. The dual sector is generated on the 
string theory side by bosonic creation operators completely symmetrized (and traceless) 
on $SO(4)_{AdS}$ vector indices. 
The simplest way to make this projection on eqn.~(\ref{fullbose11}) is to 
compute diagonal elements between the symmetrized states
\be
a_q^{( a\dag}a_r^{b\dag}a_s^{c\dag )}\Ket{J}\ ,
\ee
with $a\neq b\neq c$ (and, of course, $a,b,c \in 1,\dots,4$).  
The charges of the fermionic oscillators 
under this subgroup are $\pm 1/2$, so the three-boson state of this type 
cannot mix with one boson and two fermions (or any other state). Hence,
the above projection of eqn.~(\ref{fullbose11}) yields the 
closed sector eigenvalue correction
\be
\delta E_{AdS} (q,r,s,J) & =  &
	\frac{1}{J\omega_q\omega_r\omega_s}\biggl\{
	qs(1-qs\lambda')\omega_r
	+qr(1-qr\lambda')\omega_s
	+rs(1-rs\lambda')\omega_q
\nn\\
&&\kern+60pt	+\left[
	qr+s(q+r)\right]\lambda'
	\omega_q\omega_r\omega_s
	\biggr\}
\nn\\
	&\approx & \frac{1}{J}\biggl\{
	-2(q^2+qr+r^2)\lambda'
	-\frac{15}{8}\bigl(q^2r^2(q+r)^2\bigr){\lambda'}^3
	+ \dots
	\biggr\}\ .
\label{exactSO41}
\ee
To facilitate eventual comparison with gauge theory results,
we have performed a small-$\lambda'$ expansion in the final line
with the substitution $s\to -(q+r)$ (since the mode indices
satisfy the constraint $s+q+r=0$). The leading correction 
$-2(q^2+qr+r^2)\lambda'$ reproduces the one-loop eigenvalue
$\Lambda_{\rm BB} = -2~[280_B]$ located at level $L=4$ in the $SO(4)_{AdS}$ 
multiplet in table~\ref{3mult15}. 


The closed $\su(2)$ sector is generated by bosonic creation operators
completely symmetrized on traceless $SO(4)_{S^5}$ indices. Projection onto this
irrep is most simply achieved by choosing all mode operators in 
eqn.~(\ref{fullbose11}) to carry symmetrized, traceless $SO(4)_{S^5}$ labels (they can
also be thought of as carrying charge $+1$ under some $SO(2)$ subgroup 
of $SO(4)_{S^5}$).  Direct projection yields the $SO(4)_{S^5}$ eigenvalue
\be
\delta E_{S^5}(q,r,s,J)   &=& 
	-\frac{1}{J\omega_q\omega_r\omega_s}\biggl\{
	\left[qr+r^2+q^2(1+r^2\lambda')\right]\omega_s
\nn\\
&&	+ \left[qs+s^2+q^2(1+s^2\lambda')\right]\omega_r
\nn\\
&&	+ \left[rs+s^2+r^2(1+s^2\lambda')\right]\omega_q
	-\left[rs+q(r+s)\right]\lambda'\omega_q\omega_r\omega_s
	\biggr\}
\nn\\
&&\kern-80pt \approx 
	\frac{1}{J}\biggl\{
	-4(q^2+qr+r^2)\lambda'
	+(q^2+qr+r^2)^2{\lambda'}^2
\nn\\
&&\kern-70pt	-\frac{3}{4}\bigl(
	q^6+3q^5r+8q^4r^2+11q^3r^3
	+8q^2r^4+3qr^5+r^6\bigr)
	{\lambda'}^3 +\dots \biggr\}\ .
\label{exactSO42}
\ee
This is the all-loop formula corresponding 
to gauge theory operator dimensions in the closed $\su(2)$ subsector;
the leading-order term $-4(q^2+qr+r^2)\lambda'$ reproduces the 
one-loop eigenvalue $\Lambda_{\rm BB}=-4~[280_B]$ at level $L=4$
in the $SO(4)_{S^5}$ vector multiplet in table~\ref{3mult15}.  

The eigenvalue of the symmetrized pure-fermion irrep can be obtained 
by evaluating the exact matrix element $H_{\rm FF}$ acting on three 
symmetrized fermionic creation operators with $SO(4)\times SO(4)$ 
indices chosen to lie in the same $\Pi$ projection
(with inequivalent mode indices). The exact energy shift 
for this irrep turns out to be
\be
\delta E_{\rm Fermi}(q,r,s,J)  &=& 
	-\frac{1}{4\,J\omega_q\omega_r\omega_s}\biggl\{
	-4\bigl(rs+q(r+s)\bigr)\lambda'\omega_q\omega_r\omega_s
\nn\\
&&\kern-40pt	+\biggl[
	\omega_q
	\left(
	2 s^2
	+4r^2s^2\lambda'
	+2 r^2
	\right)
	+\bigl(s\to r,\ r\to q,\ q\to s\bigr)
	+\bigl( q\rightleftharpoons r \bigr)
	\biggr]
	\biggr\}
\nn\ee\be
&&\kern-50pt \approx 
	\frac{1}{J}\biggl\{
	-3(q^2+qr+r^2)\lambda'
	+\frac{1}{2}(q^2+qr+r^2)^2{\lambda'}^2
\nn\\
&&\kern-50pt	-\frac{3}{16}\bigl(
	2q^6+6q^5r+21q^4r^2+32q^3r^3
	+21q^2r^4+6qr^5+2r^6\bigr)
	{\lambda'}^3 +\dots \biggr\}\ .
\label{exactfermi5}
\ee
The leading-order $\lambda'$ correction $-3(q^2+qr+r^2)\lambda'$
reproduces the $\Lambda_{\rm FF}=-3~[580_F]$ eigenvalue at the $L=4$ level 
in the spinor multiplet in table~\ref{3mult15}. This and the higher
order terms in the eigenvalue will eventually be compared with 
the dimensions of operators in the closed, fermionic $\su(1|1)$ sector in the 
gauge theory. 

The argument we are making relies heavily on the claim that the perturbation
matrix is block diagonal on the closed subsectors described above: we have evaluated
the exact energy shift on these subsectors by simply taking the diagonal
matrix element of the perturbing Hamiltonian in a particular state in each
sector. We will now carry out a simple numerical test of the claimed
block diagonalization of the full perturbing Hamiltonian. The basic idea
is that, while it is impractical to algebraically diagonalize the full
$2,048\times 2,048$ perturbation matrices, it is quite easy to do a numerical
diagonalization for a specific choice of $\lambda'$ and mode indices 
$q,r,s$. One can then check that the numerical eigenvalues match
the analytic predictions evaluated at the chosen coupling and mode indices. 
For definiteness, we choose
\be
q = 1~, \qquad r=2~, \qquad s=-3~, \qquad \lambda' = 1\ .
\label{MIvals}
\ee
The predicted eigenvalue shifts of the three protected states, evaluated
at the parameter choices of (\ref{MIvals}) are given in table~\ref{centroidnum}.
These values come directly from eqns.~(\ref{exactSO41}, \ref{exactSO42}, \ref{exactfermi5})
above (with $J$ set to unity, for convenience).
\begin{table}[ht!]
\begin{eqnarray}
\begin{array}{|ccc|}
\hline
	\delta E: ~\lambda' = 1 & & q=1,\ r=2,\ s=-3 \\ 
\hline
 	\delta E_{AdS}(1,2,-3,J=1) &=& -16.255434067000426 \\ 
  	\delta E_{S^5}(1,2,-3,J=1) &=&  -20.137332508389193 \\ 
	\delta E_{\rm Fermi}(1,2,-3,J=1) &=& -18.19638328769481 \\
\hline
\end{array} \nonumber
\end{eqnarray}
\caption{Exact numerical eigenvalues of three-impurity protected sectors}
\label{centroidnum}
\end{table}
Since we want to compare these energies to a numerical diagonalization, 
we must maintain a high level of precision in the numerical computation. 
With the parameter choices of (\ref{MIvals}), the numerical diagonalization
of the full $2,048\times 2,048$ perturbation matrices 
on both the spacetime boson (table~\ref{boseblock3}) and spacetime 
fermion (table~\ref{fermiblock3}) sectors yields the spectrum and multiplicities
displayed in table~\ref{numfermi}. The multiplicities are consistent with the
superconformal multiplet structure we found in 
the one-loop analysis (given in table~\ref{3mult15}). The predicted 
closed sector eigenvalues (listed in table~\ref{centroidnum}) match, to the 
precision of the calculation, entries in the list of numerical eigenvalues. 
These energies also appear at the expected levels within the multiplets.
$E_{AdS}(1,2,-3,J)$ and $E_{S^5}(1,2,-3,J)$ appear in bosonic levels with 
multiplicity $280_B$, while energy $E_{\rm Fermi}(1,2,-3,J)$ appears as a 
fermionic level with multiplicity $560_F$; according to table~\ref{3mult15}
these are uniquely identified as the central $L=4$ levels of their respective
multiplets, exactly where the protected energy levels must lie. All of 
this is clear evidence that the closed sector states of the string theory 
do not mix with other states under the perturbing Hamiltonian, thus justifying
our method of calculating their exact eigenenergies.
\begin{table}[ht!]
\begin{eqnarray}
\begin{array}{|cc|}
\hline
\delta E(1,2,-3,J=1)\ \lambda' = 1 & {\rm Mult.}  \\
\hline
          -30.821354623065    &                     4_B\\
         -26.9394561816763    &                     4_B\\
         -26.2093998737015    &                    64_B\\
         -25.4793435657269    &                   112_B\\
         -21.5974451243382    &                   112_B\\
         -20.8673888163637    &                   448_B\\
         -20.1373325083891    &                   280_B\\
         -16.2554340670003    &                   280_B\\
         -15.5253777590258    &                   448_B\\
         -14.7953214510512    &                   112_B\\
         -10.9134230096624    &                   112_B\\
         -10.1833667016878    &                    64_B\\
          -9.4533103937133    &                     4_B\\
         -5.57141195232456    &                     4_B\\ 
\hline
\end{array} \nn
\qquad
\begin{array}{|cc|}
\hline
\delta	E(1,2,-3,J=1)\ \lambda' = 1 & {\rm Mult.}  \\
\hline
         -28.8804054023706    &                         8_F\\
          -28.150349094396    &                        32_F\\
         -24.2684506530072     &                       32_F\\
         -23.5383943450326     &                      224_F\\
          -22.808338037058     &                      224_F\\
         -18.9264395956693     &                      224_F\\
         -18.1963832876947    &                       560_F\\
         -17.4663269797201    &                       224_F\\
         -13.5844285383314    &                       224_F\\
         -12.8543722303568    &                       224_F\\
         -12.1243159223822    &                        32_F\\
         -8.24241748099347    &                        32_F\\
         -7.51236117301893    &                         8_F\\
\hline
\end{array} \nonumber
\end{eqnarray}
\caption{All-loop numerical spectrum of three-impurity states ($q=1,~r=2,~s=-3,~
\lambda'=1,~J=1$). Left panel: bosons; right panel: fermions }
\label{numfermi}
\end{table}


At one loop, we found that the three superconformal multiplets were
displaced from each other by precisely the internal level spacing.
This led to an accidental degeneracy that is lifted in the exact 
dimension formulas we have just derived. To explore this, it is useful
to have formulas for the eigenvalues of all the levels in each multiplet. 
From the discussion in Section~\ref{threeimp2}, we see that
each level in the string energy spectrum can be connected by
a simple integer shift in the angular momentum $J$.
Since we are working at $O(1/J)$ in a large-$J$ expansion, 
all contributions from this shift must come from the BMN
limit of the theory.  In other words, by sending
$J\to J+2-L/2$ in the BMN formula for the energy
\be
E = \sqrt{1+\frac{n^2 g_{YM}^2 N_c}{(J+2-L/2)^2}} + \dots\ ,	
\ee
we can generate an expansion, to arbitrary order in $\lambda'$,
for each level $L$ in the entire superconformal multiplet. 

For the vector $SO(4)_{AdS}$ multiplet, we find 
\be
\delta E_{AdS}(q,r,J,L) & \approx & 
	\frac{1}{J}\biggl\{
	(L-6)(q^2+qr+r^2)\lambda'
	-\frac{1}{2}(L-4)(q^2+qr+r^2)^2	{\lambda'}^2
\nn\\
&&	+\frac{3}{16}\Bigl[
	2(L-4)q^6
	+6(L-4)q^5r
	+5(3L-14) q^4r^2
\nn\\
&&\kern-20pt	
	+20(L-5) q^3r^3+5(3L-14) q^2r^4
	+6(L-4) qr^5
	+2(L-4)r^6
	\Bigr]{\lambda'}^3
\nn\\
&&\kern-20pt	
	-\frac{(q^2+qr+r^2)}{16}\Bigl[
	5(L-4)q^6
	+15(L-4)q^5r
	+(50L-247)q^4r^2
\nn\\
&&\kern+-0pt	
	+(75L-394)q^3r^3+(50L-247)q^2r^4
\nn\\
&&	+15(L-4)qr^5
	+5(L-4)r^6
	\Bigr]{\lambda'}^4 + \dots
	\biggr\}\ 
\ee
(for convenience in eventual comparison with the gauge theory, the
eigenvalues have been expanded to $O(\lambda'^4)$). The corresponding
result for the $SO(4)_{S^5}$ vector multiplet is
\be
\delta E_{S^5}(q,r,J,L) & \approx & 
	\frac{1}{J}\biggl\{
	(L-8)(q^2+qr+r^2)\lambda'
	-\frac{1}{2}
	(L-6)(q^2+qr+r^2)^2
	{\lambda'}^2
\nn\\
&&\kern-30pt	+\frac{3}{16}\Bigl[
	2(L-6)q^6
	+6(L-6)q^5r
	+(15L-92) q^4r^2
\nn\\
&&\kern-30pt	+4(5L-31) q^3r^3
	+(15L-92) q^2r^4
	+6(L-6) qr^5
	+2(L-6)r^6
	\Bigr]{\lambda'}^3
\nn\\
&&\kern-75pt	
	-\frac{(q^2+qr+r^2)}{16}\Bigl[
	5(L-6)q^6
	+15(L-6)q^5r
	+(50L-309)q^4r^2
\nn\\
&&\kern-0pt	+3(25L-156)q^3r^3+(50L-309)q^2r^4
\nn\\
&&	+15(L-6)qr^5
	+5(L-6)r^6
	\Bigr]{\lambda'}^4 + \dots
	\biggr\}\ .
\ee
Finally, the result for the spinor multiplet is
\be
\delta E_{\rm Fermi}(q,r,J,L) & \approx & 
	\frac{1}{J}\biggl\{
	(L-7)(q^2+qr+r^2)\lambda'
	-\frac{1}{2}
	(L-5)(q^2+qr+r^2)^2
	{\lambda'}^2
\nn\\
&&\kern-50pt
	+\frac{3}{16}\Bigl[
	2(L-5)q^6
	+6(L-5)q^5r
	+3(5L-27) q^4r^2
	+4(5L-28) q^3r^3
\nn\\
&&\kern+35pt	+3(5L-27) q^2r^4
	+6(L-5) qr^5
	+2(L-5)r^6
	\Bigr]{\lambda'}^3
\nn\\
&&\kern-100pt	
	-\frac{(q^2+qr+r^2)}{16}\Bigl[
	5(L-5)q^6
	+15(L-5)q^5r
	+2(25L-139)q^4r^2
	+(75L-431)q^3r^3
\nn\\
&&\kern-30pt	+2(25L-139)q^2r^4
	+15(L-5)qr^5
	+5(L-5)r^6
	\Bigr]{\lambda'}^4 + \dots
	\biggr\}\ .
\ee
It is important to remember that, to obtain the energies of the states as opposed
to the energy shifts $\delta E$, we must add the BMN energy of the original 
degenerate multiplet to the above results:
\be
\label{BMNen}
E_{\rm BMN} &=& \sqrt{1+\lambda' q^2}+\sqrt{1+\lambda' r^2}+\sqrt{1+\lambda' (q+r)^2}\nn\\
	&=& 3 + (q^2+r^2+qr)\lambda' - \frac{1}{4}(q^2+r^2+qr)^2\lambda'^2+ \ldots~.
\ee

We can conclude from the above formulas that all three multiplets have
a common internal level spacing given by the following function of 
$\lambda'$ and mode indices:
\be
\label{levelspacing}
\frac{\delta E}{\delta L}& \approx & 
	\frac{1}{J}\biggl\{
	(q^2+qr+r^2)\lambda'
	-\frac{1}{2}\Bigl[
	(q^2+qr+r^2)^2
	\Bigr]{\lambda'}^2
\nn\\
&&	+\frac{3}{16}\Bigl[2q^6 +6q^5r +15 q^4r^2 +20q^3r^3 
 +15q^2r^4 +6qr^5 +2r^6 \Bigr]{\lambda'}^3
\nn\\
&&\kern-00pt	-\frac{(q^2+qr+r^2)}{16}\Bigl[
	5q^6 +15q^5r +50q^4r^2 +75q^3r^3
\nn\\
&&\kern0pt
	+50q^2r^4 +15qr^5 +5(r^6 \Bigr]{\lambda'}^4 + \dots
	\biggr\}\ .
\ee
We have expanded in powers of $\lambda'$, 
but an all-orders formula can easily be constructed. The multiplets are 
displaced from one another by shifts that also depend on $\lambda'$
and mode indices. We note that the one-loop degeneracy between 
different multiplets (see table~\ref{3mult15}) is preserved to second 
order in $\lambda'$, but is broken explicitly at three loops. 
At this order and beyond, each multiplet acquires a constant overall 
($L$-independent) shift relative to the other two.

\subsection{Two equal mode indices: $(q = r = n,\ s=-2n)$}

An independent analysis is required when two mode indices are equal 
(specifically, we choose $q = r = n$, $s=-2n$). The all-loop matrix elements 
are complicated and we will refrain from giving explicit expressions for them
(though the complete formulas can be found at the following URL:
\href{http://theory.caltech.edu/~swanson/MMA1/mma1.html}{http://theory.caltech.edu/{$\sim$}swanson/MMA1/mma1.html}). 
As in the unequal mode index case, however, exact eigenvalues can easily be
extracted by projection onto certain protected subsectors. In particular,
the energy shift for states created by three bosonic mode creation operators
with symmetric-traceless $SO(4)_{AdS}$ vector indices (the $\Sl(2)$ sector)
turns out to be
\be
\delta E_{AdS}(n,J)   = &&\kern-15pt  
	-\frac{n^2\lambda'}{J(1+n^2\lambda')\sqrt{4n^2+1/\lambda'}}
	\biggl\{
	\sqrt{4n^2+\frac{1}{\lambda'}}\left( 3+4n^2\lambda'\right)
	+\omega_n\left( 4+8n^2\lambda'\right)
	\biggr\}
\nn\\
	&& \approx 
	\frac{1}{J}\biggl\{
	-7n^2\lambda' + n^4{\lambda'}^2 - \frac{17}{2}n^6{\lambda'}^3+\dots
	\biggr\}\ .
\label{nnval1}		
\ee
The leading order term in the small-$\lambda'$ expansion is the 
$-7/3~[280_B]$ level $L=4$ eigenvalue in the $\Lambda_2$ multiplet
in table~\ref{nnspec}. The energy shift of the $SO(4)_{S^5}$ partners of 
these states (belonging to the $\su(2)$ closed sector) is
\be
\delta E_{S^5}(n,J)   = &&\kern-15pt  
	-\frac{n^2\lambda'}{J(1+n^2\lambda')\sqrt{4n^2+1/\lambda'}}
	\biggl\{
	\sqrt{4n^2+\frac{1}{\lambda'}}\left( 5+4n^2\lambda'\right)
	+\omega_n\left( 6+8n^2\lambda'\right)
	\biggr\}
\nn\\
	&& \approx 
	\frac{1}{J}\biggl\{
	-11n^2\lambda' + 8n^4{\lambda'}^2 - \frac{101}{4}n^6{\lambda'}^3+\dots
	\biggr\}\ .		
\label{nnval2}
\ee
The one-loop correction corresponds to the $-11/3~[280_B]$ level
in the $\Lambda_1$ submultiplet of table~\ref{nnspec}.
As noted above, the protected symmetrized-fermion ($\su(1|1)$) sector does not
appear when two mode indices are equal. As in the previous section,
we can do a numerical diagonalization of the full perturbation matrix 
to verify that the predicted eigenvalues are indeed
exact and closed, but we will omit the details.

By invoking the angular momentum shift $J\to J+2-L/2$ in the BMN limit, 
we can use the energy shift of the $L=4$
level to recover the exact energy shifts of all other levels
in the superconformal multiplets of table~\ref{nnspec}. The energy shifts of the 
vector multiplet containing the protected $SO(4)_{AdS}$ bosonic irrep 
at level $L=4$ are given by the expression
\be
\delta E_{AdS}(n,J,L) & \approx & 
	\frac{1}{J}\biggl\{
	\frac{1}{2}(3L-19)n^2{\lambda'}
	-\frac{1}{2}(9L-38)n^4{\lambda'}^2
	+\frac{1}{8}(99L-464)n^6{\lambda'}^3
\nn\\
&&\kern+100pt	-\frac{1}{16}(645L-3160)n^8{\lambda'}^4
	+\dots
	\biggr\}\ .
\ee
The shifts of the multiplet containing the protected $SO(4)_{S^5}$ 
bosonic irrep are given by
\be
\delta E_{S^5}(n,J,L) & \approx & 
	\frac{1}{J}\biggl\{
	\frac{1}{2}(3L-23)n^2{\lambda'}
	-\frac{1}{2}(9L-52)n^4{\lambda'}^2	
	+\frac{1}{8}(99L-598)n^6{\lambda'}^3
\nn\\
&&\kern+100pt
	-\frac{1}{16}(645L-3962)n^8{\lambda'}^4
	+\dots
	\biggr\}\ .
\ee
Once again, we note that in order to get energies, rather than energy shifts, 
one must append the BMN energy of the original degenerate multiplet to these 
results. Unlike the unequal mode index case, there is no accidental degeneracy
between superconformal multiplets spanning the three-impurity
space, even at one loop in $\lambda'$. The level spacings within the two 
superconformal multiplets are the same, but the multiplets are offset from each 
other by an $L$-independent shift (but one that depends on $\lambda'$ and
mode indices).

\section{Gauge theory anomalous dimensions}
\label{threeimp4}
In the previous sections, we have given a complete analysis of the perturbed
energy spectrum of three-impurity string states. The ``data'' are internally
consistent in the sense that the perturbed energy levels organize themselves 
into proper superconformal multiplets of the classical 
nonlinear sigma model governing the string worldsheet dynamics. Since the 
quantization procedure leaves only a subgroup of the full symmetry group 
as a manifest, linearly realized symmetry, this is by itself a nontrivial 
check on the consistency of the action and quantization procedure. To address 
the issue of AdS/CFT duality, we must go further and compare the string 
energy spectrum with the anomalous dimensions of gauge theory operators 
dual to the three-impurity string states. 

As discussed in previous chapters, 
the task of finding the anomalous dimensions of BMN operators in the limit
of large $R$-charge and dimension $D$, but finite $\Delta=D-R$,
is greatly simplified by the existence of an equivalence between 
the dilatation operator of ${\cal N}=4$ SYM and the Hamiltonian of a 
one-dimensional spin chain. 
The one-loop spin chain Hamiltonian
has only nearest-neighbor interactions (in the planar large-$N_c$ limit)
and is of limited complexity.  This is tempered by the fact that the 
higher-loop gauge theory physics is encoded in increasingly long-range 
spin chain interactions that generate a rapidly growing number 
of possible terms in the Hamiltonian \cite{Beisert:2003tq}.  
Fixing the coefficients of all
these terms by comparison with diagrammatic computations would be a very
impractical approach.  Fortunately, Beisert was able to show that, at 
least for BMN operators in the $\su(2)$ closed subsector, 
general requirements (such as the existence of a well-defined BMN scaling limit) 
suffice to fix the form of the spin chain 
Hamiltonian out to three-loop order \cite{Beisert:2003ys,Beisert:2003jb}. 
In Chapter \ref{virial} we discussed the use of these higher-loop 
spin chains to generate the information
we need on the anomalous dimensions of three-impurity operators: 
we will rely on these results for our comparison with the three-impurity 
string theory predictions computed above.

We have already noted that there are three closed subsectors of BMN 
operators in which impurities taken from a subalgebra of the full 
superconformal algebra mix only with themselves: we have referred to them
as the $\Sl(2)$, $\su(2)$ (both bosonic) and $\su(1|1)$ (fermionic) sectors. We will focus
our attention on these sectors because their spin chain description is
simple and their anomalous dimensions fix the dimensions of the 
remaining three-impurity operators in the theory.  Spin chain Hamiltonians incorporating
higher-loop-order gauge theory physics have been constructed for the
$\su(2)$ and $\su(1|1)$ sectors but, as far as we know, the $\Sl(2)$ spin
chain is known only to one-loop order.\footnote{See ref.~\cite{Staudacher:2004tk}
for important progress on this problem.} 

Although these spin chains are integrable, methods such as the Bethe ansatz 
technique do not immediately yield the desired results for all multiple-impurity anomalous 
dimensions of interest. Minahan and Zarembo did use the Bethe ansatz for the one-loop $\so(6)$ 
spin chain (of which the exactly closed $\su(2)$ system is a subsector) to obtain 
approximate multi-impurity anomalous dimensions \cite{Minahan:2002ve}, but we 
need results for all sectors and for higher-loop spin chains. As mentioned above,
the $\Sl(2)$ spin chain has phenomenological applications and has been extensively
developed in that context. It is therefore possible that some of the results we need 
can be extracted from the relevant literature.\footnote{We thank A.~Belitsky for
making us aware of this literature and for helpful discussions on this point.}
In the end, since we are looking for a unified approach that can handle all 
sectors and any number of loops, we decided that numerical methods are, for the
present purposes, an effective way to extract the information we need about gauge 
theory anomalous dimensions.  Since Bethe ansatz equations exist for most of the
results that are of interest to us, the numerical results obtained here can 
be checked against the Bethe-ansatz methodology:  these exercises
were performed in Chapter \ref{virial} above (see ref.~\cite{Callan:2004dt}).

We begin with a discussion of the bosonic $\Sl(2)$ sector.
Recall from Chapter \ref{virial} that for total 
$R$-charge ${K}$ (the $R$-charge is equal to the number of lattice
sites ${K}$ in this sector), the basis for this system consists of 
single-trace operators of the form
\be
{\rm Tr}\left({\nabla}^IZ~Z^{{K}	-1} \right), 
~{\rm Tr}\left({\nabla}^{I-1}Z~{\nabla}Z~Z^{{K}	-2} \right),
	~{\rm Tr}\left({\nabla}^{I-1}Z~Z{\nabla}Z~Z^{{K}	-3} \right),~ \ldots~,
\ee
where $Z$ is the $SO(6)$ Yang-Mills boson carrying one unit of $R$-charge,
${\nabla}$ is a spacetime covariant derivative operator that scales under
the chosen $\Sl(2)$ subgroup of the Lorentz group ($\nabla \equiv \nabla_1 + i\nabla_2$), 
$I$ is the total impurity
number and the full basis contains all possible distributions of ${\nabla}$
operators among the $Z$ fields. Conservation of various $U(1)$ subgroups
of the $R$-symmetry group ensures that operators of this type mix only among
themselves to all orders in the gauge theory (as long as we work in the 't Hooft
large-$N_c$ limit). This gauge theory closed subsector corresponds to the symmetric 
traceless irrep of $SO(4)_{AdS}$ bosons in the string theory (states whose
energy shifts are given in eqns.~(\ref{exactSO41}) and (\ref{nnval1})).

To compare the results of Chapter \ref{virial} (see eqn.~(\ref{SL2pred_ch3}))
with the string theory predictions of eqns.~(\ref{exactSO41}) 
and (\ref{nnval1}), we reorganize those results as follows: we reinstate the 
BMN energy of the degenerate multiplet (\ref{BMNen}) (expanded to first order in 
$\lambda'$); we replace $\lambda'$ with $\lambda/J^2$ and replace $J$ by ${K}$. 
This gives specific string theory predictions for the large-${K}$ scaling of one-loop
anomalous dimensions of the AdS closed sector. As usual, there are
two distinct cases: for unequal mode indices $(q\ne r\ne s=-q-r)$, we have
\be
\label{adspred1}
E_{AdS}(q,r,{K}) = 3 + ({K}-2)(q^2 + r^2 + qr)\frac{\lambda}{{K}^3} 
		+ O({K}^{-4})\ .
\ee 
(Note that here we label mode indices with $q,r,s$ instead of the $k_1,k_2,k_3$
triplet used in Chapter \ref{virial}.)
For pairwise equal mode indices $(n,n,-2n)$ we have
\be
\label{adspred2}
E_{AdS}(n,{K}) = 3 + (3 {K}-7)n^2\frac{\lambda}{{K}^3} + O({K}^{-4})\ .
\ee
For convenience, we redisplay the numerical 
gauge theory predictions from the $\Sl(2)$ spin chain from eqn.~(\ref{SL2pred_ch3}):
\be
E_{\Sl(2)}^{(1,2)} = (k_1^2 +k_1 k_2+ k_2^2 )~, \qquad
E_{\Sl(2)}^{(1,3)}/E_{\Sl(2)}^{(1,2)} = -2~, \qquad k_1\neq k_2\neq k_3~,\nonumber\\
E_{\Sl(2)}^{(1,2)} = 3 n^2~, \qquad 
E_{\Sl(2)}^{(1,3)}/E_{\Sl(2)}^{(1,2)} = -7/3~, \qquad k_1=k_2=n,~ k_3=-2n\ .
\ee
The string predictions in eqns.~(\ref{adspred1}) and (\ref{adspred2}) 
match the expected virial scaling of the spin chain eigenvalues 
displayed in eqn.~(\ref{SL2pred_ch3}), with the specific identifications
\be
E_{AdS}^{(1,2)} = (q^2 + r^2 + qr)~, \qquad E_{AdS}^{(1,3)} = -2(q^2 + r^2 + qr)~, \qquad
E_{AdS}^{(1,3)}/E_{AdS}^{(1,2)} = -2
\label{expSL21}
\ee
for $q\neq r\neq s=-q-r$, or
\be
E_{AdS}^{(1,2)} = 3 n^2~, 
\qquad E_{AdS}^{(1,3)} = -7 n^2~,  \qquad
E_{AdS}^{(1,3)}/E_{AdS}^{(1,2)} = -7/3
\label{expSL22}
\ee
for $q=r=n$ and $s=-2n$.


At this point it is appropriate to say a few words about the role of integrability
in this problem. It was first argued in \cite{Bena:2003wd} 
that the complete GS action of IIB superstring theory
on $AdS_5\times S^5$ is integrable.  Integrability has since taken a central role in
studies of the AdS/CFT correspondence, as any precise non-perturbative understanding
of integrability on both sides of the duality would be extremely powerful.
Integrability on either side of the duality gives rise to an infinite tower of
hidden charges that can be loosely classified as either local (Abelian) or non-local
(non-Abelian).
In the Abelian sector, contact between the integrable structures of 
gauge theory and semiclassical string theory (a subject that was first investigated in
\cite{Arutyunov:2003rg}) has been made to two loops in $\lambda$ 
(see, e.g., \cite{Beisert:2003ea,Arutyunov:2004xy,Kazakov:2004qf,Beisert:2003xu}).
(The corresponding problem in the non-local sector was addressed to one-loop 
order in \cite{Dolan:2003uh,Dolan:2004ps}.)
One of the local gauge theory charges can be shown to anticommute
in the $\su(2)$ sector with a parity operator $P$ (to three loops in $\lambda$), 
whose action on a single-trace state in the gauge theory
is to invert the order of all fields within the trace \cite{Beisert:2003ys,Beisert:2003tq}.
Furthermore, this operator can be shown to connect states of opposite parity. 
These facts imply that all eigenstates in the spectrum
connected by $P$ must be degenerate.  
These degenerate states are known as parity
pairs and their existence can be interpreted as a necessary 
(but not sufficient) condition for integrability. The spectrum in 
table~\ref{NUM_SL(2)_1Loop3} from Chapter \ref{virial}
exhibits such a degeneracy and makes it clear
that parity pairs are simply distinct states whose lattice momenta 
(or worldsheet mode indices)
are related by an overall sign flip.  Since the net momentum of allowed 
states is zero, parity pair states can in principle scatter into each other,
and their degeneracy is a nontrivial constraint on the interactions. 
As a small caveat, we note that lattice momentum conservation implies that
mixing of parity-pair states can only occur via connected three-body (or higher)
interactions.  As the virial analysis shows, at the order to which
we are working, only two-body interactions are present and the parity pair
degeneracy is automatic.  The same remark applies to the string theory analysis
to $O(J^{-1})$ in the curvature expansion.  A calculation of the string theory spectrum
carried out to $O(J^{-2})$ is needed to see whether parity pair degeneracy 
survives string worldsheet interactions; further discussion of this point will be
given in Chapter \ref{integ2} (see also ref.~\cite{Swanson:2004mk}).


We now turn to the closed $\su(2)$ sector of gauge theory operators, 
corresponding to the symmetric-traceless bosonic $SO(4)_{S^5}$ 
sector of the string theory. The operator
basis for this sector consists of single-trace monomials built out of 
two complex scalar fields $Z$ and $\phi$, where $Z$ is the complex scalar 
carrying one unit of charge under the
$U(1)$ $R$-charge subgroup and $\phi$ is one of the two scalars with zero $R$-charge,  
transforming as an $SO(4)$ vector in the $SO(6)\simeq U(1)_R\times SO(4)$ decomposition 
of the full $R$-symmetry group of the gauge theory. The collection of operators
\be
\label{SO6basis}
\tr(\phi^I Z^{{K}-I}),\ \tr(\phi^{I-1}Z\phi Z^{{{K}-I}-1}),\ 
\tr(\phi^{I-2}Z\phi^2 Z^{{{K}-I}-1}),\ \ldots
\ee
(and all possible permutations, modulo cyclic equivalence, of the ${K}$ factors) forms 
a basis with $I$ impurities and $R$-charge equal to ${{K}-I}$. The anomalous dimension
operator simply permutes these monomials among themselves in ways that get more 
elaborate as we go to higher loop orders in the gauge theory. 
The relevant gauge theory predictions from Chapter \ref{virial} are given 
at one-loop order in eqn.~(\ref{su2num}) and at two-loop order in eqn.~(\ref{su2num2}).
For convenience, we reproduce those results here:
\be
E_{\su(2)}^{(1,2)} = (k_1^2+k_2^2+k_3^2)/2~, & \qquad & k_1+k_2+k_3=0~,\nonumber\nonumber\\
E_{\su(2)}^{(1,3)}/E_{\su(2)}^{(1,2)}  =  2~, & \qquad & (k_1\neq k_2\neq k_3)~,
\nonumber\\
E_{\su(2)}^{(1,3)}/E_{\su(2)}^{(1,2)} = \frac{7}{3}~, & \qquad &(k_1 = k_2,\ k_3=-2k_1)\ ,
\nn\\
E_{\su(2)}^{(2,4)} = -(k_1^2+k_2^2+k_3^2)^2/16~, & \qquad & k_1+k_2+k_3=0~,\nonumber\nonumber\\
E_{\su(2)}^{(2,5)}/E_{\su(2)}^{(2,3)}  =  8~, & \qquad & (k_1\neq k_2\neq k_3)~,
\nonumber\\
E_{\su(2)}^{(2,5)}/E_{\su(2)}^{(2,3)} = \frac{76}{9}~, & \qquad &(k_1 = k_2,\ k_3=-2k_1)\ .
\ee

To compare with string theory results for the bosonic symmetric-traceless
$SO(4)_{S^5}$ sector eigenvalues, we need to recast eqns.~(\ref{exactSO42}) 
and (\ref{nnval2}) as expansions in powers of $\lambda$ and ${K}^{-1}$. 
We denote by $E_{S^5}^{(n,m)}$ the coefficient of $\lambda^n {K}^{-m}$ 
in the large-${K}$ expansion of the string theory energies: they can be
directly compared with the corresponding quantities extracted from 
the numerical spin chain analysis. The string theory 
predictions for scaling coefficients, up to second order in $\lambda$, are
given in table~\ref{strscale}.
\begin{table}[ht!]
\begin{eqnarray}
\begin{array}{|c|cc|}
\hline
E_{S^5}^{(n,m)} & (q\ne r\ne s)  &  (q=r=n) \\
\hline  
E_{S^5}^{(1,2)} & (q^2 + qr + r^2) & 3n^2 \\
E_{S^5}^{(1,3)} & 2(q^2 + qr + r^2) & 7n^2 \\
E_{S^5}^{(2,4)} & -\frac{1}{4}(q^2 + qr + r^2)^2 & -\frac{9}{4} n^4 \\
E_{S^5}^{(2,5)} & -2(q^2 + qr + r^2)^2 & -19n^4 \\
\hline
\end{array} \nonumber
\end{eqnarray}
\caption{String predictions for $\su(2)$ scaling coefficients, to two loops}
\label{strscale}
\end{table}
As usual, the predictions for three-impurity states with unequal mode indices have
to be stated separately from those for states with two equal mode indices. 
We take these results as strong evidence that the string theory analysis agrees 
with the gauge theory up to $O(\lambda^2)$ in this sector.

We now turn to a discussion of gauge theory physics beyond two loops. As it happens,
the three-loop Hamiltonian can be fixed up to two unknown coefficients 
($\alpha_1$ and $\alpha_2$) by basic field theory considerations \cite{Beisert:2003jb}:
\be
H^{\so(6)}_6 & = & 
	\left(60 + 6\alpha_1 -56\alpha_2\right)\{\}
	+ \left(-104 + 14\alpha_1 +96\alpha_2\right)\{0\}
\nn\\
&&	+ \left(24 + 2\alpha_1 -24\alpha_2\right)\left( \{0,1\}+\{1,0\} \right)
	+ \left(4 + 6\alpha_1 \right) \{0,2\}
\nn\\
&&\kern-10pt	\left(-4 + 4\alpha_2 \right)\left( \{0,1,2\}+\{2,1,0\} \right)
	- \alpha_1\left( \{0,2,1\}+\{1,0,2\} \right)\ .
\label{H3SO6}
\ee 
Originally, these coefficients were determined by demanding proper BMN scaling 
in the theory and that the dynamics be integrable at three loops;
these assumptions set $\alpha_{1,2} = 0$.  By studying an $\su(2|3)$ 
spin chain model, Beisert \cite{Beisert:2003ys} was subsequently able to show that 
independent symmetry arguments, along with BMN scaling,
uniquely set $\alpha_1 = \alpha_2 = 0$  
(thus proving integrability at three loops). 

As described in Chapter \ref{virial}, the 
three-loop Hamiltonian $H^{\su(2)}_6$ can be treated as a second-order correction 
to $H^{\su(2)}_2$.  This allows us to numerically evaluate the $O(\lambda^3)$ 
contribution to the spectrum by using second-order Rayleigh-Schr\"odinger
perturbation theory (there is an intermediate state sum involved, but since we
are doing the calculation numerically, this is not a serious problem). There is
also the issue of degeneracy but the existence of a higher conserved charge once
again renders the problem effectively non-degenerate. 
The resulting three-loop data for large-${K}$ was fit in Chapter \ref{virial}
to a power series in ${K}^{-1}$ to read off the expansion coefficients
$E_{\su(2)}^{3,n}$.  It turns out that, to numerical precision, the coefficients are
non-vanishing only for $n > 5$ (as required by BMN scaling).  The results 
of this program are reproduced for convenience from Chapter \ref{virial} in 
table~\ref{NUM_SO(6)_3Loop}, where they are compared with string theory predictions derived 
(in the manner described in previous paragraphs) from eqn.~(\ref{exactSO42}).  (The
accuracy of the match is displayed in the last column of table~\ref{NUM_SO(6)_3Loop}.) 
The important point is that there is substantial disagreement 
with string results at $O(\lambda^3)$
for all energy levels: the low-lying states exhibit a mismatch ranging from roughly 
$19\%$ to $34\%$, and there is no evidence that this can be repaired by taking data 
on a larger range of lattice sizes. There is apparently a general breakdown 
of the correspondence between string theory and gauge theory anomalous dimensions at 
three loops, despite the precise and impressive agreement at first and second order.  
This disagreement was first demonstrated in the two-impurity regime \cite{Callan:2003xr}. 
It is perhaps not surprising that
the three-loop disagreement is reproduced in the three-impurity regime, 
but it provides us with more information that may help to 
clarify this puzzling phenomenon.
\begin{table}[ht!]
\begin{eqnarray}
\begin{array}{|ccc|}
\hline
 E_{\su(2)}^{(3,7)}/E_{\su(2)}^{(3,6)}  & {\rm String\ Modes\ }(q,r,s) &{\rm \% Error} \\
\hline
16.003		& (1,0,-1) & 33\%  \\   
14.07		& (1,1,-2) &   19\% \\
14.07		&  (-1,-1,2)&   19\%  \\
16.03		&  (2,0,-2)&   34\%  \\
14.37		& (1,2,-3) &   22\% \\
14.37		&  (-1,-2,3)&   22\%  \\
15.96		&  (3,0,-3)&   30\% \\
\hline
\end{array} \nonumber
\end{eqnarray}
\caption{Three-loop numerical spectrum of three-impurity 
	$\su(2)$ subsector and mismatch with string predictions }
\label{NUM_SO(6)_3Loop}
\end{table}

The same exercise can be repeated for the closed $\su(1|1)$ fermionic sector,
whose string theory dual is comprised of pure fermionic states symmetrized in
$SO(4)\times SO(4)$ indices in either the ${\bf (1,2;1,2)}$ or ${\bf (2,1;2,1)}$
irreps (projected onto $\Pi_\pm$ subspaces).
The spin chain system is embedded in Beisert's $\su(2|3)$ model, where the fermionic 
sector of the Hamiltonian has been recorded up to two-loop order \cite{Beisert:2003ys}.  
Since the relevant points of the numerical gauge/string comparison have already
been made, we will simply state the one- and two-loop results.  
(The large-${K}$ spectrum of the three-loop contribution is scrutinized 
in Chapter \ref{virial}; one again finds disagreement with string theory.)

In this sector, the $R$-charge and the lattice length are related by $J = {K} - I/2$.   
The fermionic one- and two-loop string predictions
are therefore found from eqn.~(\ref{exactfermi5}) to be
\be
&&E_{\rm Fermi}^{(1,2)} = (q^2 + qr + r^2)~, \qquad \qquad \kern-2pt
E_{\rm Fermi}^{(1,3)} = 0 \nn\\
&&E_{\rm Fermi}^{(2,4)} = -\frac{1}{4}(q^2 + qr + r^2)^2~,  \qquad
E_{\rm Fermi}^{(2,5)} = -(q^2 + qr + r^2)^2\ .
\ee
As noted above, this sector does not admit states with equivalent mode indices.

Reproducing the results from the gauge theory analysis in Chapter \ref{virial},
we find precise agreement:
\be
E_{\su(1|1)}^{(1,2)} = (k_1^2+k_1k_2+k_2^2)~, &\quad& E_{\su(1|1)}^{(1,3)} = 0~, \nn\\
E_{\su(1|1)}^{(2,4)} -\frac{1}{4}(k_1^2+k_1k_2+k_2^2)^2~,  &\quad&  
E_{\su(1|1)}^{(2,5)} = -(k_1^2+k_1k_2+k_2^2)^2 \ .
\ee
The two-loop data are obtained using the 
same first-order perturbation theory treatment described above in the 
$\su(2)$ sector (the results are recorded in table~\ref{NUM_fermi_2Loop3} above).  
The two-loop spectrum is subject to stronger ${K}^{-1}$ corrections, but the
data are still convincing and could be improved by running the extrapolation
out to larger lattice sizes.  Nonetheless, the close agreement for the 
low-lying levels corroborates 
the match between gauge and string theory up to two-loop order.

\section{Discussion}
The BMN/pp-wave mechanism has emerged as a useful proving ground
for the postulates of the AdS/CFT correspondence.  
When the full Penrose limit is lifted, a rich landscape emerges, 
even in the two-impurity regime, upon which 
the string and gauge theory sides of the duality have exhibited an intricate
and impressive match to two loops in the gauge coupling and first nontrivial
order in the curvature expansion.  
While the conditions under which agreement is obtained
are substantially more demanding in the higher-impurity problem,
we have shown that this agreement is maintained for three-impurity
string states and SYM operators.
Although the two-loop agreement survives at the three-impurity level,
we have also confirmed the previously observed mismatch at three loops
in the gauge theory coupling.  
In the end, the analyses carried out here will provide an extremely 
stringent test of any proposed solution to this vexing problem.

%
%
\chapter{$N$ impurities}                        
\label{Nimp}
In Chapters \ref{twoimp} and \ref{threeimp} we analyzed
the first curvature correction to the spectrum of string states
in the pp-wave limit of $AdS_5\times S^5$.
The string energies in this setting correspond in the gauge theory to the
difference between operator scaling dimensions and $R$-charge
($\Delta \equiv D-R$), and states are arranged into superconformal
multiplets according to the $\alg{psu}(2,2|4)$ symmetry of the
theory. The fully supersymmetric two-excitation (or two-impurity)
system, for example, is characterized by a 256-dimensional
supermultiplet of states built on a scalar primary. The complete
spectrum of this system was successfully matched to corresponding
SYM operator dimensions in Chapter \ref{twoimp}
to two loops in the modified 't~Hooft coupling $\lambda' =
\lambda/J^2$ (see also \cite{Callan:2003xr,Callan:2004uv}). 
A three-loop mismatch between the gauge and string
theory results discovered therein comprises a long-standing and
open problem in these studies, one which has appeared in several
different contexts (see, e.g.,~\cite{Serban:2004jf,Beisert:2004hm,Beisert:2003ea}). 
This was extended to the three-impurity, 4,096-dimensional supermultiplet
of string states in Chapter \ref{threeimp} (see also \cite{Callan:2004ev}), 
where precise agreement with the corresponding gauge theory was again found to two-loop
order, and a general disagreement reappeared at three loops. In
the latter study, three-impurity string predictions were compared
with corresponding gauge theory results derived both from the
virial technique described in Chapter \ref{virial} and the long-range
Bethe ansatz of \cite{Beisert:2003yb} (which overlaps at one loop
with the original $\so(6)$ system studied in \cite{Minahan:2002ve}).

In the present chapter we generalize the string side of these investigations
by computing, directly from the Hamiltonian, various $N$-impurity spectra
of IIB superstring theory at $O(J^{-1})$ in the large-$J$ curvature expansion near
the pp-wave limit of $AdS_5\times S^5$.
We focus on the bosonic $\su(2)$ and $\Sl(2)$
sectors, which are characterized by $N$ symmetric-traceless
bosonic string excitations in the $S^5$ and $AdS_5$ subspaces, respectively.
Based on calculations in these sectors,
we also formulate a conjecture for the
$N$-impurity spectrum of states in a protected $\su(1|1)$ sector composed of $N$ fermionic
excitations symmetrized in their $SO(4)\times SO(4)$ spinor indices.
We then describe the complete supermultiplet decomposition of the $N$-impurity
spectrum to two loops in $\lambda'$ using a simple generalization of the two- and
three-impurity cases.

We note here that a new Bethe ansatz for the string theory has been proposed
by Arutyunov, Frolov and Staudacher \cite{Arutyunov:2004vx} that is meant
to diagonalize the fully quantized string sigma model in the $\su(2)$ sector
to all orders in $1/J$ and $\lambda'$ (see the discussion in Chapter \ref{SYM}).
This ansatz was shown in \cite{Arutyunov:2004vx} to
reproduce the two- and three-impurity
spectra of quantized string states near the pp-wave limit detailed in
\cite{Callan:2004ev,Callan:2004uv}.  The methods developed here allow us to
check their formulas directly against the string theory for any impurity number
at $O(J^{-1})$, and we find that our general $\su(2)$ string eigenvalues
agree to all orders in $\lambda'$ with their $\su(2)$ string Bethe ansatz!
We compute the $N$-impurity energy spectra of the $\su(2)$, $\Sl(2)$
and $\su(1|1)$ closed sectors of this system in Section~\ref{ch6sec2}, 
and generalize the complete
$N$-impurity supermultiplet structure of the theory to two-loop order in
$\lambda'$ in Section~\ref{ch6sec3}.

\section{$N$-impurity string energy spectra}
\label{ch6sec2}
As described above, our string vacuum state carries the $S^5$ string angular momentum $J$ and
is labeled by $\ket{J}$; the complete Fock space of string states
is generated by acting on $\ket{J}$ with any number of the
creation operators $a_n^{A\dag}$ (bosonic) and $b_n^{\alpha\dag}$
(fermionic), where the lower indices $n,m,l,\ldots$ denote mode
numbers. The excitation number of string states (defined by the
number of creation oscillators acting on the ground state) will
also be referred to as the impurity number, and string states with
a total of $N_B+N_F = N$ impurities will contain $N_B$ bosonic and
$N_F$ fermionic impurities: \be \Ket{N_B,N_F;J} \equiv
\underset{N_B}{\underbrace{a_{n_1}^{A_1\dag}a_{n_2}^{A_2\dag}
    \ldots a_{n_{N_B}}^{A_{N_B}\dag}}}
    \underset{N_F}{\underbrace{b_{n_1}^{\alpha_{1}\dag}b_{n_2}^{\alpha_2\dag}
    \ldots b_{n_{N_F}}^{\alpha_{N_F}\dag} }}\ket{J}\ .
\ee
States constructed in this manner fall into two disjoint subsectors populated by
spacetime bosons ($N_F$ even) and spacetime fermions ($N_F$ odd).  In this notation
the pure-boson states $\ket{N_B,0;J}$ are mixed only by $H_{\rm BB}$ and
the pure-fermion states $\ket{0,N_F;J}$ are acted on by $H_{\rm FF}$.  The more
general spacetime-boson states $\ket{N_B,{\rm even};J}$ are acted on by the
complete interaction Hamiltonian $H_{\rm int}$, as are the spacetime-fermion states
$\ket{N_B,{\rm odd};J}$.  There is of course no mixing between spacetime bosons and
fermions; this block-diagonalization is given schematically in table~\ref{block}.
\begin{table}[ht!]
\begin{eqnarray}
\begin{array}{|c|cccc|}
\hline
 H_{\rm int} & \Ket{N_B,0;J} &
        \Ket{N_B,{\rm even};J} &
        \Ket{N_B,{\rm odd};J} &
        \Ket{0,{\rm odd};J}
\\   \hline
\Bra{N_B,0;J}   & H_{\rm BB}    & H_{\rm BF}                &   &   \\
\Bra{N_B,{\rm even};J} & H_{\rm BF} & H_{\rm BB}+H_{\rm BF}+H_{\rm FF}  &   &   \\
\Bra{N_B,{\rm odd};J} &  &   & H_{\rm BB}+H_{\rm BF}+H_{\rm FF} & H_{\rm BF} \\
\Bra{0,{\rm odd};J} &  &  & H_{\rm BF} & H_{\rm FF}  \\
\hline
\end{array} \nonumber
\end{eqnarray}
\caption{Interaction Hamiltonian on $N$-impurity string states $(N_B+N_F=N)$}
\label{block}
\end{table}

The full interaction Hamiltonian can be further block-diagonalized
by projecting onto certain protected sectors of string states, and
we will focus in this study on three such sectors.  Two of these
sectors are spanned by purely bosonic states $\ket{N_B,0;J}$
projected onto symmetric-traceless irreps in either the
$SO(4)_{AdS}$ or $SO(4)_{S^5}$ subspaces.  Another sector that is
known to decouple at all orders in $\lambda'$ is comprised of
purely fermionic states $\ket{0,N_F;J}$ projected onto either of
two subspaces of $SO(4)\times SO(4)$ labeled, in an
$SU(2)^2\times SU(2)^2$ notation, by $({\bf 2,1;2,1})$ and $({\bf
1,2;1,2})$, and symmetrized in spinor indices. Each of these
sectors can also be labeled by the subalgebra of the full
superconformal algebra that corresponds to the symmetry under
which they are invariant.  The bosonic $SO(4)_{AdS}$ and
$SO(4)_{S^5}$ sectors are labeled by $\Sl(2)$ and $\su(2)$
subalgebras, respectively, while the two fermionic sectors fall
into $\su(1|1)$ subsectors of the closed $\su(2|3)$ system studied
in \cite{Beisert:2003yb,Beisert:2003ys,Beisert:2003jj}.

In the large-$J$ expansion about the free pp-wave theory,
we will isolate $O(J^{-1})$ corrections to the energy eigenvalues of
$N$-impurity string states according to
\be
E(\{q_j\},N,J) = \sum_{j=1}^N\sqrt{1+q_j^2\lambda'} + \delta E(\{q_j\},N,J) + O(J^{-2})\ .
\ee
The spectrum is generically dependent upon $\lambda'$, $J$ and
the mode numbers $\{n_j\},\{q_j\},\ldots,$ where $j$ is understood
to label either the complete set of impurities ($j=1,\ldots,N$) or
some subset thereof (e.g.,~$j=1,\ldots,N_F$).
The leading order term in this expansion is the $N$-impurity free energy of states
on the pp-wave geometry, and $\delta E(\{q_j\},N,J)$ always
enters at $O(J^{-1})$.
When it becomes necessary, we will also expand the $O(1/J)$ energy shift
in the small-$\lambda'$ loop expansion:
\be
\delta E(\{q_j\},N,J) = \sum_{i=1}^\infty \delta E^{(i)}(\{q_j\},N,J) ({\lambda'})^{i}\ .
\ee
Finding the explicit form of $\delta E(\{q_j\},N,J)$
for $N$-impurity string states in certain interesting sectors of the theory will be
our primary goal.  As a side result, however, we will see that the spectrum
of \emph{all} states in the theory will be determined to two-loop order
in $\lambda'$ by the specific eigenvalues we intend to compute.

We begin by noting that the canonical commutation relations
of the bosonic fields $x^A$ and $p_A$ allow us to expand
$H_{\rm BB}$ in bosonic creation and annihilation operators using
\be
x^A(\sigma,\tau)  &=&  \sum_{n=-\infty}^\infty x_n^A(\tau ) e^{-i k_n\sigma}~,  \nn\\
x_n^A(\tau) &=& \frac{i}{\sqrt{2\omega_n}}\left(a_n^A e^{-i\omega_n\tau}
        -a_{-n}^{A\dag}e^{i\omega_n\tau}\right)\ ,
\ee
where $k_n = n$ are integer-valued, $\omega_n=\sqrt{p_-^2 + n^2}$ and
the operators $a_n^A$ and $a_n^{A\dag}$ obey the usual relation
$\left[a_m^A,a_n^{B\dag}\right] = \delta_{mn}\delta^{AB}$.
Since we are only interested in computing diagonal matrix elements of $H_{\rm BB}$
between physical string states with equal numbers of excitations, we can
restrict the oscillator expansion to contain only equal numbers of
creation and annihilation operators (all other combinations automatically
annihilate between equal-impurity string states).
Explicitly, we obtain the following expansion from the results in Chapter \ref{twoimp}:
\begin{eqnarray}
\label{Hcorrected6}
{H}_{\rm BB} & = &
    -\frac{1}{32 p_- R^2}\sum \frac{\delta(n+m+l+p)}{\xi}
\nn\\
& & 
	\times \biggl\{
    2 \biggl[ \xi^2
    - (p_-^4 - k_l k_p k_n k_m )
     +  \omega_n \omega_m k_l k_p
      +  \omega_l \omega_p k_n k_m
    + 2 \omega_n \omega_l k_m k_p
\nn\\
& &\kern-10pt
     + 2 \omega_m \omega_p k_n k_l
    \biggr]
    a_{-n}^{\dagger A}a_{-m}^{\dagger A}a_l^B a_p^B
   +4 \biggl[ \xi^2
    - (p_-^4 - k_l k_p k_n k_m )
     - 2 \omega_n \omega_m k_l k_p
     +  \omega_l \omega_m k_n k_p
\nn\\
& &   -  \omega_n \omega_l k_m k_p
    -  \omega_m \omega_p k_n k_l
    + \omega_n \omega_p k_m k_l \biggr]
    a_{-n}^{\dagger A}a_{-l}^{\dagger B}a_m^A a_p^B
     + 2  \biggl[8 k_l k_p
    a_{-n}^{\dagger i}a_{-l}^{\dagger j}a_m^i a_p^j
\nn\\
& &     + 2 (k_l k_p +k_n k_m)
    a_{-n}^{\dagger i}a_{-m}^{\dagger i}a_l^j a_p^j
    +(\omega_l \omega_p+ k_l k_p -\omega_n
    \omega_m- k_n k_m)a_{-n}^{\dagger i}a_{-m}^{\dagger i}a_l^{j'} a_p^{j'}
\nn\\
& &     -4 ( \omega_l \omega_p- k_l k_p)
    a_{-n}^{\dagger i}a_{-l}^{\dagger j'}a_m^i a_p^{j'}
    -(i,j \rightleftharpoons i',j')
    \biggr]\biggr\}~,
\end{eqnarray}
where $\xi \equiv \sqrt{\omega_n\omega_m\omega_l\omega_p}$.

\subsection{The $SO(4)_{S^5}$ ($\su(2)$) sector}
We begin in the $\su(2)$ sector spanned by symmetric-traceless
pure-boson states excited in the $S^5$ subspace.
Because we are restricting our attention to $SO(4)_{S^5}$ states
symmetric in their vector indices, we form the following oscillators:
\be
a_n = \frac{1}{\sqrt{2}}\left( a^5_n + i a^6_n \right)~, \qquad
\bar a_n = \frac{1}{\sqrt{2}}\left( a^5_n - i a^6_n \right)\ .
\label{oscdef}
\ee
By taking matrix elements of the form
\be
\bra{J} a_{n_1}a_{n_2}\ldots a_{n_{N_B}} ( H_{\rm BB} )
    a_{n_1}^\dag a_{n_2}^\dag \ldots a_{n_{N_B}}^\dag \ket{J}\ ,
\label{ME}
\ee
we can therefore select out excitations in the $(5,6)$-plane of the $S^5$ subspace and
make the symmetric-traceless projection manifest.  (More generally we
can project onto any $(n,m)$-plane, as long as $n\neq m$
and both are chosen to lie in the $S^5$ subspace.)

There are two basic oscillator structures of $H_{\rm BB}$ in eqn.~(\ref{Hcorrected6}):
one in which the creation (annihilation) operators are contracted
in their $SO(4)\times SO(4)$ indices
\be
a_{-n}^{\dag A} a_{-m}^{\dag A} a_{l}^{B}a_{p}^{B}\ , \nn
\ee
and one where pairs of creation and annihilation operators are contracted
\be
a_{-n}^{\dag A} a_{-l}^{\dag B} a_{m}^{A}a_{p}^{B}\ . \nn
\ee
In terms of the $a_n$ and $\bar a_n$ fields of eqn.~(\ref{oscdef}),
the former structure contains
\be
a_{-n}^{\dag A} a_{-m}^{\dag A} a_{l}^{B}a_{p}^{B}\Bigr|_{(5,6)} =
    \bigl( a_{-n}^\dag\,\bar a_{-m}^\dag + \bar a_{-n}^\dag\, a_{-m}^\dag \bigr)
    \bigl( a_l\,\bar a_p + \bar a_l\, a_p\bigr)\ ,
\ee
which cannot contribute to $\su(2)$ matrix elements of the form appearing in
(\ref{ME}).  The latter structure, however, contains
\be
a_{-n}^{\dag A} a_{-l}^{\dag B} a_{m}^{A}a_{p}^{B}\Bigr|_{(5,6)} =
    \bar a_{-n}^\dag\, \bar a_{-l}^\dag\, \bar a_m\, \bar a_p
    + a_{-n}^\dag\, a_{-l}^\dag\, a_m\, a_p\ ,
\label{osc2}
\ee
which will contribute to the $\su(2)$ energy spectrum.

The string states appearing in the matrix element of
eqn.~(\ref{ME}) have been written in the generic form \be
a_{n_1}^\dag a_{n_2}^\dag \ldots a_{n_{N_B}}^\dag \ket{J}\ , \nn
\ee and, as usual, they are subject to the level-matching
condition \be \sum_{j=1}^{N_B} n_j = 0\ . \label{LM} \ee The
complete set of mode indices $\{n_1,n_2,\ldots,n_{N_B}\}$ can
contain one or more subsets of indices that are equal, while still
satisfying eqn.~(\ref{LM}); this scenario complicates the
calculation of energy eigenvalues to some extent. We will
eventually compute the eigenvalues of interest for completely
general string states, but for purposes of illustration and to
introduce our strategy we will start with the simplest case in
which no two mode numbers are equal $(n_1\neq n_2 \neq \ldots\neq
n_{N_B})$. To organize the presentation of this chapter, we will use
mode numbers labeled by $\{n_j\}$ to denote those that are
inequivalent from each other, while $\{q_j\}$ will be allowed to
overlap.  Between states with completely distinct mode indices,
the oscillator structure in eqn.~(\ref{osc2}) exhibits the
following matrix element: 
\be &&\bra{J}a_{n_1}a_{n_2}\ldots
a_{N_B} ( a_{-n}^\dag a_{-l}^\dag a_m a_p )
    a_{n_1}^\dag a_{n_2}^\dag \ldots a_{N_B}^\dag \ket{J}
\nn\\
&&\kern+60pt
    = \frac{1}{2}
    \sum_{j,k=1 \atop j\neq k}^{N_B}
    \Bigl(
    \delta_{n_j+n}\,\delta_{n_k+l}\,\delta_{n_j-m}\,\delta_{n_k-p}
    +\delta_{n_j+n}\,\delta_{n_k+l}\,\delta_{n_k-m}\,\delta_{n_j-p}
\nn\\
&&\kern+70pt
    +\delta_{n_j+l}\,\delta_{n_k+n}\,\delta_{n_j-m}\,\delta_{n_k-p}
    +\delta_{n_j+l}\,\delta_{n_k+n}\,\delta_{n_k-m}\,\delta_{n_j-p}
    \Bigr)\ .
\label{delME1}
\ee
With this in hand, it is a straightforward exercise to compute the
energy eigenvalue of the $SO(4)_{S^5}$ bosonic interaction
Hamiltonian in the $N_B$-impurity symmetric-traceless irrep
(with unequal mode indices):  we simply attach the $H_{\rm BB}$ coefficient
of the oscillator structure $ a_{-n}^\dag a_{-l}^\dag a_m a_p $
to the right-hand side of eqn.~(\ref{delME1}) and carry out the
summation over mode numbers.  The result is remarkably compact:
\be
\delta E_{{S^5}}(\{n_i\},N_B,J) =&&\kern-15pt -\frac{1}{J}\sum_{j,k=1\atop j\neq k}^{N_B}
    \frac{1}{2\,\omega_{n_j}\omega_{n_k}}\left[
    n_k^2 + n_j^2\left(1+n_k^2\lambda'\right)
    + n_j n_k\left(1-\omega_{n_j}\omega_{n_k}\lambda'\right)\right]\ .
\nn\\
&&
\label{su2GEN}
\ee

This $\su(2)$ formula can be checked against
previously obtained string theory results in the two- and three-impurity regimes.
Namely, the two-impurity eigenvalue computed above (and in \cite{Callan:2003xr,Callan:2004uv})
takes the form (which is exact in $\lambda'$)
\be
\delta E_{{S^5}}(n_1,n_2,J) = -\frac{2\,n_1^2\lambda'}{J}\ ,
\label{2impsu2}
\ee
where we have set $n_2= -n_1$ using eqn.~(\ref{LM}).
This eigenvalue matches the general formula in eqn.~(\ref{su2GEN})
restricted to two impurities.  The $\su(2)$ eigenvalue for three impurities
with unequal mode indices $(n_1\neq n_2\neq n_3)$ was calculated in 
Chapter \ref{threeimp}
(and in ref.~\cite{Callan:2004ev}) and found to be
\be
\delta E_{{S^5}}(n_1,n_2,n_3,J) &=&
    -\frac{1}{J\omega_{n_1}\omega_{n_2}\omega_{n_3}}\biggl\{
    \left[n_1n_2+n_2^2+n_1^2(1+n_2^2\lambda')\right]\omega_{n_3}
\nn\\
&&\kern-20pt
    + \left[n_1n_3+n_3^2+n_1^2(1+n_3^2\lambda')\right]\omega_{n_2}
    + \left[n_2n_3+n_3^2+n_2^2(1+n_3^2\lambda')\right]\omega_{n_1}
\nn\\
&&\kern-20pt
    -\left[n_2n_3+n_1(n_2+n_3)\right]\lambda'\omega_{n_1}\omega_{n_2}\omega_{n_3}
    \biggr\}\ .
\label{3impsu2}
\ee
It is also easy to check that eqn.~(\ref{su2GEN})
reproduces this formula exactly for $N_B = 3$.

Since eqn.~(\ref{su2GEN}) matches all previously computed
results from the string theory in this sector,
it must therefore agree with corresponding
$\su(2)$ gauge theory predictions only to two-loop order in $\lambda$.
We note, however, that eqn.~(\ref{su2GEN}) is \emph{identical}
to the $N$-impurity $O(J^{-1})$ energy shift (with unequal mode numbers)
obtained from the $\su(2)$ string Bethe ansatz of \cite{Arutyunov:2004vx}.

To treat the slightly more complicated scenario of overlapping
mode indices (which can occur for three or more impurities),
we introduce the normalized eigenvectors
\be
\frac{1}{\sqrt{N_q!}}\left(a_q^\dag\right)^{N_q} a_{n_1}^\dag a_{n_2}^\dag
        \ldots a_{n_{(N_B-N_q)}}^\dag \ket{J}\ ,
\ee
which contain a single subset of $N_q$ bosonic oscillators $a_q^\dag$
that all share the same mode index $q$.  The remaining indices
$n_i \in \{n_1,n_2,\ldots,n_{N_B-N_q}\}$ are all separate from $q$
and unequal from each other,
such that the level-matching condition in eqn.~(\ref{LM}) now reads
\be
N_q\,q + \sum_{j=1}^{N_B-N_q}n_j = 0\ .
\ee
For this case we compute a matrix element analogous to that in eqn.~(\ref{delME1}):
\be
&&\kern-20pt
    \frac{1}{N_q!}\bra{J}
    \left(a_q \right)^{N_q} a_{n_1} a_{n_2}
        \ldots a_{n_{(N_B-{N_q})}}
    ( a_{-n}^\dag a_{-l}^\dag a_m a_p )
    \left(a_q^\dag\right)^{N_q} a_{n_1}^\dag a_{n_2}^\dag
        \ldots a_{n_{(N_B-{N_q})}}^\dag \ket{J}
\nn\\
&&\kern+5pt
    ={N_q}({N_q}-1)\delta_{p-q}\,\delta_{m-q}\,\delta_{n+q}\,\delta_{l+q}
\nn\\
&&\kern+00pt
    + L \sum_{j=1}^{N_B-{N_q}}
    \Bigl(
    \delta_{p-q}\,\delta_{n+q}\,\delta_{m-n_j}\,\delta_{l+n_j}
    +\delta_{m-q}\,\delta_{n+q}\,\delta_{p-n_j}\,\delta_{l+n_j}
\nn\\
&&\kern+10pt
    +\delta_{p-q}\,\delta_{l+q}\,\delta_{m-n_j}\,\delta_{n+n_j}
    +\delta_{m-q}\,\delta_{l+q}\,\delta_{p-n_j}\,\delta_{n+n_j}
    \Bigr)
\nn\\
&&\kern+00pt
    +\frac{1}{2}\sum_{j,k=1\atop j\neq k}^{N_B-{N_q}}
    \Bigl(
    \delta_{n_j+n}\,\delta_{n_k+l}\,\delta_{n_j-m}\,\delta_{n_k-p}
	+\delta_{n_j+n}\,\delta_{n_k+l}\,\delta_{n_k-m}\,\delta_{n_j-p}
\nn\\
&&\kern+10pt
    +\delta_{n_j+l}\,\delta_{n_k+n}\,\delta_{n_j-m}\,\delta_{n_k-p}
    +\delta_{n_j+l}\,\delta_{n_k+n}\,\delta_{n_k-m}\,\delta_{n_j-p}
    \Bigr)\ .
\label{delME2}
\ee
Using this result, we arrive at the $\su(2)$
energy shift for string states with $N_B$ total excitations containing
an ${N_q}$-component subset of oscillators that share the same mode index $q$:
\be
\delta E_{S^5}(\{n_i\},q,{N_q},N_B,J) & = &
    -\frac{{N_q}({N_q}-1)q^2 }{2 J \omega_q^2}
\nn\\
&&\kern-100pt
    -\sum_{j=1}^{N_B-{N_q}}
    \frac{{N_q}}{J\omega_q\omega_{n_j}}
    \left[q^2 + n_j^2(1+q^2\lambda')+q\,n_j
        \left(1-\omega_q\omega_{n_j}\lambda'\right)\right]
\nn\\
&&\kern-120pt
    -\sum_{j,k=1\atop j\neq k}^{N_B-{N_q}}
    \frac{1}{2J\,\omega_j\omega_k}\left[
    n_k^2 + n_j^2\left(1+n_k^2\lambda'\right)
    + n_j n_k\left(1-\omega_j\omega_k\lambda'\right)\right]\ .
\label{su2GENnn}
\ee

This formula can be compared with the three-impurity $\su(2)$ energy
shift with two equal mode indices ($N_q=2$) obtained in Chapter \ref{threeimp}.
For this particular case we can set the isolated mode number to $-2q$
using the level-matching condition to simplify the result:
\be
\delta E_{S^5}(q,J)  =
    -\frac{q^2}{J\omega_q^2 \omega_{2\,q} }
    \left[
    \omega_{2\,q}\left( 5+4\,q^2\lambda'\right)
    +\omega_q\left( 6+8\,q^2\lambda'\right)
    \right]\ .
\label{3impsu2nn}
\ee
It is easy to show that eqn.~(\ref{su2GENnn}) exactly reproduces
this energy shift when restricted to $N_B=3$ with a subset of two mode numbers
equal to $q$.

We now generalize the analysis completely by using eigenstates with
$M$ mode-index subsets, where all mode indices are equal within these subsets:
\be
\frac{\left( a_{q_1}^\dag\right)^{N_{q_1}}}{\sqrt{N_{q_1}!}}
\frac{\left( a_{q_2}^\dag\right)^{N_{q_2}}}{\sqrt{N_{q_2}!}}
\cdots
\frac{\left( a_{q_M}^\dag\right)^{N_{q_M}}}{\sqrt{N_{q_M}!}}\ket{J}\ .
\nn
\ee
The $j^{\rm th}$ subset contains $N_{q_j}$ oscillators with equal mode index $q_j$,
and the total impurity number is again $N_B$, such that
\be
\sum_{i=1}^M N_{q_i} = N_B~, \qquad
\sum_{i=1}^M N_{q_i} q_i = 0\ .
\ee
The matrix element of $a_{-n}^\dag\, a_{-l}^\dag\, a_m\, a_p$
between the above states,
analogous to eqns. (\ref{delME1}, \ref{delME2}), is
\be
&&\kern-30pt
\bra{J}
\frac{\left( a_{q_1} \right)^{N_{q_1}}}{\sqrt{N_{q_1}!}}
\cdots
\frac{\left( a_{q_M} \right)^{N_{q_M}}}{\sqrt{N_{q_M}!}}
\left( a_{-n}^\dag\, a_{-l}^\dag\, a_m\, a_p \right)
 \frac{\left( a_{q_1}^\dag\right)^{N_{q_1}}}{\sqrt{N_{q_1}!}}
\cdots
\frac{\left( a_{q_M}^\dag\right)^{N_{q_M}}}{\sqrt{N_{q_M}!}}
\ket{J}
\nn\\
&&\kern+00pt
    = \sum_{j=1}^M N_{q_j}(N_{q_j}-1)\,\delta_{n+n_j}\,\delta_{l+n_j}\,
    \delta_{m-n_j}\,\delta_{p-n_j}
    +\frac{1}{2}\sum_{j,k=1\atop j\neq k}^M N_{q_j} N_{q_k}
    \Bigl(
    \delta_{n+n_k}\,\delta_{l+n_j}\,\delta_{m-n_k}\,\delta_{p-n_j}
\nn\\
&&\kern-10pt
    +\delta_{n+n_j}\,\delta_{l+n_k}\,\delta_{m-n_k}\,\delta_{p-n_j}
    +\delta_{n+n_k}\,\delta_{l+n_j}\,\delta_{m-n_j}\,\delta_{p-n_k}
    +\delta_{n+n_j}\,\delta_{l+n_k}\,\delta_{m-n_j}\,\delta_{p-n_k}
    \Bigr)\ .
\ee
We thereby obtain the completely general $\su(2)$ energy shift
for $N_B$-impurity string states containing $M$ equal-mode-index subsets of oscillators:
\be
\delta E_{S^5}(\{q_i\},\{N_{q_i}\},M,J) & = & -\frac{1}{2J}\biggl\{
    \sum_{j=1}^M N_{q_j}(N_{q_j}-1)
    \left(1-\frac{1}{\omega_{q_j}^2\lambda'}\right)
\nn\\
&&\kern-50pt
    +\sum_{j,k=1\atop j\neq k}^M \frac{N_{q_j}N_{q_k}}{\omega_{q_j}\omega_{q_k}}
    \left[q_k^2 + q_j^2\omega_{q_k}^2\lambda'
    +q_j q_k(1-\omega_{q_j}\omega_{q_k}\lambda')\right]\biggr\}\ .
\label{SU2FULL}
\ee
This master formula can be used to determine the $\su(2)$ string energy spectrum to
$O(J^{-1})$ for all possible physical string states in this sector.

By taking $M=2$ and setting $N_{n_1} = N_{n_2} = 1$ (using the unequal mode indices $\{n_1,n_2\}$),
we recover from this equation the exact two-impurity result recorded in eqn.~(\ref{2impsu2}) above,
with $n_2=-n_1$.
For $M=3$ and $N_{n_1} = N_{n_2} = N_{n_3} = 1$, we get
the complete three-impurity unequal-mode-number $(n_1\neq n_2\neq n_3)$
formula found in eqn.~(\ref{3impsu2}).
Finally, the three-impurity eigenvalue with two equal mode indices $(q_1=q_2,\ q_3=-2q_1)$
given in eqn.~(\ref{3impsu2nn}) can also be extracted from eqn.~(\ref{SU2FULL})
by setting $M=2$, $N_{q_1} = 2$ and $N_{q_2}=1$.

We also note that eqn.~(\ref{SU2FULL}) agrees perfectly with the corresponding
near-pp-wave formula derived from the $\su(2)$ string Bethe ansatz of \cite{Arutyunov:2004vx}
for completely general mode-number assignment.
This successful match stands as very strong evidence that their ansatz is correct,
at least to $O(J^{-1})$.

\subsection{The $SO(4)_{AdS}$ ($\Sl(2)$) sector}
Following the derivation of eqn.~(\ref{SU2FULL}) for the energy eigenvalues
of arbitrary string states in the symmetric-traceless $SO(4)_{S^5}$ sector,
it is straightforward to find the analogous expression for symmetric-traceless
string states excited in the $SO(4)_{AdS}$ subspace, dual to operators in the
$\Sl(2)$ sector of the corresponding gauge theory.  We can define, for example,
\be
a_n = \frac{1}{\sqrt{2}}\left( a^1_n + i a^2_n \right)~, \qquad
\bar a_n = \frac{1}{\sqrt{2}}\left( a^1_n - i a^2_n \right)\ ,
\label{oscdef2}
\ee
and carry out the above calculations by computing general matrix elements
of $ a_{-n}^\dag a_{-l}^\dag a_m a_p $ defined in terms of these oscillators.
(Here we can project onto any $(n,m)$-plane in the $AdS_5$ subspace, as long
as $n\neq m$.)
General string energy eigenvalues in the $SO(4)_{AdS}$ symmetric-traceless
irrep are thus found to be
\be
\delta E_{AdS}(\{q_i\},\{N_{q_i}\},M,J) & = & \frac{1}{2J}\biggl\{
    \sum_{j=1}^M N_{q_j}(N_{q_j}-1)
    \left(1-\frac{1}{\omega_{q_j}^2\lambda'}\right)
\nn\\
&&\kern-40pt
    +\sum_{j,k=1\atop j\neq k}^M \frac{N_{q_j}N_{q_k}}{\omega_{q_j}\omega_{q_k}}
    q_j q_k \left[1-q_j q_k\lambda'+\omega_{q_j}\omega_{q_k}\lambda'  \right]\biggr\}\ .
\label{SL2FULL}
\ee
For later reference we record the limit of this equation for states with completely unequal mode
indices ($\{N_{n_i}\}=1,\ M=N_B$):
\be
\delta E_{AdS}(\{n_i\},N_B,J) &=& \frac{1}{2J}\sum_{j,k=1\atop j\neq k}^{N_B}
    \frac{n_j n_k}{\omega_{n_j}\omega_{n_k}}\left[1-n_jn_k\lambda'
    +\omega_{n_j}\omega_{n_k}\lambda'\right]\ .
\label{sl2GEN}
\ee

When $M=2$ and $N_{n_1} = N_{n_2} = 1$ in eqn.~(\ref{sl2GEN}), we find the
two-impurity eigenvalue (with $n_2=-n_1$)
\be
\delta E_{AdS}(n_1,J) = -\frac{2\,n_1^2 \lambda'}{J}\ ,
\ee
which agrees with the two-impurity result reported in Chapter 
\ref{twoimp} \cite{Callan:2003xr,Callan:2004uv}
(the $\su(2)$ and $\Sl(2)$ eigenvalues are degenerate in the two-impurity
regime).  For the three-impurity eigenvalue with three unequal mode indices
we set $M=3$ and $N_{n_1} = N_{n_2} = N_{n_3} = 1$ to obtain
\be
\delta E_{AdS} (n_1,n_2,n_3,J) & =  &
    \frac{1}{J\omega_{n_1}\omega_{n_2}\omega_{n_3}}\biggl\{
    n_1n_3(1-n_1n_3\lambda')\,\omega_{n_2}
    +n_1n_2(1-n_1n_2\lambda')\,\omega_{n_3}
\nn\\
&&\kern-60pt
    +\,n_2n_3(1-n_2n_3\lambda')\,\omega_{n_1}
    +\left[
    n_1n_2+n_3(n_1+n_2)\right]\lambda'
    \omega_{n_1}\omega_{n_2}\omega_{n_3}
    \biggr\}\ ,
\label{3impsl2}
\ee
which precisely reproduces the corresponding $\Sl(2)$ result reported in 
Chapter \ref{threeimp} \cite{Callan:2004ev}.
Finally, by setting $M=2$, $N_{q_1} = 2$, $N_{q_2}=1$ and $q_1=q_2=q,\ q_3=-2q$,
eqn.~(\ref{SL2FULL}) provides the following three-impurity eigenvalue with two equal
mode indices:
\be
\delta E_{AdS}(q,J)  =
    -\frac{q^2}{J\omega_q^2 \omega_{2\,q} }
    \left[
    \omega_{2\,q}\left( 3+4\,q^2\lambda'\right)
    +\omega_q\left( 4+8\,q^2\lambda'\right)
    \right]\ .
\label{3impsl2nn}
\ee
This again matches the corresponding three-impurity formula computed in 
Chapter~\ref{threeimp}.

\subsection{The $\su(1|1)$ sector}
Based on the above results in the bosonic $SO(4)_{AdS}$ and
$SO(4)_{S^5}$ symmetric-traceless sectors, we can easily formulate
a conjecture for the $N$-impurity eigenvalue of symmetrized
pure-fermion states in either the $({\bf 2,1;2,1})$ or $({\bf
1,2;1,2})$ of $SO(4)\times SO(4)$, labeled by the $\su(1|1)$
subalgebra. We first note that, since these states are composed of
fermionic oscillators that are symmetrized in their spinor
indices, no states in this sector can carry subsets of overlapping
mode numbers (since they would automatically vanish). Furthermore,
when restricting to states with completely unequal mode indices,
we can see that the $N$-impurity eigenvalues obtained for the
$\su(2)$ and $\Sl(2)$ sectors (eqns.~(\ref{su2GEN}) and
(\ref{sl2GEN})) are obvious generalizations of the corresponding
three-impurity formulas (eqns.~(\ref{3impsu2}) and
(\ref{3impsl2}), respectively).  Namely, if the three-impurity
eigenvalues take the generic form \be \delta E(n_1,n_2,n_3,J) =
\sum_{j,k=1\atop j\neq k}^3 F(n_j,n_k)\ , \ee the $N$-impurity
generalization is simply \be \delta E(\{n_i\},N,J) =
\sum_{j,k=1\atop j\neq k}^{N} F(n_j,n_k)\ . \ee By carrying this
over to the $\su(1|1)$ sector, we find the $N$-impurity eigenvalue
of $H_{\rm FF}$ between symmetrized $({\bf 2,1;2,1})$ or $({\bf
1,2;1,2})$ fermions (the eigenvalues of both are necessarily
degenerate): \be \delta E_{\su(1|1)}(\{n_i\},N_F,J) =
-\frac{1}{4J}\sum_{j,k=1 \atop j\neq k}^{N_F}
    \frac{1}{\omega_{n_j}\omega_{n_k}}\left[n_j^2+n_k^2+2n_j^2 n_k^2 \lambda'
        -2\,n_j n_k \omega_{n_j}\omega_{n_k}\lambda' \right]\ .
\nn\\
\label{SU11FULL}
\ee

For $N_F=2$, this formula matches the two-impurity result in
Chapter \ref{twoimp}:
\be
\delta E_{\su(1|1)}(n_1,J) = -\frac{2\,n_1^2 \lambda'}{J}\ ,
\ee
with $n_2=-n_1$ (this eigenvalue overlaps with the corresponding two-impurity
$\su(2)$ and $\Sl(2)$ values).  When $N_F=3$ we of course recover the
three-impurity eigenvalue reported in Chapter \ref{threeimp}:
\be
\delta E_{\su(1|1)}(n_1,n_2,n_3,J)  &=&
    -\frac{1}{4\,J\omega_{n_1}\omega_{n_2}\omega_{n_3}}\biggl\{
    -4\,\bigl(n_2n_3+n_1(n_2+n_3)\bigr)\lambda'\omega_{n_1}\omega_{n_2}\omega_{n_3}
\nn\\
&&\kern-110pt
    +\biggl[
    \omega_{n_1}
    \left(
    2\, n_3^2
    +4\,n_2^2n_3^2\lambda'
    +2\, n_2^2
    \right)
    +\bigl(n_3\to n_2,\ n_2\to n_1,\ n_1\to n_3\bigr)
    +\bigl( n_1\rightleftharpoons n_2 \bigr)
    \biggr]
    \biggr\}\ .
\nn\\
&&
\label{exactfermi}
\ee
It would be straightforward to check eqn.~(\ref{SU11FULL}) against
an explicit four-impurity calculation in the string theory, for example.
Better yet, one might carry out the direct $N$-impurity calculation in the
$H_{\rm FF}$ sector analogous to the above calculations for $H_{\rm BB}$.  The
latter would certainly be more technically
complicated than in the bosonic sectors, and for the moment we leave eqn.~(\ref{SU11FULL})
as it stands, withholding direct verification for a future study.

\section{Spectral decomposition }
\label{ch6sec3}
At one- and two-loop order in $\lambda'$ we can infer from basic arguments
the spectral decomposition of the extended $N$-impurity superconformal
multiplet of $O(J^{-1})$ energy corrections to the pp-wave limit.
For simplicity we will restrict the discussion to eigensystems with completely unequal
mode numbers, though the generalization to more complicated cases is straightforward.
To begin we will review the two- and three-impurity supermultiplet structures
studied in \cite{Callan:2003xr,Callan:2004uv,Callan:2004ev}.

We denote the one- and two-loop energy eigenvalue shifts as
$\Lambda^{(1)}$ and $\Lambda^{(2)}$, according to the generic formula
\be
E(\{n_j\},N,J) &=& N + \frac{\lambda'}{2}\sum_{j=1}^N
    n_j^2\left(1 + \frac{\Lambda^{(1)}}{J} + O(J^{-2})\right)
\nn\\
&&  - \frac{{\lambda'}^2}{4}\sum_{j=1}^N
    n_j^4\left(\frac{1}{2} + \frac{\Lambda^{(2)}}{J} + O(J^{-2})\right)
    + O({\lambda'}^3)\ .
\label{lambda}
\ee
The fact that these energy shifts can be expressed as coefficients
of $\sum n_j^2$ and $\sum n_j^4$ is not obvious.
In the two- and three-impurity cases this was shown to be true
by direct diagonalization of the Hamiltonian.  By expanding
eqns.~(\ref{SU2FULL}, \ref{SL2FULL}, \ref{SU11FULL}) in small $\lambda'$,
it can also be seen that the more general $N$-impurity $\su(2)$, $\Sl(2)$
and $\su(1|1)$ eigenvalues
adhere to this structure to two-loop order.
We will argue that the remaining energy shifts (those in non-protected subsectors)
can be obtained from the protected sectors
through half-integer shifts of the $S^5$ angular momentum $J$:  it will therefore
be seen that all energies considered here will appear in the form given in
eqn.~(\ref{lambda}).

As described above, the conformal invariance of
the full $\alg{psu}(2,2|4)$ symmetry algebra of the theory
guarantees that the energy eigenvalues (and hence $\Lambda^{(1)}$
and $\Lambda^{(2)}$) will be organized into conformal
(sub)multiplets built on conformal primary (or highest weight)
states.  
For the sake of continuity we will briefly review this here.
Within a given submultiplet we refer to states with lowest
energy as super-primary states, and the other conformal primaries
within the submultiplet are obtained by acting on super-primaries
with any of the eight supercharges, labeled by ${\cal Q}_\alpha$,
that increment $\Lambda^{(1)}$ or $\Lambda^{(2)}$ by a fixed
amount but leave the impurity number unchanged. In the gauge
theory these charges are understood to shift both the operator
dimension and $R$-charge such that $\Delta = D-R$ remains fixed
within the submultiplet. Acting with $L_{\rm sub}$ factors of
these supercharges on a super-primary generates nine levels within
each submultiplet labeled by $L_{\rm sub} = 0,\ldots,8$. If the
lowest energy level ($L_{\rm sub}=0$) in the submultiplet is
occupied by $p$ degenerate super-primaries, the $L_{\rm sub}^{\rm th}$
level will therefore contain $p\,C_{L_{\rm sub}}^8$ degenerate
states, where $C_n^m$ is the binomial coefficient. Furthermore, if
the super-primary in a given submultiplet is a spacetime boson,
the $L_{\rm sub}={\rm even}$ levels of the submultiplet will all
be bosonic, and the $L_{\rm sub}={\rm odd}$ levels will be
fermionic.  The opposite is true if the bottom state is fermionic.

As an example, consider the one-loop, two-impurity supermultiplet
structure studied in Chapter \ref{twoimp}. The
spectrum in this case contains only a single multiplet built on a
scalar super-primary (labeled by $1_B$, where the subscript
denotes a bosonic level) with $O(1/J)$ one-loop energy shift
$\Lambda^{(1)} = -6$. The $L_{\rm sub}=1$ level therefore has
eight degenerate states $(8_F)$ with $\Lambda^{(1)}=-5$, the
$L_{\rm sub}=2$ level contains $28_B$ states with
$\Lambda^{(1)}=-4$ and so on.  We record the two-impurity
supermultiplet structure in table~\ref{2mult1} for comparison with
higher-impurity spectra.
\begin{table}[ht!]
\begin{eqnarray}
\begin{array}{|c|ccccccccc|}
\hline
L_{\rm sub} & 0 & 1 & 2 & 3 & 4 & 5 & 6 & 7 & 8     \\  \hline
        & 1_B   & 8_F   & 28_B  & 56_F  & \fbox{$70_B$} & 56_F  & 28_B  & 8_F  & 1_B    \\ \hline
\Lambda^{(1)}(L_{\rm sub})& -6    & -5  & -4    & -3    & -2    & -1    & 0 & 1 & 2  \\     \hline
\Lambda^{(2)}(L_{\rm sub})& -4    & -3  & -2    & -1    & 0 & 1 & 2 & 3 & 4  \\     \hline
\end{array} \nonumber
\end{eqnarray}
\caption{Submultiplet breakup of the 256-dimensional two-impurity spectrum}
\label{2mult1}
\end{table}
The one-loop energies of the three protected $\Sl(2)$, $\su(2)$ and $\su(1|1)$
subsectors studied here
are degenerate in the two-impurity regime and lie in the boxed $70_B$
``centroid'' level in table~\ref{2mult1}.  We also record in table~\ref{2mult1}
the two-loop energy shifts $\Lambda^{(2)}$, which are offset from the
one-loop values by two: $\Lambda^{(2)} = \Lambda^{(1)}+2$.

In the gauge theory there are 16 operators that increment
the impurity number by one and shift the $R$-charge by certain
amounts \cite{Beisert:2002tn}. Four of these act on
single-trace operators by rotating the $SO(6)$ scalars $Z$
(carrying one unit of $R$-charge) into $\phi$ (which carry zero
$R$-charge):  they increase the operator impurity number by one
and decrease the $R$-charge by one ($N\to N+1,\ R\to R-1$).  Four
operators rotate $Z$ into $\nabla Z$, increasing $N$ by one and
leaving the $R$-charge fixed.  The remaining eight operators
are fermionic and send $N\to N+1,\ R\to R+1/2$. If
one uses these operators to generate $N$-impurity super-primaries
from those in the $(N-1)$-impurity spectrum, an immediate
implication is that, within a given $N$-impurity spectrum of
anomalous dimensions, all of the eigenvalues in the gauge theory
will be related to each other by half-integer shifts in the
$R$-charge.  Certain energy levels will therefore be common to all
of the submultiplets in the spectrum built on super-primary
operators, and this special degeneracy can be used to deduce the
overall structure of the extended supermultiplet.  This
degeneracy, however, only persists in the string theory to
two-loop order in $\lambda'$, and it is for this reason that we
are forced to limit the general superstring spectral decomposition
to two-loop order in the expansion.  (It will be shown below,
however, that a certain subset of submultiplets in the string
theory can always be determined to \emph{all} orders in
$\lambda'$.)

Sending $J\to J+A$ on the string side (dual to an $R$-charge shift in the gauge theory)
shifts $\Lambda^{(1)}$ and
$\Lambda^{(2)}$ by $-2A$:  starting from the two-impurity
super-primary $(1_B)$ with energy $\Lambda^{(1)}=-6$, the string
versions of the 16 impurity-increasing operators can be
understood to generate four (degenerate) bosonic three-impurity
super-primaries with $\Lambda^{(1)}=-8$, eight fermionic
three-impurity super-primaries with $\Lambda^{(1)}=-7$ and four
bosonic three-impurity super-primaries with $\Lambda^{(1)}=-6$. By
acting with the eight charges ${\cal Q}_\alpha$ we then generate
submultiplets based on each of these super-primaries whose levels
are populated by $p\,C_{L_{\rm sub}}^8$ degenerate states, where
$p$ here is either four (for the two four-dimensional bosonic
super-primary levels) or eight (for the eight-dimensional
fermionic super-primary level). The submultiplets themselves can
be labeled by a separate index ${L'}$, in this case running over
${L'}=0,\ldots,2$.  

The complete three-impurity multiplet structure is
recorded in table~\ref{3mult1}.  Here there are a total of 11 levels in the extended
supermultiplet, and we label these with the index $L$ such that $L=L_{\rm sub}+{L'}$.
In table~\ref{3mult1} the closed $\su(2)$ sector lies in
the boxed $280_B$ level in the ${L'}=0$ submultiplet with $\Lambda^{(1)}=-4$, the $\Sl(2)$
eigenvalue ($\Lambda^{(1)}=-2$) is in the boxed $280_B$ level of the ${L'}=2$ submultiplet
and the $\su(1|1)$ eigenvalue ($\Lambda^{(1)}=-3$) is in the $560_F$ level of the ${L'}=1$
submultiplet.  For any impurity number these protected eigenvalues will always
lie at the $L_{\rm sub}=4$ level within their respective submultiplets.
We also note that, in the ${L'}$ direction, the $\su(2)$ and $\Sl(2)$ eigenvalues
will correspond to eigenstates composed purely of $S^5$ or $AdS_5$
bosonic excitations, and will therefore fall
into the ``bottom'' and ``top'' submultiplets, respectively
(the ${L'}=0$ and ${L'}=2$ levels in the three-impurity
case).  Similarly, the $\su(1|1)$ eigenvalue will correspond to eigenstates composed
of either $({\bf 2,1;2,1})$ or $({\bf 1,2;1,2})$ excitations, and always lie in the
``centroid'' submultiplet in the ${L'}$ direction (the ${L'}=1$ level for three impurities).
The energies shared by each of the submultiplets can be collected into degenerate levels of
the complete supermultiplet.  This total level degeneracy $D(L)$ is recorded in
the bottom row of table~\ref{3mult1}.
\begin{table}[ht!]
\begin{eqnarray}
\begin{array}{|c|ccccccccccc|}
\hline
{L'} \backslash L & 0  & 1 & 2 & 3 & 4 & 5 & 6 & 7 & 8 & 9 & 10    \\  \hline\hline
0       & 4     & 32    & 112   & 224   & \fbox{$280$ }& 224    & 112   & 32    & 4 &   &   \\ \hline
1       &   & 8 & 64    & 224   & 448   & \fbox{$560$}& 448  & 224  & 64    & 8 &   \\ \hline
2       &   &   & 4     & 32    & 112   & 224   & \fbox{$280$}  & 224   & 112   & 32    & 4 \\ \hline\hline
\Lambda^{(1)}(L) & -8   & -7    & -6    & -5    & -4    & -3    & -2    & -1    & 0 & 1 & 2  \\ \hline
\Lambda^{(2)}(L) & -6   & -5    & -4   & -3 & -2    & -1    & 0 & 1 & 2 & 3 & 4  \\ \hline\hline
D(L) & 4_B  & 40_F  & 180_B & 480_F & 840_B & 1008_F& 840_B & 480_F & 180_B & 40_F  & 4_B \\
\hline
\end{array} \nonumber
\end{eqnarray}
\caption{Submultiplet breakup of the 4,096-dimensional three-impurity spectrum}
\label{3mult1}
\end{table}

It is easy to generalize this supermultiplet structure to
arbitrary impurity number based on how the complete three-impurity
spectrum is generated from the two-impurity supermultiplet above.
For $N$ impurities, the complete supermultiplet will have a total
of $16^N$ states and $5+2N$ levels: the supermultiplet level index
$L$ therefore runs over $L=0,\ldots,(4+2N)$.  The entire
supermultiplet breaks into $2N-3$ submultiplets, each of which
have nine sub-levels labeled by $L_{\rm sub}=0,\ldots,8$. The
submultiplets themselves are labeled by the index ${L'}$, which runs
over ${L'}=0,\ldots,(2N-4)$. The one-loop energy shifts
within the ${L'}^{\rm th}$ submultiplet at level $L_{\rm sub}$ are thus given by
\be
\Lambda^{(1)}_{\rm sub}({L'},L_{\rm sub},N) = {L'}+L_{\rm sub}-2(N+1)\ .
\label{lambdasub}
\ee
Equivalently, the $L^{\rm th}$ level of the entire supermultiplet has energy shift
\be
\Lambda^{(1)}(L,N)=L-2(N+1)\ .
\label{lambdaL}
\ee
The number of degenerate states at level $L_{\rm sub}$
within the ${L'}^{\rm th}$ submultiplet is
\be
D_{\rm sub}({L'},L_{\rm sub},N) = 4^{N-2}C^{2N-4}_{{L'}}C_{L_{\rm sub}}^8\ ,
\ee
so that the total dimension of the ${L'}^{\rm th}$ submultiplet is $256\times 4^{N-2}C^{2N-4}_{{L'}}$.
By summing the submultiplet degeneracies over a given supermultiplet level $L$,
the total number of degenerate states at level $L$ in the supermultiplet
is given (in terms of Euler's $\Gamma$ function) by
\be
D(L,N) = \frac{4^{N-2}\Gamma(2N+5)}{\Gamma(2N+5-L)\Gamma(1+L)}\ .
\ee
The level is bosonic when $L$ is even and fermionic when
$L$ is odd.  As a verification of this formula, we can check that the
total number of states in the $N$-impurity supermultiplet is indeed
\be
\sum_{L=0}^{4+2N}\frac{4^{N-2}\Gamma(2N+5)}{\Gamma(2N+5-L)\Gamma(1+L)} = 16^N\ .
\ee

As noted above, the one-loop $N$-impurity $\su(2)$ energy
corresponds to eigenstates that are composed purely of
symmetric-traceless $({\bf 1,1;2,2})$ excitations:
since each of these excitations increments the angular momentum $J$ by one,
the energy eigenvalue must
therefore lie within a submultiplet built on super-primary
states that exhibit the lowest possible energy in the extended supermultiplet.
In other words, the $\su(2)$ eigenvalue always lies
at level $L_{\rm sub}=4$ of the ${L'}=0$ submultiplet and,
using the general formula in eqn.~(\ref{lambdasub}),
we see that it exhibits the one-loop energy shift
\be
\Lambda^{(1)}_{S^5}(N) = \Lambda^{(1)}_{\rm sub}({L'}=0,L_{\rm sub}=4,N) = -2(N-1)\ .
\ee
As a cross-check on this result, we note that
this agrees with the one-loop limit of the general $\su(2)$ eigenvalue formula (with unequal
mode indices) in eqn.~(\ref{su2GEN}) above (with $N_B=N$):
\be
\delta E_{S^5}(\{n_i\},N,J) =&&\kern-15pt -\frac{1}{2J}
    \sum_{j,k=1\atop j\neq k}^{N} (n_j^2+n_k^2)\,\lambda' + O({\lambda'}^2)
    = -\frac{1}{J}\sum_{j=1}^{N} (N-1)\,n_j^2\, \lambda' + O({\lambda'}^2)\ .
\nn\\
&&
\ee
(Note the prefactor of $1/2$ in the definition of $\Lambda^{(1)}$ in
eqn.~(\ref{lambda}).)  At this point we also see that $\Lambda^{(1)}$ indeed
appears as a coefficient of $\sum n_j^2$, as given in eqn.~(\ref{lambda}).

The $N$-impurity $\Sl(2)$ eigenvalue, composed entirely of
$({\bf 2,2;1,1})$ excitations, must lie in the ``top'' ${L'}=2N-4$ submultiplet
at $L_{\rm sub}=4$.  This gives the one-loop energy shift
\be
\Lambda^{(1)}_{AdS}(N)
    = -2 \ .
\label{sl2oneloop}
\ee
To check this we use the general $\Sl(2)$ formula for completely unequal mode indices
in eqn.~(\ref{sl2GEN}), and again expand to one-loop order in $\lambda'$:
\be
\delta E_{AdS}(\{n_i\},N,J) &=& \frac{1}{J}\sum_{j,k=1\atop j\neq k}^{N} n_j\,n_k\,\lambda'
    + O({\lambda'}^2)\ .
\label{sl2exp}
\ee
With the level-matching condition $\sum_{j=1}^{N} n_{j} = 0$ this becomes
\be
\delta E_{AdS}(\{n_i\},N,J)=-\frac{1}{J}\sum_{j=1}^{N} n_j^2 \lambda'+ O({\lambda'}^2) \ ,
\ee
which agrees perfectly with the prediction in eqn.~(\ref{sl2oneloop})
(and again confirms that $\Lambda^{(1)}$ here is a coefficient of $\sum n_j^2$).

Finally, the $\su(1|1)$ one-loop eigenvalue,
composed of either $({\bf 2,1;2,1})$ or $({\bf 1,2;1,2})$ spinors,
lies in the ${L'}=N-2$ submultiplet at $L_{\rm sub}=4$,
exhibiting the one-loop energy shift
\be
\Lambda^{(1)}_{\su(1|1)}(N)
    = -N \ .
\label{su11oneloop}
\ee
Using eqn.~(\ref{SU11FULL}) we see that
\be
\delta E_{su(1|1)}(\{n_i\},N,J)
&=& -\frac{1}{4J}\sum_{j,k=1\atop j\neq k}^{N} (n_j-n_k)^2\lambda' + O({\lambda'}^2)
\nn\\
    & = & -\frac{1}{2J}\sum_{j=1}^{N} N\,n_j^2\,\lambda'\ ,
\label{su11exp}
\ee
where we have again invoked the level-matching condition to derive the last line.
This of course agrees with eqn.~(\ref{su11oneloop}).
For reference we present in table~6.4
the complete 65,536-dimensional four-impurity spectrum of one- and two-loop
energies.  The $\su(2)$ eigenvalue in this case lies in the
boxed $1120_B$ level with $\Lambda^{(1)} = -6$,
the $\su(1|1)$ eigenvalue is in the $6720_B$ level with $\Lambda^{(1)} = -4$,
and the $\Sl(2)$ energy lies in the $1120_B$ level with $\Lambda^{(1)} = -2$.
\begin{sidewaystable}
\begin{tabular}{|c|ccccccccccccc|}
\hline
${L'}$ $\backslash$ $L$    & 0&  1&   2&   3&    4&    5&    6&    7&    8&    9&    10&   11&   12 \\\hline\hline
0       & 16& 128& 448& 896&  \fbox{$1120$ }& 896&  448&  128&  16&   &     &     &     \\ \hline
1       & &  64&  512& 1792& 3584& 4480& 3584& 1792& 512&  64&   &     &       \\ \hline
2       & &  &   96&  768&  2688& 5376& \fbox{$6720$}& 5376& 2688& 768&  96&   &    \\ \hline
3       & &  &   &   64&   512&  1792& 3584& 4480& 3584& 1792& 512&  64&     \\ \hline
4       & &  &   &   &    16&   128&  448&  896&  \fbox{$1120$}& 896&  448&  128&  16   \\ \hline\hline
$\Lambda^{(1)}(L)$& -10 & -9    & -8    & -7    & -6    & -5    & -4    & -3    & -2    & -1    & 0 &1 &2 \\ \hline
$\Lambda^{(2)}(L)$& -8  & -7    & -6     & -5   & -4    & -3    & -2    & -1    & 0 & 1 & 2 &3 &4 \\ \hline\hline
$D(L)$&$16_B$& $192_F$& $1056_B$& $3520_F$& $7920_B$& $12672_F$& $14784_B$& $12672_F$&
        $7920_B$& $3520_F$& $1056_B$& $192_F$& $16_B$ \\ \hline
\end{tabular}
\caption{Submultiplet breakup of the 65,536-dimensional four-impurity spectrum}
\label{4mult16}
\end{sidewaystable}
\clearpage

Comparing the $\Lambda^{(2)}$ and $\Lambda^{(1)}$
spectra in tables~\ref{2mult1} and \ref{3mult1} (which are determined
directly from the string Hamiltonian), we see that the spectrum of $\Lambda^{(2)}$ is
identical to $\Lambda^{(1)}$ up to an overall shift.
The two-loop analogue of the general $N$-impurity energy shift of
eqn.~(\ref{lambdasub}) is therefore
\be
\Lambda^{(2)}_{\rm sub}({L'},L_{\rm sub},N) = {L'}+L_{\rm sub}-2N\ .
\label{lambda2}
\ee
Equivalently, we have $\Lambda^{(2)}(L,N) = L-2N$ for the entire supermultiplet
shift in terms of $L$.

Similar to the one-loop case, we can test this two-loop formula using
the $N$-impurity results derived above
in the three protected sectors.  According to eqn.~(\ref{lambda2}),
the $\su(2)$ eigenvalue in the ${L'}=0$ submultiplet at level
$L_{\rm sub}=4$ has the following two-loop energy shift:
\be
\Lambda^{(2)}_{S^5} (N)
= 4-2N\ .
\ee
Isolating the two-loop energy eigenvalue $\delta E_{S^5}^{(2)}$ from the $N$-impurity
$\su(2)$ equation
(\ref{su2GEN}), we have
\be
\delta E_{S^5}^{(2)}(\{n_i\},N,J) &=& \frac{1}{4J}\sum_{j,k=1\atop j\neq k}^N(n_j^4 + n_j^3n_k
                    +n_j n_k^3 +n_k^4){\lambda'}^2
\nn\\
    &=&-\frac{1}{4J}\sum_{j=1}^N(n_j^4)(4-2N){\lambda'}^2\ ,
\ee
which matches our prediction.
The $\Sl(2)$ eigenvalue in the ${L'}=2N-4$ submultiplet is predicted to vanish
\be
\Lambda^{(2)}_{AdS} (N)
= 0\ ,
\ee
which agrees with the two-loop expansion term in eqn.~(\ref{sl2GEN}):
\be
\delta E_{AdS}^{(2)}(\{n_i\},J) = -\frac{1}{4J}\sum_{j,k=1\atop j\neq k}^N
                    \left[n_j n_k (n_j+n_k)^2\right]{\lambda'}^2 = 0\ .
\ee
Finally, the $\su(1|1)$ pure-fermion sector in the ${L'}=N-2$ submultiplet at $L_{\rm sub}=4$
should have an energy shift of
\be
\Lambda^{(2)}_{\su(1|1)}(N)
= 2-N\ ,
\ee
which agrees with the $\su(1|1)$ formula given in eqn.~(\ref{SU11FULL}):
\be
\delta E_{\su(1|1)}^{(2)}(\{n_i\},N,J) &=& \frac{1}{8J}\sum_{j,k=1\atop j\neq k}^N
                    (n_j^2-n_k^2)^2{\lambda'}^2
\nn\ee\be       &=&-\frac{1}{4J}\sum_{j=1}^N(n_j^4)(2-N){\lambda'}^2\ .
\ee

As described in Chapter \ref{threeimp},
it should also be noted that since we know the $\su(2)$, $\Sl(2)$
and $\su(1|1)$ eigenvalues to all orders in $\lambda'$, we can easily
determine complete all-loop energy formulas for the three submultiplets
to which these eigenvalues belong.
It was previously noted that the eight supercharges
(${\cal Q}_\alpha$) that act as raising operators within each submultiplet
are known in the gauge theory to shift both the dimension and $R$-charge
by $1/2$ such that $\Delta = D-R$ is kept fixed.  Because all states within
a given submultiplet share the same $\Delta$, the string energy shift at
any level $L_{\rm sub}$ can therefore be obtained from that at some level $L'_{{\rm sub}}$
(not to be confused with $L'$) by replacing
\be
J \to J-L_{\rm sub}/2 + L'_{{\rm sub}}/2 \nn
\ee
in the energy eigenvalue evaluated at sub-level $L'_{{\rm sub}}$.
Since we are expanding to $O(J^{-1})$, however, this replacement can only affect
the eigenvalues $\delta E$ via the $O(J^0)$ BMN term in the pp-wave limit.
For the protected eigenvalues
determined above at $L_{\rm sub}=4$, we therefore find the
all-loop energy shift for the entire submultiplet by including the
appropriate $O(J^{-1})$ contribution from the BMN formula
\be
E_{\rm BMN} = \sum_{j=1}^N \sqrt{1+\frac{n_j^2\lambda}{(J+2-L_{\rm sub}/2)^2}}\ .
\ee
Explicitly, the complete level spectra of the ${L'}=0,\ {L'}=N-2$ and ${L'}=2N-4$
submultiplets are given, to all orders in $\lambda'$, by
\be
\delta E(\{n_j\},L_{\rm sub},N,J) = \frac{\lambda'}{2J}\sum_{j=1}^N
     \frac{n_j^2 (L_{\rm sub}-4)}{\sqrt{1+n_j^2\lambda'}}
    + \delta E_{L_{\rm sub}=4}(\{n_j\},J)\ ,
\ee where $\delta E_{L_{\rm sub}=4}$ is the $L_{\rm sub}=4$ energy
shift in the submultiplet of interest. Since the level degeneracy
among submultiplets is generally broken beyond two-loop order, it
is difficult to obtain similar expressions for submultiplets not
containing the $\su(2)$, $\Sl(2)$ and $\su(1|1)$ protected
eigenvalues. This can possibly be addressed by relying directly on
the commutator algebra of various impurity-increasing operators in
the string theory, and we will return to this problem in a future
study.

\section{Discussion }
In this chapter we have directly computed the near-pp-wave eigenvalues of
$N$-impurity bosonic string states with arbitrary mode-number
assignment lying in the protected symmetric-traceless irreps of
the $AdS_5$ ($\Sl(2)$) and $S^5$ ($\su(2)$) subspaces. Based on
the observation that the $\su(2)$ and $\Sl(2)$ eigenvalues are
simple generalizations of the three-impurity results obtained in
Chapter \ref{threeimp}, we have also presented a conjecture for the
$N$-impurity eigenvalues of symmetrized-fermion states in the
$\su(1|1)$ sector. This conjecture meets several basic
expectations and we believe that it is correct. (It would be
satisfying, however, to derive the $\su(1|1)$ eigenvalue formula
directly from the fermionic sector of the string theory.) We have
also found that the $\su(2)$ eigenvalues perfectly match, to all
orders in $\lambda'$, the corresponding eigenvalue predictions
given by the string Bethe ansatz of \cite{Arutyunov:2004vx}.
Along these lines, it would be very interesting to have long-range Bethe ans\"atze
analogous to \cite{Arutyunov:2004vx}
for the entire $\alg{psu}(2,2|4)$ algebra of the theory.

The supermultiplet decomposition given in Section \ref{ch6sec3} is based
on the breakup of the energy spectrum observed
between the two- and three-impurity regime, and is precisely what
is expected from the gauge theory based on how 16 particular charges are
known to act on operators that are dual to the string states
of interest \cite{Beisert:2002tn,Callan:2004ev}.
Assuming that this mechanism is not specific to the
three-impurity case, we were able to generalize the decomposition
of the $N$-impurity (unequal mode index) supermultiplet to two-loop
order in $\lambda'$.  By knowing where the eigenvalues of the $\su(2)$, $\Sl(2)$ and
$\su(1|1)$ sectors are supposed to appear in this decomposition, we
were able to provide a stringent cross-check of our results,
and we have found perfect agreement.  Given the many implicit
assumptions in this procedure, however, it would be instructive
to perform a direct diagonalization of the four-impurity Hamiltonian
to test our predictions.  While such a test is likely to be computationally
intensive, the problem could be simplified to some extent by restricting
to the pure-boson $H_{\rm BB}$ sector at one loop in $\lambda'$.
We of course expect complete agreement with the results presented in this chapter.

%
%
\chapter{Integrability in the quantum string theory}    		  
\label{integ2}
The emergence of integrable structures from planar ${\cal N}=4$ super-Yang-Mills (SYM)
theory and type IIB string theory on $AdS_5\times S^5$ has renewed hope that 
't~Hooft's formulation of large-$N_c$ QCD may eventually lead to an exact solution.
If both the gauge and string theories are in fact integrable,
each will admit infinite towers of hidden charges and,
analogous to the usual identification of the string theory Hamiltonian with 
the gauge theory dilatation generator, there will be an infinite number of mappings 
between the higher hidden charges of both theories.  
This has led to many novel tests of the AdS/CFT correspondence, particularly in the context of
the pp-wave/BMN limits \cite{Berenstein:2002jq,Metsaev:2001bj,Metsaev:2002re}. 
Barring an explicit solution, one would hope that both theories will at least be shown to 
admit identical Bethe ansatz equations, allowing us to explore a much larger 
region of the gauge/string duality.

As described above, the fact that the gauge theory harbors integrable structures 
was realized by Minahan and Zarembo when they discovered that a particular 
$SO(6)$-invariant sector of the SYM dilatation generator can be mapped,
at one-loop order in the 't~Hooft coupling $\lambda = g_{\rm YM}^2 N_c$, to the 
Hamiltonian of an integrable quantum spin chain with $SO(6)$ vector lattice sites 
\cite{Minahan:2002ve}.  
The Hamiltonian of this system can be diagonalized by solving a set of algebraic Bethe ansatz
equations:  
the problem of computing operator anomalous dimensions in this sector of the 
gauge theory was thus reduced in \cite{Minahan:2002ve} to solving the
set of Bethe equations specific to the $\so(6)$ sector of the theory.
The correspondence between operator dimensions and integrable 
spin chain systems at one loop in $\lambda$ was extended to include the complete 
$\psu(2,2|4)$ superconformal symmetry algebra of planar ${\cal N}=4$ SYM theory by Beisert and 
Staudacher in \cite{Beisert:2003yb}. 
Studies of higher-loop integrability in the gauge theory were advanced 
in \cite{Serban:2004jf,Beisert:2004hm},
where so-called long-range Bethe ansatz equations, which are understood
to encode interactions on the spin lattice that extend beyond nearest-neighbor sites,  
were developed for a closed bosonic $\su(2)$ sector of the 
gauge theory.  
The dynamics of the gauge theory therefore appear to be consistent with the 
expectations of integrability, at least to three-loop order in the 't~Hooft expansion, and 
there is convincing evidence that this extends to even higher order 
\cite{Beisert:2004ry,Beisert:2004hm}.

Concurrent with the introduction 
of the Bethe ansatz formalism in the $\so(6)$ sector of the gauge theory \cite{Minahan:2002ve}, 
related developments emerged from studies of semiclassical configurations of rotating 
string on $AdS_5\times S^5$.  This branch of investigation began
with \cite{Gubser:2002tv}, where the pp-wave limit of the string theory
was reinterpreted in the context of a semiclassical expansion about certain solitonic
solutions in the full $AdS_5\times S^5$ target space.  Using this 
semiclassical picture, Frolov and Tseytlin computed a class of two-spin string solutions
in \cite{Frolov:2002av},
demonstrating explicitly how stringy corrections in the large-spin limit
give rise to systems that can be understood as generalizations of the original pp-wave 
solution studied in \cite{Berenstein:2002jq,Metsaev:2001bj,Metsaev:2002re}.  
This work was extended by a more general study of multi-spin string solutions 
in \cite{Frolov:2003qc}, where the authors provided a detailed prescription 
for making direct comparisons with perturbative gauge theory.  
(For a more complete review of the development and current status of semiclassical
string theory and the match-up with gauge theory, see \cite{Tseytlin:2003ii} and references therein.)
Early indications of integrability in the classical limit of the string theory emerged when 
it was shown that a certain configuration of the GS superstring 
action on $AdS_5\times S^5$ admits an infinite set of classically
conserved non-local charges, and may therefore be an integrable theory 
itself \cite{Bena:2003wd} (see also \cite{Alday:2003zb} for a reduction to the pp-wave system). 
The gauge theory analogue of this non-local symmetry 
was studied in \cite{Dolan:2003uh,Dolan:2004ps},
where a direct connection with the string analysis was made to one-loop
order in $\lambda$.  Various subtleties surrounding studies of the non-local (or Yangian) algebra 
arise at higher loops, and further work is certainly warranted.

In addition to the sector of non-local charges, however, integrable systems typically 
admit an infinite tower
of local, mutually commuting charges, each of which is diagonalized by a set of Bethe 
equations \cite{Faddeev:1996iy,Faddeev:1987ph}.   
The presence of such a sector of hidden, classically conserved bosonic charges in the 
string theory was pointed out in \cite{Mandal:2002fs}.
Moreover, in accordance with the expectations
of AdS/CFT duality, various studies have been successful in matching
hidden local charges in the classical string theory to corresponding quantities
in the quantum spin chain formulation of ${\cal N}=4$ SYM theory.  
In \cite{Arutyunov:2003rg}, for example,
Arutyunov and Staudacher constructed an infinite series of conserved local charges in the 
bosonic string theory by solving the B\"acklund equations associated with certain extended 
classical solutions of the $O(6)$ string sigma model.  The local charges generated 
by the B\"acklund transformations were then matched to corresponding conserved charges obtained 
from an integrable quantum spin chain on the gauge theory side.  In fact, they were able to 
demonstrate agreement between both sides of the duality for the entire infinite tower of 
local commuting charges.   
This study was extended
in \cite{Arutyunov:2003uj,Arutyunov:2003za}, where it was shown that a general 
class of rotating classical string solutions can be mapped to solutions of a 
Neumann (or Neumann-Rosochatius) integrable system.  
More recently, a class of three-spin classical string solutions was shown in 
\cite{Engquist:2004bx} to generate hidden local charges 
(again via B\"acklund transformations) that match their gauge theory counterparts to one-loop order.  
(For a thorough review of the match-up of semiclassical string
integrable structures with corresponding structures in the gauge theory, see also 
\cite{Beisert:2004ry,Tseytlin:2003ii}; these studies were extended beyond the planar limit
in \cite{Peeters:2004pt,Peeters:2005pb}.)

The mapping between string and gauge theory integrable structures was studied from a somewhat different
perspective in \cite{Kazakov:2004qf}, where it was shown that the generator of local, classically 
conserved currents
in the string theory is related in certain sectors to a particular Riemann-Hilbert problem
that is reproduced precisely by the gauge theory integrable structure at one and two loops in $\lambda$.  
An analogous treatment of the corresponding Riemann-Hilbert problem in 
non-compact sectors of the gauge/string
duality was carried out in \cite{Kazakov:2004nh}, and an extension of these studies to $\so(6)$ and
$\su(2,2)$ sectors was recently achieved in \cite{Beisert:2004ag} and \cite{Schafer-Nameki:2004ik}, 
respectively.
The structure of the higher-loop Riemann-Hilbert problem descending from the classical string theory 
and its relationship with the corresponding gauge theory problem was used in conjunction with 
the long-range gauge theory Bethe ansatz of \cite{Beisert:2004hm} 
to develop an ansatz that, albeit 
conjecturally, is purported to interpolate between the classical and quantum regimes 
of the string theory \cite{Arutyunov:2004vx}.  Although this proposal is 
not a proof of quantum integrability on the string side, it was demonstrated in 
\cite{McLoughlin:2004dh} that the quantized string theory
in the near-pp-wave limit yields a general multi-impurity spectrum that matches the
string Bethe ansatz spectrum of \cite{Arutyunov:2004vx}.  
The intricacy of this match-up is quite remarkable, and stands as strong 
evidence that this ansatz is correct for the string theory, at least to $O(1/J)$ in the 
large angular momentum (or background curvature) expansion.  
Furthermore, the proposed string Bethe equations can accommodate the 
strong-coupling $\lambda^{1/4}$ scaling behavior predicted in \cite{Gubser:1998bc}.  The
spin chain theory implied by these Bethe equations, however, appears to disagree with that of 
the gauge theory, even at weak coupling \cite{Beisert:2004jw}.

Although the Bethe equations of \cite{Arutyunov:2004vx} reproduce several predictions of 
the string theory in a highly nontrivial way, a direct test of quantum 
integrability (beyond tree level) in the string theory is still needed:  this is the
intent of the present chapter.  Early steps in this direction were taken in \cite{Swanson:2004mk},
where the presence of a conserved local charge responsible for a certain parity degeneracy 
in the near-pp-wave string spectrum is examined at sixth-order in field fluctuations, 
or at $O(1/J^2)$ in the large-$J$ expansion.  Various subtleties of the analysis (possibly involving
the proper renormalization of the theory at $O(1/J^2)$ in the expansion) make it difficult to
reach any concrete conclusions, however.  In this chapter we take a more immediate approach,
relying primarily on a Lax representation of the classical string sigma model
and studying a semiclassical expansion about certain point-like solitonic solutions.   
The goal is to establish the existence of a series of conserved, mutually commuting
charges in the string theory that can be quantized and studied using first-order
perturbation theory.  By aligning field fluctuations with the 
finite-radius curvature expansion studied in Chapters \ref{twoimp} and \ref{threeimp}, 
we are able to study quantum corrections to quartic order, 
or to one loop beyond tree level.  We show directly that several of the low-lying 
hidden charges in the series are conserved by the 
quantum theory to this order in the expansion, and we propose a method for matching
specific eigenvalues of these charges to corresponding spectral quantities in the gauge theory.

The chapter is organized as follows:  
In Section \ref{integ2sec1} we briefly review the procedure for string quantization in 
the near-pp-wave limit developed in \cite{Callan:2003xr,Callan:2004uv}, 
with the particular goal of demonstrating how background curvature corrections to the pp-wave theory can be
interpreted as quantum corrections in a particular semiclassical expansion about
point-like classical string solutions.  In Section \ref{integ2sec2} we show how a
Lax representation of the $O(4,2)\times O(6)$ nonlinear sigma model can be 
modified to encode the string dynamics to the order of interest in this semiclassical
expansion.  We then generate a series of hidden local charges by expanding a perturbed
monodromy matrix of the Lax representation in powers of the spectral parameter.
(This method for finding higher conserved charges was related in \cite{Arutyunov:2005nk} 
to the approach based on
B\"acklund transformations in \cite{Arutyunov:2003rg}.)
In Section \ref{integ2sec3} we compute the eigenvalues of these charges in certain 
protected subsectors of the theory in the space of two-impurity string states.
The resulting spectra are then compared on the $S^5$ subspace with those of 
corresponding charges descending
from the $\su(2)$ integrable sector of the gauge theory.  We provide a prescription for
matching the spectra of local charges on both sides of the duality, and carry out 
this matching procedure to eighth order in the spectral parameter.  
To the extent that they can be compared
reliably, the gauge and string theory predictions are shown to match to this order 
(and presumably continue to agree at higher orders).  We are thus led to believe that the 
integrable structure of the classical string theory survives quantization, 
at least to the first subleading order in field fluctuations beyond tree level.

\medskip

\section{Semiclassical string quantization in $AdS_5\times S^5$}
\label{integ2sec1}
Most of the literature comparing semiclassical bosonic string theory in $AdS_5\times S^5$
to corresponding sectors of gauge theory operators has focused on classical extended string solutions
to the worldsheet sigma model in either ``folded'' or ``circular'' configurations, 
where certain components of the string angular momentum (i.e.,~certain charges of the Cartan 
subalgebra of the global symmetry group) are taken to be large 
(see, e.g.,~\cite{Frolov:2002av,Frolov:2003qc,Arutyunov:2003uj}).
The latter amounts to choosing a so-called spinning ansatz for the string 
configuration 
\cite{Arutyunov:2003uj,Arutyunov:2003za,Mandal:2002fs,Tseytlin:2003ii,Frolov:2002av,Frolov:2003qc}, 
and solutions endowed with such an ansatz
can be identified with periodic solutions of the Neumann
(or Neumann-Rosochatius) integrable system.
The standard bosonic worldsheet action is usually chosen with flat worldsheet metric
so that it is easily rewritten in terms of ${\bf R}^6$ embedding coordinates 
and identified with an $O(4,2)\times O(6)$ sigma model.  
In the present study we will modify this treatment to allow for curvature
corrections to the worldsheet metric, a complication that we are forced
to confront when moving beyond tree level in lightcone gauge \cite{Callan:2003xr,Callan:2004uv}.

We begin with a particular form of the $AdS_5\times S^5$ target space metric, 
chosen originally in Chapters \ref{twoimp} and \ref{threeimp} for the fact that it admits a simple
form for the spin connection:
\begin{equation}
\label{metric7}
ds^2_{AdS_5\times S^5}  =  {\Rhat}^2
\biggl[ -\left({1+ \frac{1}{4}z^2\over 1-\frac{1}{4}z^2}\right)^2dt^2
        +\left({1-\frac{1}{4}y^2\over 1+\frac{1}{4}y^2}\right)^2d\phi^2
    + \frac{d z_k dz_k}{(1-\frac{1}{4}z^2)^{2}}
    + \frac{dy_{k'} dy_{k'}}{(1+\frac{1}{4}y^2)^{2}} \biggr]~.
\end{equation}
While we will not address fermions in this study, we will eventually return to the 
crucial issues of supersymmetry, and the metric choice in 
eqn.~(\ref{metric7}) will undoubtedly simplify further investigations.
By defining
\be
\cosh\rho \equiv \frac{1+\frac{1}{4}z^2}{1-\frac{1}{4}z^2}~,  \qquad 
\cos\theta \equiv \frac{1-\frac{1}{4}y^2}{1+\frac{1}{4}y^2}\ ,
\ee
we may write the ${\bf R}^6\times {\bf R}^6$ embedding coordinates of $AdS_5$ and $S^5$ as
\be
	Z_{k}  =  \sinh\rho\,\frac{z_{k}}{||z||}~, &\qquad& Z_0 + iZ_5  =  \cosh\rho\, e^{it}\ ,
\nn\ee\be
        Y_{k'}  =  \sin\theta\,\frac{y_{k'}}{||y||}~, &\qquad& Y_5 +i Y_6  =  \cos\theta\, e^{i\phi}\ ,
\ee
with $||z|| \equiv \sqrt{z_{k} z_{k}}$.  The coordinates $Z_P$, with $P,Q=0,\ldots,5$, parameterize
$AdS_5$ and are contracted over repeated indices using the metric 
$\eta_{PQ}=(-1,1,1,1,1,-1)$.  The coordinates $Y_M$, with $M,N=1,\ldots,6$, encode 
the $S^5$ geometry, and are contracted with a Euclidean metric.

Decomposing the theory into $AdS_5$ and $S^5$ subspaces,
the usual conformal-gauge worldsheet action
\be
S = -\int d^2\sigma\,  h^{ab} G_{\mu\nu} \del_a x^\mu \del_b x^\nu\ 
\label{action1}
\ee
can be written as 
\be
S  &=&  \int d^2\sigma ({\cal L}_{AdS_5} + {\cal L}_{S^5} )\ ,
\nn\\
{\cal L}_{AdS_5} & = &  -\frac{1}{2} h^{ab} \eta_{PQ}\del_a Z_P \del_b Z_Q  
	+ \frac{\tilde\varphi}{2} \left(\eta_{PQ} Z_P Z_Q + 1\right)\ ,
\label{action2a}
\\
{\cal L}_{S^5} & = &  -\frac{1}{2}h^{ab} \del_a Y_M \del_b Y_M  
	+ \frac{\varphi}{2} \left(Y_M Y_M - 1\right) \ .
\label{action2b}
\ee
The quantities $\varphi$ and $\tilde \varphi$ act as 
Lagrange multipliers in the action, enforcing 
the following conditions:\footnote{Note that, in general, $\varphi$ and $\tilde \varphi$
will depend on dynamical variables.  We thank Arkady Tseytlin for clarification on this point.}
\be
\eta_{PQ} Z_P Z_Q = -1\ , \qquad Y_M Y_M = 1\ .
\ee
The action in eqn.~(\ref{action1}) must also be supplemented by the
standard conformal gauge constraints, and the worldsheet metric $h^{ab}$ (the worldsheet indices 
run over $a,b \in \tau,\sigma$) will be allowed to acquire curvature corrections in accordance with these 
constraints.

We wish to study a semiclassical expansion about the following classical point-like 
(or ``BMN-like'') solutions to the sigma model equations of motion:
\be
t = \phi = p_-\tau~, \qquad z_k = y_{k} = 0\ .
\label{soliton}
\ee
The expansion is defined in terms of quantum field fluctuations 
according to the following rescaling prescription:
\begin{eqnarray}
\label{rescalePre}
    t \rightarrow x^+~,
\qquad
    \phi \rightarrow x^+ + \frac{x^-}{\sqrt{\xi}}~,
\qquad
    z_k \rightarrow \frac{z_k}{\xi^{1/4}}~,
\qquad
    y_{k} \rightarrow \frac{y_{k}}{\xi^{1/4}}\ .
\end{eqnarray}
(A similar but notably different choice was made in \cite{Frolov:2002av}.)
This particular choice of lightcone coordinates will allow us to maintain a
constant momentum distribution on the worldsheet.  
Additionally, as noted in Chapter \ref{twoimp},  
it will have the effect of eliminating all normal-ordering 
ambiguities from the resulting worldsheet theory, an outcome that is particularly desirable in
the present study.  Furthermore, we note that if we identify $\xi \equiv {\Rhat}^4$, the proposed 
expansion about the classical solution in eqn.~(\ref{soliton}) 
is identical to the large-radius curvature expansion about the pp-wave limit
of $AdS_5\times S^5$ studied above \cite{Callan:2003xr,Callan:2004uv,Callan:2004ev}.  
In other words, we have chosen a 
perturbation to the classical point-like string geodesic that reproduces the 
target-space curvature perturbation to the pp-wave limit.  
The background metric in eqn.~(\ref{metric7}) thus yields the following large-${\Rhat}$ expansion:
\be
ds^2 & = & 2dx^+ dx^- - (x^A)^2 (dx^+)^2 + (dx^A)^2 
\nonumber \\
& & 	+ \frac{1}{{\Rhat}^2}\left[ 
	-2y^2 dx^+ dx^- + \frac{1}{2}(y^4-z^4) (dx^+)^2 + (dx^-)^2
	+\frac{1}{2}z^2 dz^2 - \frac{1}{2}y^2 dy^2 \right] 
\nonumber \\
\label{expmetric2I}
& & 	+ {O}\left({\Rhat}^{-4}\right),
\label{metricexp}
\ee
where the pp-wave geometry emerges at leading order.

The details of quantizing the string Hamiltonian in this setting
are given in Chapter \ref{twoimp}
(see also \cite{Kallosh:1998zx,Metsaev:2000yf,Kallosh:1998ji,Metsaev:1999gz,Metsaev:1998it}
for further details), 
though we will briefly review the salient points here.
The lightcone Hamiltonian $H_{\rm LC}$ is the generator of worldsheet time translations, 
and is defined in terms of the Lagrangian by
\be
-H_{\rm LC} = -p_+ = \delta {\cal L} / \delta \dot x^+\ ,
\ee 
(or $\Delta -J$ in the language of BMN), and   
this variation is performed prior to any gauge fixing.  
The non-physical lightcone variables $x^\pm$ are 
removed from the Hamiltonian by fixing lightcone gauge $x^+ = p_-\tau$ and replacing 
$x^-$ with dynamical variables by enforcing the conformal gauge constraints
\be
T_{ab} = \frac{\delta {\cal L}}{\delta h^{ab}} = 0\ .
\label{CGC}
\ee 
This procedure can be defined order-by-order in the large-${\Rhat}$ expansion.  
At leading order, for example, we obtain the following from eqn.~(\ref{CGC}):
\be
\dot x^- & = & \frac{p_-}{2}(x^A)^2 - \frac{1}{2p_-}\left[(\dot x^A)^2 + ({x'}^A)^2\right]
	+ O(1/{\Rhat}^2)\ ,
\nn\\
{x'}^- & = & -\frac{1}{p_-}\dot x^A {x'}^A + O(1/{\Rhat}^2)\ .
\ee

The conformal gauge constraints
themselves are only consistent with the equations of motion if the worldsheet
metric acquires curvature corrections (i.e.,~$h$ departs from the flat metric $h = {\rm diag}(-1,1)$), 
which we express symbolically as $\tilde h^{ab}$ according to
\be
h = \left(\begin{array}{cc}
	-1 + \tilde h^{\tau\tau}/{\Rhat}^2 & \tilde h^{\tau\sigma}/{\Rhat}^2 \\
	\tilde h^{\tau\sigma}/{\Rhat}^2 & 1 + \tilde h^{\sigma\sigma}/{\Rhat}^2
	\end{array} \right)\ .
\label{wsmetric}
\ee
The requirement that $\det h = -1$ implies $\tilde h^{\tau\tau} = \tilde h^{\sigma\sigma}$ and,
for future reference, the correction terms $\tilde h^{ab}$ are given explicitly to the order of interest
by
\be
\tilde h^{\tau\tau} &=& \frac{1}{2}(z^2-y^2) - \frac{1}{2p_-^2}\left[(\dot x^A)^2 + ({x'}^A)^2\right]\ ,
\nn\\
\tilde h^{\tau\sigma} &=& \frac{1}{p_-^2}\dot x^A {x'}^A\ .
\label{metriccorrections}
\ee
Finally, we note that the canonical momenta associated with the 
physical worldsheet excitations, defined by the variation $p_A = \delta {\cal L} / \delta x^A$, 
also acquire $O(1/{\Rhat}^2)$ corrections: consistent quantization requires that these 
corrections be taken into account.  Expressed in terms of canonical 
variables, the final bosonic Hamiltonian takes the form
\be
H_{\rm LC} &=& \frac{p_-}{2{\Rhat}^2}(x^A)^2 
	+ \frac{1}{2p_-{\Rhat}^2}\left((p_A)^2+({x'}^A)^2\right)
\nn\\
&&	+\frac{1}{{\Rhat}^4}\biggl\{
	\frac{1}{4p_-}\left[z^2(p_y^2+{y'}^2 +2{z'}^2) - y^2(p_z^2+{z'}^2+2{y'}^2)\right]
	+\frac{p_-}{8}\bigl[(x^A)^2\bigr]^2
\nn\\
&&\kern-30pt
	-\frac{1}{8p_-^3}\Bigl\{
	\bigl[(p_A)^2\bigr]^2+2(p_A)^2({x'}^A)^2 + \bigl[(x^A)^2\bigr]^2\Bigr\}
	+\frac{1}{2p_-^3}({x'}^A p_A)^2 \biggr\} +O(1/{\Rhat}^6)\ ,
\nn\\
&&
\label{HLC}
\ee
where the pp-wave Hamiltonian emerges as expected at leading order.
The lightcone momentum $p_-$ is identified (via the AdS/CFT dictionary)
with the modified 't~Hooft parameter $\lambda'$ according to 
\be
p_- = 1/\sqrt{\lambda'} = J/\sqrt{\lambda}\ .
\ee

From the point of view of the semiclassical analysis, we are
working to two-loop order in quantum corrections.  
Since the quadratic theory
can be quantized exactly, however, we can study the quartic interaction Hamiltonian using 
standard first-order perturbation theory.  
A detailed analysis of the resulting spectrum of this perturbation can be 
found above in Chapters \ref{twoimp}, \ref{threeimp} and \ref{Nimp}.  In the course of
those studies it was noticed that, analogous to the gauge theory closed
sectors studied in \cite{Beisert:2003ys,Beisert:2003tq,Beisert:2003jj,Beisert:2003jb}, certain
sectors emerged from the string analysis that decouple from the remainder 
of the theory to all orders in $\lambda'$.  One sector, which maps to the 
$\Sl(2)$ sector of the gauge theory, is diagonalized by 
bosonic string states excited in the $AdS_5$ subspace and forming symmetric-traceless
irreps in spacetime indices.  The corresponding sector of symmetric-traceless
$S^5$ string bosons maps to the closed $\su(2)$ sector in the gauge theory.
The block-diagonalization of these sectors in the string Hamiltonian will be
an important tool in the present analysis:  just as 
all higher hidden local charges in the gauge theory are simultaneously diagonalized
by a single Bethe ansatz, all of the higher hidden charges descending from the
string theory should be block-diagonalized by these particular string states as well.

\section{Lax representation}
\label{integ2sec2}
The goal is to determine whether a ladder of higher local charges can be 
computed and quantized (albeit perturbatively), analogous to the existing treatment
of the near-pp-wave Hamiltonian given in eqn.~(\ref{HLC}) above.  
To quartic order in the semiclassical expansion defined by eqn.~(\ref{rescalePre}), the difference
between the string sigma model in eqns.~(\ref{action2a}, \ref{action2b}) and that
of the $O(4,2)\times O(6)$ sigma model, defined by
\be
{\cal L}_{O(4,2)} & = &  -\frac{1}{2} \eta_{PQ}\del_a Z_P \del^a Z_Q  
	+ \frac{\tilde\varphi}{2} \left(\eta_{PQ}Z_P Z_Q + 1\right)\ , 
\nn\\
{\cal L}_{O(6)} & = &  -\frac{1}{2} \del_a Y_M \del^a Y_M  
	+ \frac{\varphi}{2} \left(Y_M Y_M - 1\right) \ ,
\label{O6action}
\ee
will essentially amount to an interaction perturbation due to curvature 
corrections to the worldsheet metric.  We therefore find it useful to rely on 
a known Lax representation of the $O(4,2)\times O(6)$ sigma model; this 
representation will define an unperturbed theory, and we will add perturbations
by hand to recover the full interaction Hamiltonian in eqn.~(\ref{HLC}).
(For a general introduction
to the Lax methodology in integrable systems, the reader is referred to \cite{Faddeev:1987ph}.)
Since worldsheet curvature corrections only appear at $O(1/{\Rhat}^2)$,
the reduction to the $O(4,2)\times O(6)$ sigma model at leading order in the 
expansion will be automatic.

For simplicity, we start from the four-dimensional Lax representation given 
for the $O(6)$ sigma model in \cite{Arutyunov:2003za} (see also \cite{Pohlmeyer:1975nb} 
for details), and work only to leading order in the semiclassical expansion.
The complexified coordinates 
\be
{\cal Y}_1 = Y_1 + i\,Y_2~, \qquad {\cal Y}_2 = Y_3 + i\,Y_4~,
	\qquad {\cal Y}_3 = Y_5 + i\,Y_6~, 
\label{complexY}
\ee
are used to form a unitary matrix $S_{S^5}$
\be
S_{S^5} = 
{\scriptsize
	\left(
	\begin{array}{cccc}
	0 & {\cal Y}_1 & -{\cal Y}_2 & \bar {\cal Y}_3 \\
	-{\cal Y}_1 & 0 &{\cal Y}_3 & \bar{\cal Y}_2 \\
	{\cal Y}_2 & - {\cal Y}_3 & 0 & \bar{\cal Y}_1 \\
	-\bar{\cal Y}_3 & -\bar {\cal Y}_2 & -\bar{\cal Y}_1 & 0 
	\end{array}
\right)}\ ,
\label{smat}
\ee
in terms of which one may form the following $SU(4)$-valued currents:
\be
A_a = S_{S^5}\del_a {S_{S^5}}^\dag\ .
\ee
The equations of motion of the $O(6)$ sigma model 
\be
\del_a \del^a Y_M + \varphi Y_M = 0
\label{geneom}
\ee
are then encoded by
the auxiliary system of linear equations 
\be
(\del_\sigma - U)X = (\del_\tau - V)X = 0\ ,
\label{laxpair1}
\ee
where the Lax pair $U$ and $V$ are defined by
\be
U = \frac{1}{1+\gamma}A_- - \frac{1}{1-\gamma}A_+\ , \qquad 
\label{laxmat1}
V = -\frac{1}{1+\gamma}A_- - \frac{1}{1-\gamma}A_+\ .
\label{laxmat2}
\ee
The constant $\gamma$ is a free spectral parameter,
and $A_{\pm}$ are defined by $A_{\pm} \equiv \frac{1}{2}(A_\tau\pm A_\sigma)$.  
Note that on the $SO(4)$ subspace spanned by $y_{k'}$, eqn.~(\ref{geneom})
reduces to the pp-wave equations of motion on $S^5$:
\be
\ddot y_{k'} - {y''}_{k'} + p_-^2 y_{k'} = 0\ .
\label{ppeom}
\ee

The utility of the Lax representation arises from the fact that $U$ and $V$ may be 
considered as local connection coefficients, and a consistency equation for the
auxiliary linear problem can be reinterpreted as a flatness condition for the $(U,V)$-connection:
\be
\del_\tau U - \del_\sigma V + \left[U,V\right] = 0\ .
\label{ZCC1}
\ee
Parallel transport along this flat connection is defined by the path-ordered exponent
\be
\Omega_C(\gamma) = {\cal P}\exp \int_{\cal C} (U\,d\sigma + V\,d\tau) \ ,
\ee
where ${\cal C}$ is some contour in ${\bf R}^2$.
Restricting to transport along the contour defined by $\tau=\tau_0$ and $0\leq\sigma \leq 2\pi$
yields a monodromy matrix:
\be
T(2\pi,\gamma) = {\cal P}\exp \int_0^{2\pi} d\sigma\, U\ .
\label{monodromy}
\ee
The flatness condition in eqn.~(\ref{ZCC1}) admits an infinite number
of conservation laws, which translates to the fact that the trace of the monodromy matrix yields 
an infinite tower of local, mutually commuting charges $\Qo_n^{S^5}$ when expanded 
in powers of the spectral index about the poles of $U$ 
($\gamma = \pm 1$, in this case):\footnote{In general, an expansion around some $\gamma$
that is finitely displaced from a singularity of $U$ 
will yield combinations of local and non-local quantities.  One is of course free
to redefine $\gamma$ such that the expansion about $\gamma=0$ in eqn.~(\ref{chargedef})
is local. }
\be
\tr T(2\pi,\gamma) = \sum_n \gamma^n \widehat Q_n^{S^5}\ .
\label{chargedef}
\ee
The first nonvanishing charge $\Qo_2^{S^5}$, for example, is the Hamiltonian of the 
theory (on the $S^5$ subspace).

Moving beyond leading order in the semiclassical expansion, 
the essential difference between the $O(6)$ sigma model defined in 
eqn.~(\ref{O6action}) and the string action given in eqn.~(\ref{action2b}) 
is, as noted above, that worldsheet indices are contracted in the 
latter case with a non-flat worldsheet metric.  Keeping the components of
$h^{ab}$ explicit, the lightcone Hamiltonian derived from the string sigma model in 
eqn.~(\ref{action2b}) appears at leading order as
\be
H_{\rm LC}^{S^5} = -\frac{1}{2p_-\Rhat^2}\left[
	h^{\tau\tau}( p_-^2 y^2 + {y'}^2 + \dot y^2 )
	+2 h^{\tau\sigma} \dot y \cdot y' \right] + O(1/\Rhat^4)\ ,
\label{LOham}
\ee
where $h^{\tau\tau} = -1 + \tilde h^{\tau\tau}/\Rhat^2$ 
and $h^{\tau\sigma} = \tilde h^{\tau\sigma}/\Rhat^2$.
The prescription will be to find a perturbation to the 
$(U,V)$-connection such that the Hamiltonian in eqn.~(\ref{LOham})
emerges in an appropriate limit from the charge $\Qo_2^{S_5}$ defined by eqn.~(\ref{chargedef}).  
Such a perturbation is achieved by transforming the $U$ matrix according to 
\be
U\to U = \frac{1}{1+\gamma}(1+u_-/\Rhat^2 )A_- - \frac{1}{1-\gamma}(1+ u_+/\Rhat^2)A_+\ ,
\label{Udef}
\ee
where $u_\pm$ are given by
\be
u_\pm \equiv -\frac{1}{2} \tilde h^{\tau\tau} \mp \frac{1}{3} \tilde h^{\tau\sigma}\ . 
\ee
These perturbations should be treated as constants, to 
be replaced in the end with dynamical variables by fixing conformal gauge according to 
eqn.~(\ref{CGC}).  
The remaining quartic perturbations to the pp-wave theory will be naturally encoded in the
semiclassical expansion of the underlying $O(6)$ (likewise, $O(4,2)$) sigma model.
The matrix $V$ can be transformed in a similar way:
\be
V\to V = -\frac{1}{1+\gamma}(1+v_-/\Rhat^2)A_- - \frac{1}{1-\gamma}(1+v_-/\Rhat^2 )A_+\ ,
\ee
where $v_\pm$ may be chosen such that the perturbed Lax pair satisfies the flatness condition
in eqn.~(\ref{ZCC1}).  
Given that the intent is simply to determine whether the higher local charges generated by the 
perturbed monodromy matrix are conserved when quantum fluctuations are included, fixing $V$ to
satisfy the flatness condition is not really necessary:  the complicated formulas for $v_\pm$ 
that do satisfy eqn.~(\ref{ZCC1}) will therefore not be needed.

The perturbation in eqn.~(\ref{Udef}) can be obtained by a slightly different method.  When the path-ordered 
exponent defining the monodromy matrix is expanded, it can be seen that all odd products of 
the Lax matrix $U$ will not contribute to the final expression.  By replacing all even products
of $U$ according to the rule
\be
U(\sigma_1) U(\sigma_2) 
	&\to& \frac{1}{(\gamma^2-1)^2}\Bigl[
	h^{\sigma\sigma} A_\sigma(\sigma_1)A_\sigma(\sigma_2)
	-\gamma^2 h^{\sigma\tau} A_\sigma(\sigma_1)A_\tau(\sigma_2)
\nn\\
&&	-\gamma^2 h^{\tau\sigma} A_\tau(\sigma_1)A_\sigma(\sigma_2)
	-\gamma^2 h^{\tau\tau} A_\tau(\sigma_1)A_\tau(\sigma_2)
	\Bigr]\ ,
\label{UUrule}
\ee
the Hamiltonian in eqn.~(\ref{LOham}) is again obtained at leading order in the expansion.
Computationally, this latter method seems to be much more efficient, and
we will use eqn.~(\ref{UUrule}) in what follows.
At leading order in the $1/\Rhat$ expansion, the first nonvanishing integral of 
motion descending from the monodromy matrix is thereby found to be
\be
{Q}_2^{S^5} 
	= \frac{4\pi}{\Rhat^2}\int_0^{2\pi} d\sigma\,
	\left[
	h^{\tau\tau}( p_-^2 y^2 + {y'}^2 + \dot y^2 )
	+2 h^{\tau\sigma} \dot y \cdot y'\right] + O(1/\Rhat^4)\ ,
\label{Q20}
\ee
which, by construction, matches the desired structure in eqn.~(\ref{LOham}).

The same construction may be carried out for the $AdS_5$ system.
In fact, to make matters simple, we may borrow the Lax structure of the $O(6)$ 
model defined in eqns.~(\ref{smat}-\ref{laxmat2}), replacing the
$O(6)$ coordinates in eqn.~(\ref{complexY}) with the following Euclideanized 
$O(4,2)$ complex embedding coordinates:
\be
{\cal Z}_1 = Z_1 + i\,Z_2~, \qquad {\cal Z}_2 = Z_3 + i\,Z_4~,
	\qquad {\cal Z}_3 = i\,Z_0 - Z_5\ .
\ee
In this case, however, the Lax matrix $S_{AdS_5}$ will obey $S_{AdS_5}^\dag S_{AdS_5} = -1$.
Otherwise, the analysis above applies to the $AdS_5$ sector by direct analogy: 
expanding the perturbed $O(4,2)$ monodromy matrix in the spectral parameter yields 
a set of charges labeled by $\widehat Q_n^{AdS_5}$.  The local charges
for the entire theory are then given by
\be
\widehat Q_n \equiv \widehat Q_n^{S^5} - \widehat Q_n^{AdS_5}\ .
\ee
The corresponding currents will be labeled by ${\cal Q}_n$.

It turns out that the expansion in the spectral parameter
$\gamma$ is arranged such that the path-ordered exponent defining
the monodromy matrix can be computed explicitly to a given order in
$\gamma$ by evaluating only a finite number of worldsheet integrals.  
The procedure for extracting local, canonically quantized currents is then
completely analogous to that followed in computing the
lightcone Hamiltonian described above.  All gauge fixing is done after the currents 
are evaluated, all occurrences of $x^-$ are replaced 
with dynamical variables by solving the conformal gauge constraints, and worldsheet 
metric corrections $\tilde h^{ab}$ are evaluated according to eqns.~(\ref{metriccorrections}) 
above.  We note, however, that previous studies 
involving the matching of integrable structures between gauge and string
theory have found it necessary to invoke certain redefinitions of $\gamma$ to obtain agreement
\cite{Beisert:2004hm,Kazakov:2004qf}.  
It would be straightforward to allow for rather general redefinitions of the spectral parameter
in the present calculation.
When we turn to computing spectra and comparing with gauge theory, however,
such redefinitions can lead to unwanted ambiguity.
We will therefore be primarily interested in finding ratios of 
eigenvalue coefficients for which arbitrary redefinitions of $\gamma$ are 
irrelevant, and for simplicity we will simply retain the original definition of $\gamma$
given by eqn.~(\ref{laxmat2}) above.

As previously noted, the first current ${\cal Q}_1$ defined by eqn.~(\ref{chargedef}) vanishes.  
In fact, all ${\cal Q}_n$ vanish for odd values of $n$, 
and this property of the integrable structure is mirrored 
on the gauge theory side.  The first nonvanishing current emerging from the monodromy matrix
is given by
\be
{\cal Q}_2 & = & \frac{4\pi }{\Rhat^2} \left( (\dot x^A)^2
    + ({x'}^A)^2 + p_-^2(x^A)^2 \right) 
\nonumber \\
& & \kern-20pt
    +\frac{\pi }{\Rhat^4}\biggl\{
     {2z^2} \left[{y'}^2
    + 2{z'}^2- \dot y^2 \right]
    - {2y^2} \left[{z'}^2 + 2{y'}^2
    - \dot z^2 \right]
	-\frac{4}{p_-^2}(\dot x^A{x'}^A)^2
\nonumber \\
& & \kern-20pt
	+\frac{1}{p_-^2}\left[3(\dot x^A)^2-({x'}^A)^2\right]
	\left[(\dot x^A)^2+({x'}^A)^2\right]
	+{p_-^2}\left[(x^A)^2\right]^2 \biggr\} + O(1/\Rhat^6)\ .
\label{Q2pre}
\ee
The leading-order term is the quadratic pp-wave Hamiltonian, as expected, and the perturbation
is strictly quartic in field fluctuations.  All occurrences of $x^-$ and all curvature corrections 
to the worldsheet metric $\tilde h^{ab}$ have been replaced with physical variables as described above.
The final step is to express eqn.~(\ref{Q2pre}) in
terms of canonically conjugate variables determined by directly varying the Lagrangian in 
eqn.~(\ref{action1}).  We obtain
\be
\label{HppwaveI}
{\cal Q}_2 & = & 
	\frac{4\pi }{\Rhat^2}\left({p_-^2}(x^A)^2 + (p_A)^2 + ({x'}^A)^2\right)
\nn\\ 
&&	\kern-10pt
	+\frac{\pi }{\Rhat^4}\biggl\{
	{2}\left[ -y^2\left( p_z^2 + {z'}^2 + 2{y'}^2\right)
	+ z^2\left( p_{y}^2 + {y'}^2 + 2{z'}^2 \right)\right]
	+ {p_-^2}\left[ (x^A)^2 \right]^2
\nn\\
& &\kern-10pt
 	- \frac{1}{p_-^2}\Bigl\{  \left[ (p_A)^2\right]^2 + 2(p_A)^2({x'}^A)^2 
	+ \left[ ({x'}^A)^2\right]^2 \Bigr\}
	 + \frac{4}{p_-^2}\left({x'}^A p_A\right)^2
	\biggr\} + O(1/\Rhat^6)\ .
\nn\\
&&
\label{hamQ2}
\ee
Comparing this with eqn.~(\ref{HLC}) above, we see that, to the order of interest,
\be
{\cal Q}_2 = 8\pi\, p_- H_{\rm LC}\ .
\ee
As expected, the perturbed monodromy matrix precisely reproduces the structure of 
the lightcone Hamiltonian to quartic order in the semiclassical expansion.  
(Note that ${\cal Q}_2$ is only expected to be identified with the lightcone Hamiltonian
up to an overall constant.)

Computationally, the expansion of the monodromy matrix becomes increasingly time consuming
at higher orders in the spectral index.  The situation can be mitigated to some extent 
by projecting the theory onto $AdS_5$ or $S^5$ excitations, eliminating all interaction
terms (from the quartic perturbation) that mix fluctuations from both subspaces.  
We will eventually want to compute eigenvalue spectra in the block-diagonal subsectors
discussed above (which require such a projection), so this maneuver will not affect the 
outcome.  

The next nonvanishing $S^5$ current in the series is given by
\be
{\cal Q}_4^{S^5} & = & \frac{8\pi}{3\Rhat^2}\left(3 - \pi^2 p_-^2 \right)
	\left(p_-^2 y^2 + p_y^2 + {y'}^2 \right)
\nn\\
&&	+\frac{2\pi}{3p_-^2 \Rhat^4}\biggl\{
	-3 
	( p_y^2 - 2p_y\cdot y' + {y'}^2 )
	( p_y^2 + 2p_y\cdot y' + {y'}^2 )
\nn\\
&&	-  \pi^2 p_-^2 \Bigl[ 4 (p_y\cdot y')^2 + (p_y^2 + {y'}^2)^2 \Bigr]
	- 12 p_-^2 {y'}^2 y^2
\nn\ee\be
&&	- p_-^4 y^2 \Bigl[
		4 \pi^2 p_y^2 - 3  y^2\Bigr]
	-3 \pi^2 p_-^6 (y^2)^2 \biggr\}+ O(1/\Rhat^6)\ .
\label{hamQ4}
\ee
Although the quadratic interaction of ${\cal Q}_4^{S^5}$ is proportional to the pp-wave Hamiltonian on the
$S^5$, the structure of the perturbing quartic interaction differs from 
that obtained for ${\cal Q}_2$.  
The corresponding $AdS_5$ current takes the form
\be
{\cal Q}_4^{AdS_5} & = & \frac{8\pi}{3\Rhat^2}\left(3 - \pi^2 p_-^2 \right)
	\left(p_-^2 z^2 + p_z^2 + {z'}^2 \right)
\nn\\
&&
	+\frac{2\pi}{3p_-^2 \Rhat^4}\biggl\{ 
 	-3 
	( p_z^2 - 2p_z\cdot z' + {z'}^2 )
	( p_z^2 + 2p_z\cdot z' + {z'}^2 )
\nn\\
&&	-  \pi^2 p_-^2 \Bigl[ 4 (p_z\cdot z')^2 + (p_z^2 + {z'}^2)^2 \Bigr]
	+ 12 p_-^2 {z'}^2 z^2
\nn\\
&&	+ p_-^4 z^2 \Bigl[
		-4 \pi^2 {z'}^2 + 3  z^2\Bigr]
	+ \pi^2 p_-^6 (z^2)^2 \biggr\}+ O(1/\Rhat^6)\ ,
\label{hamQ4Z}
\ee
where the quadratic sector is again proportional to the pp-wave Hamiltonian, 
projected in this case onto the $AdS_5$ subspace.
Continuing on to sixth order in the spectral index, 
we find the $S^5$ current
\be
{\cal Q}_6^{S^5} & = & \frac{1}{15\Rhat^2}\biggl\{
	4 \pi 
	\Bigl[
	45 -40 \pi^2 p_-^2 + 2\pi^4 p_-^4 \Bigr]
	\left(p_-^2 y^2 + p_y^2 + {y'}^2 \right) \biggr\}
\nn\\
&& 	+ \frac{\pi}{15p_-^2 \Rhat^4}\biggl\{
	 -45
	(p_y^2-2p_y\cdot y' +{y'}^2)( p_y^2+2 p_ y\cdot y' +{y'}^2)
\nn\\
&&	-20 p_-^2 \Bigl[
	2\pi^2 \left(4 (p_y\cdot y')^2 + (p_y^2+{y'}^2)^2\right)
	+9  {y'}^2 {y}^2 \Bigr]
\nn\\
&&	+ p_-^4\Bigl[
	2 \pi^4 \Bigl( 4 (p_y\cdot y')^2+3(p_y^2 +{y'}^2)^2\Bigr)
	-160 \pi^2  p_y^2 y^2 + 45 (y^2)^2 \Bigr]
\nn\\
&&\kern-10pt
	+8\pi^2 p_-^6 y^2\Bigl[
	(2\pi^2 p_y^2 + \pi^2 {y'}^2) - 15 y^2 \Bigr]
	+10 \pi^4 p_-^8 (y^2)^2
	\biggr\}+ O(1/\Rhat^6)\ .
\label{hamQ6}
\ee
The quadratic piece of ${\cal Q}_6^{S^5}$ is again identical in structure to the pp-wave Hamiltonian.
The analogous current in the $AdS_5$ subspace is arranged in a similar fashion:
\be
{\cal Q}_6^{AdS_5} & = & \frac{1}{15\Rhat^2}\biggl\{
	4 \pi \Bigl[
	45 -40 \pi^2 p_-^2 + 2\pi^4 p_-^4 \Bigr]
	\left(p_-^2 z^2 + p_z^2 + {z'}^2 \right) \biggr\}
\nn\\
&& 	+ \frac{\pi}{15p_-^2 \Rhat^4}\biggl\{
	 -45
	(p_z^2-2p_z\cdot z' +{z'}^2)( p_z^2+2 p_ z\cdot z' +{z'}^2)
\nn\\
&&	-20 p_-^2 \Bigl[
	2\pi^2  \left(4 (p_z\cdot z')^2 + (p_z^2+{z'}^2)^2\right)
	-9 {z'}^2 {z}^2 \Bigr]
\nn\\
&&	+ p_-^4\Bigl[
	2\pi^4 \Bigl( 4 (p_z\cdot z')^2+3(p_z^2 +{z'}^2)^2\Bigr)
	-160 \pi^2  {z'}^2 z^2 + 45 (z^2)^2 \Bigr]
\nn\\
&&\kern-10pt
	+8\pi^2 p_-^6 y^2
	(\pi^2 {z'}^2 + 5 z^2 )
	-6\pi^4 p_-^8 (z^2)^2
	\biggr\}+ O(1/\Rhat^6)\ .
\label{hamQ6Z}
\ee
While we will not present explicit formulas for the resulting currents, it is easy to carry this
out to eighth order in $\gamma$.

Taken separately, each current can be viewed as a free pp-wave Hamiltonian
plus a quartic interaction.  This is particularly useful, as it allows us to quantize each 
charge exactly at leading order and express the perturbation in terms of 
free pp-wave oscillators.  More explicitly, we quantize the quadratic sectors of these currents
by expanding the fluctuation fields in their usual Fourier components:
\be
x^A(\sigma,\tau) &=& \sum_{n=-\infty}^\infty x_n^A(\tau)e^{-ik_n\sigma}\ , \nn\\
x^A_n(\tau) &=& \frac{i}{\sqrt{2\omega_n}}\left( a_n^A e^{-i\omega_n\tau}
					+{a_n^{A\dag}} e^{i\omega_n\tau} \right)\ .
\ee
The quadratic (pp-wave) equations of motion
\be
\ddot x^{A} - {x''}^{A} + p_-^2 x^{A} = 0
\ee
are satisfied by setting $k_n = n$ (integer), and $\omega_n = \sqrt{p_-^2 + k_n^2}$, where 
the operators $a^A_n$ and ${a_n^{A\dag}}$ obey the commutation relation
$\bigl[ a_m^A,{a_n^{B\dag}}\bigr] = \delta_{mn}\delta^{AB}$.

In accordance with integrability, we expect that the local charges in 
eqns.~(\ref{hamQ2}--\ref{hamQ6Z}) 
should all be mutually commuting.  
Expressed in terms of quantum raising and lowering operators, we can check 
the commutators of the hidden local charges directly.  
To avoid mixing issues, we will need to select out closed subsectors of each charge that completely
decouple from the remaining terms in the theory.  We have already noted that the 
Hamiltonian $\Qo_2$ is known to be closed under $AdS_5$ and $S^5$ 
string states forming symmetric-traceless irreps in their spacetime indices.  
The equivalent gauge theory statement is that the dilatation generator is closed in
certain $\Sl(2)$ and $\su(2)$ projections.  Since the complete tower of corresponding charges in the 
gauge theory (including the dilatation generator) can be diagonalized by a single set of 
$\Sl(2)$ or $\su(2)$ Bethe equations, it is a reasonable guess that the full tower
of local string charges decouples under corresponding projections.  (A similar conjecture
is made, for example, in \cite{Arutyunov:2003rg,Kazakov:2004qf}.)  Following the 
treatment in Chapter \ref{Nimp}, we therefore define the following $AdS_5$ oscillators
\be
a_n = \frac{1}{\sqrt{2}}\left(a_n^j + i a_n^k\right)~, \qquad 
\bar a_n = \frac{1}{\sqrt{2}}\left(a_n^j - i a_n^k\right)~, \qquad (j\neq k)\ ,
\label{oscads}
\ee
which satisfy the standard relations
\be
\left[ a_n, a_m^\dag \right] = \left[ \bar a_n, \bar a_m^\dag \right] = \delta_{nm}~, 
\qquad \left[ a_n, \bar a_m^\dag \right] = \left[ \bar a_n, a_m^\dag \right] = 0\ .
\ee
When restricted to these oscillators, the symmetric-traceless projection in the 
$AdS_5$ subspace is achieved by setting all $\bar a_n,\ \bar a_n^\dag$ to zero
(see \cite{McLoughlin:2004dh} for details).  
A corresponding definition on the $S^5$ takes the form
\be
a_n = \frac{1}{\sqrt{2}}\left(a_n^{j'} + i a_n^{k'}\right)~, \qquad 
\bar a_n = \frac{1}{\sqrt{2}}\left(a_n^{j'} - i a_n^{k'}\right)~, \qquad (j'\neq k')\ ,
\label{oscs5}
\ee
where the symmetric-traceless projection is again invoked by setting 
$\bar a_n,\ \bar a_n^\dag$ to zero.  In other words, we can test the commutativity
of the local charges in the $AdS_5$ and $S^5$ symmetric-traceless projections
by rewriting their oscillator expansions according to eqns.~(\ref{oscads}, \ref{oscs5}) 
and setting all $\bar a_n,\ \bar a_n^\dag$ to zero.

Since the currents are expanded to $O(1/\Rhat^4)$, we only require that the commutators vanish 
to $O(1/\Rhat^6)$.  This simplifies the problem somewhat, since we only need to
compute commutators involving at most six oscillators.  On the subspace of symmetric-traceless 
$AdS_5$ string states, we obtain
\be
\left[ \widehat Q_n^{AdS_5},\widehat Q_m^{AdS_5} \right] 
= O(1/\Rhat^6)~, \qquad n,m \in 2,\ldots,8\ .
\ee
The corresponding projection on the $S^5$ yields
\be
\left[ \widehat Q_n^{S^5},\widehat Q_m^{S^5} \right] 
= O(1/\Rhat^6)~, \qquad n,m \in 2,\ldots,8\ .
\ee
We therefore find evidence for the existence of a tower of mutually 
commuting charges (within these particular closed sectors) that are 
conserved perturbatively by the quantized theory.

\section{Spectral comparison with gauge theory}
\label{integ2sec3}
Given the freedom involved in redefinitions of the
spectral parameter, it may seem that any spectral agreement between the string charges computed above 
and corresponding quantities in the gauge theory would be rather arbitrary.  
We therefore seek a comparison of integrable structures on both sides of the
duality that avoids this ambiguity.  It turns out that such a test is indeed possible 
in the symmetric-traceless sector of $S^5$ excitations, which
will map in the gauge theory to the closed $\su(2)$ sector.  
We will further restrict ourselves to computing spectra associated with  
the following two-impurity string states:
\be
 {a_q^{j'\dag}}{a_{-q}^{k'\dag}} \ket{J}\ .
\nn
\ee
The analysis for three- or higher-impurity states would require an accounting of interactions
between $AdS_5$ and $S^5$ string excitations;  as noted above, however, this dramatically 
complicates the computational analysis.  (We intend to return to the question
of higher-impurity string integrability in a future study.)
The ground state $\ket{J}$ is understood to carry $J$ units of angular momentum on the $S^5$,
and the two-impurity $SO(4)$ subspace above comprises a $16\times 16$-dimensional sub-block of the Hamiltonian. 
In addition, the mode indices (labeled here by $q$) of physical string states must sum to zero to
satisfy the usual level-matching condition
(the Virasoro constraint is understood to be satisfied by the leading-order solution
to the equations of motion; any higher-order information contained in the $T_{01}$ component
of eqn.~(\ref{CGC}) is irrelevant).

To simplify the analysis, and for comparison
with previous chapters, 
we will also rescale each of the charges computed above by a factor of $\Rhat^2$: 
\be
\widehat Q_n \to \Rhat^2 \widehat Q_n\ .
\ee
The two-impurity matrix elements of the charge $\widehat Q_2^{S^5}$ are then given by:
\be
\braket{J|a_q^{a'}a_{-q}^{b'}(\widehat Q_2^{S^5})a_{-q}^{c'\dag}a_{q}^{d'\dag} |J} & = & 
	{16\pi \omega_q} \delta^{a'd'}\delta^{b'c'}
\nn\\
&&\kern-140pt
	-\frac{8\pi q^2}{J \sqrt{\lambda'} \omega_q^2}
	\left[(3+2q^2\lambda)\delta^{a'd'}\delta^{b'c'} 
		-\delta^{a'c'}\delta^{b'd'} 
		+ \delta^{a'b'}\delta^{c'd'} \right]+ O(1/J^2)\ ,
\label{matq2}
\ee
The radius $\Rhat$ has been replaced with the angular momentum $J$, and 
$p_-$ has been replaced with $1/\sqrt{\lambda'}$ via 
\be
J/p_- = \Rhat^2 = \sqrt{\lambda}\ .
\ee
As expected, contributions to the pp-wave limit of eqn.~(\ref{matq2}) all lie on the diagonal. 
Up to an overall factor, one may further check that the correction terms at $O(1/J)$ agree 
with those computed in Chapter \ref{twoimp}, projected onto the $S^5$ subspace.  
The next higher charges in the series yield matrix elements given by
\be
\braket{J|a_q^{a'}a_{-q}^{b'}(\widehat Q_4^{S^5})a_{-q}^{c'\dag}a_{q}^{d'\dag} |J} & = & 
	\frac{32\pi\omega_q}{3\lambda'}( 3\lambda' - \pi^2  )\delta^{a'd'}\delta^{b'c'}
\nn\\
&&\kern-20pt
	-\frac{16\pi}{3{\lambda'}^{5/2}\omega_q^2 J}\biggl\{
	\Bigl[
	\pi^2(2+q^2\lambda') + 3 q^2 {\lambda'}^2(3+2q^2{\lambda'})\Bigr]\delta^{a'd'}\delta^{b'c'}
\nn\\
&&\kern-20pt
	+\Bigl[
	\pi^2(2+q^2\lambda')-3q^2{\lambda'}^2\Bigr]\delta^{a'c'}\delta^{b'd'}
\nn\\
&&\kern-20pt	
	+ q^2\lambda'(3\lambda' - \pi^2 ) \delta^{a'b'}\delta^{c'd'} 
	\biggr\}
	+ O(1/J^2)\ ,
\\
\braket{J|a_q^{a'}a_{-q}^{b'}(\widehat Q_6^{S^5})a_{-q}^{c'\dag}a_{q}^{d'\dag} |J} & = & 
	\frac{16\pi}{15\omega_q{\lambda'}^2}(2\pi^4 - 40\pi^2{\lambda'} + 45{\lambda'}^2 )
	\omega_q^2 \delta^{a'd'}\delta^{b'c'}
\nn\\
&&\kern-80pt
	+\frac{8\pi}{15{\lambda'}^{7/2}\omega_q^2}\biggl\{
	\Bigl[
	2  \pi^4 (4+q^2\lambda'(5+q^2\lambda'))
	-40\pi^2\lambda'(2+q^2\lambda')
\nn\\
&&\kern-80pt	
	-45 q^2{\lambda'}^3(3+2q^2\lambda') 
	\Bigr]\delta^{a'd'}\delta^{b'c'}
	+\Bigl[
	2 \pi^4(4+q^2\lambda')
	-40\pi^2\lambda'(2+q^2\lambda')
\nn\\
&&\kern-100pt
	+45q^2{\lambda'}^3  \Bigr]\delta^{a'c'}\delta^{b'd'}
	+q^2\lambda'\Bigl[
	\lambda'(40\pi^2  - 45\lambda' )-2\pi^4 \Bigr]\delta^{a'b'}\delta^{c'd'}
	\biggr\}+ O(1/J^2)\ .
\nn\\
&&
\ee

We will again project onto 
symmetric-traceless irreps of $SO(4)\times SO(4)$, transforming
as $({\bf 1,1;3,3})$ in an $SU(2)^2\times SU(2)^2$ notation. 
Although it is not necessarily
guaranteed that the symmetric-traceless states will diagonalize the higher charges 
$\widehat Q_4$ and $\widehat Q_6$ at quartic order,
this can be checked directly at one-loop order in $\lambda'$ by computing the eigenvectors of the charges above
(the higher-loop version of this check is much more 
difficult because the above charges are no longer completely block diagonal under 
the $SO(4)$ projection, a fact that can be seen in the structure of ${\cal Q}_2$ above).
The $\widehat Q_2^{S^5}$ eigenvalue between symmetric-traceless $({\bf 1,1;3,3})$ 
$S^5$ states (denoted by $Q_2^{S^5}$) is then found to be
\be
Q_2^{S^5} = 16\pi  \left(\omega_q - \frac{q^2\sqrt{\lambda'}}{J}\right) + O(1/J^2)\ .
\label{Q2eig}
\ee
Up to an overall constant, this is just the two-impurity energy shift 
computed in Chapter \ref{twoimp}.
The corresponding eigenvalues of the higher charges $\Qo_4$ and $\Qo_6$ can be
computed in an analogous fashion:
\be
Q_4^{S^5}  &= & \frac{32 \pi}{3}\biggl\{
	\frac{\omega_q}{\lambda'}
	\Bigl[ 3\lambda' - \pi^2  \Bigr]
\nn\\
&&\kern+00pt
	-\frac{\pi}{{\lambda'}^{5/2} \omega_q^2 J}\Bigl[
	\pi^2 (2+q^2\lambda') + 3q^2{\lambda'}^3\omega_q^2 \Bigr]
	\biggr\}+ O(1/J^2)\ ,
\label{Q4eig}
\nn\\
Q_6^{S^5} & = & \frac{16\pi}{15}\biggl\{
	\frac{\omega_q}{{\lambda'}^2}(2\pi^4 - 40 \pi^2\lambda' + 45{\lambda'}^2)
	-\frac{1}{\omega_q^2 {\lambda'}^{7/2} J}
	\Bigl[ 40 \pi^2\lambda'(2+q^2\lambda') 
\nn\\
&&
	+ 45 q^2{\lambda'}^4 \omega_q^2 - 2 \pi^4 (4+q^2\lambda' (3+q^2\lambda'))\Bigr]
	\biggr\} + O(1/J^2)\ .
\label{Q6eig}
\ee
Similar formulas can be extracted for the $AdS_5$ charges $Q_2^{AdS_5}$, 
$Q_4^{AdS_5}$ and $Q_6^{AdS_5}$, which are diagonalized by symmetric-traceless $({\bf 3,3;1,1})$
string states excited in the $AdS_5$ subspace.  Though we have not given explicit formulas,
it is also straightforward to obtain the corresponding eigenvalues for  $Q_8^{AdS_5}$ and $Q_8^{S^5}$.

By modifying the Inozemtsev spin chain of \cite{Serban:2004jf} to exhibit higher-loop BMN scaling, 
Beisert, Dippel and Staudacher were able to formulate a long-range
Bethe ansatz for the gauge theory in the closed $\su(2)$ sector \cite{Beisert:2004hm} 
(we will simply state their results here, referring the reader
to \cite{Beisert:2004hm} for further details).  In essence, the Bethe ansatz encodes 
the interactions of pseudoparticle excitations on a spin lattice and, in terms
of pseudoparticle momenta $p_k$, the ansatz given in \cite{Beisert:2004hm} diagonalizes the entire tower of local 
gauge theory $\su(2)$ charges.  The eigenvalues of these charges, which we label here 
as $D_n$, are given by
\be
D_n = \sum_{k=1}^{I} q_n (p_k)\ , &\qquad& 
q_n(p) = \frac{2\sin(\frac{p}{2}(n-1))}{n-1}
	\left( \frac{ \sqrt{1+8 g^2\sin^2(p/2)} -1}{2 g^2\sin^2(p/2)} \right)^{n-1}\ ,
\nn\\
&&
\label{Ddef}
\ee
where $g^2 \equiv {\lambda}/{8\pi^2}$, and the index $k$ runs over the total number ${I}$ of 
pseudoparticle excitations (or $R$-charge impurities) on the spin lattice.
These eigenvalues can then be expanded perturbatively in
inverse powers of the gauge theory $R$-charge $({R})$ by approximating the pseudoparticle
momenta $p_k$ by the expansion
\be
p_k = \sum_j \frac{f_j(n_k)}{{R}^{j/2}}\ ,
\ee
where $f_j$ are functions of the integer mode numbers $n_k$, determined by solving the
Bethe equations explicitly to a given order in $1/{R}$. 

In general, we wish to identify the local string charges with linear combinations of 
corresponding charges in the gauge theory.  From eqn.~(\ref{Ddef}), however, it is easy to see that
as one moves up the ladder of higher charges in the gauge theory, the eigenvalues $D_n$ of these
charges have leading contributions at higher and higher powers of $g^2/{R}^2$
in the large-${R}$, small-$\lambda$ double-scaling expansion.  This is puzzling because the
string eigenvalues computed above do not exhibit similar properties.
The difference in scaling behavior therefore motivates the following prescription 
for identifying the eigenvalues of the higher local charges on both sides of the correspondence:
\be
Q_{n} - N = C \left(\frac{n}{2} D_{n}\right)^{2/n}\ .
\label{QdefD}
\ee
$N$ here counts the number of string worldsheet impurities and $C$ is an arbitrary constant.  
Fractional powers of the gauge theory charges $D_n$ 
are well defined in terms of the double-scaling expansion, so that the right-hand side of 
eqn.~(\ref{QdefD}) is in fact just a linear combination of conserved quantities 
in the gauge theory. 

A potential subtlety arises when matching $Q_n$ and 
$D_n$ in this fashion for $n>2$ beyond one-loop order in $\lambda$.  
The problem is that, under the identification in eqn.~(\ref{QdefD}), information 
from string energy eigenvalues at $O(1/J^2)$ and higher
is required to completely characterize the higher-loop (in $\lambda$) coefficients
of the gauge theory charges $D_n$.  
The essential reason for this is that the string loop expansion is in powers 
of the modified 't~Hooft coupling, which, in terms of the gauge theory $R$-charge ${R}$, is
\be
\lambda' = \lambda/J^2 = \lambda/{R}^2\ .
\label{RJid}
\ee
In other words, under eqn.~(\ref{QdefD}), it is
impossible to disentangle higher-order $1/J$ contributions to the string charges 
$Q_n$ from higher-order $\lambda$ corrections to $D_n$.
The prescription given in eqn.~(\ref{QdefD}) therefore holds only 
to one-loop order in $\lambda$, where knowing the $1/J$ corrections in the string
theory is sufficient.

Furthermore, since the local charges in the string and gauge theories are only 
identified up to an overall multiplicative constant, directly comparing the spectra 
of each theory is not especially rigorous.  A convenient quantity to work 
with, however, is the ratio of the $O(1/J)$ eigenvalue correction to the pp-wave coefficient:  
at first-loop order in $\lambda$ this ratio eliminates all ambiguity 
associated with overall constants and $\gamma $ redefinitions, and 
thus provides a meaningful comparison with gauge theory. 
(The analogous quantity computed for charges in the $AdS_5$ subspace is not free from such
ambiguities.)
We therefore arrange the one-loop, two-impurity 
eigenvalues of local $S^5$ string theory charges according to
\be
Q_n^{S^5} = 2 + q^2 \lambda'\left(\Lambda_{n,0} + \frac{\Lambda_{n,1}}{J} \right) + O({\lambda'}^2) + O(1/J^2)\ ,
\ee
where the numbers $\Lambda_{n,0}$ and $\Lambda_{n,1}$ characterize eigenvalue coefficients in the
pp-wave limit and at $O(1/J)$, respectively, and $q$ is the mode number associated with the two-impurity
string states defined above.  
On the gauge theory side we make a similar arrangement:
\be
\left(\frac{n}{2} D_{n}\right)^{2/n} 
	=  \frac{q^2 \lambda}{{R}^2}\left(\bar\Lambda_{n,0} + \frac{\bar\Lambda_{n,1}}{{R}} \right) 
							+ O({\lambda}^2) + O(1/{R}^4)\ ,
\ee
where the integer $q$ is a mode number associated with the momenta of pseudoparticle
excitations on the spin lattice (which, in turn, correspond to roots of the $\su(2)$ Bethe equations).
The $R$-charge ${R}$ is understood to be identified with the string angular momentum $J$ via eqn.~(\ref{RJid}).

\begin{table}[ht!]
\begin{eqnarray}
\begin{array}{|c|cc|}
\hline
n & \Lambda_{n,1}/\Lambda_{n,0} & \bar\Lambda_{n,1}/\bar\Lambda_{n,0}  \\
\hline
2 & -2 & -2 \\
3 & 0 & 0 \\
4 & -2 & -2 \\
5 & 0 & 0 \\
6 & -2 & -2 \\
7 & 0 & 0 \\
8 & -2 & -2 \\
\hline
\end{array} \nonumber
\end{eqnarray}
\caption{Ratios of $O(1/J)$ (or $O(1/{R})$) 
	corrections to pp-wave/BMN coefficients in string and gauge theory local charges}
\label{tab1}
\end{table}

\ \\

The quantities $\Lambda_{2,0}$ and $\Lambda_{2,1}$ 
for the string Hamiltonian $Q_2$ can be computed from the eigenvalue formula
in eqn.~(\ref{Q2eig}) (or, alternatively, 
retrieved from the two-impurity string results reported in Chapter \ref{twoimp}).
We find the following ratio:
\be
\Lambda_{2,1}/\Lambda_{2,0} = -2\ .
\ee
As shown in Chapter \ref{twoimp}, this agrees with the corresponding gauge theory
prediction at one-loop order in $\lambda$:
\be
\bar\Lambda_{2,1}/\bar\Lambda_{2,0} = -2\ .
\ee
The ratio of $O(1/J)$ eigenvalue corrections to pp-wave coefficients is in fact 
$-2$ for all of the nonvanishing string charges.
Under the matching prescription in eqn.~(\ref{QdefD}), this agrees with the gauge theory perfectly.  
(The odd charges vanish altogether on both sides of the correspondence.) 
We summarize the results of this comparison for the first eight charges in the series in Table~\ref{tab1}. 
It would be satisfying to test this agreement at higher loop-orders in $\lambda$.  The corresponding
computation at two-loop order, however, would require evaluating the local string theory charges
at $O(1/\Rhat^6)$ in the semiclassical expansion, where several subtleties of perturbation theory 
(and, for that matter, lightcone quantization) would need to be addressed.
This emphasizes the need to understand the quantum string theory at higher orders in the
expansion away from the pp-wave limit.

\section{Discussion}
In this chapter we have provided evidence that an infinite tower of local,
mutually commuting bosonic charges of type IIB string theory on $AdS_5\times S^5$,
known to exist in the classical theory, can also be identified in the quantum theory.  
In addition, we have provided a prescription
for matching certain eigenvalues of these charges in a protected subsector
of the string theory to corresponding eigenvalues in the closed $\su(2)$ sector
of the gauge theory.  
The fact that the spectra of local string charges computed here can only be matched 
to corresponding quantities in the gauge theory via the matching prescription in
eqn.~(\ref{QdefD}), however, indicates that the monodromy matrix used 
to derive the local string charges is substantially different from that which would give rise
to the proposed quantum string Bethe ansatz of \cite{Arutyunov:2004vx} (or, since they are equivalent at
one-loop order, the corresponding $\su(2)$ Bethe ansatz in the gauge theory).  
In other words, we expect that there is a Lax representation for the string 
sigma model that gives rise to hidden local charges that can be compared directly 
with the gauge theory, without having to take fractional powers or linear combinations.  

There are a number of additional tests of integrability in the quantum 
string theory that, in the context of the present calculation,
should be relatively straightforward.  By computing the quartic interactions
among fluctuations in the $AdS_5$ and $S^5$ subspaces for each of the 
higher local charges studied here, it would be easy, for example, to find the
resulting spectra of three- or higher-impurity string states.  
Apart from the difficulty of actually computing the mixing interactions, 
this would provide a simple check on the methodology employed here.
A more difficult problem would be to address whether the integrable structure
of the string theory respects supersymmetry.  By formulating a supersymmetric
Lax representation that generates the complete interaction Hamiltonian computed
in Chapter \ref{twoimp}, one might be able to show that each of the higher
local charges are individually supersymmetric, and a comparison with gauge theory
could be carried out in the closed $\su(1|1)$ sector studied in Chapters \ref{threeimp}
and \ref{Nimp} 
(the corresponding sector of the string theory would be comprised of symmetrized
fermionic excitations in the $({\bf 3,1;3,1})$ or $({\bf 1,3;1,3})$ of
$SO(4)\times SO(4)$).  

Ultimately, the hope is that the arsenal of techniques associated with 
integrable systems can be employed to find an exact solution 
to the string formulation of large-$N_c$ Yang-Mills theory.  
Alternatively, a proof that both sides of the duality are diagonalized by 
identical Bethe equations should be obtainable.  At present, the major obstacle preventing such a
proof is the disagreement between gauge and string theory at three-loop order in the
't~Hooft coupling.   The fact that the integrable
systems of both theories seem to agree in certain limited cases, however, 
stands as strong evidence that they are likely to be equivalent.

%
%
\chapter{Conclusions and outlook}   					  
\label{con}
Since the advent of the BMN/pp-wave limit, the AdS/CFT correspondence has 
been subject to a new class of rigorous and detailed tests.  These
studies have not only provided novel verifications of the validity 
of Maldacena's conjecture, but they have given a much more detailed
understanding of how holographic dualities are realized.
In this dissertation we have focused on the correspondence between
type IIB superstring theory on $AdS_5\times S^5$ and
${\cal N}=4$ SYM theory in four dimensions, widely viewed as
the simplest and most striking example of AdS/CFT duality. 
Given the large number of symmetries on either side of the correspondence,
this is, in many respects, the easiest system to study.  
It has been a longstanding problem, however, that
string quantization is not well understood in the presence of a curved, 
RR background.  Together with the difficulty of computing 
non-BPS operator dimensions in the strong-coupling limit
of the gauge theory, the obstacles preventing a direct test of the 
proposal have been formidable.  Of course, this situation changed dramatically
when BMN discovered a large $R$-charge limit of the gauge theory that 
matched the pp-wave limit of the string theory, a limit that was shown by Metsaev 
to render the string theory exactly soluble.  
Their insights were a tremendous success because 
the match-up between the string and gauge theory in this limit marked
the first direct comparison of string energy spectra with a corresponding
set of anomalous dimensions in the gauge theory.

At this level of the analysis, however, one is limited to dealing 
with spectra on either side of the duality that are highly degenerate.
We demonstrated that this degeneracy can 
be lifted by including worldsheet interactions associated with 
background curvature corrections to the Penrose limit.  
Including such corrections is a difficult task, as it reintroduces the
puzzle of quantizing the string theory in a curved background with 
RR flux.  The essential point is that this vexing problem
can be circumvented by treating the curvature corrections in a 
purely perturbative setting.  The corrections to the spectrum can be 
controlled by the expansion in inverse (squared) powers of the scale radius 
$(1/\Rhat^{2n})$ or, equivalently, in inverse powers of the $S^5$ angular 
momentum $(1/J)$.
We have focused in this dissertation 
on the leading set of these corrections, appearing at $O(1/J)$ 
in the expansion.  

Chapter~\ref{twoimp} was dedicated to computing the
interaction Hamiltonian and analyzing the resulting spectrum in the Fock space 
of two-impurity string states, formed by acting with two raising operators
on the ground state $\ket{J}$.  The resulting spectrum was composed of
256 distinct states that sort themselves into a nine-level supermultiplet
whose multiplicity structure matches that which is expected from 
the structure of ${\cal N}=4$ supersymmetry.  Furthermore, the string theory provides energies
that are exact in the gauge theory coupling $\lambda = g_{\rm YM}^2 N_c$.
When compared with higher-order $\lambda$ corrections to anomalous 
dimensions computed directly from the gauge theory, we found a perfect match at
both one- and two-loop order, but this remarkable agreement with the gauge theory
breaks down at third order.

By extending this analysis to the 4,096-dimensional Fock space of 
three-impurity string states, we were able to show in Chapter~\ref{threeimp}
that the string and gauge theories again agree in this perturbative setting 
at one- and two-loop order.  The agreement again breaks down at three loops, however.  
When compared with the conjectured Bethe equations that provide anomalous
dimension formulas in the gauge theory, this pattern was discovered in Chapter~\ref{Nimp}
to exist for the generic $N$-impurity case, albeit restricted to certain protected
subsectors of the theory.  

The BMN/pp-wave limit of the AdS/CFT correspondence provides
an immensely powerful testing ground for holographic gauge/string dualities.
We have provided direct spectral comparisons of the gauge and string theories, 
along with convincing evidence that, in this setting, one can expect to see an impressive match-up 
that holds to two-loop order in the 't~Hooft coupling, but 
breaks down at three loops.\footnote{This pattern has been found elsewhere, 
particularly in the study of the duality 
between semiclassical extended string configurations and corresponding sectors of
the gauge theory (see also \cite{Tseytlin:2003ii} for a general review of this program, and
\cite{Beisert:2005mq,Frolov:2004bh,Kruczenski:2004kw,Stefanski:2004cw,Kruczenski:2004cn,Park:2005ji,Alday:2005gi} 
for more recent developments).  Extended string configurations typically give rise
to additional conserved charges that provide an intuitive generalization of 
the BMN/pp-wave picture.}
Whether one can perform the similar analyses 
in the string theory at even higher orders in the $1/\Rhat$ expansion is still an open
question.  (Early steps in this direction are taken in \cite{Swanson:2004mk}.)
It is unclear whether lightcone methods will be helpful in this context, because,
dimensionally speaking, the theory becomes nonrenormalizable at $O(1/\Rhat^4)$ in the 
expansion.  
In addition, it still remains to be seen whether the diverse set of integrable structures 
underlying the duality will be useful for actually solving certain sectors of
the gauge theory or the string theory.

In the end we stand to learn a great deal about the non-perturbative aspects of
Yang-Mills theories by relying on their string theory counterparts.  As noted above, 
the next landmark achievement will likely be proving the complete equivalence of some sector
of IIB string theory on $AdS_5\times S^5$ and ${\cal N}=4$ SYM theory, perhaps at the
level of the Bethe equations.  For this to happen, however, we need to rectify the
higher-loop disagreement.  Given the huge amount of symmetry on 
both sides of the duality, it may be possible to reach the strong-coupling regime of
the gauge theory and verify the form of the conjectured wrapping interactions therein.  
If current trends continue, we will undoubtedly uncover fascinating new realms of physics
in the process.

\renewcommand{\thefigure}{A-\arabic{equation}} 
\setcounter{equation}{0}
\numberwithin{equation}{chapter}
\appendix

%
%
\chapter{Notation and conventions}					  
For convenience we record in this appendix the most common symbols used in the
text.  In the following list we collect quantities defined on the CFT side of the 
correspondence:
\begin{tabbing}
$\qquad$ \=  $D$  $\qquad \qquad \quad $ \= Operator dimension 	\\
\>$R$ \>			$U(1)_R$ component of the $SU(4)$ $R$-symmetry (the $R$-charge);
				\\ \>\> is mapped to and used interchangeably with the string theory
				\\ \>\> angular momentum $J$ (see below) \\
\>$K$	\>			Naive dimension; counts the total number of fields in 
				\\ \>\> an operator, 
				 or the number of sites on the corresponding spin \\ \>\>
				chain; maps to 
				$J+I$ on the string side, where \\ \>\>
				$I$ is the impurity number \\
\>$\Delta$\>			$D-R$; maps to $P_+$ on the string side \\ 
\>$g_{\rm YM}$\>		Yang-Mills coupling \\ 	
\>$\lambda$\>			$g_{YM}^2 N_c$, 't~Hooft coupling \\
\>$N_c$	\>			Rank of the Yang-Mills gauge group \\
\>$\overline T^{(+)}_K$\>		Symmetric-traceless, rank-two $SO(4)$ tensor operator
				in the \\ \>\> $(2,K-4,2)$ $SU(4)$ irrep  \\
\>$\overline T^{(-)}_K$\> 		Antisymmetric, rank-two $SO(4)$ tensor operator
				in the \\ \>\> $(0,K-3,2)$ $+(2,K-3,0)$ of $SU(4)$  \\
\>$\overline T^{(0)}_K$\>		Trace part of the set of rank-two $SO(4)$ tensor operators,
				\\ \>\> in the $(0,K-2,0)$ of $SU(4)$  \\
\>$D-K$\>			Anomalous dimension \\
\>$\Delta_0$\>			$K-R$, counts the number of $R$-charge impurities (can be half-integer 
				\\ \>\> valued when fermionic impurities are present) \\
\>$L$\>				The level of a supermultiplet, reached by acting on a primary level
				\\ \>\> with some SUSY generator $L$ times \\
\>$\suphi$\>			SYM scalars, rank-two antisymmetric $SU(4)$ tensor in the 
				\\ \>\> six-dimensional $(0,1,0)$  ($({\bf 2,1})$ of $SL(2,C)$) \\
\>$\suchi_{~a}$ \>		SYM gluino, rank-one $SU(4)$ tensor in the four-dimensional
				\\ \>\> fundamental $(1,0,0)$ ($({\bf 1,2})$ of $SL(2,C)$) \\
\>$\suchib_{~\dot a}$ \>	SYM gluino, rank-three antisymmetric $SU(4)$ tensor in the 
				\\ \>\>four-dimensional antifundamental $(0,0,1)$ \\
\>$\nabla_\mu$\>		Spacetime covariant derivative \\
\>$Z$\>				Scalar field with $R=1$: $\phi^{\,\Yboxdim5pt\tiny\young(1,2)}$ \\
\>$A,B$\>			Scalar impurity fields with $R=0$: 
				$\phi^{\,\Yboxdim5pt\tiny\young(1,3)},\
    				\phi^{\,\Yboxdim5pt\tiny\young(1,4)},\
    				\phi^{\,\Yboxdim5pt\tiny\young(2,3)},\
    				\phi^{\,\Yboxdim5pt\tiny\young(2,4)}$ (or simply $\phi^A$, $\phi^B$) \\
\>$\bar Z$\>			Scalar field with $R=-1$: $\phi^{\,\Yboxdim5pt\tiny\young(3,4)}$ \\
\>$P_{i,j}$\>			Permutation operator that exchanges spins on the $i^{\rm th}$ and $j^{\rm th}$ 
				\\ \>\> lattice sites of a spin chain  \\
\>$\{ n_1,n_2,\dots\}$\>	Shorthand for the following series of permutation 
				\\ \>\> operators: $\sum_{k=1}^L P_{k+n_1,k+n_1+1} P_{k+n_2,k+n_2+1}\cdots$\\
\>$I$\>				Spin chain impurity number (Chapter~\ref{virial}) \\
\>$n_i$\>			Mode number of the $i^{\rm th}$ pseudoparticle excitation on the 
				\\ \>\>spin lattice (Chapter~\ref{virial}) \\
\>$b_j^\dag,b_j$\>		Position space raising and lowering operators for magnon 
				\\ \>\> excitations on the spin lattice  \\
\>$\tilde b_p^\dag,\tilde b_p$\>  Momentum space raising and lowering operators for magnon 
				\\ \>\> excitations \\
\>$\ket{L}$\>			Ground state of the length-$L$ spin chain; corresponds to
				\\ \>\> the BPS operator $\tr(Z^L)$ \\
\> $\bar\Lambda$ \>		$O(1/R)$ shift of the anomalous dimension \\
\end{tabbing}
On the string side we use the following symbols:
\begin{tabbing}
$\qquad$ \= $g_s$ $\qquad \qquad \quad $ \= String coupling, equal to $g_{\rm YM}^2$ \\
\> $\Rhat$ \>			Curvature radius of $AdS_5\times S^5$ \\
\>$\lambda'$\>			Modified 't~Hooft coupling, $g_{\rm YM}^2 N_c / J^2$ \\
\> $J$ \>			Angular momentum of string states along an equatorial geodesic
				\\ \>\> in the $S^5$ subspace; maps to $L-I$, or simply $R$, in the CFT \\
\> $\omega $ \>			String excitation energy \\
\> $\ket{J}$ \>			The string ground state, carrying $J$ units of angular momentum
				\\ \>\> on the $S^5$  \\
\> $P_+$ \> 			String lightcone Hamiltonian $\omega -J$, maps to $\Delta$ 
				on the CFT side  \\
\> $L$ \>			Supermultiplet level of a string energy spectrum \\
\>$a_n^\dag,a_n$ \>		Bosonic raising and lowering operators with mode number $n$  \\
\>$b_n^\dag,b_n$ \>		Fermionic raising and lowering operators with mode number $n$ \\
\> $z^k$ \>			$SO(4)$ vector in $AdS_5$ \\
\> $y^{k'}$ \>			$SO(4)$ vector in $S^5$ \\
\> $x^A$ \>			Vector in a transverse $SO(8)$ subspace of $AdS_5\times S^5$ \\
\> $G_{\mu\nu}$ \>		Metric tensor of $AdS_5\times S^5$ \\
\> $h_{ab}$ \>			Worldsheet metric tensor \\
\> $s^{IJ}$ \>			$2\times 2$ matrix $s \equiv {\rm diag}(1,-1)$ \\
\> $G$ \>			Coset space representative \\
\> $L^\mu$ \>			Cartan one-forms \\
\> $L^\alpha$ \>		Cartan superconnections \\
\> $H_{\rm LC}$\>		The full lightcone string Hamiltonian \\
\> $H_{\rm int}$ \>		Interaction sector of the string Hamiltonian, appearing at
				\\ \>\> $O(1/\Rhat^2)$ in the curvature expansion \\
\> $H_{\rm BB}$ \>		Purely bosonic sector of the interaction Hamiltonian  \\
\> $H_{\rm FF}$ \>		Purely fermionic sector of the interaction Hamiltonian  \\
\> $H_{\rm BF}$ \>		Bose-fermi mixing sector of the interaction Hamiltonian  \\
\> $\Lambda$ \>			String energy shift at $O(1/J)$; the shifts denoted by 
				\\ \>\> $\Lambda_{\rm BB}$, $\Lambda_{\rm FF}$ and $\Lambda_{\rm BF}$
				are associated with the corresponding  
				\\ \>\> sectors of $H_{\rm int}$ \\
\> $\omega_n$ \>		String energy at mode number $n$: $\sqrt{n^2 + p_-^2}$ \\
\> $k_n$ \>			String mode function $k_n = n$ (integer) \\
\> $p_-$ \>			Worldsheet momentum in the $x^-$ direction, equal to $1/\sqrt{\lambda'}$ \\
\> $L$ \>			Marks the overall level within a (super)multiplet of energy states \\
\> $L_{\rm sub}$ \>		Marks the level within an energy submultiplet (Chapter \ref{Nimp}) \\
\> $L'$ \>			Index labeling submultiplets within a supermultiplet of 
				\\ \>\> energy levels $(L = L_{\rm sub} + L')$
\end{tabbing}
The various indices on the string side are chosen to represent the following:
\begin{tabbing}
$\qquad$ \= $\mu,\nu,\rho = 0,\dots,9$   $\qquad\qquad$	\=  $SO(9,1)$ vectors~,  
\\
\> $\alpha,\beta,\gamma,\delta = 1,\dots,16$  \>  $SO(9,1)$ spinors ~,
\\
\> $A,B = 1,\dots,8$ \> 	$SO(8)$ vectors~,
\\
\> $i,j,k = 1,\dots,4$  \>	$SO(4)$ vectors~,
\\
\> $i',j',k' = 5,\dots,8$  \> 	$SO(4)'$ vectors~,
\\
\> $a,b = 0,1$ 	\> 	Worldsheet coordinates  $(\tau,\sigma)$~,
\\
\> $I,J,K,L = 1,2$	\> 	Label two Majorana-Weyl spinors of equal chirality~.
\end{tabbing}

The $32\times 32$ Dirac gamma matrices are decomposed into a $16\times 16$
representation according to
\begin{eqnarray}
(\Gamma^\mu)_{32\times 32} = \left( \begin{array}{cc}
    0   &   \gamma^\mu  \\
    \bar\gamma^\mu &    0
    \end{array}  \right)\ ,
&  \qquad &
\gamma^\mu \bar\gamma^\nu
+ \gamma^\nu \bar\gamma^\mu = 2\eta^{\mu\nu}~,
\nonumber \\
	\gamma^\mu = (1,\gamma^A, \gamma^9)~,
&  \qquad  &
	\bar\gamma^\mu = (-1,\gamma^A, \gamma^9)~,
\nonumber \\
	\gamma^+ = 1+\gamma^9 ~,
& \qquad & 
	\bar\gamma^+ = -1 + \gamma^9~.
\end{eqnarray}
In particular, the notation $\bar \gamma^\mu$ lowers the $SO(9,1)$ spinor indices
$\alpha,\beta$:
\be
\gamma^\mu = (\gamma^\mu)^{\alpha\beta}~, \qquad 
\bar \gamma^\mu = (\gamma^\mu)_{\alpha\beta}~.
\ee
These conventions are chosen to match those of Metsaev in \cite{Metsaev:2001bj}.
By invoking $\kappa$-symmetry,
\be
\bar\gamma^+\theta = 0 & \Longrightarrow & \bar\gamma^9\theta = \theta~, \\
\bar\gamma^- = 1+\bar\gamma^9 & \Longrightarrow & \bar\gamma^-\theta = 2\theta~.
\ee
The antisymmetric product $\gamma^{\mu\nu}$ is given by
\be
(\gamma^{\mu\nu})^\alpha_{\phantom{\alpha}\beta} & \equiv &
	\frac{1}{2}(\gamma^\mu \bar\gamma^\nu)^\alpha_{\phantom{\alpha}\beta}
	- (\mu \rightleftharpoons \nu)~, 
\nonumber \\
(\bar \gamma^{\mu\nu})^\alpha_{\phantom{\alpha}\beta} & \equiv &
	\frac{1}{2}(\bar \gamma^\mu \gamma^\nu)_\alpha^{\phantom{\alpha}\beta}
	- (\mu \rightleftharpoons \nu)~.
\ee
We form the matrices $\Pi$ and $\tilde\Pi$ according to:  
\be
\Pi & \equiv & \gamma^1 \bar\gamma^2 \gamma^3 \bar\gamma^4 ~,
\nn \\
\tilde \Pi & \equiv & \gamma^5 \bar\gamma^6 \gamma^7 \bar\gamma^8\ .
\ee
These form the projection operators $(\Pi^2 = \tilde\Pi^2 = 1)$
\be
\Pi_+ \equiv \frac{1}{2}(1+\Pi)~, & \qquad & \Pi_- \equiv \frac{1}{2}(1-\Pi)~, \nn\\
\tilde\Pi_+ \equiv \frac{1}{2}(1+\tilde\Pi)~, & \qquad & \tilde\Pi_- \equiv \frac{1}{2}(1-\tilde\Pi)~.
\ee

The spinors $\theta^I$ represent two 32-component Majorana-Weyl spinors
of $SO(9,1)$ with equal chirality.  The 32-component Weyl condition is
$\Gamma_{11}\theta = \theta$, with 
\be
\Gamma_{11} = \Gamma^0\dots \Gamma^9 = \left(
\begin{array}{cc}
{ 1} & 0 \\
0 & -{ 1} 
\end{array}  \right)_{32\times 32}~.
\ee
The Weyl condition is used to select the top 16 components of $\theta$ to form 
the 16-component spinors
\begin{eqnarray}
\theta^I & = & \left( {\theta^\alpha \atop 0} \right)^I\ .
\end{eqnarray}
It is useful to form a single complex 16-component spinor $\psi$ from 
the real spinors $\theta^1$ and $\theta^2$:
\be
\psi & = & \sqrt{2} (\theta^1 + i \theta^2 )\ .
\ee
The 16-component Weyl condition $\gamma^9 \theta = \theta $ selects the upper
eight components of $\theta $, with
\be
\gamma^9 = \gamma^1\dots \gamma^8 = \left(
\begin{array}{cc}
{ 1} & 0 \\
0 & -{ 1} 
\end{array}  \right)_{16\times 16}~.
\ee
The 16-component Dirac matrices $\gamma^\mu$ can, in turn, be constructed
from the familiar Spin(8) Clifford algebra, wherein (in terms of $SO(8)$ vector indices)
\be
(\gamma^A)_{16\times 16} =  \left(
	\begin{array}{cc}
	0 & \gamma^A \\
	(\gamma^A)^T & 0 
	\end{array} \right)~,
\ee
and
\be
\left\{ \gamma^A,\gamma^B \right\}_{16\times 16} = 2\delta^{AB}~, &\qquad &
\left( \gamma^A (\gamma^B)^T + \gamma^B(\gamma^A)^T = 2\delta^{AB}\right)_{8\times 8}~.
\ee
The Spin(8) Clifford algebra may be constructed explicitly in terms of eight real matrices
\be
\label{cliffmat}
\gamma^1 = \epsilon\times\epsilon\times\epsilon~, & \qquad & 
	\gamma^5 = \tau_3\times\epsilon\times 1~, \nn\\
\gamma^2 = 1\times \tau_1\times\epsilon~, & \qquad & 
	\gamma^6 = \epsilon\times 1\times\tau_1~, \nn\ee\be
\gamma^3 = 1\times \tau_3\times\epsilon~, & \qquad & 
	\gamma^7 = \epsilon\times 1\times\tau_3~, \nn\\
\gamma^4 = \tau_1\times\epsilon\times 1~, & \qquad & 
	\gamma^8 = 1\times 1\times 1~,
\ee
with
\be
\epsilon = \left( \begin{array}{cc}
		0 & 1 \\
		-1 & 0 \end{array}\right)~, \qquad 
\tau_1 = \left( \begin{array}{cc}
		0 & 1 \\
		1 & 0 \end{array}\right)~, \qquad 
\tau_3 = \left( \begin{array}{cc}
		1 & 0 \\
		0 & -1 \end{array}\right)~.
\ee
 
\bibliographystyle{utcaps}
\bibliography{main_doc}

\end{document}